\documentclass[rmp,aps,preprint,nofootinbib,endfloats]{revtex4}
\usepackage{graphics}
\usepackage{epsfig}

\begin{document}

\title{Half-metallic ferromagnets: From band structure to many-body effects}
\author{M.I. Katsnelson}
\email{katsnel@science.ru.nl} \affiliation{Institute for Molecules
and Materials, Radboud University of Nijmegen, NL-6525 ED
Nijmegen,, Netherlands}
\author{V.Yu. Irkhin}
\affiliation{Institute of Metal Physics, 620219 Ekaterinburg,
Russia}
\author{L. Chioncel}
\affiliation{Institute of Theoretical Physics, Graz University of
Technology, A-8010 Graz, Austria} \affiliation{Department of Physics, University of Oradea,
410087 Oradea, Romania}
\author{A.I.~Lichtenstein}
\affiliation{Institute of Theoretical Physics, University of Hamburg,  \\
20355 Hamburg, Germany}
\author{R.A. de Groot}
\affiliation{Institute for Molecules and Materials, Radboud
University of Nijmegen, The Netherlands}
\affiliation{Zernicke Institute for
Advanced Materials, NL-9747 AG
Groningen, The Netherlands}

\begin{abstract}
A review of new developments in theoretical and experimental
electronic structure investigations of half-metallic ferromagnets
(HMF) is presented. Being semiconductors for one spin projection
and metals for another ones, these substances are promising
magnetic materials for applications in spintronics (i.e.,
spin-dependent electronics). Classification of HMF by the
peculiarities of their electronic structure and chemical bonding is
discussed. Effects of electron-magnon interaction in HMF and their
manifestations in magnetic, spectral, thermodynamic, and transport
properties are considered. Especial attention is paid to
appearance of non-quasiparticle states in the energy gap, which
provide an instructive example of essentially many-body features
in the electronic structure. State-of-art electronic calculations
for correlated $d$-systems is discussed, and results for specific
HMF (Heusler alloys, zinc-blende structure compounds, CrO$_{2},$
Fe$_{3}$O$_{4}$) are reviewed.
\end{abstract}

\maketitle \tableofcontents


\newpage

\section{Introduction}

\label{sec:intro} Twenty-five years ago the unusual magnetooptical
properties of several Heusler alloys motivated the study of their
electronic structure. It yielded an unexpected result: some of
these alloys showed the properties of metals as well as insulators
at the \textit{same} time in the \textit{same} material depending
on the spin direction. This property was baptized half-metallic
magnetism \cite{deGroot:2024}. Although it is not exactly clear
how many half-metals are known to exist at this moment,
half-metallic magnetism as phenomenon has been generally accepted.
Formally the expected $100\%$ spin polarization of the charge
carriers in a half-metallic ferromagnet (HMF) is a hypothetical
situation that can only be approached in the limit of vanishing
temperature and neglecting spin-orbital interactions. However, at
low temperatures (as compared with the Curie temperature which
exceeds 1000K for some HMF's) and minor spin-orbit interactions a
half-metal deviates so markedly from a normal material, that the
treatment as a special category of materials is justified. The
confusion on the number of well-established half-metals originates
from the fact that there is no \textquotedblleft smoking
gun\textquotedblright\ experiment to prove or disprove
half-metallicity. The most direct measurement is spin-resolved
positron annihilation \cite{Hanssen:5009}, but this is a tedious,
expensive technique requiring very dedicated equipment. The only
proven HMF so far is NiMnSb to the precision of the experiment,
which was better than one hundredth of an electron
\cite{Hanssen:1533}. This number also sets the scale for concerns
of temperature-induced depolarization and spin-orbit effects,
detrimental for half-metallicity.

The half-metallicity in a specific compound should not be confused
with the ability to pick up $100 \% $ polarized electrons from a
HMF. The latter process involves electrons crossing a surface or
interface into some medium where their degree of spin polarization
is analyzed. This is clearly not an intrinsic materials property.
The richness but also the complications of surfaces and interfaces
are still not fully appreciated.

Because of all these experimental complications, it is not
surprising that electronic structure calculations continue to play
an important role in the search for new HMF as well as the
introduction of new concepts like half-metallic
antiferromagnetism. However, electronic structure calculations
have weaknesses as well. Most of the calculations are based on
density functional theory in the LDA or GGA approximation. It is
well known that these methods underestimate the band gap for many
semiconductors and insulators, typically by $30 \%$. It has been
assumed that these problems do not occur in half-metals since
their dielectric response is that of a metal. This assumption was
disproved recently. A calculation on the HMF
La$_{0.7}$Sr$_{0.3}$MnO$_3$ employing the GW approximation (that
gives a correct description of band gaps in many semi-conductors)
leads to a half-metallic band gap 2~eV in excess to the DFT value
\cite{Kino:858}. The consequences of this result are possibly
dramatic: if it were valid in half-metallic magnetism in general,
it would imply that many of the materials, showing band gaps in
DFT based calculations of insufficient size to encompass the Fermi
energy, are actually in reality \textit{bona fide} half-metals.
Clearly much more work is needed in this area.

The strength of a computational approach is that it does not need
samples: even nonexistent materials can be calculated! But in such
an endeavour a clear goal should be kept in mind.
Certainly, computational studies can help in the design of new
materials, but the challenge is not so much in finding exotic
physics in materials that have no chance of ever being realized.
Such studies \textit{can} serve didactical purposes, in which case
they will be included in this review. However, the main attention
will be devoted to materials that either exist or are (meta)stable
enough to have a fair chance of realization.

This review will cover half-metals and will not discuss the
thriving area of magnetic semiconductors. Some overlap exists,
however. The much older field of magnetic semiconductors started
with semiconductors like the europium monochalcogenides and
cadmium-chromium chalcogenides \cite{nagaev:1983}. Later, the
attention changed to the so-called diluted magnetic semiconductors
\cite{Delves:549}. These are regular (i.e. III-V or II-VI)
semiconductors, where magnetism is introduced by partial
substitution of the cation by some (magnetic) 3$d$ transition
element. The resulting Curie temperatures remained unsatisfactory,
however. The next step in the development was the elimination of
the non-magnetic transition element altogether. This way HMF's can
be realized \textit{provided that} the remaining transition-metal
pnictides could be stabilized in the zinc-blende or related
structures. The review will  treat not the (diluted) magnetic
semiconductors as such, but some aspects of metastable zinc-blende
HMF's.

HMF's form a quite diverse collection of materials with very
different chemical and physical properties and even the origin of
the half-metallicity can be quite distinct. For this reason it is
required to discuss the origin of the band gap in terms of two
ingredients that define a solid: the crystal structure and the
chemical composition. Two aspects are of importance in this
context. The first one is ``strong magnetism'' versus ``weak
magnetism''. In a strong magnet, the magnetic moment will not
increase if the exchange splitting is increased hypothetically.
Thus the size of the magnetic moment is not determined by the
strength of the exchange interaction, but is limited instead by
the availability of electron states. In practice this implies that
either the minority spin sub shell(s) responsible for the
magnetism is (are) empty or the relevant majority channel(s) is
(are) completely filled. In the case of weak magnetism the
magnetic moment is determined by a subtle compromise between the
energy gain of an increase in magnetic moment (the exchange
energy) and the (band) energy the increase of the magnetic moment
costs. To avoid misunderstanding, we emphasize that this
definition of ``weak'' and ``strong'' magnets differs from that
used in Moriya's book \cite{Moriya:1985} and most theoretical
works on itinerant-electron magnetism. According to Moriya,
``strong'' magnets are those with well-defined magnetic moments
which means, e.g., the Curie-Weiss behavior of the
wavevector-dependent magnetic susceptibility $\chi (\mathbf{q},T)$
in the whole Brillouin zone. In this sense, all HMF's (e.g.,
containing Mn ions) are strong magnets. However, within this group
of materials we may introduce a more fine classification based on
the sensitivity of magnetic moment with respect to small variation
of parameters.

Any combination of ``weak'' or ``strong'' magnetism with majority
or minority spin band gaps is known today. Thus "weak" magnets
with minority spin band gaps are found in the Heusler alloys and
artificial zinc-blendes, examples of weak magnets with majority
spin gaps are the double perovskites and magnetite. The colossal
magnetoresistance materials are examples of strongly magnetic
half-metals with minority spin band gaps as is for example,
chromium dioxide, while the anionogenic ferromagnets like rubidium
sesquioxide are examples of a strongly magnetic half-metal with a
majority spin band gap.

An interesting and relatively new development is the work on
half-metallic sulphides. The HMF state in oxides with the spinel
structure is relatively rare. The prime example of course is
magnetite. However any substitution into the transition-metal
sublattices leads practically invariable to a Mott insulating
state, as magnetite itself possesses below the Verwey transition
at 120K. On the other hand the electrons in sulphides, are
substantially less correlated. Hence a wealth of substitutions is
possible in order to optimize properties, design half-metallic
ferro- or antiferromagnets etc. without the risk of losing the
metallic properties for the second spin direction as well. There
is a prize to be paid however: since the cation-cation distances
are larger in sulphides, the Curie and Neel temperatures are lower
in comparison with the oxides. Nevertheless, the work on
half-metallic sulphides deserves strong support.

In all the metallic ferromagnets, interaction between conduction
electrons and spin fluctuation is of crucial importance for
physical properties. In particular, the scattering of charge
carriers by magnetic excitations determines transport properties
of itinerant magnets (temperature dependences of resistivity,
magnetoresistivity, thermoelectric power, anomalous Hall etc.).
From this point of view, HMF, as well as ferromagnetic
semiconductors, differ from ``normal'' metallic ferromagnets by
the absence of spin-flip (one-magnon) scattering processes. This
difference is also important for magnetic excitations since there
no Stoner damping, and spin waves are well defined in the whole
Brillouin zone, as well as in magnetic insulators
\cite{Auslender:301,Irkhin:705}.

\begin{figure}[tbh]
\begin{center}
\includegraphics[width=0.75\columnwidth, angle=0]{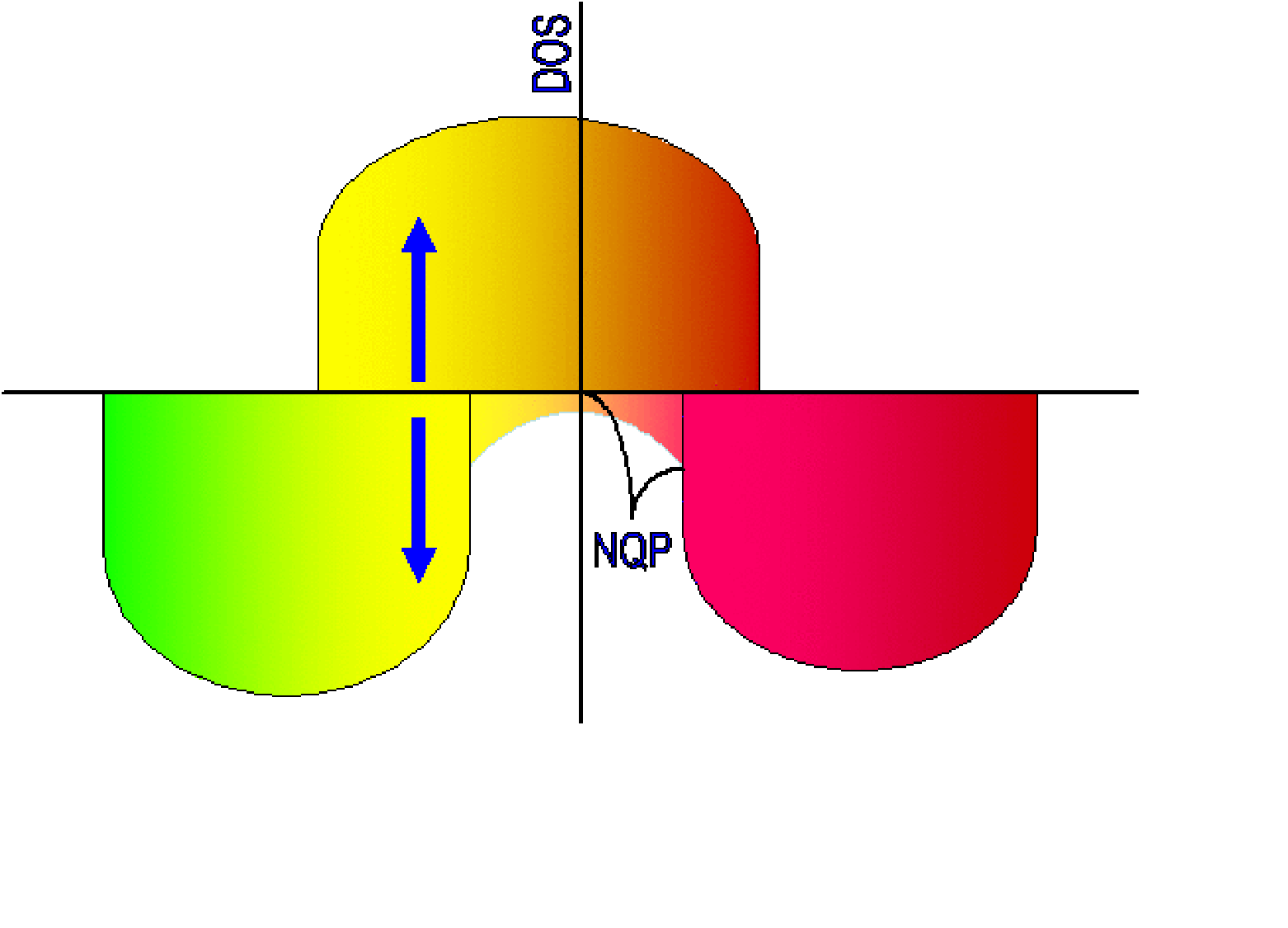}
\end{center}
\caption{(color online) Density of states of non-quasiparticles, for half-metallic
ferromagnets, possessing the gap in the minority spin channel. NQP states
are the dominant many-body feature around $E_F$ in comparison other
mean-field effects, such as spin-orbit or non-collinearity as it will
be discussed in the following sections.} \label{pict-nqp}
\end{figure}

Electron-magnon interaction modifies also considerably electron
energy spectrum in HMF. These effects take place both in usual
ferromagnets and in HMF. However, peculiar band structure of HMF
(the energy gap for one spin projection) results in important
consequences. In generic itinerant ferromagnets the states near
the Fermi level are quasiparticles for both spin projections. On
the contrary, in HMF and important role belongs to incoherent
(non-quasiparticle, NQP) states which occur near the Fermi level
in the energy gap \cite{Irkhin:705}. The appearance of the NQP
states in \cite{edwards:2191,Irkhin:1947} is one of the most
interesting correlation effects typical for HMF. The origin of
these states is connected with \textquotedblleft
spin-polaron\textquotedblright\ processes: the spin-down
low-energy electron excitations, which are forbidden for HMF in
the one-particle picture, turn out to be possible as a
superposition of spin-up electron excitations and virtual magnons.
The density of the NQP states vanishes at the Fermi level but
increases drastically at the energy scale of the order of a
characteristic magnon frequency $\overline{\omega }$. These states
are important for spin-polarized electron spectroscopy
\cite{irkhin:401,irkhin:104429}, NMR \cite{Irkhin:479}, and subgap
transport
in ferromagnet-superconductor junctions (Andreev reflection) \cite%
{Tkachov:024519}. Recently, the density of NQP states has been
calculated from first principles for a prototype HMF, NiMnSb
\cite{chioncel:144425}, as well as for other Heusler alloys
\cite{chioncel:137203}, zinc-blend structure compounds
\cite{chioncel:085111,chioncel:197203} and CrO$_{2}$
\cite{chioncel:cro2}. Fig \ref{pict-nqp} shows in a pictorial way the
NQP contribution to the density of states.

Therefore,  HMF are very interesting conceptually as
a class of materials which may be convenient to treat many-body
solid state physics that is essentially beyond band theory. It is
accepted that usually many-body effects lead only to
renormalization of the quasiparticle parameters in the sense of
Landau's Fermi liquid (FL) theory, the electronic
\textit{liquid} being \textit{qualitatively }similar to the electron \textit{%
gas} (see, e.g., \cite{Nozieres:1964}). On the other hand, NQP
states in HMF are not described by the FL theory. As an example of
highly unusual
properties of the NQP states, we mention that they can contribute to the $T$%
-linear term in the electron heat capacity \cite%
{irkhin:7151,Irkhin:1733,Irkhin:397}, despite their density at
$E_{F}$ is zero at temperature $T=0$~K. Some developments
concerning physical effects of NQP states in HMF are considered in
the present review.

\section{Classes of half-metallic ferromagnets}

\subsection{Heusler alloys and zinc-blende structure compounds}

In this chapter we will treat HMF with the Heusler C1$_b$ and
L2$_1$ structures. Although being not Heusler alloys in the strict
sense, artificial half-metals in the zinc-blende structure will be
also discussed because of their close relation with the Heusler
C1$_b$ ones. Zinc blende has a face centered
cubic (\textit{fcc}) Bravais lattice with a basis of $(0,0,0)$ and $%
(1/4,1/4,1/4)$, both species coordinating each other
tetrahedrally. The Heusler C1$_b$ structure consists of the
zinc-blende structure with an additional occupation of the
$(1/2,1/2,1/2)$ site. Atoms at the latter position, as well as
those in the origin, are tetrahedrally coordinated by the third
constituent, which itself has a cube coordination consisting of
two tetrahedra. The Heusler L2$_1$ structure is obtained by an
additional occupation of the $(3/4,3/4,3/4)$ by the same element
already present on the $(1/4,1/4,1/4)$ site. This results in
occurrence of an inversion center which is not present in the
zinc-blende and Heusler C1$_b$ structures. This difference has
important consequences for the half-metallic band gaps.
Electronic structure of the Heusler alloys has been reviewed 
recently (from a bit different positions)
by Galanakis and Mavropoulos \cite{Galanakis:315213}.

\subsubsection{Heusler C1$_b$ alloys}

Interest in fast, non-volatile mass storage memory sparked many
activities in the area of magnetooptics in general and the
magnetooptic Kerr effect specifically in the beginning of the 80-s
of last century. All existing
magnetic solids were investigated leading to a record MOKE rotation of $%
1.27^{\circ }$ for PtMnSb \cite{engen:202}. The origin of these
properties remained an unsolved problem. This formed the
motivation for the study of the electronic structure of the
isoelectronic Heusler C1$_b$ compounds NiMnSb, PdMnSb and PtMnSb
and the subsequent discovery of half-metallic magnetism.
Interestingly enough, there seems to be still no consensus on the
origin of the magnetooptical properties. The original simple and
intuitive explanation \cite{deGroot:45} was complementary to the
production of spin-polarized electrons by optical excitation in
III-V semiconductors. In that case, the top of the valence band is
split by the the spin-orbit coupling, and the photoexcitation of
electrons from the very top of the band by circularly polarized
light leads to 50\% spin polarization. 
Vice versa,
excitations from a 100
valenceband is possible for only one of the two components of circular
light, as in the case of PtMnSb,
this should result in a strong difference of refraction
and absorption for two opposite polarizations. In PtMnSb this
difference is maximal for the visible light, and for NiMnSb the
maximum of off-diagonal optical conductivity is shifted to the
ultraviolet region. The main contribution to this shift comes from
scalar relativistic interactions in the final state
\cite{Wijngaard:9318}, which are much weaker for Ni than for Pt
due to difference in nuclear charges. Further the magnetooptical
properties of the Heusler alloys were calculated by Antonov et al. \cite%
{Antonov:13012} in a good agreement with experimental data, but
the physical
explanation was not given in this paper. Recently, Chadov et al \cite%
{chadov:140411} demonstrated that an agreement between calculated
and experimental values of the Kerr rotation and ellipticity in
NiMnSb can be further improved by taking into account correlation
effects within so-called LDA+DMFT approach (see below
Sect.\ref{sec:diffunc}).

Since NiMnSb is the most studied HMF (at least within the Heusler C1$%
_{b}$'s) we will concentrate on it here. The origin of its
half-metallic properties has an analogy with the electronic
structure of III-V zinc-blende semiconductors. Given the magnetic
moment of $4\mu _{B}$ manganese is trivalent for the minority
spin direction, and antimony is pentavalent. The Heusler C1$_{b}$
structure is the zinc-blende one with an additional site
$(1/2,1/2,1/2)$ being occupied. The role of  nickel is to supply
both Mn and Sb with the essential tetrahedral coordination and to
stabilize MnSb in the cubic structure (MnSb in the zinc-blende
structure is half-metallic, but not stable). Thus a proper site
occupancy is essential:
nickel has to occupy the double tetrahedrally coordinated site \cite%
{Helmholdt:249,Orgassa:13237}. The similarity in chemical bonding
between NiMnSb and zinc-blende semiconductors also explains why it
is a \textquotedblleft weak\textquotedblright\ magnet, in the
sense discussed in the Introduction: the presence of occupied
manganese minority $d $ states is essential for the band gap.
These states play the same role as the metal $p$ states in
zinc-blende semiconductors, a situation which is possible only
because of the absence of inversion symmetry. The similarity of
chemical bonding in Heusler's and zinc-blende in the original
paper \cite{deGroot:2024} was illustrated by "removing the nickel {\it
d} states from the Hamiltonian". This phrasing has lead to
considerable confusion. Actually, the coupling of manganese states
and antimony states through non-diagonal matrix elements of nickel
{\it d} states was maintained in this calculation.

Several explanations of the band gap have been given in terms of a
Ni-Mn interaction only \cite{galanakis:134428}. While this
interaction is certainly present, it is not sufficient to explain
the band gap in NiMnSb. These analyses are based on calculations
of NiMnSb excluding the antimony, but keeping the volume fixed.
This is a highly inflated situation with a volume more than twice
the equilibrium one \cite{Egorushkin:61}. Under expansion
bandwidths in metals decrease, leading eventually to a Mott
insulating state. But even before this transition a band gap
appear, simply due to the inflation itself. This is not a
hypothetical scenario: a solid as simple as elemental lithium
becomes a half-metallic ferromagnet under expansion
\cite{Min:324}, yet there is no evidence for half-metallic
magnetism under equilibrium conditions for this element. Also, it
is not clear from these considerations why NiMnSb is half-metallic
only in the case of tetrahedrally coordinated manganese. Probably
the chemical bonding in relation to the band gap is best
summarized by in Ref. \cite{Kuebler2000}: a nickel induced Mn-Sb
covalent interaction.

Surfaces of NiMnSb do not show the $100\%$ spin polarization as
determined by positron annihilation for the bulk
\cite{bona:391,soulen:4589}. Part of the reason is their tendency
of showing surface segregation of manganese \cite{ristoiu:2349}.
Also, surfaces of NiMnSb are quite reactive and are easily
oxidized. But even without contaminations none of the surfaces of
NiMnSb are genuinely half-metallic
\cite{dewijs:020402,Galanakis:6329}. This is just another example
of the sensitivity of the half-metallic properties in NiMnSb on
the correct crystal structure. But this does not necessarily imply
that \textit{interfaces} of NiMnSb with, for example,
semiconductors cannot be completely spin-polarized. For example,
it was shown that at the $111b$ interface of NiMnSb with CdS or
InP the HMF properties are completely conserved, if the
semiconductors are anion terminated at the interface
\cite{dewijs:020402}. This anion-antimony bond may look exotic,
but such a coordination is quite common in minerals like costobite
and paracostobite (minerals are stable on a geological timescale)
No experimental verification is at hand at this moment, partially
because experimentalists tend to prefer the easier $100$ surfaces
in spite of the fact that calculations show that no half-metallic
properties are possible here.

Several photoemission measurements have been reported on NiMnSb\cite%
{correa:125316} as well as the closely related PtMnSb
\cite{Kisker:21}. We concentrate on the latter one, because it is
the first angular-resolved measurement using single crystalline
samples. A very good agreement with the calculated band structure
was obtained. This is a remarkable result. In calculations based
on density functional theory eigenvalues depend on the occupation.
These occupations deviate from the ground state in a photoemission
experiment. To very good precision the dependence of the
eigenvalues on the occupation numbers is given by the Hubbard
parameter $U$. The consequence is that the effective $U~$value in
alloys like NiMnSb and PtMnSb is much smaller then, e.g., in Ni
metal where photoemission experiments indicate the satellite
structure related to the so-called Hubbard bands
\cite{lichtenstein:067205}.

The transport properties of NiMnSb were studied extensively \cite%
{otto:2351,Hordequin:287,Borca:052409}
(a theoretical discussion
of transport properties in HMF is given in Sect.
\ref{sec:transp}).
At low temperatures, the
temperature dependence of the resistivity follows a $T^{2}$ law.
However, the $T^{2}$ law at low temperatures is absent in thin
films \cite{moodera:6101}. At around 90K a transition takes place,
beyond which the temperature dependence is $T^{1.65}$. The nature
of this phase transition is unknown. One possibility is the effect
of thermal excitations provided the Fermi energy is positioned
close to the top of the valence band or the bottom of the
conduction band. For example, in the latter case thermal
excitations are possible from the metallic majority spin direction
to the empty states in the conduction band of the minority spin
direction. Such excitation reduces the magnetic moment, which in
its turn will reduce the exchange splitting resulting in an even
further reduction of the energy difference between Fermi energy
and bottom of the conduction band. This positive feed-back will
lead to a collapse of the half-metallic properties at a certain
temperature. An analogous situation exists for $E_{F}$ close to
the valence band maximum. Numerical simulations indicate that this
scenario is highly unlikely in the case of NiMnSb because of its
unusual low density of states at the Fermi energy. Another
explanation has been put forward based on the crossing of a magnon
and a phonon branch at an energy corresponding to 80K
\cite{Hordequin:602,Hordequin:605}. It is unclear how this phonon-magnon
interaction influences the electronic properties of NiMnSb.

Local magnetic moments were studied experimentally as a function of
temperature with polarized neutron scattering. The manganese
moment decreases slightly with temperature from 3.79 $\mu _{B}$ at
15K to 3.55 $\mu _{B}$ at 260K, while the nickel moment remains
constant at 0.19 $\mu _{B}$ in the same temperature range. On the
other hand, magnetic circular dichroism shows a reduction of both
the manganese and nickel moments around 80K. Borca et al.
\cite{Borca:052409} conclude that at the phase transition the
coupling of the manganese and nickel moments is lost. A
computational study \cite{lezaic:026404} shows vanishing of the
moment of nickel at the transition temperature. None of these
anomalies are reflected in the spontaneous magnetization
\cite{otto:2351} of bulk NiMnSb.

Two Heusler C1$_{b}$ alloys exist, isoelectronic with NiMnSb:
PdMnSb and PtMnSb. Their electronic structures are very similar,
but we will discuss the differences. The calculated DFT band
structure for PdMnSb is not half-metallic. The minority spin
direction does show a band gap very similar as NiMnSb but the
Fermi energy intersects the top of the valence band. Reliable
calculations (e.g., based on the GW approximation) are needed here
to settle the issue whether PdMnSb is half-metallic or not. PtMnSb
is very similar to NiMnSb, the largest differences being in the
empty states just above the Fermi energy. The direct band gap (at
the gamma point) is between the triplet top of the valence band
(neglecting spin-orbit interactions) and a total symmetrical
singlet state in PtMnSb. This singlet state is positioned at much
higher energy in NiMnSb. These differences have been attributed to
the much stronger mass-velocity and Darwin terms in platinum
\cite{Wijngaard:9318}. Platinum does not carry a magnetic moment
in PtMnSb. Consequently, no 90K anomaly like in NiMnSb is to be
expected and none has been reported so far.

Let us consider whether half-metals in the Heusler C1$_{b}$
structure exist when substituting other than iso-electronic
elements for Ni in NiMnSb. Since NiMnSb is a weak magnet,
substitutions with elements reducing the total magnetic moment are
possible only while maintaining the half-metallic properties. Thus
cobalt, iron manganese and chromium will be considered. The case
of Co was studied by K\"{u}bler \cite{Kubler:257}. The
half-metallic
properties are conserved, consequently the magnetic moment is reduced to $%
3\mu _{B}$. Calculations on FeMnSb \cite{deGroot:330}, MnMnSb \cite%
{Wijngaard:5395} as well as CrMnSb \cite{deGroot:45} all show the
preservation of the half-metallic properties. In the case of
FeMnSb this implies a reduction of the total magnetic moment per
formula unit to $2\mu _{B}$, which is an unusually small moment to
be shared by iron and manganese. The way out is that FeMnSb orders
\textit{anti}ferromagnetically. Thus the $2\mu _{B}$ total
magnetic moment corresponds to the difference in moments of iron
and manganese rather than to their sum implied by ferromagnetic
ordering. This way to preserve a band gap (energetically favorable
from a chemical-bonding point of view) together with the
maintenance of sizable magnetic moments (favorable for the
exchange energy) determine the magnetic ordering here. Both these
effects are usually larger than the exchange-coupling energies.
The antiferromagnetic ordering is maintained in MnMnSb with a
total moment of $1\mu _{B}$. In the case of CrMnSb the
antiferromagnetic coupling leads to a half-metallic solution with
a zero net moment. This is a really exotic state of matter. It is
genuinely half-metallic, implying $100\%$ spin polarization of the
conduction-electrons, yet it lacks a net magnetization
\cite{deGroot:45}. The stability of such a solution depends
sensitively on the balance between the energy gain of the band gap
and the energy gain due to the existence of magnetic moments: if
the first one dominates a nonmagnetic semiconducting solution will
be more stable (remember that both spin directions are
isoelectronic here).

In reality, the situation is more complex. CoMnSb does exist, but
it crystallizes in a tetragonal superstructure with Co partially
occupying the empty sites \cite{Senateur:226}. The magnetic
moments deviate from the ones expected for a half-metallic
solution. FeMnSb does not exist, but part of the nickel can be
substituted by iron. Up to 10$\%$ the Heusler C1$_{b}$ structure
is maintained, from 75$\%$ to 95$\%$ a structure comparable with
CoMnSb is stable and between 10$\%$ and 95$\%$ both phases coexist \cite%
{deGroot:330}. MnMnSb exists, orders antiferromagnetically and has
a net moment of 1$\mu _{B}$. It does not crystallize in the
Heusler C1$_{b}$ structure and consequently is not half-metallic.
CrMnSb exists, is antiferromagnetic at low temperatures, and shows
a transition at room temperature to a ferromagnetic phase.

A different substitution is the replacement of Mn by another
transition metal. An interesting substitution is a rare earth
element R. Because of the analogy of half-metals with C1$_b$
structure and III-V semiconductors one expects NiRSb compounds to
be nonmagnetic semiconductors.

Several of these compounds do exist in the C1$_b$ structure,
examples are Sc, Y and heavy rare-earth elements from the second
half of the lanthanide series. All of them are semiconductors
indeed \cite{Pierre:845,Pierre:74}. Doping of NiMnSb by rare earth
elements has been suggested as a way to improve the
finite-temperature spin polarization in NiMnSb. These
substitutions do not influence the electronic band structure much,
(see also Sect. \ref{sec:la_nimnsb}),
the band gap for the minority spin direction remaining completely
intact. However, the random substitution of nonmagnetic (Y, Sc) or
very different (Ho-Lu)
magnetic elements for manganese will modify the magnon spectrum \cite%
{attema:S5517,Chioncel:thesis}. This could be beneficial to
increase spin polarization in some temperature range.

\subsubsection{Half-metals with zinc-blende structure}

The Curie temperatures of diluted magnetic semiconductors remain
somewhat disappointing. A solution is to replace all the main
group metals by transition metals. But there is a heavy prize to
be paid: These systems can only be prepared as metastable states
-- if at all -- on a suitable chosen substrate. An alternative to
come to the same conclusion is to consider Heusler C1$_{b}$'s with
larger band gaps. This is most easily accomplished by replacement
of the antimony by arsenic or phosphorous. No stable Heusler
C1$_{b}$ alloys exist with these lighter pnictides, however. An
alternative is to try to grow them as metastable systems on a
suitable chosen substrate. This makes the nickel superfluous:
since it fails in the case of lighter pnictides to play the role
it does so well in the NiMnSb. The bottleneck in this quest is not
so much in predicting systems that are good half-metals, but to
design combinations of half-metals and substrates that are
meta-stable enough to have a chance of being realized
experimentally.

Shirai et al. \cite{shirai:1383} were the first to relate
concentrated magnetic semiconductor with half-metallic magnets in
their study of MnAs in the zinc-blende structure. The experimental
realization showed an increase of the Curie temperature indeed:
400K was reported on for CrAs grown on GaAs. Xie et al.
\cite{xie:2003} calculated the stability of all 3$d$
transition-metal chalcogenides in the zinc-blende with respect to
the ground state structure. Chromium telluride and selenide, as
well as vanadium telluride, are good half-metals which are stable
towards a tetrahedral and rhombohedral distortions. Zhao and
Zunger \cite{zhao:132403} consider the stability of an epitaxial
layer as a function of the lattice parameter of the substrate
allowing for relaxation in the growth direction. The result is
that while the bulk zinc-blende phase is always unstable with
respect to the (equilibrium) NiAs structure, there exist lattice
constants where the epitaxial zinc-blende phase is more stable as
compared with the epitaxial nickel arsenide structure. This is
realized (computationally) for half-metallic CrSe.

An alternative to the concentrated III-V magnetic semiconductors
is given by delta doped III-V semiconductors. Here the magnetic
properties are not introduced by a more or less homogeneous
replacement of main group metals by magnetic transition metals.
Instead, a very thin transition-metal layer is sandwiched between
undoped III-V semiconductor material \cite{nazmul:3120}. A clear
increase in Curie temperature results \cite{chiba:3020}. This is
not unrelated to the interface-half-metallicty introduced before
\cite{deGroot:45}.

\subsubsection{Heusler L2$_1$ alloys}

The crystal structure of the Heusler L2$_1$ alloys is closely related with that 
of the C1$_b$ alloys. In the L2$_1$ structure the $(1/2, 1/2, 1/2)$ position, empty in the 
C1$_b$ structure, is occupied by the same element that occupies the $(0, 0, 0)$ position. 
The similarity in structure suggests a similarity in interactions and physical 
properties, but on the contrary, the interactions and the physical properties of 
the two classes are actually quite distinct. The introduction of the fourth atom 
in the unit cell introduces an inversion centre in the crystal structure. The bandgap 
in the C1$_b$ compounds resulted from an interaction very similar to that in III-V 
semiconductors, where the manganese $t_{2g}$ $d$ electrons play the role of the $p$ electrons in 
the III-V semiconductor. This is no longer possible in the presence of an inversion 
centre. Consequently bandwidths are reduced and usually Van Hove singularities occur 
in the vicinity of the Fermi energy. The smaller bandwidth leads to several (pseudo) gaps. 
Correlation effects are expected to become better observable here. 

Another difference is the occurrence of defects. Experimentally it was noted that 
``The strong effect of cold work on Heusler alloys (L2$_1$ structure) contrasts with almost 
unobservable effects in the C1$_b$ structure alloy NiMnSb'' \cite{sc.da.83}. 
But also here there are indications 
that defects that destroy the bandgap are energetically less favourable.

Experimental work goes back to Heuβler in the beginning of the last century. The motivation 
of his work was the possibility of preparing magnetic alloys out of non-magnetic elements 
\cite{heus.03} (A material 
was only considered magnetic in that period if it possessed a spontaneous net magnetisation). 
More recently the landmark work of Ziebeck and Webster on neutron-diffractions investigations 
\cite{zi.we.74} deserves mentioning as well as the 
NMR work in the Orsay group of Campbell. 

The first bandstructure calculations were by Ishida and coworkers 
\cite{Ishida:1570,Ishida:1239,Ishida:814,Ishida:1111}, as well as 
Kuebler, Williams and Sommers \cite{ku.wi.83}. 
The latter paper contains a clue to half-metallic properties in the L2$_1$ compounds: the authors remark 
that ``The minority state densities at the Fermi energy for ferromagnetic Co$_2$MnAl and Co$_2$MnSn nearly vanish. 
This should lead to peculiar transport properties in these two Heusler alloys''. 

Calculations that explicitly addressed the question of half-metallic properties in the full 
heuslers appeared not earlier than in 1995 \cite{fu.is.85,Ishida:2152}. 
A systematic study of the electronic structure 
of Heusler L2$_1$ compounds was undertaken by Galanakis, Dederichs and Papanikolaou \cite{ga.de.02}. 
This paper also reviews the work on half-metallic magnetism in full heusler compounds till 2002. For this reason 
we refer to it for details and concentrate on subsequent developments here. 

The heusler L2$_1$ compounds take a unique position in the spectrum of halfmetals because of 
their Curie temperatures. High Curie temperatures are important in the application of 
halfmetals at finite temperature, since many of the depolarisation mechanisms scale with 
the reduced temperature $T/T_c$. Curie temperatures approach 1000K: Co$_2$MnSn shows a Curie 
temperature of 829K, the Germanium analogue 905K while Co$_2$MnSi was a record holder for 
some time with a Curie temperature of 985K \cite{br.ne.00}. 
A further increase was realized in Co$_2$FeSi. Experimentally it shows an integer magnetic 
moment of 6 $\mu_B$ and a Curie temperature of 1100K \cite{wurmehl:184434}. 
This result was not reproduced in calculations employing the LDA approximation. The magnetic 
moment of 6 Bohr-magnetons could only be reproduced by the application of U in excess of 7.5 
electron-volt. This is an unusual high number and alternative explanations should also be considered. 
The question of lattice defects has been studied. On the basis of neutron-diffraction, Co-Fe disorder 
could be excluded, but no data are available for the degree of Fe-Si interchange. A calculation 
of the magnetic saturation moment as function of the iron-silicon disorder seems a logical next 
step in the understanding of this fascinating compound.

Whereas the investigations of the bulk electronic structures of full heuslers has advanced 
comparable as with the half-heuslers, the situation with respect of the preservation of 
halfmetallic properties at surfaces and interfaces clearly still lacks behind. Two important 
results were obtained recently. One result is the preservation of halfmetallic properties of 
an Co$_2$MnSi (001) surface provided it is purely manganese terminated. This is the only surface 
of this half-metal showing this property \cite{hashemifar:096402}. 

No genuine halfmetallic interfaces between full-heuslers and semiconductors are reported yet. 
But the results for Co$_2$CrAl/GaAs look promising. For an (110) interface a spin-polarization 
of $\approx 90\%$ was obtained \cite{na.ma.04}. Although 
this is clearly not a genuine half-metallic interface yet it should provide a good basis for 
analyses why half-metallic behaviour is lost at in interface in analogy with the successful 
work for the C1$_b$ case.

An interesting development in half-metallic-magnetism is in electron-deficient full-Heusler 
alloys.  Reduction of the number of valence electrons to 24 per formula unit leads to either 
a non-magnetic semiconductor or a halfmetallic anti-ferromagnet. But remarkable enough, the 
reduction or the number of valence-electrons can be continued here, re-entering a range of 
half-metals but with a bandgap for the majority spin-direction now. This is best exemplified 
for the case of Mn$_2$VAl. It is a halfmetallic ferrimagnet of calculated with the generalized 
gradient exchange-correlation potential \cite{ru.pi.99}. 
Halfmetals with a bandgap for the majority spin-direction hardly occur. The search for new 
candidates should strongly be supported.

\begin{figure}[tbh]
\begin{center}
\includegraphics[width=0.75\columnwidth, angle=0]{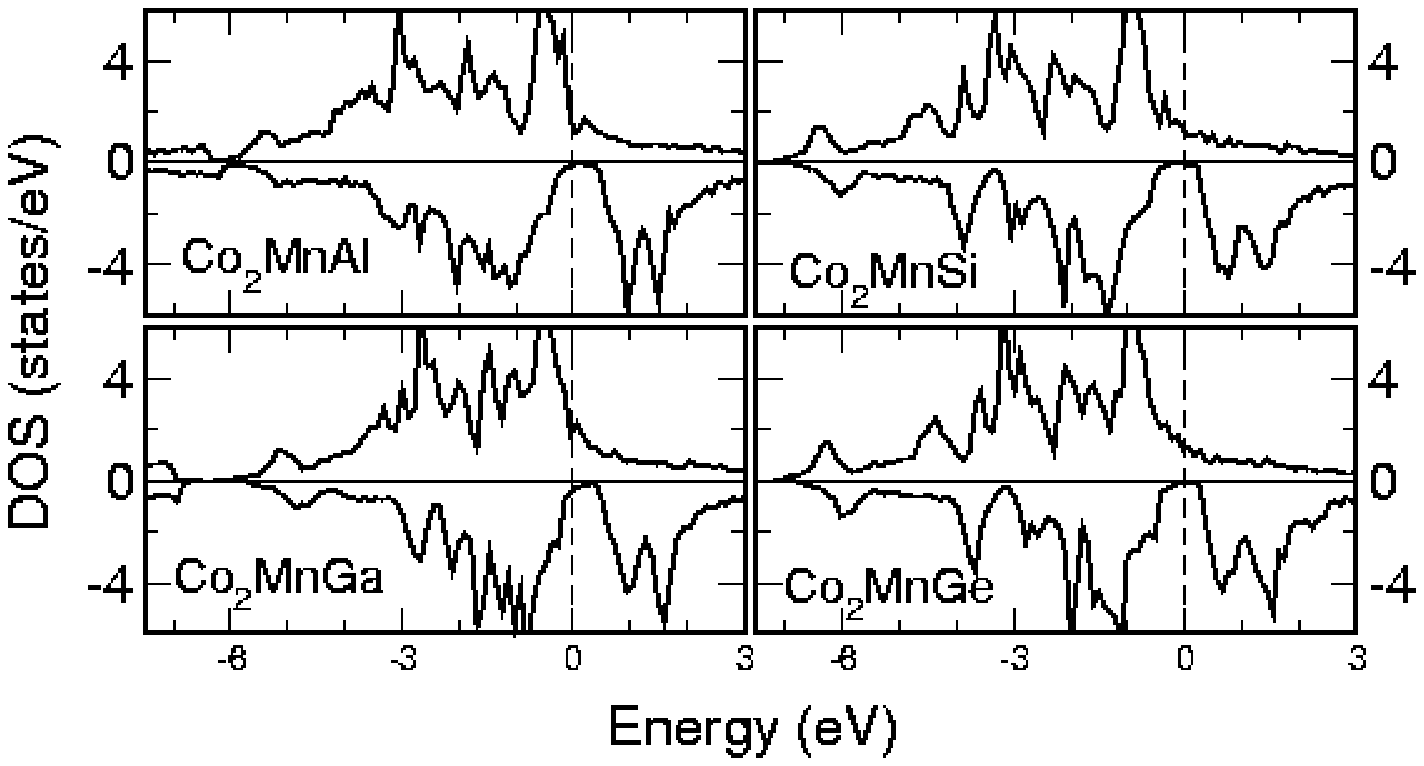}
\end{center}
\caption{Atom-resolved DOS for the Co$_{2}$MnZ compounds with
Z=Al, Si, Ge, Sn \protect\cite{Galanakis:765}.} \label{L21}
\end{figure}

\subsection{Strongly magnetic half-metals with minority spin gap}

\subsubsection{Chromium dioxide}

Chromium dioxide is the only metallic oxide of chromium. It orders
ferromagnetically with a Curie temperature of about 390 K. Its
half-metallic
state was discovered by band structure calculations \cite%
{schwarz:L211,Matar:315}. The origin of the half-metallicity is
straightforward: in an ionic picture the chromium is in the form of a Cr$%
^{4+}$ ion. The two remaining $d~$electrons occupy the majority
$d~$states. The crystal field splitting is that of a (slightly)
deformed octahedron. The valence band for the majority-spin
direction is 2/3 filled, hence the metallic properties. The
minority-spin $d$ states are at a significant higher energy due to
the exchange splitting. For this reason the Fermi level falls in a
band gap between the (filled) oxygen 2$p$ states and the (empty)
chromium $d$ states. Thus the HMF properties of chromium dioxide
are basically a property of chromium and its valence and, as long
as the crystal-field splitting is not changed too drastically, the
half-metallic properties are conserved. This implies that the
influence of impurities should not be dramatic and a number of
surfaces retain the half-metallicity of the bulk. As a matter of
fact, all the surfaces of low index are
half-metallic with a possible exception of one of the (101) surfaces \cite%
{vanLeuken:7176,Attema:793}. Although initial measurements did not
confirm these expectations \cite{Kamper:2788}, they were confirmed
later by the
experiments like tunneling \cite{Bratkovsky:2344}, Andreev reflection \cite%
{ji:5585} on well-characterized surfaces. Recently, the flow of a
triplet-spin supercurrent has been realized in CrO$_{2}$
sandwiched employing two superconducting contacts
\cite{keizer:825}.

As mentioned before, an interesting question is the origin of the
metallic ferromagnetism in CrO$_{2}$. This was explained in terms
of the double
exchange (Zener) model by Korotin et al. \cite%
{korotin:4305,schlottmann:174419}. The octahedral coordination in
the rutile
structure is slightly distorted. This leads to splitting of the degenerate $%
t_{2g}$ state into a more localized $d_{xy}$ state and more delocalized $%
d_{xz}$ and $d_{yz}$ states (or linear combinations of these). The
localized filled $d_{xy}$ state plays the same role as the filled
$t_{2g}$ majority
spin state in the Zener double-exchange model, while the partially occupied $%
d_{xy} \pm d_{yz}$ majority states in CrO$_{2}$ the role of the
partially occupied $e_{g}$ states. The transport properties of
CrO$_{2}$ were investigated in detail \cite{wa.wi.00} and
interpreted in terms of a two-band model, very much in line with
the double exchange model for CrO$_{2}$.

The importance of explicit electron-electron interactions in
CrO$_{2}$ remains a subject of active research. On one hand,
Mazin, Singh and Ambrosch-Draxl \cite{mazin:411} compared LSDA
calculations with experimental optical conductivities and found no
indications for strong correlations related exotic phenomena. On
the other hand, Craco, Laad and M{\"u}ller-Hartman
\cite{craco:237203,laad:214421} considered photoemission results
and conductivity (both DC and optical) and concluded the
importance of dynamical correlation effects. The ferromagnetic
correlated state was investigated also in a combined local and
non-local approach \cite{chioncel:cro2} which demonstrates that
the $d_{xy}$ orbital is not completely filled and localized
as described by LDA+U or model calculations \cite%
{korotin:4305,schlottmann:174419,to.ko.05}. More recently,
Toropova, Kotliar, Savrasov and Oudovenko \cite{to.ko.05}
concluded that the low-temperature experimental data are best
fitted without taking into account the Hubbard $U$ corrections.
Chromium dioxide will clearly remain an area of active research.

\subsubsection{The colossal magnetoresistance materials}

The interest in ternary oxides of manganese with di- or trivalent
main group metals goes back to Van Santen and Jonker
\cite{jonker:337}. The occurrence of ferromagnetism in transition
metal oxides, being considered unusual at that time, was explained
by Zener \cite{Zener:403} by the introduction of \textquotedblleft
double exchange\textquotedblright\ mechanism. In 70th and 80th
these systems were investigated theoretically in connection with
the problem of phase separation and \textquotedblleft
ferron\textquotedblright\
(magnetic polaron) formation in ferromagnetic semiconductors \cite%
{Auslender:436,nagaev:1983}. The interest in spintronics fifteen
years ago revived the interest in the ternary manganese
perovskites, generally referred to as colossal magnetoresistance
materials. A wealth of interesting
physics is combined in a single phase diagram of, for example, La$_{1-x}$Sr$%
_{x}$MnO$_{3}$. From a \textquotedblleft
traditional\textquotedblright\ antiferromagnetic insulator for
$x=1$, the reduction of $x$ results in a ferromagnetic metallic
state, while finally at $x=0$ a Mott insulating antiferromagnet is
found. Some of the transitions are accompanied with charge and or
orbital ordering. Finite temperatures and applied magnetic fields
complicate the phase diagram substantially. The ferromagnetic
metallic phase for intermediate values of $x$ is presumably
half-metallic \cite{Pickett:1146}. We will concentrate on this
phase here and refer to
other reviews for a more complete overview of the manganites \cite%
{Salamon:583,nagaev:387,ziese:143,Dagotto2003}.

Once the occurrence of a ferromagnetic magnetic ordering is
explained, the occurrence of half-metallic magnetism is rather
straightforward. Manganese possesses around $3.5$ $d$ electrons in
the metallic high-spin state; its rather localized majority spin
$t_{2g}$ state is filled, the majority, much more dispersive,
$e_{g}$ state is partially occupied and the minority $d$ states
are positioned at higher energy, thus being empty. Hence a rather
large band gap exists for the minority spin at the Fermi energy
and the manganites are strong magnets. Correlation effects are
expected to be much stronger here. Notice that no reference has
been made to the actual crystal structure: subtleties like in the
Heusler structure are absent here. The half-metallic properties
are basically a property of the valence of the manganese alone.
Surface sensitivity of the HMF properties is not to be expected as
long as the valence of the manganese is maintained: This is easily
accomplished in the layered manganites \cite{Boer:10758}.

Experimental verification of the half-metallic properties has not
been without debate. The origin of the controversy is that the
calculated position of the Fermi energy in the energy gap is
invariably very close to the bottom of the conduction band. The
experimental conformation of the HMF behavior by photoemission
\cite{Park:794}, was contested on the basis of Andreev reflection
measurements, that did show minority-spin $d$ states at the Fermi
energy \cite{Nadgorny:184433,Nadgorny:315209}. Also, tunnelling
experiments initially casted doubt on the half-metallic properties \cite%
{jo:3803,jo:R14905,Viret:8067}. Mazin subsequently introduced the
concept of transport half-metal: the Fermi energy may straddle the
bottom of the minority spin $t_{2g}$ band, but since these states
are localized this does not influence the half-metallic
properties, as far as transport is concerned \cite{mazin:411,Nadgorny:315209}.
Recent magnetotransport measurements on better
samples support the HMF picture of the CMR materials \cite%
{bowen:233}. The recent GW calculations by Kino et al.
\cite{Kino:858} shed a different light on this matter. In these
calculations the half-metallic band gap is increased by as much as
2~eV with respect to the DFT value. This implies that the minority
spin $d~$band is not even close to the Fermi energy and the CMR
materials should be considered as genuine, real HMF's.

\subsection{Weakly magnetic half-metals with majority spin gap}

\subsubsection{The double perovskites}

\begin{figure}[tbh]
\begin{center}
\includegraphics[width=0.75\columnwidth, angle=0]{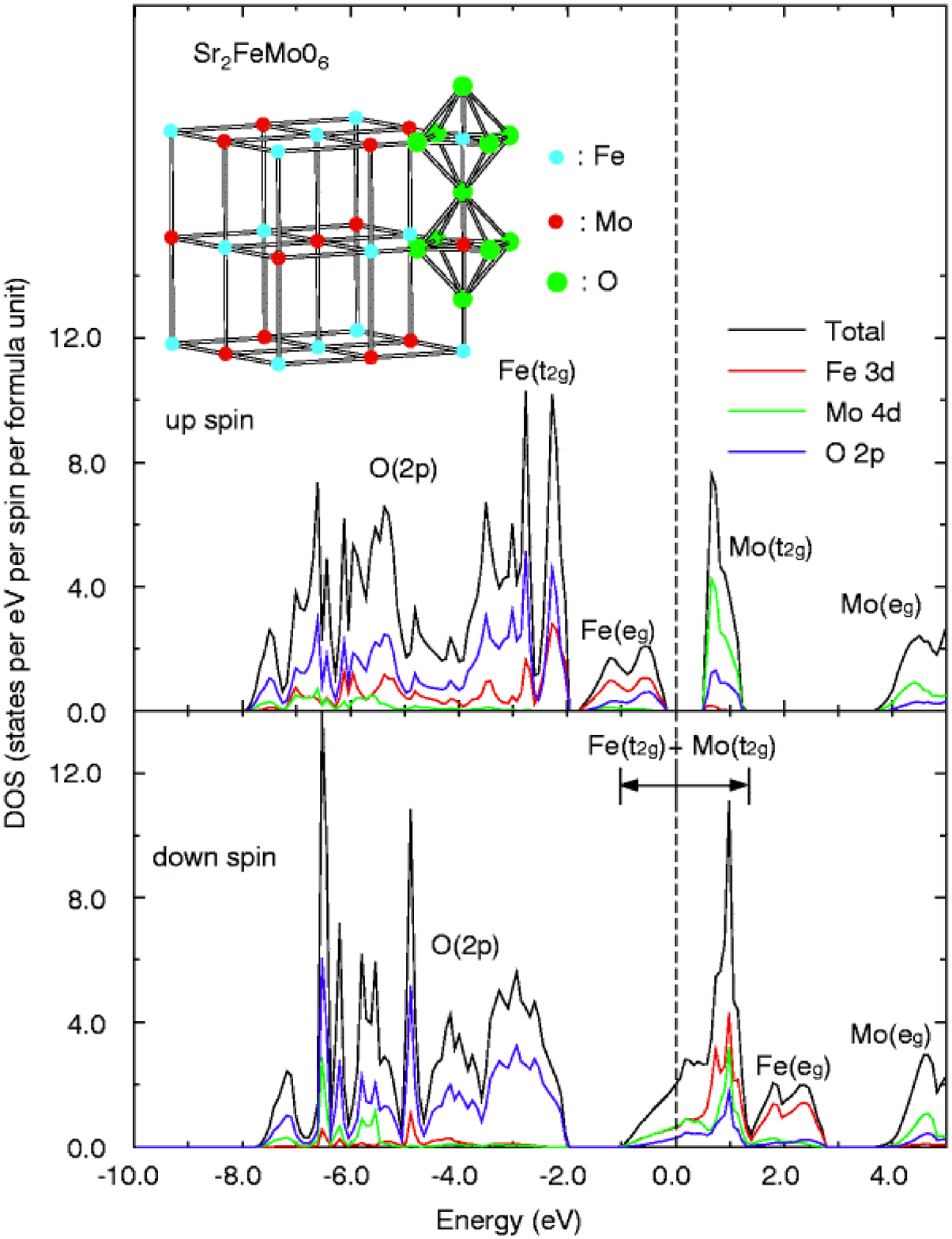}
\end{center}
\caption{The density of states (DOS) of Sr$_2$FeMoO$_6$
\cite{Kobayashi:677}} \label{double_per}
\end{figure}

The double perovskites have a unit cell twice the size of the
regular perovskite structure. The two transition-metal sites are
occupied by different elements. Double perovskites are interesting
for two reasons.
Half-metallic antiferromagnetism has been predicted to occur for La$_{2}$VMnO%
$_{6}$ \cite{Pickett:10613} (we will return to this question later
in Sect.\ref{spinel}). The second reason the double perovskites
are important is that the high Curie temperatures can be obtained
in them as compared with the regular perovskites.
Sr$_{2}$FeMoO$_{6}$ was the first example to be studied in this
respect by means of band structure calculations
\cite{Kobayashi:677}. The density of states is shown in figure
\ref{double_per}. In the majority spin direction the valence band
consists of filled oxygen 2$s$ and 2$p$ states, as well as a
completely filled Fe 3$d$ band, showing the usual crystal field
splitting. A band gap separates the conduction band, which is
primarily formed by molybdenum $d$ states. The minority spin
direction shows an occupied oxygen derived valence band and a
hybridized $d$ band of mixed iron and molybdenum character. It
intersects the Fermi energy.

The Curie temperature is in the range of 410 to 450K. More
recently, a similar behavior was found for Sr$_{2}$FeReO$_{6}$
\cite{Kobayashi:11159}. Optical measurements did show excitations
across the half-metallic band gap of 0.5 eV \cite{Tomioka:422} . A
substantial higher Curie temperature is found in
Sr$_{2}$CrReO$_{5}$, $T_{C}=635$K \cite{kato:328}, but band
structure calculations show that the band gap for the majority
spin direction is closed by the spin-orbit interaction
\cite{vaitheeswaran:032513}.

\subsubsection{Magnetite}

Magnetite Fe$_{3}$O$_{4}~$ is one of most wide-spread natural iron
compounds and the most ancient magnetic material known to
humanity. Surprisingly, we still have no complete explanation of
its magnetic, electronic and even structural properties, many
issues about this substance remaining controversial. At room
temperature magnetite has inverted cubic spinel structure with
tetrahedral A-sites occupied by Fe$^{3+}$ ions, whereas octahedral
B-sites are randomly occupied by Fe$^{2+}$ and Fe$^{3+}$ ions with
equal concentrations. Fe$_{3}$O$_{4}$ is a ferrimagnet with a high
Curie temperature $T_{C}\simeq 860$ K. As discovered by Verwey \cite%
{Verwey:327}, at $T_{V}\simeq 120$ K magnetite undergoes a
structural distortion and metal-insulator transition. Usually the
Verwey transition is treated as a charge ordering of Fe$^{2+}$ and
Fe$^{3+}$ states in octahedral sites (for a review, see
\cite{Mott1974,Mott:327}). The nature of the Verwey transition and
low-temperature phase of Fe$_{3}$O$_{4}$ is a subject of numerous
investigations which are beyond our topic, see, e.g., recent
reviews \cite{Walz:R285,Garsia:R145}. As demonstrated by the
band-structure calculation \cite{Yanase:312}, magnetite in the
cubic spinel structure is a rather rare example of HMF with
majority-spin gap. This means a saturated state of itinerant
$\mathit{3}$\textit{d} electrons propagating over B-sites, the
magnetic moment being close to 4$\mu _{B}$ per formula unit.
Recently this picture was questioned by the x-ray magnetic
circular dichroism (XMCD) data \cite{huang:077204} which were
interpreted as an evidence of large orbital contribution to the
magnetization and non-saturated spin state. However, later XMCD
experiments \cite{Goering:97} confirm the purely spin saturated
magnetic state. Direct measurements of
spin polarization by spin-polarized photoemission spectroscopy \cite%
{Mortonx:L451} yield the value about $-40$\% (instead of $-100$\%
predicted by naive band picture), which can be due to both surface
effects and electron correlations in the bulk (see
Sect.\ref{pol}). Transport properties of Fe$_{3}$O$_{4}$-based
films are now intensively studied (see, e.g., Refs.
\cite{Zhao:2595,Eerenstein:247204}). In particular, a large
magnetoresistance owing to electron propagation through antiphase
boundaries was found \cite{Eerenstein:247204}.

Unlike the Heusler alloys, magnetite is a system with a narrow
\textit{3d} band and therefore strong correlation effects. The
fact of the metal-insulator transition itself can be already
considered as an evidence of strong electron-electron interaction
\cite{Mott1974}. The influence of these effects on the electronic
structure of Fe$_{3}$O$_{4}$ has been recently considered in Refs.
\cite{craco:064425,leonov:165117}.

\subsection{Strongly magnetic half-metals with majority spin gap}

\subsubsection{Anionogenic ferromagnets}

Until recently, strongly magnetic HMF with a majority spin band
gap were absent. The chemical composition of the compounds
calculated to be half-metallic in this category was quite
unexpected: heavy alkali oxides \cite{Attema:16325}. The magnetic
moment is carried by complex oxygen ions, hence the above name.
Besides the oxygen molecule, that has two unpaired
electrons, the O$^{2-}$ ion occurs in the so-called hyperoxides like RbO$%
_{2}$ and CsO$_{2}$. These are antiferromagnetic insulators with
rather low Neel temperatures. Another molecular ion of interest is
the non-magnetic peroxide ion O$_{2}^{2-}$. In the series
molecular oxygen -- hyperoxide ion -- peroxide ion the antibonding
$\pi $ orbital is progressively filled, leading to the vanishing
of the magnetic moment for the peroxides. Also sesquioxides exist
which have composition between peroxide and hyperoxide. They are
rather thermally stable, but do react with atmospheric water and
carbondioxide. The analogy between the holes in the antibonding
double-degenerate $\pi $ level and the electrons in the
double-degenerate antibonding $e_{g}$ level of the colossal
magnetoresistance materials motivated a computational study. This
yielded a HMF state with surprisingly high Curie temperatures
(300K). Partial  explanation is the absence of superexchange in
these oxides, since the mediator for it, the alkali ions, do not
possess the required electron states in the vicinity of the Fermi
level. Direct experimental evidences are
 unfortunately lacking. An indirect evidence is the cubic crystal
structure measured down to 5K (unlike peroxides and hyperoxides),
the crystallographic equivalence of the
molecular oxygen ions, the occurrence of charge fluctuations down to 5K \cite%
{Jansen:591}, the opaque optical properties and indications of
unusual widths of the stability regions of the sequioxides in the
oxygen-rubidium and oxygen-caesium phase diagrams
\cite{Rengade:348}.

\subsection{Sulphides}

The spectacular developments in the area of high temperature
superconductivity succeeded by the interest in colossal
magnetoresistance materials have pushed the interest in sulphides
and selenides somewhat to the background. These materials have
some advantages over oxides, however. Two main differences, both
due to the increased metal-anion covalence as compared with
oxides, are of importance here: the more correlated behavior of
the oxides as well as their preference for a high-spin
configuration. Sulphides often prefer a low-spin configuration,
which make their behavior less predictable without careful
computation. So, the sulphur analogue of magnetite, the mineral
Greigite, has a magnetic moment of $2\mu _{B}$ only, compared with
the $4\mu _{B}$ of magnetite. Consequently, it  is not
half-metallic. In the widespread mineral pyrite FeS$_{2}$, iron
has a non-magnetic $d^{6}$ configuration, unimaginable in oxides.
As mentioned before, magnetite shows half-metallic properties, but
is at the brink of Mott localization: cooling down below 120K
suffices to accomplish this.  On the other hand, the much less
correlated behavior of sulphospinels allows the occupation of a
broad range of different transition metals on the octahedral and
tetrahedral cation sites without the risk of a Mott
insulating state. This does not hold for all the pyrites, however. Thus FeS$%
_{2}$ is a non-magnetic semiconductor. The excellent agreement
between LDA calculations and the photoemission spectra indicate
negligible correlation effects \cite{Folkerts:4135}. CoS$_{2}$ is
a ferromagnetic metal with a Curie temperature of 122K. The
magnetic moments were calculated function of the Hubbard U and
comparison with experimental data indicated the importance of $U$
(of less than 1~eV). NiS$_{2}$ is a Mott insulator. NiSe$_{2}$ is
metallic while in NiSe$_{2x}$S$_{2(1-x)}$ the strength of the
correlation effects can be adjusted by variations in the
composition.

Magnetic ordering temperatures, important for maintaining the
polarization of charge carriers at finite temperature, of oxides
are usually superior to those of sulphides and selenides.

\subsubsection{Pyrites}

Saturated itinerant ferromagnetism in the pyrite-structure system Fe$_{1-x}$%
Co$_{x}$S$_{2}$ was discovered experimentally in Ref.
\cite{Jarrett:617} and
discussed from the theoretical-model point of view in Ref. \cite%
{Auslender:5521}. Half-metallic ferromagnetism in pyrites was
first
considered in band calculations by Zhao, Callaway and Hayashibaran \cite%
{Zhao:15781}. Their results for CoS$_{2}$ show near the Fermi
energy two completely filled $t_{2g}$ bands for the two spin
directions, a partial filled $e_{g}$ majority band as well as a
minority $e_{g}$ band just overlapping the Fermi energy. At
slightly higher energy the antibonding sulphur 3$p$ states are
found. Clearly, cobalt disulphide is an almost half-metallic
ferromagnet. Also it is suggested that half-metallic magnetism can
be obtained in the ternary system Fe$_{x}$Co$_{1-x}$S$_{2}$, an
idea further worked out by Mazin \cite{mazin:3000}. He calculated
the expected HMF region in the phase diagram to extend from $0.2$
to $0.9$. A detailed study, both computational and experimental
\cite{wang:056602}, reveals a strong dependence of the spin
polarization at the Fermi level on
the composition. Theoretically, $100\%$ spin polarization is obtained for $%
x=0.25$, whereas the maximal polarization ($85\%$) determined with
Andreev reflection at 4.2K is obtained at $x=0.15$. The
polarization drops for higher concentrations of iron. The Fermi
level is located very close to the bottom of the conduction band.
This can lead to thermal instabilities of the half-metallicity as
discussed for NiMnSb.
Recently half-metallic properties of pyrite-structure compounds 
have been reviewed by Leighton et. al. \cite{Leighton:315219}.

\subsubsection{Spinels}
\label{spinel}

The activities in the area of half-metallicity are somewhat
underrepresented. A complication in this class of compounds is
that of cation ordering. The application of high temperatures
leads to disproportionation, so long annealing at lower
temperatures may be required. The type of cation ordering depends
on the preparation conditions. On the other hand, once being
controlled, the cation occupancy can form a degree of freedom to
achieve HMF materials.

One of the compounds considered in a study on chromium
chalcogenides is of interest here CuCr$_2$S$_4$. It shows an
almost HMF band structure: the Fermi level is positioned 50 meV
below the top of the valence band \cite{Antonov:14552}.

Sulphospinels were also considered in detail in the quest for the
elusive half-metallic antiferromagnet
\cite{Park:100403,Min:S5509}. Mn(CrV)S$_{4}$, with chromium and
vanadium occupying the octahedral sites is calculated to fulfill
all the requirements. It shows a band gap of approximately 2eV,
while the Fermi level intersects a band of primarily vanadium
character. The Mn moment is compensated by the moments of chromium
and vanadium on the octahedral sites. Another sulphospinel with
predicted half-metallic properties is
(Fe$_{0.5}$Cu$_{0.5}$)(V$_{0.5}$Ti$_{1.5}$)$_{2}$S$_{4}$. In this
case, the metallic behavior is attributed to the atoms at the
tetrahedral site; their magnetic moments are exactly canceled by
those at the octahedral site.


\subsection{Miscellaneous}

\label{sec:miscel}

\subsubsection{Ruthenates}

The 3$d$ transition-elements and their compounds have been studied
in much more detail as compared with their 4$d$ and 5$d$
analogues. Part of the reason is that magnetism is expected to be
favored more in the 3$d$ series where no d core is present.
Ruthenium is a perfect example of the contrary. The binary and
ternary oxides of this 4d transition metal show a rich variety of
physical properties like ferromagnetism in SrRuO$_{3}$,
unconventional superconductivity in Sr$_2$RuO$_4$
\cite{Maeno:532}. Here we consider the case of SrRuO$_{3}$.
Ruthenium is tetravalent in this compound, just as in RuO$_{2}$.
The latter compound is a non-magnetic metal with 4$d$ electrons in
the slightly split $t_{2g}$ subband. In SrRuO$_{3}$, a magnetic,
low-spin state occurs with a filled $t_{2g}$ majority spin band
and a partially filled $t_{2g}$ minority spin band. Thus all the
ingredients seem to be present for a half-metal. Calculations show
that the exchange and crystal field splitting is not sufficient to
create a band gap large enough to encompass the Fermi energy.
Recently, it was shown that the application of the LDA+U method
leads to a substantial increase in band gap in conjunction with
orbital ordering. Thus a half-metallic solution is obtained.
Comparison with experiment does not lead to a definite conclusion.
No experimental determination of $U$ is available. The measured
magnetic moment is more in line with the LDA results ($0.8\div
1.6$ $\mu _{B}$), but extrapolation to the high-field limit could
lead to an integer magnetic moment. There is no experimental
evidence for orbital ordering, however.

The research on ruthenates is relatively recent and especially
ternary compounds have not yet been investigated exhaustively.


\subsubsection{Organic half-metals}

Conducting organic materials have been an area of active research
since the
discovery of electrical conduction in doped polyacethylene \cite%
{Shirakawa:578}. A surge of activities has resulted in
applications generally referred to as plastic electronics.
Recently, attempts have started to develop HMF suitable for these
applications. Originally, the focus was on carbon nanotubes where
magnetism was achieved by the introduction of 3d metals.
Calculations were performed for 3,3 single wall carbon nanotubes
with a linear iron nanowire inside \cite{Rahman:S5755}. Structure
optimization resulted in a slightly asymmetrical position of the
iron wire in the nanotube. The results were somewhat
disappointing: the iron looses its magnetic moment and the overall
system is semiconducting. A subsequent investigation of the 3,3
single wall carbon nanotube with a linear cobalt wire inside
resulted in a HMF band structure with the band gap for the
minority spin direction of order 1 eV. This band structure is very
much like that of the iron system. The metallic properties are
caused by the extra electron of the cobalt system, that is
completely absorbed by the majority spin band structure.

Another series of materials investigated is inspired by the
molecule ferrocene. This is a so-called sandwich complex with an
Fe ion between two cyclopentadienyl anions. Ferrocene can be
considered to be the first member of a series of so-called
multiple decker sandwhich structures. They are formed by adding
additional pairs of iron atoms and cyclopentadienyl molecules.
Thus the chemical structure is fundamentally different from the
nanotubes discussed above: The latter can be thought of as two
interacting wires in parallel, one of organic and one of metal
nature. The former is characterized by a parallel stacking of
cyclopentadienyl (or benzene) rings coupled together by
transition-metal atoms. The syntheses of these systems
was shown to be possible for various vanadium benzene clusters \cite%
{Hoshino:3053}. The most promising candidate at the moment is the
one-dimensional manganese-benzene polymer \cite{xiang:243113}. It
is a genuine HMF with a moment of 1 $\mu _{B}$. The ferromagnetic
ordering is much more stable than the antiferromagnetic one.
This large difference (0.25 eV) can be traced back to the
coexistence of a rather narrow and a rather dispersive band for
the metallic spin direction, a scenario very reminiscent of the
double-exchange model.

\section{Model theoretical approaches}

\subsection{Electron spectrum and strong itinerant ferromagnetism in the
Hubbard model}

\label{sec:Hubbard}

To investigate the spectrum of single-particle and spin-wave
excitations in metallic magnets we use many-electron models which
permit to describe effects of inter-electron correlations. The
simplest model of such a type is the Hubbard model. In the case of
a non-degenerate band its Hamiltonian reads

\begin{eqnarray}
\mathcal{H} &=&\sum_{\mathbf{k}\sigma
}t_{\mathbf{k}}c_{\mathbf{k}\sigma
}^{\dagger }c_{\mathbf{k}\sigma }+\mathcal{H}_{\mathrm{int}},  \nonumber \\
\mathcal{H}_{\mathrm{int}} &=&U\sum_{i}c_{i\uparrow }^{\dagger
}c_{i\uparrow }c_{i\downarrow }^{\dagger }c_{i\downarrow }
\label{Hub}
\end{eqnarray}
with $U$ being the on-site repulsion parameter, $t_{\mathbf{k}}$
the bare electron spectrum. The Hubbard model was widely used to
consider itinerant electron ferromagnetism since this takes into
account the largest term of the Coulomb interaction -- the
intra-atomic one. Despite apparent simplicity, this model contains
a very reach physics, and its rigorous investigation is a very
difficult problem.

The simplest Hartree-Fock (Stoner) approximation in the Hubbard model (\ref%
{Hub}), which corresponds formally to first-order perturbation theory in $U$%
, yields the electron spectrum of the form

\begin{equation}
E_{\mathbf{k}\sigma }=t_{\mathbf{k}}+Un_{-\sigma}=
t_{\mathbf{k}}+U(\frac n2-\sigma \langle S^z\rangle )\equiv
t_{\mathbf{k}\sigma }
\end{equation}
so that we have for the spin splitting $\Delta
=U(n_{\uparrow}-n_{\downarrow })=2U\langle S^{z}\rangle$ and $U$
plays the role of the Stoner parameter.

Consider more strictly the case of a half-metallic (saturated)
ferromagnet where $n_{+}=n=1-n_{0},$ $n_{-}=0$ (note that for
realistic HMF the saturated ferromagnetic behavior is described by
the generalized Slater-Pauling rule, see Sect.\ref{magn1}). Then
the spin up electrons behave at $T$ =0~K as free ones,
\begin{equation}
G_{\mathbf{k}\uparrow }(E)=(E-t_{\mathbf{k}})^{-1}
\end{equation}
For spin-down states the situation is non-trivial. Writing down
the sequence of equations of motion for $G_{\mathbf{k}\downarrow
}(E)$ and for the Green's function
\begin{equation}
\Phi _{\mathbf{kp}}^{{}}(E)=\langle \langle S_{\mathbf{p}}^{+}c_{\mathbf{k}-%
\mathbf{p\uparrow }}|c_{\mathbf{k}\downarrow }^{\dagger }\rangle
\rangle
_{E},S_{\mathbf{q}}^{+}=\sum_{\mathbf{k}}c_{\mathbf{k}\uparrow
}^{\dagger }c_{\mathbf{k}+\mathbf{q}\downarrow }
\end{equation}
and performing decoupling in spirit of a ladder approximation we
obtain for the self-energy
\begin{equation}
\Sigma _{\mathbf{k}\downarrow
}=\frac{Un_{+}}{1-UR_{\mathbf{k}}(E)} \label{sig}
\end{equation}
where $n_{\mathbf{k}}=f(t_{\mathbf{k}})$ is the Fermi function,
\begin{equation}
R_{\mathbf{k}}(E)=\sum_{\mathbf{q}}\frac{1-n_{\mathbf{k-q}}}{E-t_{\mathbf{k-q%
}}-\omega _{\mathbf{q}}}  \label{Edwards:L327}
\end{equation}
describes the electron-magnon scattering. This result corresponds
to the Edwards-Hertz approximation \cite{edwards:2191}.

In a more general non-saturated situation one obtains for the
self-energy to second order in $U$ \cite{irkhin:7151}
\begin{equation}
\Sigma _{\mathbf{k}\sigma }(E)=U^{2}\sum_{\mathbf{q}}\int_{-\infty
}^{\infty
}\frac{d\omega }{\pi }\text{\textrm{Im~}}\langle \langle S_{\mathbf{q}%
}^{\sigma }|S_{-\mathbf{q}}^{-\sigma }\rangle \rangle _{\omega }\frac{%
N_{B}(\omega )+n_{\mathbf{k+q,}-\sigma
}}{E-t_{\mathbf{k+q,}-\sigma }+\omega }
\end{equation}%
where $N_{B}(\omega )$ is the Bose function. Retaining only the
magnon pole contribution to the spin spectral density (i.e.
neglecting the spin-wave damping) we have
\begin{eqnarray}
\Sigma _{\mathbf{k\uparrow }}(E) &=&U\Delta \sum_{\mathbf{q}}\frac{N_{%
\mathbf{q}}+n_{\mathbf{k+q\downarrow }}}{E-t_{\mathbf{k+q\downarrow }%
}+\omega _{\mathbf{q}}}  \label{s1} \\
\Sigma _{\mathbf{k\downarrow }}(E) &=&U\Delta \sum_{\mathbf{q}}\frac{1+N_{%
\mathbf{q}}-n_{\mathbf{k-q\uparrow }}}{E-t_{\mathbf{k-q\uparrow }}-\omega _{%
\mathbf{q}}}  \label{s2}
\end{eqnarray}%
where $\omega _{\mathbf{q}}$ is the magnon energy, $N_{\mathbf{q}%
}=N_{B}(\omega _{\mathbf{q}})$. These results are valid in the
$\mathit{s-d}$
model ($U\rightarrow I$, see below) to first order in the small parameter $%
1/2S$. Taking into account the relation
\begin{equation}
\langle S^{z}\rangle =S_{0}-\sum_{\mathbf{p}}N_{\mathbf{p}}
\end{equation}%
where $S_{0}$ is the saturation magnetization one obtains for the
spin-wave
correction to the electron energy 
\begin{equation}
\delta E_{\mathbf{k}\sigma }(T)=\sum_{\mathbf{q}}A_{\mathbf{kq}}^{\sigma }N_{%
\mathbf{q}}=\frac{v_{0}}{2\langle S^{z}\rangle }\frac{\xi (5/2)}{32\pi ^{3/2}%
}\left( \frac{T}{D}\right) ^{5/2}\left[ \frac{\partial ^{2}t_{\mathbf{k}}}{%
\partial k_{x}^{2}}-\frac{\sigma }{U\langle S^{z}\rangle }\left( \frac{%
\partial t_{\mathbf{k}}}{\partial k}\right) ^{2}\right]
\end{equation}%
%
%
%
where
\begin{equation}
\mathcal{A}_{\mathbf{kq}}^{\sigma }=\sigma U\frac{t_{\mathbf{k+q}}-t_{%
\mathbf{k}}}{t_{\mathbf{k+q}}-t_{\mathbf{k}}+\sigma \Delta }
\label{amp}
\end{equation}%
The $T^{5/2}$-dependence of the electron spectrum owing to magnons
is weaker than the $T^{3/2}$-dependence of the magnetization. This
fact is due to vanishing of electron-magnon interaction amplitude
$\mathcal{A}$ at zero magnon wavevector, which is connected with
the symmetry of exchange interaction. Such a weakening of
temperature dependence of the spin splitting was observed in iron
\cite{Lonzarich:225}. The one-electron
damping in the half-metallic situation was calculated in Ref.\cite%
{Auslender:301}, a Fermi-liquid-type behavior (small damping near
the Fermi level containing high powers of temperature) being
obtained.

The problem of ferromagnetic ordering in narrow energy bands is up
to now extensively discussed. To stabilize the ferromagnetic
solution within the Hubbard model is yet another difficult
problem. It was proved recently, that the necessary conditions for
ferromagnetism is a density of state with large spectral weight
near the band edges \cite{ulmke:301} and the Hund's rule coupling
for the degenerate case \cite{Vollhardt:383}. Real examples of
saturated ferromagnetic ordering are provided by pyrite structure systems Fe$%
_{1-x}$Co$_{x}$S$_{2}$ with itinerant-electron ferromagnetism in
double-degenerate narrow $e_{g}$ band \cite%
{Jarrett:617,Auslender:5521,ramesha:214409}. CMR manganites, magnetite Fe$%
_{3}$O$_{4}$ above the Verwey transition temperature,
\textquotedblleft anionic\textquotedblright\ half-metallic
ferromagnets are another examples (see Section
\ref{sec:elstr_hmf}). Recently, a model of $sp$~electron magnetism
in narrow impurity bands has been proposed \cite{Edwards:7209}
which may be applicable to some carbon- or boron-based systems
such as doped CaB$_{6}$. In this model, the magnon excitations
turn out to have higher energy than the Stoner ones. Also,
$T$-matrix renormalization of the Stoner exchange parameter which
decreases its value essentially in a typical itinerant-electron
magnets is much less relevant. For these reasons, the narrow-band
$sp$~systems can provide an example of real \textquotedblleft
Stoner\textquotedblright\ magnets which can have rather high Curie
temperatures at small enough magnetization value
\cite{Edwards:7209}. According to that model, these ferromagnets
also should be saturated.

Systems with strong inter-electron correlations are the most
difficult case for standard approaches in the itinerant electron
magnetism theory (band calculations, spin-fluctuation theories).
Physically, the magnetism picture in this case differs
substantially from the Stoner picture of a weak itinerant
magnetism~\cite{Moriya:1985} since correlations lead to a radical
reconstruction of the electron spectrum~--- formation of the
Hubbard's subbands \cite{Hubbard:238} which are intimately
connected with the local magnetic moments \cite{Auslender:5521}.

In the limit $U\rightarrow \infty $, considering the case where
the number of electron $n=1-\delta <1$ ($\delta $ is the hole
concentration), the Hubbard Hamiltonian reads
\begin{equation}
\mathcal{H}=\sum_{\mathbf{k}\sigma }\varepsilon _{\mathbf{k}}X_{-\mathbf{k}%
}^{0\sigma }X_{\mathbf{k}}^{\sigma 0},  \label{HHM}
\end{equation}
where $X_{\mathbf{k}}^{\alpha \beta }$ is the Fourier transform of
the Hubbard operators $X_{i}^{\alpha \beta }=|i\alpha \rangle
\langle i\beta |$,
$\varepsilon _{\mathbf{k}}=-t_{\mathbf{k}}$. According to Nagaoka~\cite%
{Nagaoka:392}, the ground state for simple lattices is a saturated
ferromagnetic state for a low density $\delta $ of current
carriers (\textquotedblleft doubles\textquotedblright\ or
\textquotedblleft
holes\textquotedblright\ in an almost half-filled band). Roth~\cite%
{Roth:451,Roth:428} applied a variational principle to this
problem and obtained two critical concentrations. The first one,
$\delta _{\text{c}}$, corresponds to instability of saturated
ferromagnetic state, and the second one, $\delta
_{\text{c}}^{\prime }$, to the transition from non-saturated
ferromagnetic into paramagnetic state. For the simple cubic (sc)
lattice, the values $\delta _{\text{c}}=0.37$ and $\delta
_{\text{c}}^{\prime }=0.64$ were obtained. Next, the stability of
ferromagnetism was investigated within various approximations and
methods. Most calculations for a number of lattices yield the
value of $\delta _{\text{c}}$ which is close to $0.3$. In
particular, the Gutzwiller method~ \cite{Fazekas:69}, $t/U$ expansion~\cite%
{Zhao:2321}, density matrix renormalization group approach and
Quantum
Monte-Carlo (QMC) method~\cite{Liang:173} yielded $\delta _{\text{c}%
}=0.2-0.35$.

At the same time, for the critical concentrations~$\delta _{\text{c}%
}^{\prime }$ the interval of values is broader and varies from~$0.38$ to~$%
0.64$. Irkhin and Zarubin \cite{irkhin:035116,irkhin:246} have
obtained the density-of-states (DOS) pictures in a Hubbard
ferromagnet with account of the \textquotedblleft
Kondo\textquotedblright\ scattering and spin-polaron contributions
and calculated the values of the critical concentrations of
current carriers. This approach yields a rather simple
interpolational description of saturated and non-saturated
ferromagnetism.

The simplest \textquotedblleft Hubbard-I\textquotedblright\
approximation for the electron spectrum \cite{Hubbard:238}
corresponds to the zeroth order in the inverse nearest-neighbor
number $1/z$ (\textquotedblleft mean-field\textquotedblright\
approximation in the electron hopping). This approximation is
quite non-satisfactory at describing ferromagnetism (in
particular, ferromagnetic solutions are absent, except for
peculiar models of bare density of states). Therefore, to treat
the problems connected with the ferromagnetism formation in narrow
bands the one-particle Green's functions were calculated to first
order in~$1/z$ and in the corresponding self-consistent
approximations.

The retarded anticommutator Green's functions $G_{\mathbf{k}\sigma
}(E)=\langle \!\langle X_{\mathbf{k}}^{\sigma
0}|X_{-\mathbf{k}}^{0\sigma }\rangle \!\rangle _{E}$ can be
calculated by using the equation-of motion-approach of
Refs.~\cite{Irkhin:41,irkhin:035116,irkhin:246} with
account of spin fluctuations. In the locator representation one obtains \cite%
{irkhin:035116}
\begin{equation}
G_{\mathbf{k}\sigma }(E)=[F_{\mathbf{k}\sigma }(E)-\varepsilon _{\mathbf{k}%
}]^{-1},\ F_{\mathbf{k}\sigma }(E)=\frac{b_{\mathbf{k}\sigma }(E)}{a_{%
\mathbf{k}\sigma }(E)},  \label{eq:GF:1}
\end{equation}%
%
%
%
with
\begin{eqnarray}
a_{\mathbf{k}\sigma }(E) &=&n_{0}+n_{\sigma }  \nonumber \\
&+&\sum_{\mathbf{q}}\varepsilon _{\mathbf{k-q}}\frac{\chi _{\mathbf{q}%
}^{\sigma -\sigma }+n_{\mathbf{k-q}-\sigma }}{n_{0}+n_{-\sigma }}G_{\mathbf{%
k-q}-\sigma }^{0}(E-\sigma \omega _{\mathbf{q}})+\sum_{\mathbf{q}%
}\varepsilon _{\mathbf{k-q}}\frac{\chi _{\mathbf{q}}^{-\sigma -\sigma }}{%
n_{0}+n_{\sigma }}G_{\mathbf{k-q}\sigma }^{0}(E)  \nonumber \\
b_{\mathbf{k}\sigma }(E) &=&E+\sum_{\mathbf{q}}\varepsilon _{\mathbf{k-q}%
}^{2}\frac{n_{\mathbf{k-q}-\sigma }}{n_{0}+n_{-\sigma }}G_{\mathbf{k-q}%
-\sigma }^{0}(E-\sigma \omega _{\mathbf{q}}),  \label{eq:GF:3}
\end{eqnarray}%
where $\chi _{\mathbf{q}}^{\sigma -\sigma }=\langle
S_{\mathbf{q}}^{\sigma }S_{-\mathbf{q}}^{-\sigma }\rangle =\langle
X_{\mathbf{q}}^{\sigma -\sigma }X_{\mathbf{-q}}^{-\sigma \sigma
}\rangle $ and $\chi _{\mathbf{q}}^{-\sigma
-\sigma }=\langle \delta X_{\mathbf{q}}^{-\sigma -\sigma }\delta X_{-\mathbf{%
q}}^{-\sigma -\sigma }\rangle $ are the correlation functions for
spin and charge densities, $n_{\mathbf{k}\sigma }=\langle
X_{\mathbf{-k}}^{0\sigma }X_{\mathbf{k}}^{\sigma 0}\rangle $. To
simplify numerical calculations, the
long-wavelength dispersion law $\omega _{\mathbf{q}}=\mathcal{D}q^{2}$ ($%
\mathcal{D}$ is the spin-wave stiffness constant) was used with
the magnon spectral function $K_{\mathbf{q}}(\omega )$ being
average this in~$\mathbf{q} $. Following to Ref.~
\cite{Nagaoka:392} the value $\mathcal{D}=0.66\delta |t|$ was
taken for the cubic lattice and the same $\overline{K}(\omega )$
adopted for other lattices (the choice of $\mathcal{D}$ influences
weakly
the critical concentration). Then $a(E)$ and $b(E)$ do not depend on~$%
\mathbf{k}$ and can be expressed in terms of the bare electron
density of
states. In the case of the saturated ferromagnetism the expressions (\ref%
{eq:GF:1}) reduce approximately to the result (\ref{Edwards:L327}) for $%
U\rightarrow \infty $,
\begin{equation}
\Sigma _{\mathbf{k}\downarrow }(E)=-(1-n_{0})\left( \sum_{\mathbf{q}}\frac{%
n_{\mathbf{k-q}}}{E-\varepsilon _{\mathbf{k-q}}+\omega
_{\mathbf{q}}}\right) ^{-1}.
\end{equation}

To write down the self-consistent approximation one has to replace in (\ref%
{eq:GF:3})  $G_{\mathbf{k}\sigma }^{0}(E)\rightarrow
G_{\mathbf{k}\sigma }(E) $ and calculate $n_{\mathbf{k}\sigma }$
via spectral representation. In such an approach, large electron
damping is present which smears the \textquotedblleft
Kondo\textquotedblright\ peak.

The $1/z$-corrections lead to a non-trivial structure of the
one-particle density of states. In the non-self-consistent
approach the integral with the Fermi functions yields, similar to
the Kondo problem, the logarithmic singularity~\cite{Irkhin:463}.
For very low $\delta $ a significant logarithmic singularity
exists only in the imaginary part of the Green's
function, which corresponds to a finite jump in the density of states~\cite%
{Irkhin:522}. However, when $\delta $ increases, it is necessary
to take into account the resolvents in both the numerator and
denominator of the Green's function, so that the real and
imaginary parts are \textquotedblleft mixed\textquotedblright\ and
a logarithmic singularity does appear in DOS. The magnon
frequencies in the denominators of Eqs.~(\ref{eq:GF:3}) result in
that the singularity is spread out over the interval~$\omega
_{\text{max}}$ and the peak becomes smoothed. In the
self-consistent approximation the form of~$N_{\downarrow }(E)$
approaches the bare density of states and the peak is smeared,
even neglecting spin dynamics.

\begin{figure}[htb]
\begin{center}
\includegraphics[width=0.75\columnwidth, angle=-90]{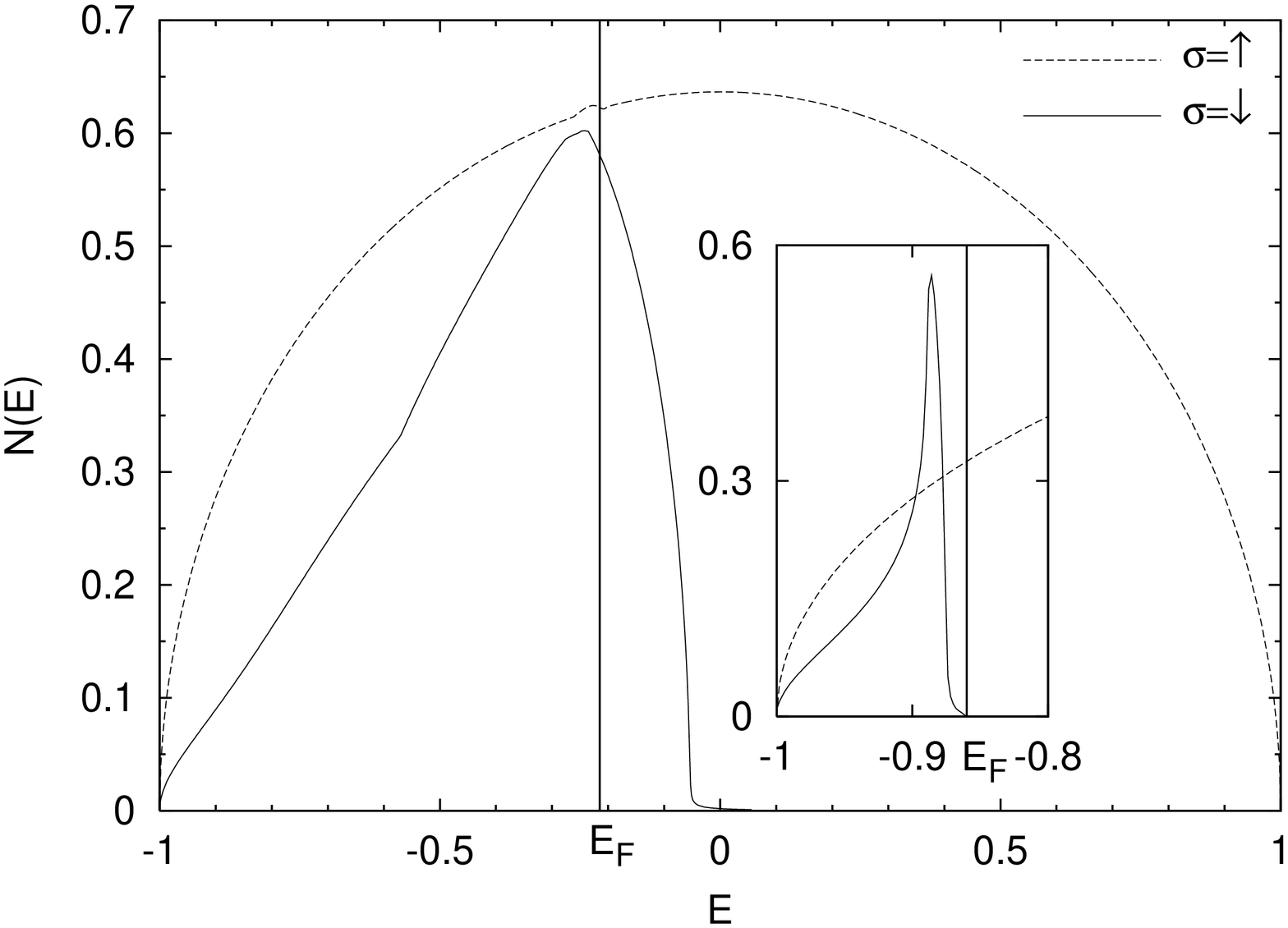}
\end{center}
\caption{Density of states for the semielliptic DOS at
concentration of
carriers current~$\protect\delta =0.35$ ($\protect\delta_{\text{c}}<\protect%
\delta <\protect\delta _{\text{c}}^{\prime }$) and $\protect\delta
=0.02< \protect\delta _{\text{c}}$ (inset) in the self-consistent
approximation.} \label{fig:11}
\end{figure}

There are no poles of the Green's function for~$\sigma =\downarrow
$ above the Fermi level at small $\delta $, i.e. the saturated
ferromagnetic state is preserved. Unlike most other analytical
approaches, the results of Ref. \cite{irkhin:035116,irkhin:246}
for the one-particle Green's function describe formation of
non-saturated ferromagnetism too, the account of longitudinal spin
fluctuations $\chi _{\mathbf{q}}^{-\sigma -\sigma }$ being
decisive for obtaining the non-saturated solution and calculating
the second critical concentration $\delta _{\text{c}}^{\prime }$
where the magnetization vanishes. For $\delta >\delta _{\text{c}}$
this dependence deviates from the linear one, $\langle
S^{z}\rangle =(1-\delta )/2$. The
calculations of Ref.\cite{irkhin:035116,irkhin:246} yield the $\delta _{%
\text{c}}^{\prime }$ values which are considerably smaller than
the results of the spin-wave
approximation~\cite{Roth:451,Roth:428}. In the non-saturated state
a spin-polaron pole occurs, so that quasiparticle states with
$\sigma =\downarrow $ occur above the Fermi level
(Fig.~\ref{fig:11}).

\begin{figure}[htb]
\begin{center}
\includegraphics[width=0.75\columnwidth, angle=-90]{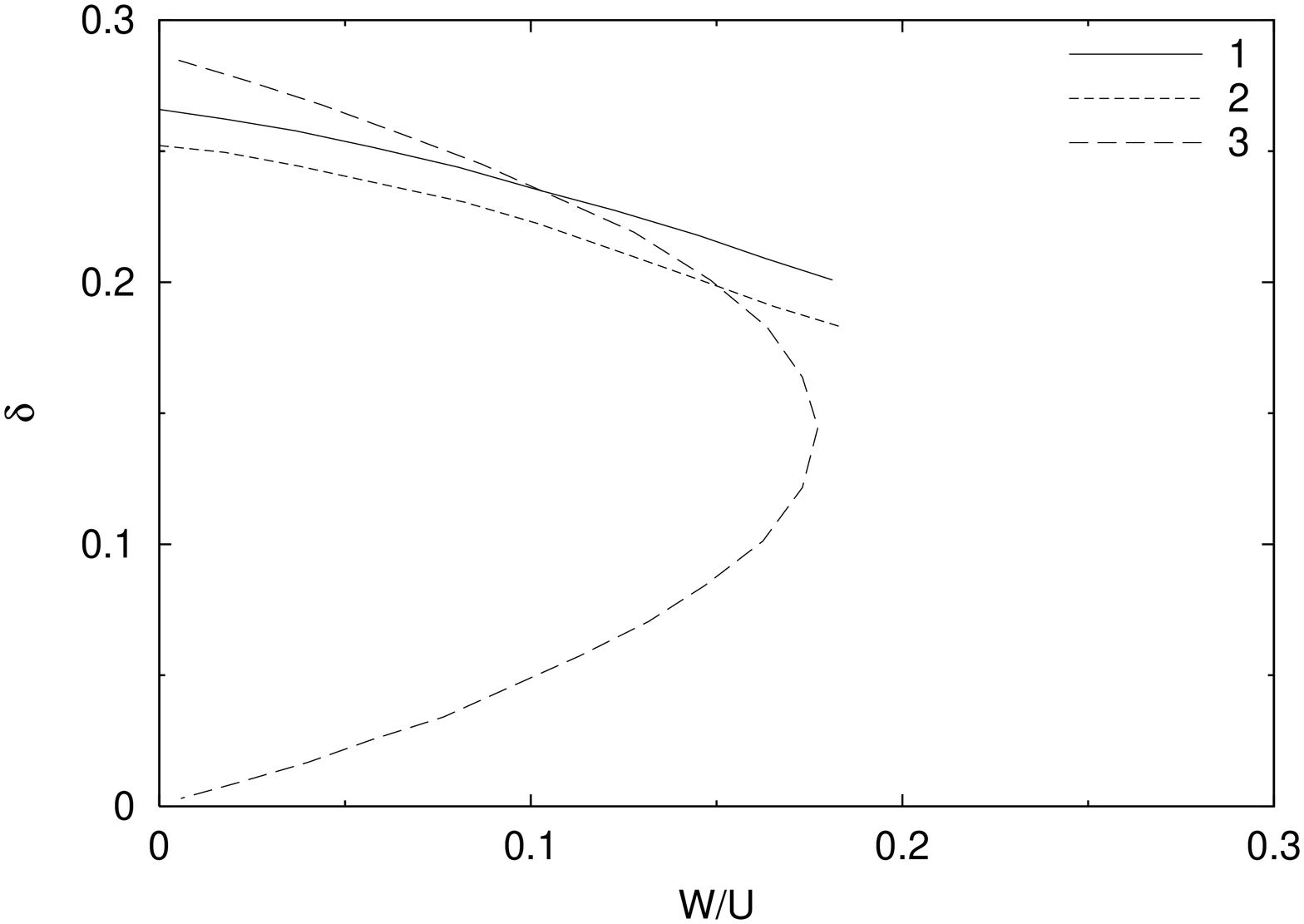}
\end{center}
\caption{The boundary of the saturated ferromagnetic region in the
approximation (\protect\ref{sig}) for the semielliptic band (solid
line) and square lattice (short-dashed line), $W$ being bandwidth.
The results of Ref.~ \protect\cite{Linden:4917} for the square
lattice are shown by long-dashed line.} \label{fig:22}
\end{figure}

The finite-$U$ case can be also treated within the Green's
function methods discussed. The Edwards-Hertz approximation
(\ref{sig}) enables one to investigate stability of the saturated
ferromagnetic state only, i.e. calculate $\delta _{\text{c}}$. The
corresponding results are presented in
Fig.~\ref{fig:22}. For comparison, the variational results of Refs.\cite%
{Linden:4917} are shown which yield a strict upper boundary for
the saturated state. An agreement takes place for large $U$ (far
from antiferromagnetic or phase-separation instability which are
not taken into account in the calculations). It should be noted
that DMFT yields qualitatively similar
results~\cite{Obermeier:8479}. One can see that saturated
ferromagnetism can occur for large $U/|t|,$ and its existence at
realistic $U$ is, generally speaking, a not too simple problem.

Now we treat the orbital-degenerate case which is more realistic
for transition-metal compounds. Consider the many-electron system
with two
ground terms of the $d^{n}$ and $d^{n+1}$ configurations, $\Gamma _{n}=\{SL{%
\}}$ and $\Gamma _{n+1}=\{S^{\prime }L^{\prime }{\}}$. It is
suitable to use the representation of the Fermi operators in terms
of the many-body atomic quantum numbers
\cite{Irkhin:9,Irkhin:2007}
\begin{equation}
c_{il\sigma m}^{\dagger }=(n+1)^{1/2}\sum G_{SL}^{S^{\prime
}L^{\prime }}C_{L\mu ,lm}^{L^{\prime }\mu ^{\prime
}}C_{SM,\frac{1}{2}\sigma }^{S^{\prime }M^{\prime }}X_{i}\left(
S^{\prime }L^{\prime }M^{\prime }\mu ^{\prime },SLM\mu \right)
\label{pro}
\end{equation}
where $G_{SL}^{S^{\prime }L^{\prime }}$ are the fractional
parentage coefficients. We can introduce a further simplification
assuming that only one of the competing configurations has
non-zero orbital moment $L=l$. This assumption holds for the
$d^{5}$ and $d^{6}$ ground-state configurations of Fe$^{3+}$ and
Fe$^{2+}$ respectively, the first configuration having zero
orbital moment. A similar situation takes place for the CMR
manganites (with $d^{3}$ and $d^{4}$ configuration for Mn$^{4+}$
and Mn$^{3+}$: due to the relevance of $t_{2g}-e_{g}$
crystal-field splitting the former configuration corresponds to
the completely filled $t_{2g}$ band with $L=0$.

We treat the narrow-band case which should be described by a
two-configuration Hubbard model where both conduction electrons
and local moments belong to the same $d$-band, the states with
$n+1$ electrons playing the role of current-carrier states. After
performing the procedure of mapping onto the corresponding state
space, the one-electron Fermi operators for the strongly
correlated states $c_{il\sigma m}^{\dagger }$ are replaced by
many-electron operators according to Eq.(\ref{pro}). Taking into
account the values of the Clebsh-Gordan coefficients which
correspond to the coupling of momenta $S$ and 1/2 we obtain the
\textquotedblleft double-exchange\textquotedblright\ Hamiltonian
\begin{equation}
\mathcal{H}=\sum_{\mathbf{k}\sigma
m}t_{\mathbf{k}m}g_{\mathbf{k}\sigma m}^{\dagger
}g_{\mathbf{k}\sigma m}.  \label{eq:H}
\end{equation}
Here we have redefined the band energy by including the
many-electron renormalization factor,
$t_{\mathbf{k}m}(n+1)(G_{SL}^{S^{\prime }0})^{2}/(2l+1)\rightarrow
t_{\mathbf{k}m}$, and
\begin{eqnarray}
g_{i\sigma m}^{\dagger } &=&\sum_{M=-S}^{S}\sqrt{\frac{S-\sigma M}{2S+1}}%
X_{i}(S-1/2,M+\frac{\sigma }{2};SMm),\;\;\;S^{\prime }=S-1/2,  \nonumber \\
g_{i\sigma m}^{\dagger } &=&\sum_{M=-S}^{S}\sqrt{\frac{S+\sigma M+1}{2S+1}}%
X_{i}(S+1/2,M+\frac{\sigma }{2};SMm),\;\;\;S^{\prime }=S+1/2
\end{eqnarray}
where $|SMm\rangle $ are the empty states with the orbital index $m,$ $%
|S^{\prime }M^{\prime }\rangle $ are the singly-occupied states
with the
total on-site spin $S^{\prime }=S\pm 1/2$ and its projection $M^{\prime }$, $%
\sigma =\pm $. We see that the two-configuration Hamiltonian is a
generalization of the narrow-band $s-d$ exchange model with
$|I|\rightarrow \infty $ (double-exchange model)
\cite{Irkhin:9,Irkhin:2007,Irkhin:522}: in the case
where the configuration $d^{n+1}$ has larger spin than the configuration $%
d^{n}$, we have the effective \textquotedblleft $s-d$ exchange
model\textquotedblright\ with ferromagnetic coupling, and in the
opposite case with antiferromagnetic coupling. In the absence of
orbital degeneracy the model (\ref{eq:H}) is reduced to the
narrow-band Hubbard model.

For $S=1/2$ the narrow-band $s-d$ exchange model with
$|I|\rightarrow
-\infty $ is equivalent to the Hubbard model with the replacement $t_{%
\mathbf{k}}\rightarrow t_{\mathbf{k}}/2$, so that the
ferromagnetism picture corresponds to that described above. In a
general case the criteria for spin and orbital instabilities
\cite{irkhin:054421} are different. It turns out that the
saturated spin ferromagnetism is more stable than the orbital one
in the realistic case $S>1/2$ (e.g., for magnetite and for
colossal magnetoresistance manganites). This means that the
half-metallic ferromagnetic phases both with saturated and
non-saturated orbital moments
can arise. The phase diagram at finite temperatures is discussed in Ref.\cite%
{Edwards:1259}.

In contrast with usual itinerant-electron ferromagnets, additional
collective excitation branches (orbitons) occur in the model.
Also, mixed excitations with the simultaneous change of spin and
orbital projections exist (\textquotedblleft optical
magnons\textquotedblright ). All these excitations can be well
defined in the whole Brillouin zone, the damping due
to the interaction with current carriers being small enough \cite%
{irkhin:054421}.

The XMCD data \cite{huang:077204} suggest large orbital
contributions to magnetism in Fe$_{3}$O$_{4}$. However, more
recent experimental XMCD data \cite{Goering:97} yield very small
orbital moments in Fe$_{3}$O$_{4}$ and confirm HMF behavior of
magnetite. In any case, the model of orbital itinerant
ferromagnetism \cite{irkhin:054421} is of a general physical
interest and can be applied, e.g., to CMR manganites.

\subsection{Electron spectrum in the \textit{s-d} exchange model: The
non-quasiparticle density of states}
\label{nqp}

Besides the Hubbard model, it is often convenient to use for
theoretical
description of magnetic metals the $\mathit{s-d(f)}$ exchange model. The $%
\mathit{s-d}$ exchange model was first proposed for transition
\textit{d} metals to consider peculiarities of their electrical
resistivity (Vonsovsky 1971)\cite{Vonsovsky:1974}. This model
postulates the existence of two electron subsystems: itinerant
\textit{s} electrons which play the role of current carriers, and
localized \textit{d }electrons which give the main contribution to
magnetic moments. Such an assumption can be hardly justified
quantitatively for $d$ metals, but is useful at qualitative
consideration of some physical properties, especially of transport
phenomena. At the same time, the $\mathit{s-d}$ model provides a
good description of magnetism in rare-earth metals and their
compounds with well-localized \textit{4f}
states. Now this model is widely used in the theory of anomalous \textit{f }%
systems (intermediate valence compounds, heavy fermions...) as the
\textquotedblleft Kondo-lattice\textquotedblright\ model
\cite{Hewson:1993}.

The Hamiltonian of the $s-d$ exchange model in the case of
arbitrary inhomogeneous potential reads
\begin{eqnarray}
\mathcal{H} &=&\int d\mathbf{r}\left( \sum_{\sigma }\Psi _{\sigma
}^{\dagger
}(\mathbf{r})\mathcal{H}_{0}^{\sigma }\Psi _{\sigma }(\mathbf{r}%
)-I\sum_{\sigma \sigma ^{\prime }}\delta \mathbf{S(r)}\Psi
_{\sigma }^{\dagger }(\mathbf{r})\mbox {\boldmath
$\sigma$}_{\sigma \sigma ^{\prime
}}\Psi _{\sigma ^{\prime }}(\mathbf{r})\right) +\mathcal{H}_{d}  \nonumber \\
\mathcal{H}_{0}^{\sigma } &=&-\frac{\hbar ^{2}}{2m}\nabla ^{2}+V_{\sigma }(%
\mathbf{r})  \label{hamilt}
\end{eqnarray}
where $V_{\sigma }(\mathbf{r})$ is the potential energy (with
account of the electron-electron interaction in the mean field
approximation) which is supposed to be spin dependent, $\Psi
_{\sigma }(\mathbf{r})$ is the field operator for the spin
projection $\sigma ,$ $\mathbf{S(r)}$ is the spin
density of the localized-moment system, $\delta \mathbf{S(r)=S(r)-}%
\left\langle \mathbf{S(r)}\right\rangle $ is its fluctuating part,
the
effect of the average spin polarization\ $\left\langle \mathbf{S(r)}%
\right\rangle $ being included into $V_{\sigma }(\mathbf{r})$. We
use an approximation of contact electron-magnon interaction
described by the $s-d$ exchange parameter $I$,
\begin{equation}
\mathcal{H}_{d}=-\sum_{\mathbf{q}}J_{\mathbf{q}}\mathbf{S}_{\mathbf{q}}%
\mathbf{S}_{-\mathbf{q}}
\end{equation}
(for simplicity we neglect the inhomogeneity effects for the
magnon
subsystem), $\mathbf{S}_{\mathbf{q}}$ are operators for localized spins, $J_{%
\mathbf{q}}$ are the Fourier transforms of the exchange parameters
between localized spins. In rare earth metals the latter
interaction is usually the indirect RKKY exchange via conduction
electrons which is due to the same \textit{s-d} interaction.
However, at constructing perturbation theory, it is convenient to
include this interaction in the zero-order Hamiltonian.

Although being more complicated in its form, the $\mathit{s-d}$
model turns out to be in some respect simpler than the Hubbard
model (\ref{Hub}) since
it permits to construct the quasiclassical expansion in the small parameter $%
1/2S$. Within simple approximations, the results in the
$\mathit{s-d(f)}$ and Hubbard models differ as a rule by the
replacement $I\rightarrow U$ only. To describe effects of
electron-magnon interaction we use the formalism of the exact
eigenfunctions \cite{Irkhin:3055,irkhin:104429}. Passing to the
representation of the exact eigenfunctions of the Hamiltonian
$\mathcal{H}_{0}^{\sigma },$%
\begin{eqnarray}
\mathcal{H}_{0}^{\sigma }\psi _{v\sigma } &=&\varepsilon _{\nu
\sigma }\psi
_{\nu \sigma },  \nonumber \\
\Psi _{\sigma }(\mathbf{r}) &=&\sum\limits_{\nu }\psi _{\nu \sigma
}\left( \mathbf{r}\right) c_{\nu \sigma },
\end{eqnarray}
one can rewrite the Hamiltonian (\ref{hamilt}) in the following
form:
\begin{equation}
\mathcal{H}=\sum_{\nu \sigma }\varepsilon _{\nu \sigma }c_{\nu
\sigma
}^{\dagger }c_{\mathbf{\nu }\sigma }-I\sum_{\mu \nu \alpha \beta \mathbf{q}%
}\left( \nu \alpha ,\mu \beta |\mathbf{q}\right) \delta
\mathbf{S_{q}}c_{\nu
\alpha }^{\dagger }\mbox {\boldmath $\sigma $}_{\alpha \beta }c_{\mu \beta }+%
\mathcal{H}_{d}  \label{eigen}
\end{equation}
where
\begin{equation}
\left( \nu \sigma ,\mu \sigma ^{\prime }|\mathbf{q}\right)
=\left\langle \mu \sigma ^{\prime }\right\vert
e^{i\mathbf{qr}}\left\vert \nu \sigma \right\rangle .
\end{equation}
We take into account again the electron-spectrum spin splitting in
the mean-field approximation by keeping the dependence of the
eigenfunctions on the spin projection.

In the spin-wave region one can use for the spin operators the
magnon (e.g., Dyson-Maleev) representation. Then we have for the
one-electron Green's function
\begin{equation}
G_{\nu }^{\sigma }(E)=\left[ E-\varepsilon _{\mathbf{\nu }\sigma }-\Sigma _{%
\mathbf{\nu }}^{\sigma }(E)\right] ^{-1},  \label{dys}
\end{equation}
with the self-energy $\Sigma _{\mathbf{\nu }}^{\sigma }(E)$
describing correlation effects.

We start with the perturbation expansion in the electron-magnon
interaction. To second order in $I$ one has
\begin{equation}
\Sigma _{\mathbf{\nu }}^{\uparrow }(E)=2I^{2}S\sum_{\mu
\mathbf{q}}|\left(
\nu \uparrow ,\mu \downarrow |\mathbf{q}\right) |^{2}\frac{N_{\mathbf{q}}+n_{%
\mathbf{\mu }}^{\downarrow }}{E-\varepsilon _{\mathbf{\mu \downarrow }%
}+\omega _{\mathbf{q}}},\Sigma _{\mathbf{\nu }}^{\downarrow
}(E)=2I^{2}S\sum_{\mu \mathbf{q}}|\left( \nu \downarrow ,\mu \uparrow |%
\mathbf{q}\right) |^{2}\frac{1+N_{\mathbf{q}}-n_{\mathbf{\mu }}^{\uparrow }}{%
E-\varepsilon _{\mu \mathbf{\uparrow }}-\omega _{\mathbf{q}}}
\label{q}
\end{equation}
where $n_{\mathbf{\mu }}^{\sigma }=f(\varepsilon _{\mathbf{\mu
}\sigma })$ (discussion of a more general \textquotedblleft
ladder" \ approximation is given below). Using the expansion of
the Dyson equation (\ref{dys}) we obtain for the spectral density
\begin{eqnarray}
\mathcal{A}_{\nu \sigma }\left( E\right) &=&-\frac{1}{\pi }\text{\textrm{Im}}%
G_{\mathbf{\nu }}^{\sigma }(E)=\delta (E-\varepsilon _{\mathbf{\nu
}\sigma })
\nonumber \\
&&\ \ \ \ -\delta ^{\prime }(E-\varepsilon _{\mathbf{\nu }\sigma })\text{%
\textrm{Re}}\Sigma _{\mathbf{\nu }}^{\sigma }(E)-\frac{1}{\pi }\frac{\text{%
\textrm{Im}}\Sigma _{\mathbf{\nu }}^{\sigma }(E)}{(E-\varepsilon _{\mathbf{%
\nu }\sigma })^{2}}  \label{N(E)}
\end{eqnarray}
The second term in the right-hand side of Eq.(\ref{N(E)}) gives
the shift of quasiparticle energies. The third term, which arises
from the branch cut of the self-energy, describes the incoherent
(non-quasiparticle) contribution owing to scattering by magnons.
One can see that this does not vanish in the energy region,
corresponding to the \textquotedblleft alien\textquotedblright\
spin subband with the opposite projection $-\sigma $.

Neglecting temporarily in Eq.(\ref{q}) the magnon energy $\omega _{\mathbf{q}%
}$ in comparison with typical electron energies and using the
identities
\begin{equation}
\sum_{\mu \mathbf{q}}\frac{|\left( \nu \mu |\mathbf{q}\right) |^{2}}{%
E-\varepsilon _{\mathbf{\mu }}}F\left( \varepsilon _{\mathbf{\mu }}\right) =-%
\frac{1}{\pi }\int dE^{\prime }\frac{F\left( E^{\prime }\right) }{%
E-E^{\prime }}\text{\textrm{Im}}\left\langle \nu \right\vert (E^{\prime }-%
\mathcal{H}_{0}+i0)^{-1}\left\vert \nu \right\rangle  \nonumber
\end{equation}
one derives at zero temperature
\begin{eqnarray}
\Sigma _{\nu }^{\uparrow }(E) &=&2I^{2}S\int dE^{\prime }\frac{f(E^{\prime })%
}{E-E^{\prime }}\left\langle \nu \uparrow \right\vert \delta
\left( E-E^{\prime }-\mathcal{H}_{0}^{\downarrow }\right)
\left\vert \nu \uparrow
\right\rangle  \label{sigmaup} \\
\Sigma _{\nu }^{\downarrow }(E) &=&2I^{2}S\int dE\frac{1-f(E^{\prime })}{%
E-E^{\prime }}\left\langle \nu \downarrow \right\vert \delta
\left( E-E^{\prime }-\mathcal{H}_{0}^{\uparrow }\right) \left\vert
\nu \downarrow \right\rangle  \label{sigmadn}
\end{eqnarray}
Using the tight-binding model for the ideal-crystal Hamiltonian
one obtains in the real-space representation
\begin{eqnarray}
\Sigma _{\mathbf{R,R}^{\prime }}^{\uparrow }(E) &=&2I^{2}S\int
dE^{\prime
}f(E^{\prime })\left( -\frac{1}{\pi }\text{\textrm{Im}}G_{\mathbf{R,R}%
}^{\downarrow }(E^{\prime })\right) \delta _{\mathbf{R,R}^{\prime
}}
\label{sig_up} \\
\Sigma _{\mathbf{R,R}^{\prime }}^{\downarrow }(E) &=&2I^{2}S\int dE^{\prime }%
\left[ 1-f(E^{\prime })\right] \left( -\frac{1}{\pi }\text{\textrm{Im}}G_{%
\mathbf{R,R}}^{\uparrow }(E^{\prime })\right) \delta
_{\mathbf{R,R}^{\prime }}  \label{sig_dn}
\end{eqnarray}
where $\mathbf{R,R}^{\prime }$ are lattice site indices, and
therefore
\begin{equation}
\Sigma _{\mathbf{\nu }}^{\sigma }(E)=\sum_{\mathbf{R}}\left\vert
\psi _{\nu
\sigma }\left( \mathbf{R}\right) \right\vert ^{2}\Sigma _{\mathbf{R,R}%
}^{\sigma }(E).  \label{sig11}
\end{equation}
One can generalize the above results to the case of arbitrary
$s-d$ exchange
parameter. Simplifying the sequence of equations of motion (cf. Ref. \cite%
{Irkhin:3055}) we have for the operator Green' function
\begin{equation}
G^{\sigma }\left( E\right) =\left[ E-\mathcal{H}_{0}^{\sigma }+\sigma I(%
\mathcal{H}_{0}^{\sigma }-\mathcal{H}_{0}^{-\sigma
})\frac{1}{1+\sigma IR^{\sigma }(E)}R^{\sigma }(E)\right] ^{-1}
\label{gm1}
\end{equation}
If we consider spin dependence of electron spectrum in the
simplest rigid-splitting approximation $\varepsilon _{\mathbf{\nu
}\sigma }=\varepsilon _{\mathbf{\nu }}-\sigma I\left\langle
S^{z}\right\rangle $ and thus neglect spin-dependence of the
eigenfunctions $\psi _{\nu \sigma }\left( \mathbf{R}\right) $ the
expressions (\ref{sigmaup}),(\ref{sigmadn}) are drastically
simplified. Then the self-energy does not depend on $\nu $:
\begin{eqnarray}
\Sigma ^{\sigma }\left( E\right) &=&\frac{2I^{2}SR^{\sigma
}(E)}{1+\sigma
IR^{\sigma }(E)},  \label{sss} \\
R^{\uparrow }(E) &=&\sum_{\mu }\frac{n_{\mathbf{\mu }}^{\downarrow }}{%
E-\varepsilon _{\mathbf{\mu \downarrow }}},\,R^{\downarrow }(E)=\sum_{\mu }%
\frac{1-n_{\mathbf{\mu }}^{\uparrow }}{E-\varepsilon _{\mu \mathbf{\uparrow }%
}}
\end{eqnarray}
If $\mathcal{H}_{0}^{\sigma }$ is just the crystal Hamiltonian ($\nu =%
\mathbf{k},\varepsilon _{\mathbf{\nu }\sigma }=t_{\mathbf{k}\sigma }$, $t_{%
\mathbf{k}\sigma }$ being the band energy), the expression
(\ref{gm1})
coincides with the result for the Hubbard model after the replacement $%
I\rightarrow U$ (see Sect.\ref{sec:Hubbard}). The imaginary part
of $\Sigma ^{\sigma }\left( E\right) $ determines the NQP states.
Description of such states in the Hubbard model with arbitrary $U$
by the dynamical mean-field theory will be presented in Section
\ref{sec:diffunc}

The expression (\ref{gm1}) can be also represented in the form
\begin{equation}
G^{\sigma }\left( E\right) =\left[ E-\mathcal{H}_{0}^{-\sigma }-(\mathcal{H}%
_{0}^{\sigma }-\mathcal{H}_{0}^{-\sigma })\frac{1}{1+\sigma IR^{\sigma }(E)}%
\right] ^{-1}  \label{gm2}
\end{equation}
The equation (\ref{gm2}) is convenient in the narrow-band case. In
this limit where spin splitting is large in comparison with the
bandwidth of
conduction electrons we have $\mathcal{H}_{0}^{\uparrow }-\mathcal{H}%
_{0}^{\uparrow }=-2IS$ and we obtain for the ``lower'' spin subband with $%
\sigma =-\mathrm{sign}I$
\begin{equation}
G^{\sigma }\left( E\right) =\left[ E-\mathcal{H}_{0}^{-\sigma }+\frac{2S}{%
R^{\sigma }(E)}\right] ^{-1}  \label{nb}
\end{equation}

For a periodic crystal Eq.(\ref{nb}) takes the form
\begin{equation}
G_{\mathbf{k}}^{\sigma }\left( E\right) =\left[ E-t_{\mathbf{k}-\sigma }+%
\frac{2S}{R^{\sigma }(E)}\right] ^{-1}  \label{nb1}
\end{equation}
This expression yields exact result in the limit $I\rightarrow
+\infty $,
\begin{equation}
G_{\mathbf{k}}^{\downarrow }\left( E\right) =\left[ \epsilon -t_{\mathbf{k}}+%
\frac{2S}{R(\epsilon )}\right] ^{-1}, \\
R(\epsilon )=\sum_{\mathbf{k}}\frac{1-f(t_{\mathbf{k}})}{\epsilon -t_{%
\mathbf{k}}}  \label{nb22}
\end{equation}
with $\epsilon =E+IS$. In the limit $I\rightarrow -\infty $
Eq.(\ref{nb1}) gives correctly the spectrum of spin-down
quasiparticles,
\begin{equation}
G_{\mathbf{k}}^{\downarrow }\left( E\right) =\frac{2S}{2S+1}\left[
\epsilon -t_{\mathbf{k}}^{\ast }\right] ^{-1}  \label{t*}
\end{equation}
with $\epsilon =E-I(S+1),t_{\mathbf{k}}^{\ast
}=[2S/(2S+1)]t_{\mathbf{k}}$. However, it does not describe the
NQP states quite correctly, so that more
accurate expressions can be obtained by using the atomic representation \cite%
{irkhin:054421},
\begin{equation}
G_{\mathbf{k}}^{\uparrow }\left( E\right) =\frac{2S}{2S+1}\left[
\epsilon
-t_{\mathbf{k}}^{\ast }+\frac{2S}{R^{\ast }(\epsilon )}\right] ^{-1}, \\
R^{\ast }(\epsilon )=\sum_{\mathbf{k}}\frac{f(t_{\mathbf{k}}^{\ast })}{%
\epsilon -t_{\mathbf{k}}^{\ast }}  \label{nb2}
\end{equation}

The Green's functions (\ref{nb1}), (\ref{nb22}), (\ref{nb2}) have
no poles, at least for small current carrier concentration, and
the whole spectral weight of minority states is provided by the
branch cut (non-quasiparticle states)
\cite{Irkhin:1947,Irkhin:4173,irkhin:7151}. For surface states
this result was obtained in Ref.\cite{Katsnelson:3289} in a
narrow-band Hubbard model. Now we see that this result can be
derived in an arbitrary inhomogeneous case. For a HMF with the gap
in the minority-spin subband NQP states occur above the Fermi
level, and for the gap in the majority-spin subband below the
Fermi level.

In the absence of spin dynamics (i.e., neglecting the magnon
frequencies) the NQP density of states has a jump at the Fermi
level. However, the magnon frequencies can be restored in the
final result, in analogy with the case of ideal crystal, which
leads to a smearing of the jump on the energy scale of a
characteristic magnon energy $\overline{\omega }$. It should be
mentioned once more that we restrict ourselves to the case of the
usual three-dimensional magnon spectrum and do not consider the
influence of surface states on the spin-wave subsystem. The
expressions obtained enable us to investigate the energy
dependence of the spectral density.

An analysis of the electron-spin coupling yields different DOS
pictures for two possible signs of the $s-d$ exchange parameter
$I$. For $I>0$ the spin-down NQP scattering states form a
\textquotedblleft
tail\textquotedblright\ of the upper spin-down band, which starts from $%
E_{F} $ (Fig.1) since the Pauli principle prevents electron
scattering into
occupied states. 

\begin{figure}[tbp]
\includegraphics[clip]{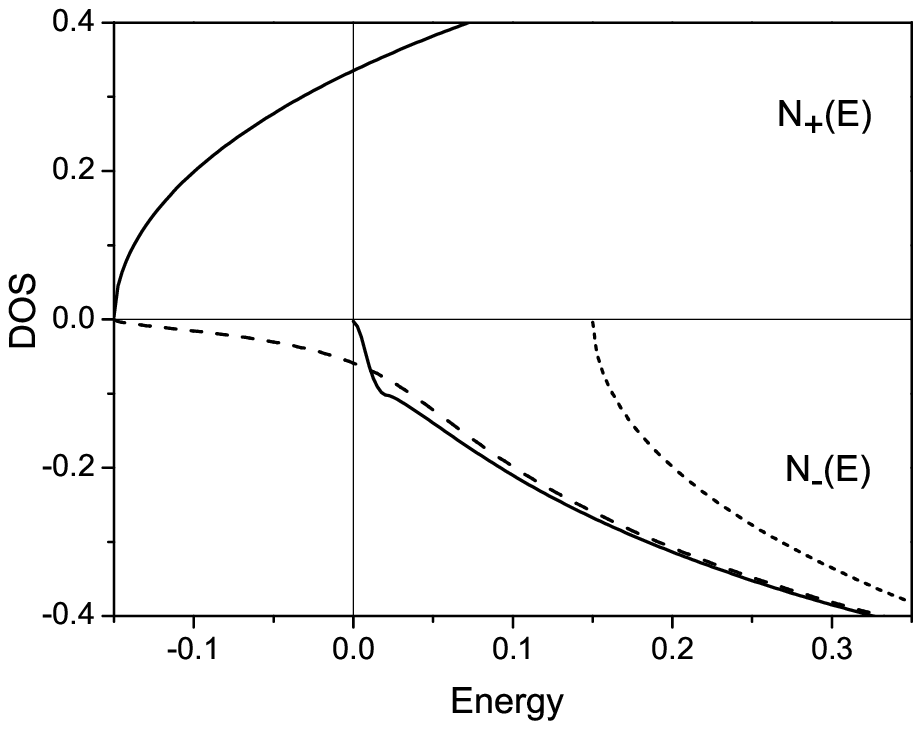}
\caption{ Density of states in the $s-d$ exchange model of a
half-metallic ferromagnet with $S=1/2,I=0.3$ for the semielliptic
bare band with the width of $W=2$. The Fermi energy calculated
from the band bottom is 0.15 (the energy is referred to $E_F$).
The magnon band is also assumed semielliptic with the width of
$\protect\omega_{\mathrm{max}}=0.02$. The non-quasiparticle tail
of the spin-down subband (lower half of the figure) occurs above
the Fermi level. The corresponding picture for the empty
conduction band is shown by dashed line; the short-dashed line
corresponds to the mean-field approximation.} \label{fig:1}
\end{figure}

\begin{figure}[tbp]
\includegraphics[clip]{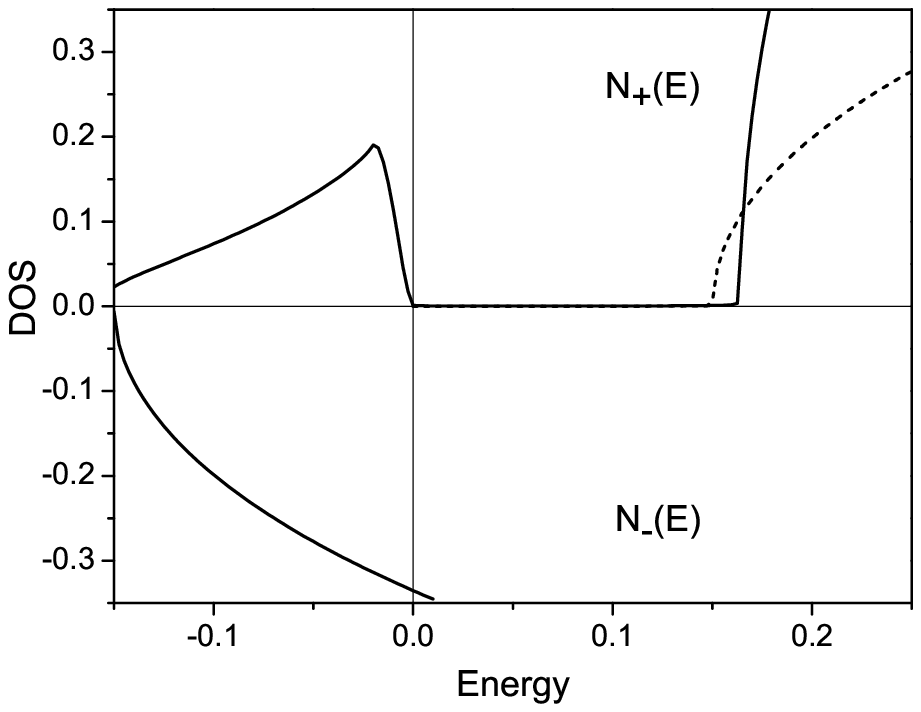}
\caption{Density of states in a half-metallic ferromagnet with
$I=-0.3<0$, other parameters being the same as in Fig.1. The
spin-down subband (lower half of the figure) nearly coincides with
the bare band shifted by $IS$. Non-quasiparticle states in the
spin-up subbands (upper half of the figure) occur below the Fermi
level; the short-dashed line corresponds to the mean-field
approximation.} \label{fig:2}
\end{figure}

For $I<0$ spin-up NQP states are present below the Fermi level as
an isolated region (Fig.2): occupied states with the total spin
$S-1$ are a
superposition of the states $|S\rangle |\downarrow \rangle $ and $%
|S-1\rangle |\uparrow \rangle $. The entanglement of the states of
electron and spin subsystems which is necessary to form the NQP
states is a purely quantum effect formally disappearing at
$S\rightarrow \infty $. To understand better why the NQP states
are formed only below $E_F$ in this case we can treat the limit
$I=-\infty .$ Then the current carrier is really a many-body state
of the occupied site as a whole with total spin $S-1/2,$ which
propagates in the ferromagnetic medium with spin $S$ at any site.
The
fractions of the states $|S\rangle |\downarrow \rangle $ and $%
|S-1\rangle|\uparrow \rangle $ in the current carrier state are
$1/(2S+1)$ and $2S/(2S+1)$, respectively, so that the first number
is just a spectral weight of occupied spin-up electron NQP states.
At the same time, the density of empty states is measured by the
number of electrons with a given spin projection which one can add
to the system. It is obvious that one cannot put any spin-up
electrons in the spin-up site at $I=-\infty .$ Therefore the
density of NQP states should vanish above $E_F$.

It is worthwhile to note that in the most of known HMF the gap
exists for minority-spin states \cite{Irkhin:705}. This is similar
to the case $I>0$, so that the NQP states should arise above the
Fermi energy. For exceptional
cases with the majority-spin gap such as a double perovskite Sr$_{2}$FeMoO$%
_{6}$ \cite{Kobayashi:677} and magnetite one should expect
formation of the NQP states below the Fermi energy.

The presence of space inhomogeneity (e.g., surfaces, interfaces,
impurities) does not change qualitatively the spectral density
picture, except smooth matrix elements. Further in this section we
consider, for simplicity, the case of clean infinite crystal; all
the temperature and energy dependences of the spectral density
will be basically the same, e.g., for the surface layer.

The second term in the right-hand side of Eq. (\ref{N(E)})
describes the renormalization of quasiparticle energies. The third
term, which arises from the branch cut of the self-energy $\Sigma
_{\sigma }(E)$, describes the incoherent (non-quasiparticle)
contribution owing to scattering by magnons. One can see that this
does not vanish in the energy region, corresponding to the
\textquotedblleft alien\textquotedblright\ spin subband with the
opposite projection $-\sigma $. Consider for definiteness the case
$I>0$ (the case $I<0$ differs, roughly speaking, by a
particle-hole
transformation). On summing up Eq.(\ref{N(E)}) to find the total DOS $%
N_{\sigma }\left( E\right) $ and neglecting the quasiparticle
shift we get
\begin{eqnarray}
N_{\uparrow }(E) &=&\sum_{\mathbf{kq}}\left[ 1-\frac{2I^{2}SN_{\mathbf{q}}}{%
(t_{\mathbf{k+q\downarrow }}-t_{\mathbf{k\uparrow }})^{2}}\right]
\delta
(E-t_{\mathbf{k}\uparrow })  \nonumber \\
N_{\downarrow }(E) &=&2I^{2}S\sum_{\mathbf{kq}}\frac{1+N_{\mathbf{q}}-n_{%
\mathbf{k\uparrow }}}{(t_{\mathbf{k+q\downarrow }}-t_{\mathbf{k\uparrow }%
}-\omega _{\mathbf{q}})^{2}}\delta (E-t_{\mathbf{k}\uparrow }-\omega _{%
\mathbf{q}})  \label{DOS1}
\end{eqnarray}
%
%
%
%
%
%
%
%
The $T^{3/2}$-dependence of the magnon contribution to the residue
of the Green's function, i.e. of the effective electron mass in
the lower spin subband, and an increase with temperature of the
incoherent tail from the upper spin subband result in a strong
temperature dependence of partial densities of states $N_{\sigma
}(E)$, the corrections being of opposite sign. At the same time,
the temperature shift of the band edge for the quasiparticle
states is proportional to $T^{5/2}$ rather than to magnetization
\cite{Irkhin:1947,Irkhin:4173,Irkhin:3055}.

It is worthwhile to note that there exists a purely
single-particle mechanism of the gap filling in HMF which is due
to relativistic interactions. Specifically, one should take into
account spin-orbit coupling effects which connect the spin-up and
spin-down channels through the angular momentum $\mathbf{l}$. The
strength of this interaction is proportional to
the spatial derivatives of the crystal potential $V(\mathbf{r})$: $%
V_{SO}\propto \nabla V~(\mathbf{l}\cdot \mathbf{s}),$ off-diagonal elements $%
V_{SO}^{\sigma ,\sigma ^{\prime }}$being non-zero. For HMF with a
gap in the minority-spin (spin-down) channel one could construct
the wave function for spin-down electrons within perturbation
theory, so that the DOS in the gap has a square dependence on the
spin-orbit coupling strength, $\delta
n_{\downarrow }^{SO}(E)\propto ({V}_{SO}^{\downarrow ,\uparrow })^{2}$ \cite%
{Mavropoulos:S5759}. There is an obvious qualitative distinction
between the many-body and  spin-orbit contribution in the minority
spin channel; besides that, that the latter is orders in magnitude
smaller and very weakly temperature-dependent.
For further discussions of the
spin-orbit effects in HMF, see Ref. \cite{Pickett:315203}.

The exact solution in the atomic limit (for one conduction
electron), which is valid not only in the spin-wave region, but
for arbitrary temperatures, reads \cite{Auslender:309}
\begin{equation}
G^{\sigma }\left( E\right) =\frac{S+1+\sigma \left\langle
S^{z}\right\rangle
}{2S+1}\frac{1}{E+IS}+\frac{S-\sigma \left\langle S^{z}\right\rangle }{2S+1}%
\frac{1}{E-I\left( S+1\right) }.  \label{atom}
\end{equation}
In this case the energy levels are not temperature dependent at
all, whereas the residues are strongly temperature dependent via
the magnetization.

Now we consider the case $T=0~$K for a finite band filling. The picture of $%
N(E)$ in HMF (or degenerate ferromagnetic semiconductor)
demonstrates strong energy dependence near the Fermi level (Figs.
1,2). If we neglect magnon frequencies in the denominators of
Eq.(\ref{DOS1}), the partial density of
incoherent states should occur by a jump above or below the Fermi energy $%
E_{F}$ for $I>0$ and $I<0$ respectively owing to the Fermi
distribution
functions. An account of finite magnon frequencies $\omega _{\mathbf{q}}=%
\mathcal{D}q^{2}$ ($\mathcal{D}$ is the spin wave stiffness
constant) leads
to smearing of these singularities, $N_{-\alpha }(E_{F})$ being equal to zero. For $%
|E-E_{F}|\ll \overline{\omega }$ we obtain

\begin{equation}
\frac{N_{-\alpha }(E)}{N_{\alpha }(E)}=\frac{1}{2S}\left\vert \frac{E-E_{F}}{%
\overline{\omega }}\right\vert ^{3/2}\theta (\alpha (E-E_{F})),\,\alpha =%
\mathrm{sign}I  \label{alpha1}
\end{equation}
($\alpha =\pm $ corresponds to the spin projections $\uparrow ,\downarrow $%
). With increasing $|E-E_{F}|,\,N_{-\alpha }/N_{\alpha }$ tends to
a constant value which is of order of $I^{2}$ within the
perturbation theory.

In the strong coupling limit where $|I|\rightarrow \infty $ we have from (%
\ref{DOS1})
\begin{equation}
\frac{N_{-\alpha }(E)}{N_{\alpha }(E)}=\frac{1}{2S}\theta (\alpha
(E-E_{F})),|E-E_{F}|\gg \overline{\omega }  \label{alpha2}
\end{equation}
In fact, this expression is valid only in the framework of the $1/2S$%
-expansion, and in the narrow-band quantum case we have to use
more exact expressions (\ref{nb22}),(\ref{nb2}).

To investigate details of the energy dependence of $N\left(
E\right) $ in the broad-band case we assume the simplest isotropic
approximation for the majority-spin electrons,
\begin{equation}
t_{\mathbf{k}\uparrow }-E_{F}\equiv \xi _{\mathbf{k}}=\frac{k^{2}-k_{F}^{2}}{%
2m^*}.  \label{DOSbareup}
\end{equation}
Provided that we use the rigid splitting approximation $t_{\mathbf{k}%
\downarrow }=t_{\mathbf{k}\uparrow }+\Delta $ ($\Delta =2IS,I>0$),
the half-metallic situation (or, more precisely, the situation of
degenerate ferromagnetic semiconductor) takes place for $\Delta
>E_{F}$. Then qualitatively the equation (\ref{alpha1}) works to
accuracy of a prefactor. It is worthwhile to note that, formally
speaking, the NQP contribution to DOS occurs also for an ``usual"
metal where $\Delta <E_{F}$. In the case of small $\Delta $ there
is a crossover energy (or temperature) scale
\begin{equation}
T^{\ast }=\mathcal{D}\left(m^* \Delta/ k_F\right) ^{2}
\label{crossover}
\end{equation}
which is the magnon energy at the boundary of Stoner continuum,
$T^{\ast
}\simeq \overline{\omega }\left( \Delta /E_{F}\right) ^{2}\ll \overline{%
\omega }.$ At $|E-E_{F}|\ll \overline{\omega }$ the equation
(\ref{DOS1}) for the NQP contribution reads
\begin{equation}
\delta N_{\downarrow }(E)\propto \left[ \frac{1}{2}\ln \left| \frac{1+\sqrt{%
\left( E-E_{F}\right) /T^{\ast }}}{1-\sqrt{\left( E-E_{F}\right) /T^{\ast }}}%
\right| -\sqrt{\left( E-E_{F}\right) /T^{\ast }}\right] \theta
(E-E_{F}). \label{DOS111}
\end{equation}
At $|E-E_{F}|\ll T^{\ast }$ this gives the same results as above.
However, at $T^{\ast }$ $\ll |E-E_{F}|\ll \overline{\omega }$ this
contribution is
proportional to $-\sqrt{\left( E-E_{F}\right) /T^{\ast }}$ and is \textit{%
negative} (of course, the total DOS is always positive). This
demonstrates that one should be very careful when discussing the
NQP states for the systems which are not half-metallic.

The model of rigid spin splitting used above is in fact not
applicable for
the real HMF where the gap has a hybridization origin \cite%
{deGroot:2024,Irkhin:705}. The simplest model for HMF is as
follows: a \textquotedblleft normal\textquotedblright\ metallic
spectrum for majority electrons (\ref{DOSbareup}) and the
hybridization gap for minority ones,
\begin{equation}
t_{\mathbf{k}\downarrow }-E_{F}=\frac{1}{2}\left( \xi _{\mathbf{k}}+\mathrm{%
sign}\left( \xi _{\mathbf{k}}\right) \sqrt{\xi _{\mathbf{k}}^{2}+\Delta ^{2}}%
\right)  \label{DOSbaredn}
\end{equation}
Here we assume for simplicity that the Fermi energy lies exactly
in the
middle of the hybridization gap (otherwise one needs to shift $\xi _{\mathbf{%
k}}\rightarrow \xi _{\mathbf{k}}+E_{0}-E_{F}$ in the last
equation, $E_{0}$
being the middle of the gap). One can replace in Eq.(\ref{DOS1}) $\xi _{%
\mathbf{k+q}}$ by $\mathbf{v}_{\mathbf{k}}\mathbf{q}$, $\mathbf{v}_{\mathbf{k%
}}=\mathbf{k}/m^{\ast }$. Integrating over the angle between the vectors $%
\mathbf{k}$ and $\mathbf{q}$ we derive
\begin{equation}
\left\langle \left( \frac{1}{t_{\mathbf{k+q\downarrow }}-t_{\mathbf{%
k\uparrow }}-\omega _{\mathbf{q}}}\right) ^{2}\right\rangle =\frac{8}{%
v_{F}q\Delta }\left( \frac{2}{3}\left[
X^{3}-(X^{2}+1)^{3/2}+1\right] +X\right)  \label{angle}
\end{equation}
where angular brackets stand for the average over the angles of the vector $%
\mathbf{k}$, $X=k_{F}q/m^{\ast }\Delta .$ Here we do have the
crossover with the energy scale $T^{\ast }$ which can be small for
small enough hybridization gap. For example, in NiMnSb the
conduction band width is about 5~eV and the distance from the
Fermi level to the nearest gap edge (i.e. indirect energy gap
which is proportional to $\Delta ^{2}$) is smaller than 0.5~eV, so
that $(\Delta /E_{F})^{2}\leq 0.1$.

For the case $0<E-E_{F}\ll \overline{\omega }$ one has
\begin{equation}
N_{\downarrow }(E)\propto b\left( \frac{E-E_{F}}{T^{\ast }}\right) , \\
b(y)==\frac{2}{5}\left[ y^{5/2}-\left( 1+y\right) ^{5/2}+1\right]
+y+y^{3/2}\simeq \left\{
\begin{array}{cc}
y^{3/2}, & y\ll 1 \\
y, & y\gg 1%
\end{array}
\right.  \label{ss}
\end{equation}
Thus the behavior $N_{\downarrow }(E)\propto \left( E-E_{F}\right)
^{3/2}$ takes place only for very small excitation energies
$E-E_{F}\ll T^{\ast }$, whereas in a broad interval $T^{\ast }$
$\ll E-E_{F}\ll \overline{\omega }$ one has the linear dependence
$N_{\downarrow }(E)\propto E-E_{F}.$

\subsection{The problem of spin polarization}
\label{pol}

Functionality of devices that exploit charge as well as spin
degrees of freedom depends in a crucial way on the behavior of the
spin polarization of current carriers \cite{Prinz:1660}.
Unfortunately, many potentially promising  half-metallic systems
exhibit dramatic decrease in the spin polarization. Crystal
imperfections \cite{ebert:4627}, interfaces \cite{dewijs:020402},
and surfaces \cite{Galanakis:6329} constitute important examples
of static perturbations of the ideal periodic potential which
affect the states in the half-metallic gap.

In addition, several other depolarization mechanisms were
suggested that are based on magnon and phonon excitations
\cite{dowben:7948,skomski:544,dowben:7453,Skomski:315202}. These papers extend
the view of spin-disorder as random inter-atomic exchange fields
and claim that  disorder rotates locally the spin direction and
thus modifies the local magnetic moment and spin polarization
\cite{orgassa:5870,Orgassa:13237,ma.ju.98,it.oh.00}. The coupling
between atomic moments can be treated in terms of the
Heisenberg-type interactions (see Sect.\ref{subsec:GF} and
\ref{subsec:FM.SS}). The sign and magnitude of the exchange
constants determines whether the spin structure is collinear on
not \cite{sand.01}.

Simple qualitative considerations \cite{Edwards:L327}, as well as
direct Green's functions calculations
\cite{Auslender:301,Auslender:1003} for ferromagnetic
semiconductors, demonstrate that  spin polarization of conduction
electrons in spin-wave region is proportional to magnetization
\begin{equation}
P\equiv \frac{N_{\uparrow }-N_{\downarrow }}{N_{\uparrow }+N_{\downarrow }}%
=P_{0}\langle S^{z}\rangle/S  \label{polar1}
\end{equation}
A weak ground-state depolarization $1-P_{0}$ occurs in the case
for the empty conduction band where $I>0$. As discussed in the
previous section, in the case of the Fermi statistics of charge
carriers (degenerate ferromagnetic semiconductor and HMF) the NQP
states at $T=0$ exist only below $E_{F}~$for majority-spin gap
($I<0$ for the case of semiconductors) and only above $E_{F}$ for
minority-spin gap ($I>0$ for the case of semiconductors).

Spin-resolved photoelectron spectra for magnetite slightly above
the Verwey transition point have been measured in
Ref.\cite{Mortonx:L451}, a negative polarization about $-40$\%
being found near the Fermi energy. The strong deviation from
$-100$\% polarization can be, at least partially, related to NQP
states. Since according to the electronic structure calculations
\cite{Yanase:312} magnetite is a HMF with majority-spin gap, the
NQP should exist below the Fermi energy and thus be relevant for
photoelectron spectroscopy. Since electron correlations in
Fe$_{3}$O$_{4}$ are quite strong, the spectral weight of the NQP
states should be considerable. Of course, the photoemission is a
surface-sensitive method, and it is not quite clear to what degree
these data characterize the electronic structure of bulk
Fe$_{3}$O$_{4}$.

An instructive limit is  the Hubbard ferromagnet with infinitely
strong correlations $U=\infty $ (\ref{HHM}) and electron
concentration $n<1$ The DOS calculations yield
\cite{Irkhin:1947,Irkhin:4173}
\begin{equation}
N_{\downarrow }(E)=\sum_{\mathbf{k}\sigma }f(t_{\mathbf{k+q}})\delta (E-t_{%
\mathbf{k+q}}+\omega _{\mathbf{q}})=\left\{
\begin{array}{ll}
N_{\uparrow }(E) & ,\qquad E-E_{F}\gg \omega _{\max } \\
0 & ,\qquad E<E_{F}%
\end{array}
\right.  \label{ultim}
\end{equation}
A schematic density of states is shown in Fig. \ref{model_cras}a
(a more realistic picture is presented in Fig.~\ref{fig:11}, see
also Ref.\cite{irkhin:054421}). The result (\ref{ultim}) has a
simple physical meaning. Since the current carriers are spinless
doubles (doubly occupied sites), the electrons with spins up and
down may be picked up with an equal probability from the states
below the Fermi level of doubles, so that these states are fully
depolarized. On the other hand, according to the Pauli principle,
only the spin down electrons may be added in the singly occupied
states in the saturated ferromagnet.

The behavior $P(T)\sim \langle S^{z}\rangle $ is qualitatively
confirmed by  experimental data on field emission from
ferromagnetic semiconductors \cite{kisker:2256} and transport
properties of half-metallic Heusler alloys \cite{otto:2351}. An
attempt was used \cite{dowben:7948,skomski:544} to generalize the
result (\ref{polar1}) on the HMF case (in fact, using qualitative
arguments which are valid only in the atomic limit, see
Eq.(\ref{atom})). However, the situation for HMF is  more
complicated. We focus on the magnon contribution to DOS
(\ref{DOS1}) and calculate the function
\begin{equation}
\Lambda =\sum_{\mathbf{kq}}\frac{2I^{2}SN_{\mathbf{q}}}{(t_{\mathbf{%
k+q\downarrow }}-t_{\mathbf{k\uparrow }}-\omega
_{\mathbf{q}})^{2}}\delta (E_{F}-t_{\mathbf{k}\uparrow })
\label{phi}
\end{equation}
Using the parabolic electron spectrum $t_{\mathbf{k\uparrow }%
}=k^{2}/2m^{\ast }$ and averaging over the angles of the vector
$\mathbf{k}$ we obtain
\begin{equation}
\Lambda =\frac{2I^{2}Sm^{2}}{k_{F}^{2}}\rho \sum_{\mathbf{q}}\frac{N_{%
\mathbf{q}}}{\left( q^{\ast }\right) ^{2}-q^{2}}  \label{phi1}
\end{equation}
where $\rho =N_{\uparrow }(E_{F},T=0),$ we have used the condition
$q\ll k_{F},\,q^{\ast }=m^{\ast }\Delta /k_{F}=\Delta /v_{F},$
where $\Delta =2\left\vert I\right\vert S$ is the spin splitting.
In the ferromagnetic semiconductor we have, in agreement with the
qualitative considerations presented above:
\begin{equation}
\Lambda =\frac{S-\langle S^{z}\rangle }{2S}\rho \propto \left( \frac{T}{T_{C}%
}\right) ^{3/2}\rho  \label{phi3}
\end{equation}

Further on we consider the spectrum model (\ref{DOSbareup}),
(\ref{DOSbaredn}) where the gap has a hybridization origin. At
$T\ll T^{\ast }$ we reproduce the result (\ref{phi3}) which is
actually universal for this temperature region. \ At $T^{\ast }$
$\ll T\ll \overline{\omega }$ we derive
\begin{equation}
\Lambda =\sum_{\mathbf{kq}}2I^{2}SN_{\mathbf{q}}\delta (\xi _{\mathbf{k}})%
\frac{16}{3v_{F}q\Delta }\propto q^{\ast }\sum_{\mathbf{q}}\frac{N_{\mathbf{q%
}}}{q}\propto \frac{T^{\ast 1/2}}{T_{C}^{1/2}}T\ln
\frac{T}{T^{\ast }} \label{phi11}
\end{equation}
This result distinguishes HMF like the Heusler alloys from
ferromagnetic semiconductors and narrow-band saturated
ferromagnets. In the narrow-band case the spin polarization
follows the magnetization up to the Curie temperature $T_{C}$.

For finite temperatures the density of NQP states at the Fermi
energy is proportional to
\begin{equation}
N(E_{F})\propto \int_{0}^{\infty }d\omega \frac{K(\omega )}{\sinh (\omega /T)%
}  \label{atFermi}
\end{equation}
The filling of the energy gap is very important for possible
applications of HMF in spintronics: in fact HMF  have deciding
advantages only provided that $T\ll T_{C}$. Since a
single-particle Stoner-like theory leads to much less restrictive
(but unfortunately completely wrong) inequality $ T\ll \Delta ,$
the many-body treatment of the spin-polarization problem
(inclusion of collective spin-wave excitations) is crucial.
Generally, for temperatures which are comparable with the Curie
temperature $T_{C}$ there are no essential difference between
half-metallic and \textquotedblleft ordinary" ferromagnets since
the gap is filled.

The corresponding symmetry analysis was performed in Ref.
\cite{Irkhin:1733,Irkhin:397} for a model of conduction electrons
interacting with \textquotedblleft pseudospin\textquotedblright\
excitations in ferroelectric semiconductors. The symmetrical (with
respect to $E_{F}$) part of $N(E)$ in the gap can be attributed to
smearing of electron states by electron-magnon scattering; the
asymmetrical (\textquotedblleft Kondo-like") one is the density of
NQP states owing to the Fermi distribution function.

Skomski and Dowben \cite{dowben:7948,skomski:544,dowben:7453,Skomski:315202}
investigated spin mixing effects for NiMnSb  by using a simple
tight-binding approximation.

\begin{figure}[tbh]
\begin{center}
\includegraphics[width=0.30\columnwidth, ]{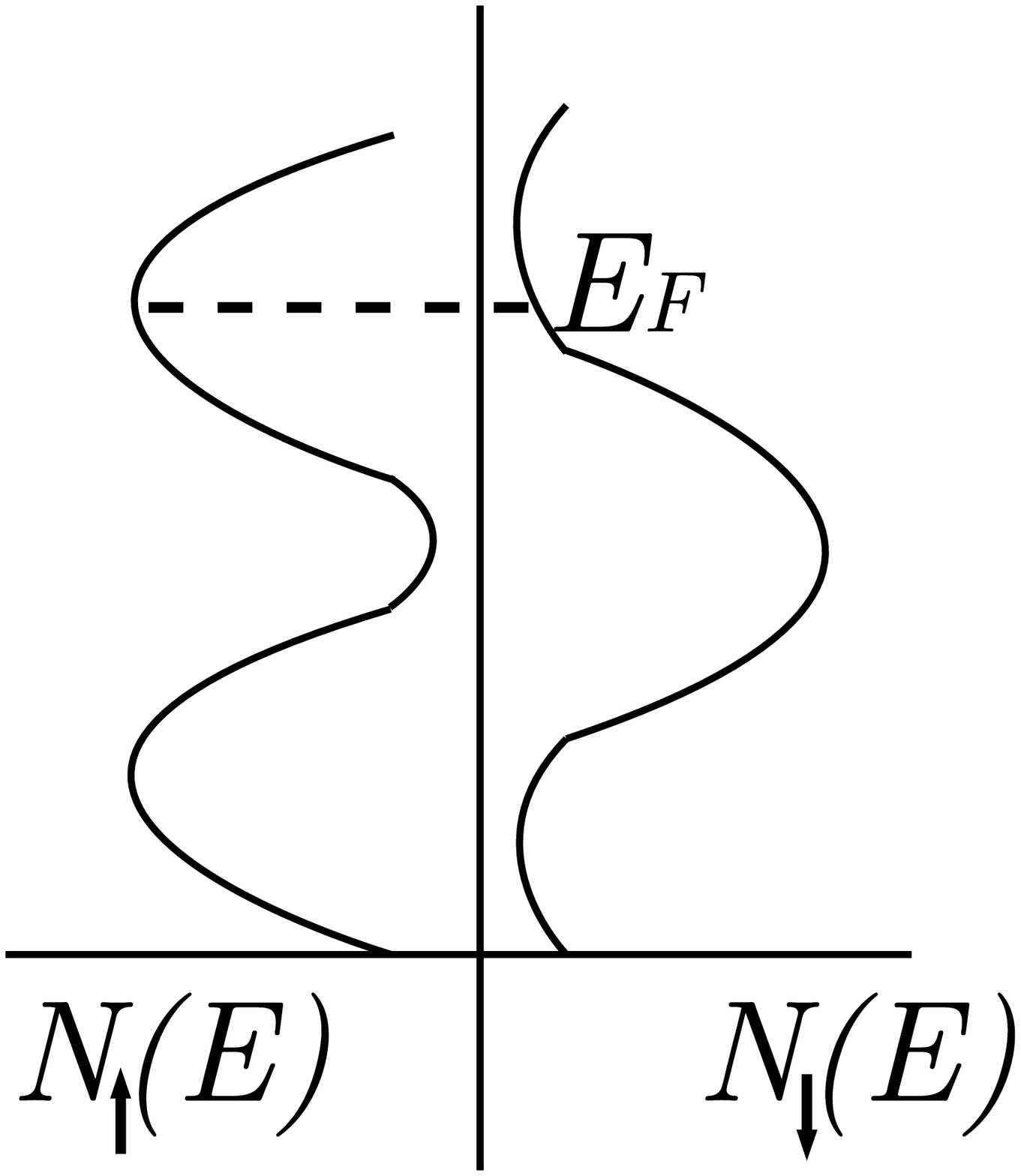}
\includegraphics[width=0.45\columnwidth, angle=0]{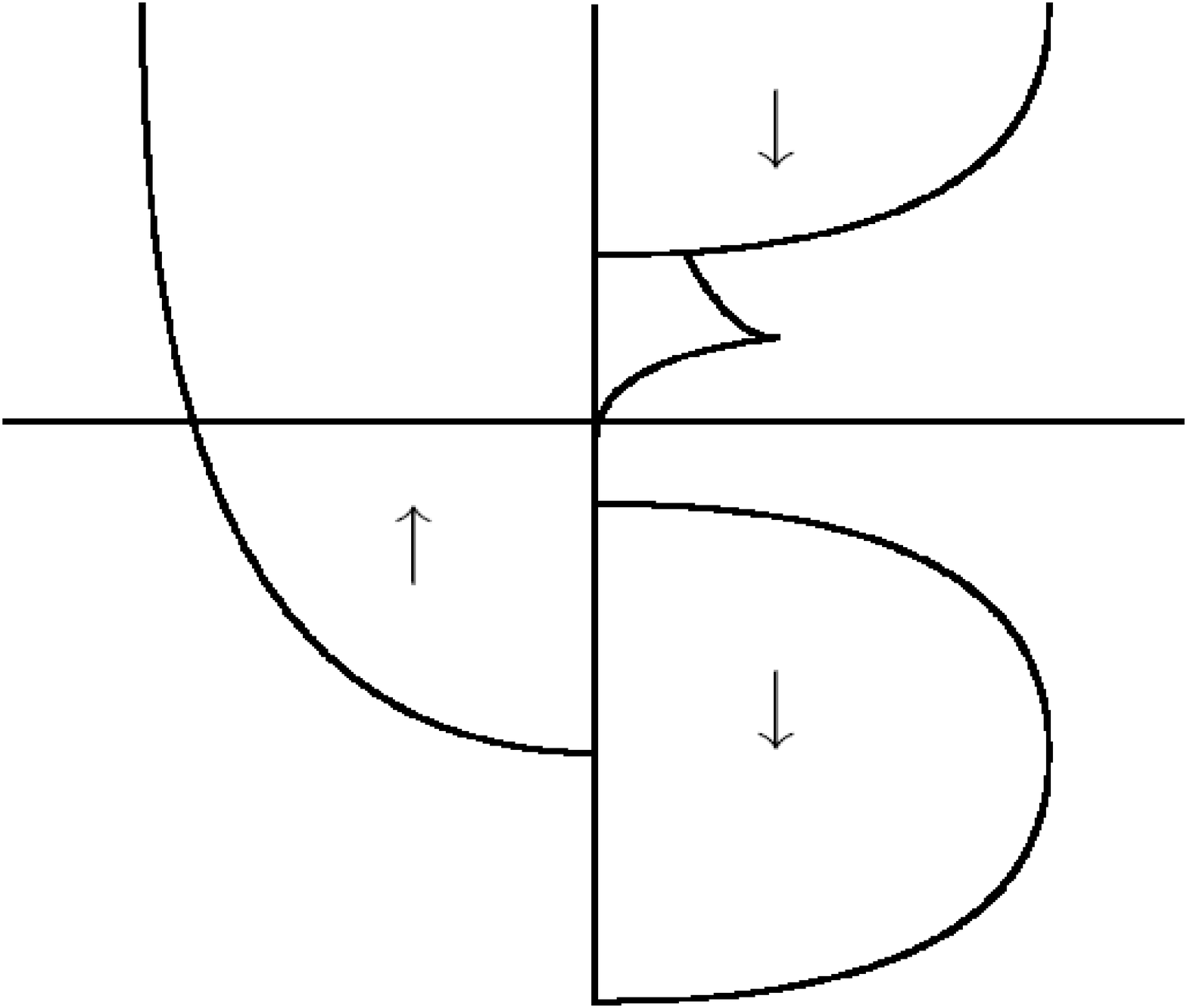}
\end{center}
\caption{Left: Schematic half-metallic DOS with and without the
spin mixing \protect\cite{dowben:7453}. Right: Schematic
half-metallic DOS, and the presence of NQP states, a genuine
many-body effect.} \label{tb_dos}
\end{figure}

Fig. \ref{tb_dos} shows also a schematic comparison between this
approximation and many-body results. In the tight-binding
approach, the distortion of the spin-up and spin-down DOS is
presented by the dark regions. The spin mixing gives a non-zero
symmetric $N_\downarrow(E)$, DOS being weakly modified by thermal
fluctuations.

Itoh et al. \cite{it.oh.00} calculated the polarization for a
ferromagnet-insulator magnetic tunnel junction with and without
spin fluctuations in a thermally randomized atomic potential. The
results indicate that the effect of spin fluctuations is
significant. The idea of spin fluctuations was further developed
by Lezaic et al. \cite{lezaic:026404} by considering the
competition between hybridization and thermal spin fluctuation in
the prototype HMF NiMnSb.

\begin{figure}[tbh]
\begin{center}
\includegraphics[width=0.40\linewidth, angle=-90]{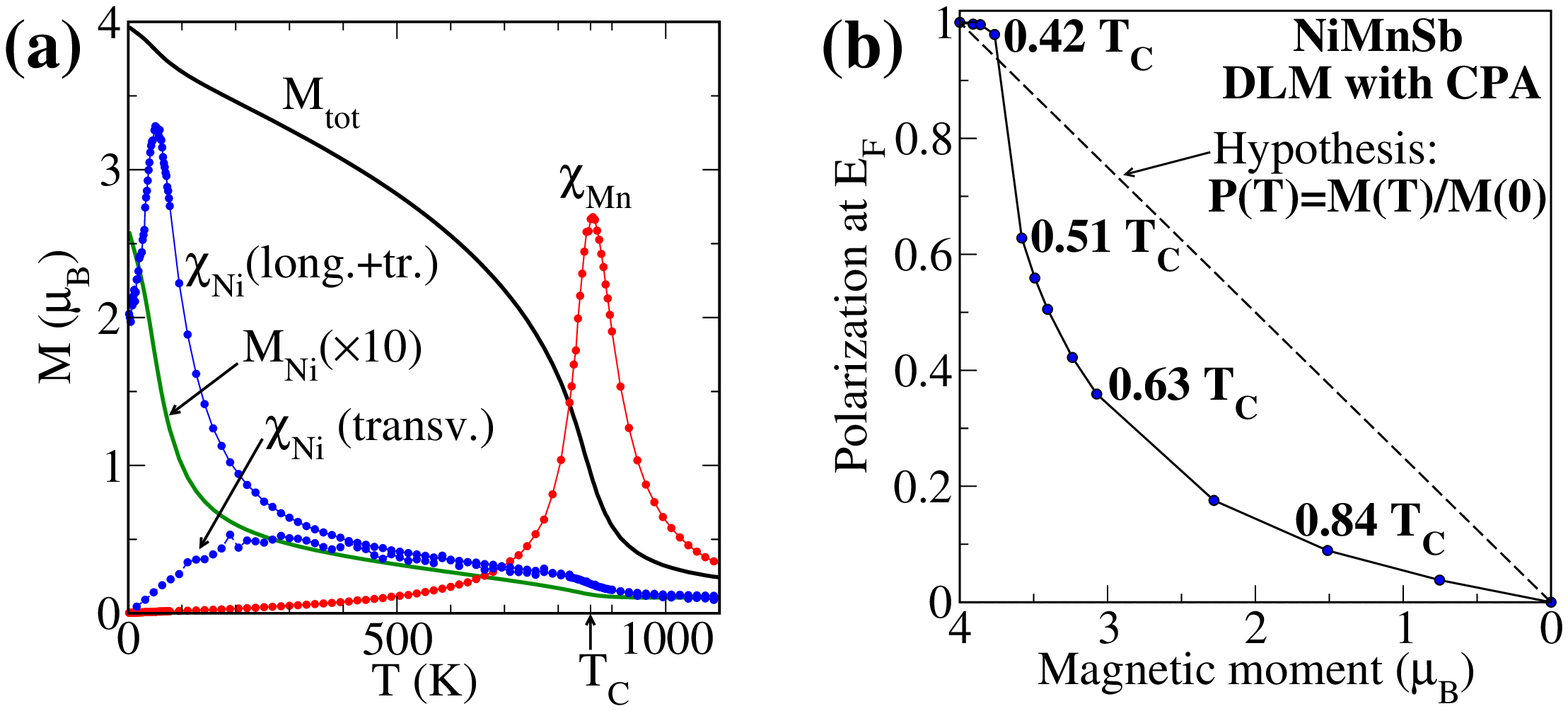}
\end{center}
\caption{Left: Monte-Carlo results for finite temperature magnetic
properties.  Right: Polarization at E$_F$ as a function of total
spin moment in the DLM picture for
NiMn$_{1-x}^{\uparrow}$Mn$_x^{\downarrow}$Sb calculated with CPA.
\cite{lezaic:026404}}. \label{dlm_dos}
\end{figure}

Figure \ref{dlm_dos} shows the sublattice susceptibilities
computed within a generalized Heisenberg-type Hamiltonian. These
results demonstrate that the Ni-sublattice magnetic order is lost
already at 50K (this effect is a consequence of  weakness of the
coupling between the Ni moments and the neighbor atoms), however,
neutron diffraction data \cite{Hordequin:602,Hordequin:605}
does not support this scenario. The right
hand side of Fig. \ref{dlm_dos} presents the polarization $P(T)$
calculated in a disordered local moment (DLM) approach
\cite{gy.pi.85}, representing the system at finite temperatures in
a mean-field way. It is concluded that the thermal collapse of the
polarization is connected with a change in hybridization due to
the moment fluctuation, the effect of non-collinearity being much
milder.

The issue of finite-temperature spin polarization in half-metals
remains an open question. Magnons play a crucial role,
independently of the theoretical approach. Besides that, a role of
phonon modes in the the many-sublattice half-metals is not
excluded. Nevertheless, a non-trivial contributions to physics
of half-metallic ferromagnets comes from
genuine many-body effects. 
The corresponding first-principle calculations will be
presented in Sect.\ref{sec:elstr_hmf}.

\subsection{The tunneling conductance and spin-polarized STM}

An useful tool to probe the spin polarization and
non-quasiparticle states in HMF is provided by tunneling phenomena
\cite{auslender:701,bratkovsky:2334,auth:024403,gercsi:082512,sakuraba:052508,sa.ha.06,sa.ha.07},
in particular, by Andreev reflection spectroscopy for a
HMF-superconductor tunnel junction \cite{Tkachov:024519}. The most
direct way is probably the measurement of a tunnel current between
two pieces of HMF with the opposite magnetization directions.

Let us explain in a simple qualitative way why the NQP states are
important for the tunneling transport. To this aim we consider the
above-discussed narrow-band saturated Hubbard ferromagnet where
the current carriers are the holes in the lowest Hubbard band and
the NQP states provide \textit{all} the spectral weight for the
minority spin projection . Suppose we have a tunnel
junction with two pieces of this ferromagnet with either parallel (Fig. \ref%
{model_cras}b) or antiparallel (Fig. \ref{model_cras}c)
magnetization directions. From the one-particle point of view, the
spin-conserving tunneling is forbidden in the latter case.
However, in the framework of many-particle picture the charge
current is a transfer process between an empty site and a
single-occupied site rather than the motion of the electron
irrespective to the site like in the band theory and therefore the
distinction between these two cases (see Fig. \ref{model_cras}),
is due only to the difference in the densities of states. It means
that the estimations of the tunneling magnetoresistance based on
simple one-electron picture is too optimistic; even for
antiparallel spin orientation of two pieces of the half-metallic
ferromagnets in the junction for zero temperature the current is
not zero, due to the non-quasiparticle states. More exactly, it
vanishes for zero bias since the density of NQP states at the
Fermi energy equals to zero. However, it grows with the bias
sharply, having the scale of typical \textit{magnon} energies,
i.e., millivolts.

Formally, we can consider a standard tunneling Hamiltonian (see,
e.g., Ref. \cite{mahan:1990})

\begin{equation}
\mathcal{H}=\mathcal{H}_{L}+\mathcal{H}_{R}+\sum_{\mathbf{kp}}(T_{\mathbf{kp}%
}c_{\mathbf{k}\uparrow }^{\dagger }c_{\mathbf{p}\downarrow }+h.c.)
\end{equation}%
where $\mathcal{H}_{L,R}$ are the Hamiltonians of the left (right)
half-spaces, respectively, $\mathbf{k}$ and $\mathbf{p}$ are the
corresponding quasimomenta, and spin projections are defined with
respect to the magnetization direction of a given half-space (spin
is supposed to be conserving in the \textquotedblleft
global\textquotedblright\ coordinate
system). Carrying out standard calculations of the tunneling current $%
\mathcal{I}$ in the second order in $T_{\mathbf{kp}}$ one has (cf. \cite%
{mahan:1990}) 
\begin{equation}
\mathcal{I}\propto \sum_{\mathbf{kqp}}|T_{\mathbf{kp}}|^{2}[1+N_{\mathbf{q}%
}-f(t_{\mathbf{p-q}})][f(t_{\mathbf{k}})-f(t_{\mathbf{k}}+eV)]\delta (eV+t_{%
\mathbf{k}}-t_{\mathbf{p-q}}+\omega _{\mathbf{q}})
\end{equation}%
where $V$ is the bias voltage.

A very efficient new experimental method is spin-polarized
scanning tunneling microscopy (STM)
\cite{Heinze:1805,Wiesendanger:247,Kleiber:4606} which enables one
to probe directly the spectral density with spin resolution in
magnetic systems. The spin-polarized STM should be able to probe
the NQP states via their contribution to the differential
tunneling conductivity $d\mathcal{I}_{\sigma }/dV\propto N_{\sigma
}\left( eV\right) $
\cite{irkhin:104429}. Note that the value $N_{\sigma }(eV)$ vanishes for $%
|eV|<\hbar \omega _{0}$, where $\hbar \omega _{0}$ is an
anisotropy gap in the magnon spectrum \cite{irkhin:481}, which is
small, but could be changed by suitable substitution
\cite{attema:S5517}. Keeping in mind that ferromagnetic
semiconductors can be considered as a peculiar case of HFM
\cite{Irkhin:1947}, an account of NQP states can be important for
proper description of spin diodes and transistors
\cite{flatte:1273,Tkachov:024519}.

The above formulas are derived for the usual one-electron density
of states at $E_{F}$, which is observed, say, in photoemission
measurements. However, the factors which are present in the
expression for the tunneling current do not influence the
temperature dependence, and therefore these results are valid for
spin polarization from tunneling conductance at zero bias in STM
\cite{Ukraintsev:11176,irkhin:104429}. Unlike the photoemission
spectroscopy which probes only occupied electron states, STM
detects the states both above and below $E_{F}$, depending on the
sign of bias.

One should keep in mind that sometimes the surface of HMF is not
half-metallic; in particular, this is the case of a prototype HMF,
NiMnSb \cite{dewijs:020402}. In such a situation, the tunneling
current for minority electrons is due to the surface states only.
However, the NQP states can be still visible in the tunneling
current via the hybridization of the bulk states with the surface
one. This leads to the Fano antiresonance picture which is usually
observed in STM investigations of the Kondo effect at metallic
surfaces. In such cases the tunneling conductance will be
proportional to a mixture of $N_{\sigma }$ and the real part of
the on-site Green's function, $L_{\sigma }$. Surprisingly, in this
case the effect of NQP states on the tunneling current can be even
more pronounced in comparison with the ideal crystal. The reason
is that the analytical continuation of the jump in $N_{\sigma
}\left( E\right) $ is logarithm; both singularities are cut at the
energy $\overline{\omega };$ nevertheless, the
energy dependence of $L_{\sigma }\left( E\right) $ can be pronounced \cite%
{irkhin:104429}.

STM measurements of electron DOS give also an opportunity to probe \textit{%
bosonic} excitations interacting with the conduction electrons.
Due to electron-phonon coupling, the derivative $dN_{\sigma
}\left( E\right) /dE$ and thus $d^{2}\mathcal{I}_{\sigma }\left(
V\right) /dV^{2}$ at $eV=E$ have peaks at the energies $E=\pm
\omega _{i}$ corresponding to the peaks in the phonon DOS.
According to above treatment (see, e.g., Eq.(\ref{DOS1}), the same
effect should be observable for the case of electron-magnon
interaction. However, in the latter case these peaks are
essentially asymmetric with respect to the Fermi energy (zero
bias) due to asymmetry of the non-quasiparticle contributions.
This asymmetry can be used to distinguish phonon and magnon peaks.

Thermoelectric power in the tunnel situation was theoretically
investigated in Refs.\cite{McCann:134424,mccann:172404}. The
relative polarizations of ferromagnetic layers can be manipulated
by an external magnetic field, and a large difference occurs for a
junction between two ferromagnets with antiparallel and parallel
polarizations. This magnetothermopower effect becomes giant in the
extreme case of a junction between two half-metallic ferromagnets,
since the thermopower is inversely proportional to the area of the
maximal cross section of the Fermi surface of minority electrons
in the plane parallel to the interface. One has a strong
polarization dependence of the thermopower
\begin{equation}
\mathcal{Q}_{AP}=0.64kB/e,\mathcal{Q}_{P}\propto
k_{B}^{2}T/(eE_{F})
\end{equation}
This result is independent of temperature and of the specific
half-metallic material, and it represents a giant
magnetothermopower effect, $\Delta \mathcal{Q}\simeq
\mathcal{Q}_{AP}=-55\mu $V/K.

\subsection{Spin waves}

\label{magn}

Unlike the Stoner theory, the Hubbard model and other model with
electron correlations enable one to describe spin-wave excitations
in an itinerant ferromagnet. This was already done in the old
approaches based on the random phase approximation (RPA)
\cite{Herring:1966}. To discuss related approaches we present the
interaction Hamiltonian in terms of the spin density operators
\begin{equation}
\mathcal{H}_{\mathrm{int}}=\frac{U}{2}\sum_{\mathbf{k}\sigma }c_{\mathbf{k}%
\sigma }^{\dagger }c_{\mathbf{k}\sigma }-\frac{U}{2}\sum_{\mathbf{q}}(S_{-%
\mathbf{q}}^{-}S_{\mathbf{q}}^{+}+S_{\mathbf{q}}^{+}S_{-\mathbf{q}}^{-})
\label{G.5}
\end{equation}%
where $S_{\mathbf{q}}^{\alpha }$ are the Fourier components of
spin density operators. The first term in (\ref{G.5}) yields a
renormalization of the chemical potential and may be omitted.
Writing down the sequence of equations of motion for the spin
Green's function
\begin{equation}
G_{\mathbf{q}}(\omega )=\langle \langle S_{\mathbf{q}}^{+}|S_{-\mathbf{q}%
}^{-}\rangle \rangle _{\omega }  \label{G.9}
\end{equation}%
one derives \cite{irkhin:7151}
\begin{equation}
G_{\mathbf{q}}(\omega )=\frac{\langle S^{z}\rangle -\Omega _{\mathbf{q}%
}(\omega )/U}{\omega -\Omega _{\mathbf{q}}(\omega )-\pi
_{\mathbf{q}}(\omega )}  \label{G.10}
\end{equation}%
where
\begin{equation}
\Omega _{\mathbf{q}}(\omega )=U\sum_{\mathbf{k}}\frac{t_{\mathbf{k+q}}-t_{%
\mathbf{k}}}{t_{\mathbf{k+q}}-t_{\mathbf{k}}+\Delta -\omega }(n_{\mathbf{%
k\uparrow }}-n_{\mathbf{k+q\downarrow }})
\end{equation}%
and the self-energy $\pi $ describes corrections to RPA. Unlike
the standard form of RPA,
\begin{equation}
G_{\mathbf{q}}(\omega )=\frac{\Pi _{\mathbf{q}}(\omega )}{1-U\Pi _{\mathbf{q}%
}(\omega )},
\end{equation}

\begin{equation}
\Pi _{\mathbf{q}}(\omega )=\sum_{\mathbf{k}}\frac{n_{\mathbf{k\uparrow }}-n_{%
\mathbf{k+q\downarrow }}}{\omega +t_{\mathbf{k\uparrow }}-t_{\mathbf{%
k+q\downarrow }}},
\end{equation}
the equivalent representation (\ref{G.10}) yields explicitly the
magnon (spin-wave) pole
\begin{equation}
\omega _{\mathbf{q}}\simeq \Omega _{\mathbf{q}}(0)=\sum_{\mathbf{k}\sigma }%
\mathcal{A}_{\mathbf{kq}}^{\sigma }n_{\mathbf{k}\sigma }
\end{equation}
where $\mathcal{A}_{\mathbf{kq}}^{\sigma }$ is given by
(\ref{amp}). Expanding in $q$ we get $\omega
_{\mathbf{q}}=\mathcal{D}_{\alpha \beta }q_{\alpha }q_{\beta }$
where
\begin{equation}
\mathcal{D}_{\alpha \beta }=\frac{U}{\Delta }\sum_{\mathbf{k}}\left[ \frac{%
\partial ^{2}t_{\mathbf{k}}}{\partial k_{\alpha }\partial k_{\beta }}(n_{%
\mathbf{k}\uparrow }+n_{\mathbf{k}\downarrow })-\frac{1}{\Delta }\frac{%
\partial t_{\mathbf{k}}}{\partial k_{\alpha }}\frac{\partial t_{\mathbf{k}}}{%
\partial k_{\beta }}(n_{\mathbf{k}\uparrow }-n_{\mathbf{k}\downarrow })%
\right]  \label{G.15}
\end{equation}
Eq. \ref{G.15} are spin-wave stiffness tensor components. For a
weak ferromagnet ($\Delta \ll E_{F},U$) we have
$\mathcal{D}\propto \Delta $. The magnon damping in the RPA is
given by
\begin{equation}
\gamma _{\mathbf{q}}^{(1)}(\omega )=-\mathrm{Im~}\Omega
_{\mathbf{q}}(\omega
)=\pi U\Delta \omega \sum_{\mathbf{k}}\left( -\frac{\partial n_{\mathbf{k}%
\uparrow }}{\partial t_{\mathbf{k}\uparrow }}\right) \delta (\omega -t_{%
\mathbf{k+q\downarrow }}+t_{\mathbf{k\uparrow }})  \label{G.17}
\end{equation}

\begin{equation}
\gamma _{\mathbf{q}}^{(1)}\equiv \gamma _{\mathbf{q}}^{(1)}(\omega _{\mathbf{%
q}})\simeq \pi U\Delta \omega _{\mathbf{q}}N_{\uparrow
}(E_{F})N_{\downarrow }(E_{F})\theta (\omega _{\mathbf{q}}-\omega
_{-})  \label{G.18}
\end{equation}
with $\theta (x)$ being the step function. Here $\omega
_{-}=\omega (q^{\ast })$ is the threshold energy which is
determined by the condition of entering into the Stoner continuum
(decay into the Stoner excitations, i.e. electron-hole pairs),
$q^{\ast }$ being the minimal (in $\mathbf{k}$) solution to the
equation
\begin{equation}
t_{\mathbf{k+q}^{\ast }\downarrow }=t_{\mathbf{k\uparrow }}=E_{F}
\end{equation}
The quantity $\omega _{-}$ determines a characteristic energy
scale separating two temperature regions: the contributions of
spin waves (poles of the Green's function \ref{G.10} dominate at
$T<\omega _{-}$, and those of Stoner excitations (its branch cut)
at $T>\omega _{-}$.

Although the formal expressions in the $s-d$ exchange model are
similar, presence of two electron subsystems leads to some new
effects, in particular, to possible occurrence of the
\textquotedblleft optical mode\textquotedblright\ pole $\omega
\simeq 2|I|S$. The problem of the optical mode formation and its
damping was investigated in application to degenerate
ferromagnetic semiconductors \cite{Auslender:301,Auslender:129,
Irkhin:522,irkh.41}. Kaplan \textit{et al} \cite{Kaplan:3634}
performed exact diagonalization studies of the double exchange
model which indicate the existence of continuum states in the
single-spin-flip channel that overlap the magnons at very low
energies (of order 10$^{-2}$~eV) and extend to high energies. This
picture differs dramatically from the prevalent view, where there
are the magnons, plus the Stoner continuum at the high-energy
scale, with nothing in between. Peculiarities of magnons in HMF, 
especially, in the collossal magnetoresistanse materials, have been
recently reviewed in Ref. \cite{Zhang:315204}.

In the case of weak ferromagnets, the contribution of the branch
cut of the spin Green's function may be approximately treated as
that of a paramagnon pole at imaginary $\omega $, and we obtain

\begin{equation}
q^{\ast }=k_{F\uparrow }-k_{F\downarrow
}\,,\,\,\,\,\,\,\,\,\,\,\,\,\omega _{-}=\mathcal{D}(k_{F\uparrow
}-k_{F\downarrow })^{2}\sim \Delta ^{3}\sim T_{C}^{2}/E_{F}
\label{G.20}
\end{equation}
Since $q^{\ast }$ is small, we have at small $q>q^{\ast }$, instead of Eq. (%
\ref{G.18}),

\begin{equation}
\gamma _{\mathbf{q}}^{(1)}(\omega _{\mathbf{q}})\simeq
\frac{U\Delta \omega }{q}\frac{\Omega _{0}}{4\pi }(m^{\ast
})^{2}\equiv A/q  \label{G.21}
\end{equation}
with $\Omega _{0}$ the lattice cell volume. The estimation
(\ref{G.20}) holds also for the $\mathit{s-d(f)}$ exchange model
with the indirect RKKY-interaction where $\mathcal{D}\sim
T_{C}/S\sim I^{2}S/E_{F}$.

The damping at very small $q<q^{\ast }$ (where (\ref{G.17})
vanishes) is due to the two-magnon scattering processes. To
consider these we have to calculate the function $\pi $ to leading
order in the fluctuating part of the Coulomb interaction. Writing
down the equation of motion for the Green's function (\ref{G.9})
we obtain

\begin{eqnarray}
\pi _{\mathbf{q}}(\omega )
&=&\sum_{\mathbf{pk}}(A_{\mathbf{kq}}^{\uparrow
})^2[B(\mathbf{k\uparrow ,k+q-p\uparrow ,\omega }_{\mathbf{p}}-\omega )+B(%
\mathbf{k+p\downarrow ,k+q\downarrow ,\omega }_{\mathbf{p}}-\omega
)
\nonumber \\
&-& B(\mathbf{k+p\downarrow ,} \mathbf{k\uparrow ,\omega }_{\mathbf{p}})-B(%
\mathbf{k+q\downarrow ,k+q-p\downarrow ,p,\omega }_{\mathbf{p}})]
\label{G.23}
\end{eqnarray}
where

\begin{equation}
B(\mathbf{k}^{\prime }\mathbf{\sigma }^{\prime },\mathbf{k\sigma },\omega )=%
\frac{N_{\mathbf{p}}(n_{\mathbf{k}^{\prime }\mathbf{\sigma }^{\prime }}-n_{%
\mathbf{k\sigma }})+n_{\mathbf{k}^{\prime }\mathbf{\sigma }^{\prime }}(1-n_{%
\mathbf{k\sigma }})}{\omega -t_{\mathbf{k}^{\prime }\mathbf{\sigma
}^{\prime }}+t_{\mathbf{k\sigma }}}
\end{equation}
The magnon damping needed is given by the imaginary part of
(\ref{G.23}),
\begin{equation}
\gamma _{\mathbf{q}}^{(2)}(\omega )=\pi \sum_{\mathbf{kp\sigma }}(A_{\mathbf{%
kq}}^{\uparrow })^{2}(n_{\mathbf{k\sigma }}-n_{\mathbf{k+q-p\sigma
}})\left[
N_{\mathbf{p}}-N_{B}(\omega _{\mathbf{p}}-\omega )\right] \delta (\omega +t_{%
\mathbf{k}}-t_{\mathbf{k+q-p}}-\omega _{\mathbf{p}})
\end{equation}
%
%
Integration for the isotropic electron spectrum gives
\cite{irkhin:7151}
\begin{equation}
\gamma _{\mathbf{q}}^{(2)}(\omega )=\frac{\Omega _{0}^{2}}{12\pi ^{3}}\frac{%
q^{4}}{4\langle S^{z}\rangle ^{2}}\sum_{\sigma }k_{F\sigma
}^{2}\times \left\{
\begin{array}{ll}
\omega _{\mathbf{q}}/35 & ,\qquad T\ll \omega _{\mathbf{q}} \\
(T/4)\left( \ln (T/\omega _{\mathbf{q}})+\frac{5}{3}\right) &
,\qquad T\gg
\omega _{\mathbf{q}}%
\end{array}
\right.
\end{equation}

These results were obtained by Silin and Solontsov within the
phenomenological Fermi-liquid theory \cite{Silin:1080} and
Auslender and Irkhin \cite{Auslender:301,Auslender:129} within the
$s-d$ exchange model.
Golosov \cite{Golosov:3974} reproduced the results of Refs.\cite%
{Auslender:301,Auslender:129,Irkhin:522} within the
$1/2S$-expansion and performed numerical investigations of the
magnon spectrum and damping in the limit of large $|I|$
(double-exchange situation) in application to colossal
magnetoresistance compounds.

Real part of (\ref{G.23}) describes the temperature dependence of
the spin stiffness owing to two-magnon processes (besides the
simplest $T^{2}$ -contribution which occurs from the temperature
dependence of the Fermi distribution functions in (\ref{G.10}).
The spin-wave contribution connected with the magnon distribution
functions is proportional to $T$. More interesting is the
non-analytical many-electron contribution owing to the Fermi
functions:

\begin{eqnarray}
\delta \mathcal{D}_{\alpha \beta } &=&\frac{1}{4\langle S^{z}\rangle ^{2}}%
\sum_{\mathbf{pk}}\frac{\partial t_{\mathbf{k}}}{\partial k_{\alpha }}\frac{%
\partial t_{\mathbf{k}}}{\partial k_{\beta }}\left[ \frac{n_{\mathbf{k}%
\downarrow }(1-n_{\mathbf{k-p}\uparrow })}{t_{\mathbf{k}}-t_{\mathbf{k-p}%
}-\omega _{\mathbf{p}}}+\right.  \nonumber \\
&&\ \ \left. +\frac{n_{\mathbf{k+p}\downarrow }(1-n_{\mathbf{k}\downarrow })%
}{t_{\mathbf{k+p}}-t_{\mathbf{k}}-\omega _{\mathbf{p}}}-\frac{n_{\mathbf{k+p}%
\downarrow }(1-n_{\mathbf{k}\uparrow })}{t_{\mathbf{k+p\downarrow }}-t_{%
\mathbf{k\uparrow }}-\omega
_{\mathbf{p}}}-\frac{n_{\mathbf{k}\downarrow
}(1-n_{\mathbf{k-p}\downarrow })}{t_{\mathbf{k\downarrow }}-t_{\mathbf{%
k-p\uparrow }}-\omega _{\mathbf{p}}}\right]  \label{G.26}
\end{eqnarray}
Performing integration for parabolic spectra of electrons and
magnons yields

\begin{equation}
\delta \mathcal{D}(T) =\left( \frac{\pi \Omega _{0}T}{12\langle
S^{z}\rangle m^{\ast }}\right) ^{2}\frac{1}{D}\left[ \sum_{\sigma
}N_{\sigma}^{2} (E_{F})\ln \frac{T}{\omega _{+}} - 2N_{\uparrow
}(E_{F})N_{\downarrow }(E_{F})\ln \frac{\max (\omega_{-},T)}
{\omega _{+}}\right]  \label{G.27}
\end{equation}
with
\begin{equation}
\omega _{\pm }=\mathcal{D}(k_{F\uparrow }\pm k_{F\downarrow
})^{2}\,,\,\,\,\,\,\,\,\,\,\,\,\,\,\,\,N_{\sigma }(E_{F})=m^{\ast
}\Omega _{0}k_{F}/2\pi ^{2}  \label{G.28}
\end{equation}
It should be noted that the correction (\ref{G.27}) dominates at
low temperatures over the above-mentioned $T^{2}$-correction,
which demonstrates an important role of corrections to the RPA
approximation. Unfortunately, the $T^{2}\ln T$-term has not yet to
be considered at analyzing magnon spectra of ferromagnetic metals.
We see that temperature dependences of spin-wave characteristics
in conducting magnets differ considerably from those in the
Heisenberg model.

\subsection{Magnetization and local moments}

\label{magn1}

To treat the problem of magnetic moments in the Hubbard model we
consider corrections to the magnetization $\langle S^{z}\rangle $.
We have
\begin{equation}
\langle S^{z}\rangle =\frac{n}{2}-\sum_{\mathbf{q}}\langle S_{\mathbf{q}%
}^{-}S_{\mathbf{q}}^{+}\rangle -\langle \widehat{n}_{i\uparrow }\widehat{n}%
_{i\downarrow }\rangle   \label{G.41}
\end{equation}%
The first average involved in (\ref{G.41}) is calculated from the
spectral representation of the RPA Green's function \ref{G.10}:
\begin{equation}
\langle S_{-\mathbf{q}}^{-}S_{\mathbf{q}}^{+}\rangle \,=2S_{0}N_{\mathbf{q}%
}\,\,\,\,\,\,\,\,\,\,\,\,\,\,\,\,\,\,\,\,\,\,\,\,\,(q<q^{\ast })
\label{G.42}
\end{equation}%
\begin{equation}
\langle S_{-\mathbf{q}}^{-}S_{\mathbf{q}}^{+}\rangle \,=\frac{1}{\pi }%
\int_{-\infty }^{\infty }d\omega \frac{N_{B}(\omega )\gamma _{\mathbf{q}%
}^{(1)}(\omega )(\Delta -\omega )/U}{[\omega -\mathrm{Re~}\Omega _{\mathbf{q}%
}(\omega )]^{2}+[\gamma _{\mathbf{q}}^{(1)}(\omega )]^{2}}%
\,\,\,\,\,\,\,\,\,\,\,\,\,(q>q^{\ast })  \label{G.43}
\end{equation}%
In contradiction with (\ref{G.41}), (\ref{G.42}), in the true
Bloch
spin-wave contribution to magnetization every magnon should decrease $%
\langle S^{z}\rangle $ by unity. The agreement may be restored by
allowing
not only the magnon pole, but also branch cut contributions \cite%
{irkhin:7151}. In the semi-phenomenological manner, it is
convenient to introduce \textquotedblleft
magnon\textquotedblright\ operators which satisfy on the average
the Bose commutation relations:
\begin{equation}
b_{\mathbf{q}}=(2S_{0})^{-1/2}S_{\mathbf{q}}^{+}\,,\,\,b_{\mathbf{q}%
}^{\dagger }=(2S_{0})^{-1/2}S_{-\mathbf{q}}^{-}\,  \label{G.47}
\end{equation}%
Then we have

\begin{equation}
\delta \langle S^z\rangle \,=-\sum_{\mathbf{q}}\langle \,\,b_{\mathbf{q}%
}^{\dagger }b_{\mathbf{q}}\rangle \,=\frac
1{(2S_0)}\sum_{\mathbf{q}}\langle
S_{-\mathbf{q}}^{-}S_{\mathbf{q}}^{+}\rangle
\end{equation}
Performing integration over $\omega $ in (\ref{G.43}) at $T$=0~K
we obtain

\begin{equation}
\delta \langle S^{z}\rangle \,=-\frac{1}{\pi
}\sum_{\mathbf{q}}\frac{\gamma _{\mathbf{q}}^{(1)}}{\omega
_{\mathbf{q}}}\ln \frac{W}{\omega _{\mathbf{q}}} \label{G.49}
\end{equation}
with $W$ being the bandwidth. This contribution of the order of
$U^{2}\ln (W/\omega _{+})$ describes zero-point decrease of the
magnetization due to the ground-state magnon damping which is
owing to the Stoner excitations. For parabolic electron and magnon
spectra, neglecting the damping in the denominator of (\ref{G.43})
we obtain at low temperatures $T<\omega _{-}$ the dependence
$\delta \langle S^{z}\rangle _{cl}\propto U^{2}(T/\omega
_{+})^{2}$. For a weak ferromagnet, the temperature correction is
proportional to $(T/T_{C})^{2}$, in agreement with the
self-consistent
renormalization theory \cite{Moriya:1985}. It should be stressed that the $%
T^{2}$-correction obtained is much larger than the Stoner
contribution of the order of $(T/E_{F})^{2}$. The spin-wave
corrections to the local magnetic moment at a site $\langle
\mathbf{S}^{2}\rangle =(3/4)(n-2\langle \widehat{n}_{i\uparrow
}\widehat{n}_{i\downarrow }\rangle )$ at low $T\ll \omega _{-}$
have a weak dependence $-(T/T_{C})^{5/2}$ \cite{irkhin:7151}. This
justifies their neglecting in the above discussion of the
magnetization (\ref{G.41}).

At high $T>\omega _{-}$ the damping in the denominator dominates
at small $q$ in the case of a weak ferromagnet. Taking into
account (\ref{G.21}) we obtain from (\ref{G.43})
\begin{equation}
\delta \langle S_{-\mathbf{q}}^{-}S_{\mathbf{q}}^{+}\rangle =\frac{\Delta }{%
\pi U}\int_{-\infty }^{\infty }d\omega N_{B}(\omega )\int_{0}^{\infty }\frac{%
\omega Aqdq}{(\mathcal{D}q^{2})^{2}+A^{2}\omega ^{2}/q^{2}}\propto
\left( \frac{T}{E_{F}}\right) ^{4/3}  \label{G.51}
\end{equation}
Thus we get from (\ref{G.41}) the $T^{4/3}$-contribution to the
magnetization, which agrees with the result of the phase
transition scaling theory near $T=T_{C}$. For a ferromagnet with
well-localized magnetic
moments the damping may be neglected and we derive a small correction $%
\delta \langle S^{z}\rangle _{el}\propto -I^{2}\omega _{-}\ln
(T/\omega _{-})~$\cite{Irkhin:14008}.

Now we discuss  a more realistic situation in HMF, in particular
in  the Heusler alloys. These compounds demonstrate high values of
the saturation magnetization and Curie temperature (see
\cite{Irkhin:705,Dederichs2005,Ozdogan:2905,fecher:08J106}). The
strong ferromagnetism is mainly due to local moments of
well-separated Mn atoms. On the other hand, the highest magnetic
moment ($6\mu_B$) and Curie-temperature (1100K) in the classes of
Heusler compounds as well as half-metallic ferromagnets was
revealed for Co$_2$FeSi \cite{wurmehl:184434}. It was found
empirically that the Curie temperature of Co$_2$-based Heusler
compounds can be estimated from a nearly linear dependence on the
magnetic moment  \cite{fecher:08J106}.

High spin polarization and magnetic moment of half-metallic
ferromagnets can be treated within the generalized Slater-Pauling
rule \cite{ga.de.02,fecher:08J106}. According to the original
formulation by Slater and Pauling, the magnetic moments $m$ of 3d
elements and their binary compounds can be described by the mean
number of valence electrons $n_V$ per atom. A plot of $m$ versus
magnetic valence $m(n_M)$ is called the generalized Slater-Pauling
rule, as described by K\"ubler \cite{Kubler:257}. According to
Hund's rule it is often favorable that the majority d states are
fully occupied ($n_{d \uparrow}=5$). Starting from
$m=2n_{\uparrow} - n_V$, this leads to the definition of the
magnetic valence as $n_M=10 - n_V$, so that the magnetic moment
per atom is given by $m=n_M+2n_{sp\uparrow}$.

\begin{figure}[tbh]
\begin{center}
\includegraphics[width=0.75\columnwidth, angle=0]{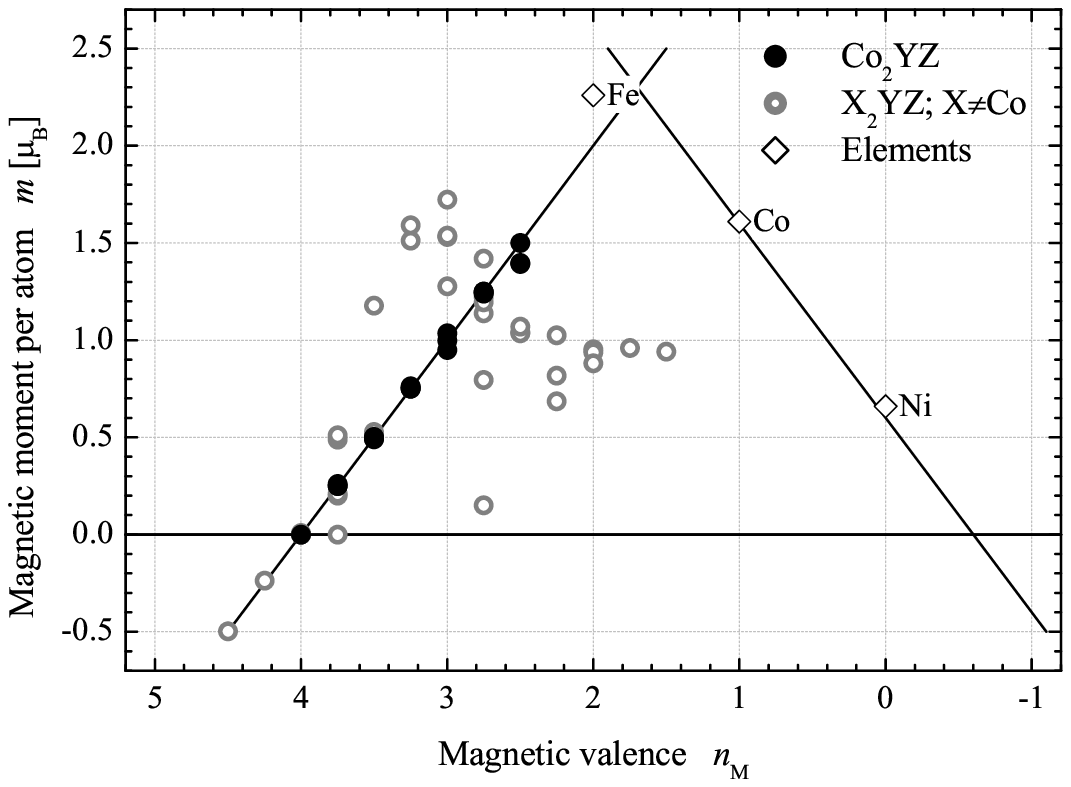}
\end{center}
\caption{Slater Pauling graph for Heusler compounds \cite{fecher:08J106}.}
\label{fig:sl-paul}
\end{figure}

In the case of localized moments, the Fermi energy is pinned in a
deep valley of the minority electron density. This constrains
$n_{d \downarrow}$ to be approximately 3, and $m = n_V - 6 -
2n_{sp\uparrow}$. HMF are supposed to exhibit a real gap in the
minority density of states where the Fermi energy is pinned. Then
the number of occupied minority states has to be an integer. Thus,
the Slater-Pauling rule will be strictly fulfilled with the spin
magnetic moment per atom $m = n_V - 6$. The situation for the HMF
and non-HMF full Heusler alloys is shown in Fig.\ref{fig:sl-paul}.

For ordered compounds with different kinds of atoms it may be more
convenient to consider the total spin magnetic moment  $M_t$ of
all atoms of the unit cell. This quantity scales with the number
of valence electron $Z_t$: $M_t= Z_t - 18$ for the half-Heusler
and $M_t= Z_t - 24$ for the full-Heusler alloys. Thus in both
types of compounds  the spin magnetic moment per unit cell is
strictly integer for the HMF situation. On the other hand, for
alloys with non-integer site occupancies like the quaternaries
X$_2$Y$_{1-x}$Y$_x$Z the moment may become non-integer depending on
the composition, even for the HMF state. The first-principle
calculations of the quaternary Heusler alloys
 Co$_2$[Cr$_{1-x}$Mn$_x$]Al, Co$_2$Mn[Al$_{1-x}$Sn$_x$] and
[Fe$_{1-x}$Co$_x$]$_2$MnAl \cite{Galanakis:3089}  demonstrated the
Slater-Pauling behavior and half-metallic properties. Moreover,
this behavior was investigated theoretically in V-based Heusler
alloys Mn$_2$VZ (Z$=$Al, Ga, In, Si, Ge, Sn) which are predicted
to demonstrate half-metallic ferrimagnetism
\cite{ru.pi.99,Ozdogan:2905,sa.ga.05}.

An interesting feature of the half-metallic Heusler alloys is that
the Rhodes-Wolfarth ratio $p_{C}/p_{s}$ ($p_{C}$ are the effective
moments, $ p_{s}$ the saturation moments) can be essentially
smaller than unity \cite {Irkhin:705}. Moreover, the effective
moment in the paramagnetic state, determined from the paramagnetic
susceptibility, decreases appreciably with temperature. We recall
that for the Heisenberg magnets (atomic-like magnetic states) we
have $p_{C}/p_{s}=1$ and for weak itinerant ferromagnets such as
ZrZn$_{2}$ $p_{C}/p_{s}\gg 1$, in accordance with the concept of
\textquotedblleft thermally induced local magnetic
moments\textquotedblright\ \cite{Moriya:1985}. In the case of
conventional strong itinerant ferromagnets (for example, Fe and
Ni) this ratio is also greater than unity (for detailed
discussion, see \cite{lichtenstein:067205} ). It therefore follows
that the inequality $p_{C}/p_{s}<1$ is a striking property of
HMF's, which could be used in their preliminary experimental
identification.

This behavior may be explained by that the change of electronic
structure. The temperature dependence of magnetic moment in the
paramagnetic state may be due to short-range magnetic order (local
densities of states are similar to those in the ferromagnetic
state). The numerical calculations demonstrate that the reduction
in the moments is a consequence of a change in the electron
structure as a results of rotation of the magnetic moments as it
was demonstrated for Fe and Ni in Ref.\cite{Turzhevskii:1952}. One
would expect such changes to be particularly large in the case of
HMF's and they should be of qualitative nature (smearing out of
the hybridization gap because of spin disorder, for a review, see
\cite{Irkhin:705}). From the many-electron model point of view,
the decrease of the local moment with increasing temperature is
connected with the absence of corrections to ground state
magnetization of the type (\ref{G.49}). However, such corrections
do occur at high temperatures.

\subsection{Nuclear magnetic relaxation}

\label{sec:nmr}

Nuclear magnetic resonance (NMR), which is one of most powerful
tools for investigating various physical properties, has a number
of peculiarities for magnetically ordered materials and especially
for HMF. The localized-spin Heisenberg model is inadequate to
describe the most interesting systems mentioned above where the
role of conduction electrons is essential in magnetic properties.
Usually the data on the longitudinal nuclear magnetic relaxation
rate $1/T_{1}$ are discussed within itinerant-electron models
such as Hubbard model or phenomenological spin-fluctuation theories \cite%
{Ueda:32,Moriya:1985,Ishigaki:3402,Moriya:1994,Millis:167}. On the
other hand, in the $s-d(f)$ exchange model (well-separated
localized and itinerant subsystems) magnetic properties differ
essentially from those in the paramagnon regime. We discuss the
contributions to $1/T_{1}(T)~$owing to electron-magnon interaction
for three- and two-dimensional ($3D$ and $2D$) metallic
ferromagnets with well-defined local magnetic moments with
especial attention to HMF case \cite{irkhin:401}.

The standard Hamiltonian of the hyperfine interaction, $\mathcal{H}_{hf}=%
\mathbf{hI,}$ ($h_{\alpha }=A_{\alpha \beta }S_{\beta }$,
$\widehat{A}$ is the hyperfine interaction matrix) contains the
Fermi (contact) and dipole-dipole contributions, $A_{\alpha \beta
}=A^{F}\delta _{\alpha \beta }+A_{\alpha \beta }^{dip}.$ According
to \cite{Abragam:1961} we have
\begin{eqnarray}
h^{-} &=&(A^{F}+\frac{1}{3}aF^{(0)})S^{-}+aF^{(2)}S^{+}+2aF^{(1)}S^{z}, \\
h^{z} &=&(A^{F}-\frac{2}{3}aF^{(0)})S^{z}+a (F^{(1)} S^{+} +
F^{(1)\ast} S^{-})
\end{eqnarray}
where
\begin{eqnarray}
F^{(0)} &=&\langle (1-3\cos ^{2}\theta )/r^{3}\rangle
,F^{(1)}=\langle \sin
\theta \cos \theta \exp (-i\phi )/r^{3}\rangle ,  \nonumber \\
F^{(2)} &=&\langle \sin ^{2}\theta \exp (-2i\phi )/r^{3}\rangle ,a=-\frac{3}{%
2}\gamma _{e}\gamma _{n},
\end{eqnarray}
$\langle ...\rangle $ is the average over the electron subsystem states, $%
\gamma _{e}$ and $\gamma _{n}$ are gyromagnetic ratios for
electron and nuclear moments, respectively. In the case of the
\textit{local} cubic symmetry we have $F^{(a)}=0.$ The Fermi
hyperfine interaction is proportional to the electron density at
the nucleus and therefore only $s$ states participate in it, the
contribution of core $s$ states (which are polarized due to local
magnetic moments) being much larger than of conduction electrons.
It is just the consequence of considerably smaller localization
area (and therefore higher density on nuclei) for the core states.
It is obvious that magnetic $f$ or $d$ electrons dominate also in
dipole interactions because of large spin polarization. Hence the
direct interaction of nuclear spins with that of conduction
electrons can be neglected in magnets with well-defined local
magnetic moments. Nevertheless, conduction electrons do effect
nuclear relaxation via their influence on the local-moment system;
besides that, as we shall see below, such contributions possess
large exchange enhancement factors.

Using the expressions for $1/T_{1}~$and linewidth $1/T_{2}$ in
terms of the Green's functions \cite{Moriya:516}
\begin{eqnarray}
\frac{1}{T_{1}} &=&-\frac{T}{2\pi }\text{\textrm{Im}}\sum_{\mathbf{q}%
}\langle \langle h_{\mathbf{q}}^{+}|h_{-\mathbf{q}}^{-}\rangle
\rangle
_{\omega _{n}}/\omega _{n}, \\
\frac{1}{T_{2}} &=&\frac{1}{2T_{1}}-\frac{T}{2\pi }\lim_{\omega
\rightarrow
0}\text{\textrm{Im}}\sum_{\mathbf{q}}\langle \langle h_{\mathbf{q}}^{z}|h_{-%
\mathbf{q}}^{z}\rangle \rangle _{\omega }/\omega
\end{eqnarray}
($\omega _{n}=\langle h^{z}\rangle \ll T$ is the NMR frequency) we
derive
\begin{eqnarray}
\frac{1}{T_{1}} &=&\frac{T}{2}\{[(A^{F}+\frac{1}{3}%
aF^{(0)})^{2}+a^{2}|F^{(2)}|^{2}]K^{+-}+4a^{2}|F^{(1)}|^{2}K^{zz}\}
\label{FK} \\
\frac{1}{T_{2}} &=&\frac{1}{2T_{1}}+\frac{T}{2}\{(A^{F}-\frac{2}{3}%
aF^{(0)})^{2}K^{zz}+a^{2}[2|F^{(1)}|^{2}K^{+-}]\}  \label{FK1}
\end{eqnarray}
\begin{equation}
K^{\alpha \beta }=-(1/\pi )\lim_{\omega \rightarrow 0}\text{\textrm{Im}}%
\sum_{\mathbf{q}}\langle \langle S_{\mathbf{q}}^{+}|S_{-\mathbf{q}%
}^{-}\rangle \rangle _{\omega }/\omega
\end{equation}

Passing to the magnon representation we obtain
\begin{equation}
\langle \langle S_{\mathbf{q}}^{+}|S_{-\mathbf{q}}^{-}\rangle
\rangle _{\omega }=2S/[\omega -\omega _{\mathbf{q}}+i\gamma
_{\mathbf{q}}(\omega )] \label{fgf}
\end{equation}
where $\omega _{\mathbf{q}}=2S(J_{\mathbf{q}}-J_{0})+\omega _{0}$
is the magnon frequency, $\gamma _{\mathbf{q}}(\omega )\propto
\omega $ is the magnon damping. Then we have
\begin{equation}
K^{+-}=\frac{2S}{\pi \omega _{n}}\sum_{\mathbf{q}}\frac{\gamma _{\mathbf{q}%
}(\omega _{n})}{\omega _{\mathbf{q}}^{2}}  \label{gfm}
\end{equation}
The damping owing to the one-magnon decay processes can be
represented as
\begin{equation}
\gamma _{\mathbf{q}}^{(1)}(\omega )=2\pi I^{2}S\omega \sum_{\mathbf{k}%
}\delta (t_{\mathbf{k\uparrow }})\delta (t_{\mathbf{k-q\downarrow
}}). \label{G1F}
\end{equation}
where the energy is referred to the Fermi level. The linearity of
spin fluctuation damping in $\omega $ is a characteristic property
of metals. According to (\ref{FK}) this leads to $T$-linear
contributions to $1/T_{1}$ which is the Korringa law. It is
important that the simplest expression for the Korringa relaxation
\begin{equation}
1/T_{1}\simeq 1/T_{2}\simeq A^{2}N_{\uparrow }(E_{F})N_{\downarrow
}(E_{F})T, \label{kor}
\end{equation}
($A$ is an effective hyperfine interaction constant) is
practically never applicable for magnetic metals: exchange
enhancement factors can change even the order of magnitude of
$1/T_{1}$ \cite{Moriya:1985,Irkhin:705}. Accurate expression for
the \textquotedblleft Korringa\textquotedblright\ contribution in
the case under consideration can be derived by the substitution
(\ref{gfm}) and (\ref{G1F}) into (\ref{FK}).

Apart from three-dimensional case we can consider also
two-dimensional HMF keeping in mind, e.g., layered CMR compounds
like LaSr$_{2}$Mn$_{2}$O$_{7}$
\cite{nagaev:387,Boer:10758}. According to (\ref{G.18}), the damping (\ref%
{G1F}) has the threshold value of $q,$ which is determined by the
spin splitting $\Delta =2|I|S$, $q^{\ast }=\Delta /v_{F}$ ($v_{F}$
is the electron velocity at the Fermi level), corresponding
characteristic
temperature and energy scale being $\omega _{-}\sim (\Delta /v_{F})^{2}T_{C}$%
. After integration for the parabolic electron spectrum the
one-magnon damping contribution to (\ref{gfm}) takes the form
\begin{equation}
\delta ^{(1)}K^{+-}=\frac{N_{\uparrow }(E_{F})N_{\downarrow }(E_{F})}{%
\mathcal{D}^{2}m^{2}}\times \left\{
\begin{tabular}{ll}
$1/4,$ & $D=3$ \\
$1/(\pi q^{\ast }),$ & $D=2$%
\end{tabular}
\ \ \right.  \label{t1f}
\end{equation}
with $m$ the electron effective mas. Thus in the $3D$ case the factor of $%
I^{2}$ is canceled, and the factor of $I^{-1}$ occurs in the $2D$
case, so that we obtain a strongly enhanced $T$-linear
Korringa-type term (remember that $\mathcal{D}\sim J\sim I^{2}/W$
for the RKKY interaction). This means that the contribution of
conduction electrons to $T$-linear relaxation rate via their
interaction with localized spins is indeed much more important
than the \textquotedblleft direct\textquotedblright\ contribution:
perturbation theory in the $s-d$ exchange coupling parameter $I$
turns out to be singular.

In HMF the one-magnon decay processes are absent and
electron-magnon (two-magnon) scattering processes should be
considered (Sect.\ref{magn}). Substituting the corresponding
damping into (\ref{gfm}) yields for $D=3$
\[
\delta ^{(2)}K^{+-}=\frac{\Omega _{0}T^{1/2}}{128\pi ^{2}Sm^{2}\mathcal{D}%
^{7/2}}\sum_{\sigma }N_{\sigma }^{2}(E_{F})\times \left\{
\begin{tabular}{ll}
$3\pi ^{1/2}\zeta (\frac{3}{2})T,$ & $T\ll \omega _{-}$ \\
$5.2\omega _{,}$ & $T\gg \omega _{-}$%
\end{tabular}
\ {}\right.
\]
where $\zeta (z)$ is the Riemann function, $\Omega _{0}$ the
lattice cell volume. This contribution can also modify
considerably the temperature dependence of $1/T_{1}$ in
\textquotedblleft usual" ferromagnets, a crossover from $T^{5/2}$
to $T^{3/2}$ dependence of the correction taking place.

For $D=2$ and $T,\omega _{-}\gg \omega _{0},$ small magnon momenta
of order of $\left( \omega _{0}/\mathcal{D}\right) ^{1/2}$ make
the main contribution
to (\ref{gfm}). On using the high-temperature expression $N_{\mathbf{p}%
}=T/\omega _{\mathbf{p}}$ one gets
\begin{equation}
\delta ^{(2)}K^{+-}=1.23\frac{\Omega _{0}^{3}k_{F}}{8\pi ^{4}S\mathcal{D}%
^{5/2}\omega _{0}^{1/2}}T
\end{equation}
Thus in the $2D$ FM case, in contrast with $3D$ one, the relaxation rate $%
1/T_{1}$ is strongly dependent on the anisotropy gap. It is
worthwhile to note an important difference between relaxation
processes via phonons and via magnons. The main difference is due
to the gap in magnon spectrum. Usually $\omega _{0}>\omega _{n}$
and therefore one-magnon processes contribute to the relaxation
rate due to magnon damping only (cf. discussion of the
phonon-induced relaxation processes in Ref.\cite{Abragam:1961}).
However, the mechanisms of magnon damping in magnetic dielectrics
(magnon-magnon interactions) are different from those in magnetic
metals and degenerate semiconductors
\cite{Auslender:301,Auslender:129}.

\subsection{Thermodynamic properties}

\label{sec:thermo}

Consider the renormalization of electronic specific heat in an
itinerant
ferromagnet due to interaction with spin fluctuations. Integration in (\ref%
{s1}), (\ref{s2}) at $T=0~$ gives
\[
\mathrm{Re}\Sigma _{\sigma }(k_{F\sigma },E)=-\frac{U\Delta
}{\omega _{+}-\omega _{-}}N_{-\sigma }(E_{F})\sum_{\alpha =\pm
}\alpha (E-\omega _{\alpha })\ln \frac{|E-\omega _{\alpha }|}{W}
\]%
Then the inverse residue of the electron Green's function, $1/Z_{\mathbf{{k}%
\sigma }}(E)=1-(\partial /\partial E)\mathrm{Re}\Sigma
_{\mathbf{k}\sigma }(E)$, which determines the renormalization of
the electron effective mass owing to the electron-magnon
interaction, contains a logarithmic factor. We obtain for the
coefficient at the linear term in the electronic specific heat at
$T\ll \omega _{-}$
\begin{equation}
\gamma _{\sigma }=\gamma _{\sigma }^{(0)}/Z_{\sigma }(k_{F\sigma },E_{F})=%
\frac{\pi ^{2}}{3}N_{\sigma }(E_{F})\left[ 1+\frac{U\Delta
}{\omega
_{+}-\omega _{-}}N_{-\sigma }(E_{F})\ln \frac{\omega _{+}}{\omega _{-}}%
\right]   \label{G.55}
\end{equation}%
For weak itinerant ferromagnets we have
\begin{equation}
\ln \frac{\omega _{+}}{\omega _{-}}\simeq -2\ln (UN(E_{F})-1)
\label{G.56}
\end{equation}%
so that the we have a paramagnon enhancement of the specific heat,
the numerical factor in (\ref{G.55}) being inexact in this limit
because of neglecting longitudinal spin fluctuations (see
\cite{Moriya:1985}). On the other hand, we have a considerable
enhancement of specific heat owing to spin fluctuations strong
ferromagnets which is really observed in a number of systems.

Other thermodynamic properties may be treated by calculating the
free energy of the system. The spin-wave contribution to the free
energy has the form, usual for the Bose excitations with the
square dispersion law and is proportional to $(T/T_{C})^{5/2}$. At
low $T<\omega _{-}$ the many-electron (branch cut) contribution
reads

\begin{equation}
F_{\mathrm{el}}=\frac{1}{2S_{0}}\sum_{\mathbf{q>q}^{\ast }}\omega _{\mathbf{q%
}}\langle S_{-\mathbf{q}}^{-}S_{\mathbf{q}}^{+}\rangle \,\simeq
U\Delta
\sum_{\mathbf{kk}^{\prime }}\frac{n_{\mathbf{k}^{\prime }\downarrow }(1-n_{%
\mathbf{k\uparrow }})}{t_{\mathbf{k\uparrow
}}-t_{\mathbf{k}^{\prime }\downarrow }+\omega
_{\mathbf{k-k}^{\prime }}}  \label{G.57}
\end{equation}%
%
%
%
%
%
Differentiating (\ref{G.57}) over $T$ one obtains

\begin{eqnarray}
\delta C_{\mathrm{el}} &=&-\frac{\partial }{\partial T}\delta
F_{el}(T)
\nonumber \\
&=&U^{2}\frac{2\langle S^{z}\rangle }{\omega _{+}-\omega
_{-}}N_{\uparrow
}(E_{F})N_{\downarrow }(E_{F})\frac{2\pi ^{2}}{3}T\ln \frac{\omega _{+}}{%
\max (\omega _{-},T)}
\end{eqnarray}%
Thus at $T\gg \omega _{-}$ we have instead of (\ref{G.55}) the $T\ln T$%
-dependence of specific heat.

As one can see from (\ref{G.55}), the enhancement of effective
mass and electronic specific heat owing to spin fluctuations is
absent in the half-metallic state. We shall demonstrate that the
specific heat of a conducting ferromagnet may contain
spin-fluctuation contributions of another nature. Write down a
general expression for the specific heat in the $s-d$ exchange
model in terms of the total energy

\begin{eqnarray}
C(T) &=&\frac{\partial \langle \mathcal{H}\rangle }{\partial T}=\frac{%
\partial }{\partial T}\int dEEf(E)N_{t}(E)  \nonumber \\
&=&\frac{\pi ^{2}}{3}N_{t}(E)T+\int dEEf(E)\frac{\partial }{\partial T}%
N_{t}(E,T)  \label{G.65}
\end{eqnarray}
where

\[
N_{t}(E)=-\frac{1}{\pi }\sum_{\mathbf{k}\sigma }\text{\textrm{Im}}G_{\mathbf{%
k}\sigma }(E)
\]
The first term in the right-hand side of (\ref{G.65}) yields the
standard result of the Fermi-liquid theory. The second term is due
to the energy dependence of the density of states. Such a
dependence occurs in the conducting ferromagnet owing to
non-quasiparticle (incoherent) states. Using again the expressions
for the self-energies (\ref{s1}), (\ref{s2}) we derive
\cite{irkhin:7151}
\begin{equation}
\delta C_{\sigma }(T)=2\sigma I^{2}\langle S^{z}\rangle \sum_{\mathbf{kq}}%
\frac{f(t_{\mathbf{k+q},-\sigma }-\sigma \omega _{\mathbf{q}})}{(t_{\mathbf{%
k+q},-\sigma }-t_{\mathbf{k},\sigma })^{2}}\frac{\partial }{\partial T}n_{%
\mathbf{k+q},-\sigma }
\end{equation}
Since at low temperatures $f(t_{\mathbf{k+q},\downarrow }-\omega _{\mathbf{q}%
})=1,f(t_{\mathbf{k+q},\uparrow }-\omega _{\mathbf{q}})=0$, the
non-quasiparticle states with $\sigma =\downarrow $ do not
contribute to linear specific heat since they are empty at
$T$=0~K. In the half-metallic state the non-quasiparticle
contributions (\ref{G.65}) with $\sigma =\uparrow $ are present
for $I<0$ only, and we obtain
\begin{equation}
\delta C_{\uparrow }(T)=\frac{2\pi ^{2}}{3}I^{2}\langle
S^{z}\rangle
N_{\downarrow }(E_{F})T\sum_{\mathbf{k}}\frac{1}{(t_{\mathbf{k\uparrow }%
}-E_{F})^{2}}
\end{equation}
To avoid misunderstanding, it should be stressed that presence of
such contributions to specific heat means inapplicability of the
Fermi-liquid description in terms of dynamical quasiparticles
only, which are determined by poles of Green's functions. It may
be shown rigorously that the entropy of interacting Fermi systems
at low $T$ is expressed in terms of Landau quasiparticles with the
energies, determined as variational derivatives of the total
energy with respect to occupation numbers \cite{Carneiro:1106}.
Thus, even in the presence of non-pole contributions to the
Green's functions, the description of thermodynamics in terms of
statistical quasiparticles \cite{Carneiro:1106} holds. (However,
the quasiparticle description is insufficient for spectral
characteristics, e.g., optical and emission data.) The anomalous
$\gamma T$-term is determined by the difference of the spectra of
statistical and dynamical quasiparticles.

Similar contributions to specific heat in the Hubbard model with
strong correlations are discussed in the paper \cite{irkhin:7151}
too. They dominate in the enhancement of specific heat for
half-metallic ferromagnets and may be important, besides the
effective mass enhancement (\ref{G.55}), for \textquotedblleft
usual\textquotedblright\ magnets with well-defined local moments.

\subsection{Transport properties}

\label{sec:transp}

Transport properties of HMF are the subject of numerous
experimental investigations (see, e.g., recent works for CrO$_2$
\cite{rabe:7} and NiMnSb
\cite{Borca:052409}, and the reviews \cite{Irkhin:705,ziese:143,nagaev:387}%
). At the same time, the theoretical interpretation of these
results is still a problem. As for electronic scattering
mechanisms, the most important difference between HMF and
``standard" itinerant electron ferromagnets like iron or nickel is
the absence of one-magnon scattering processes in the former case
\cite{Irkhin:705}.

Since the states with one spin projection only exist at the Fermi
level and one-magnon scattering processes are forbidden in the
whole spin-wave region, the corresponding $T^{2}$-term in
resistivity is absent in the
case of a half-metallic ferromagnets. This seems to be confirmed
by comparing experimental data on
resistivity of Heusler alloys TMnSb (T = Ni, Co, Pt, Cu, Au) and PtMnSn \cite%
{otto:2351} (see also discussion in Sect. \ref{sec:NiMnSb1}). 
The $T^{2}$-contribution from one-magnon processes to
resistivity for half-metallic systems (T = Ni, Co, Pt) was really
not picked out, whereas the dependences $\rho (T)$ for
\textquotedblleft usual\textquotedblright\ ferromagnets were
considerably steeper.

Two-magnon scattering processes have been considered many years
ago, the
temperature dependence of resistivity obtained being $T^{7/2}$ \cite%
{Roesler:K31,Hartman:114}. The obtained temperature dependence of
the resistivity has the form $T^{7/2}$. At low enough temperatures
the first result fails and should be replaced by $T^{9/2}$
\cite{lutovinov:707}; the reason is the compensation of the
transverse and longitudinal contributions in the long-wavelength
limit which is a consequence of the rotational symmetry of the
$s-d$ exchange Hamiltonian
\cite{Grigin:65,nagaev:1983,Auslender:309}. We discuss effects of
interaction of current carriers with local moments are
investigated in the standard $s-d$ exchange model in the spin-wave
region.

\begin{eqnarray}
\mathcal{H} &=&\mathcal{H}_{0}-I(2S)^{1/2}\sum_{\mathbf{kq}}(c_{\mathbf{k}%
\uparrow }^{\dagger }c_{\mathbf{k}+\mathbf{q}\downarrow }b_{\mathbf{q}%
}^{\dagger }+h.c.)  \nonumber \\
&&\ \ \ \ \ +I\sum_{\mathbf{kqp}\sigma }\sigma c_{\mathbf{k}\sigma
}^{\dagger }c_{\mathbf{k+q-p}\sigma }b_{\mathbf{q}}^{\dagger
}b_{\mathbf{p}}
\end{eqnarray}
The zero-order Hamiltonian includes non-interacting electrons and
magnons,
\begin{eqnarray}
\mathcal{H}_{0} &=&\sum_{\mathbf{k}\sigma }E_{\mathbf{k}\sigma }c_{\mathbf{k}%
\sigma }^{\dagger }c_{\mathbf{k}\sigma }+\sum_{\mathbf{q}}\omega _{\mathbf{q}%
}b_{\mathbf{q}}^{\dagger }b_{\mathbf{q}}, \\
E_{\mathbf{k}\sigma } &=&t_{\mathbf{k}}-\sigma \Delta /2,\omega _{\mathbf{q}%
}=2S(J_{0}-J_{\mathbf{q}}),  \nonumber
\end{eqnarray}
with $\Delta =2IS$ being the spin splitting which is included in $\mathcal{H}%
_{0}$. In the half-metallic case the spin-flip processes do not
work in the second order in $I$ since the states with one spin
projection only are present at the Fermi level. At the same time,
we have to consider the renormalization of the longitudinal
processes in higher orders in $I$ (formally, we have to include
the terms up to the second order in the quasiclassical small
parameter $1/S$). To this end we can eliminate from the
Hamiltonian the terms which are linear in the magnon operators by
using the canonical transformation \cite{Grigin:65,nagaev:1983}.
Then we obtain the effective Hamiltonian
\begin{equation}
\widetilde{\mathcal{H}}=\mathcal{H}_{0}+\frac{1}{2}\sum_{\mathbf{kqp}\sigma
}(\mathcal{A}_{\mathbf{kq}}^{\sigma
}+\mathcal{A}_{\mathbf{k+q-p,q}}^{\sigma
})c_{\mathbf{k}\sigma }^{\dagger }c_{\mathbf{k+q-p}\sigma }b_{\mathbf{q}%
}^{\dagger }b_{\mathbf{p}}^{{}}  \label{hef}
\end{equation}
Here $\mathcal{A}_{\mathbf{kq}}^{\sigma }$ is the $s-d$ scattering
amplitude which is defined by (\ref{amp}) ($U\rightarrow I$). More
general interpolation expression for the effective amplitude,
which does not assume the smallness of $|I|$ or $1/2S$ was
obtained in Ref.\cite{Auslender:309} by a variational approach.

The most general and rigorous method for calculating the transport
relaxation time is the use of the Kubo formula for the
conductivity $\sigma _{xx}$ \cite{kubo:570} (see details in
Ref.\cite{irkhin:481}):

\begin{equation}
\sigma _{xx}=\beta \int_0^\beta d\lambda \int_0^\infty dt\exp
(-\varepsilon t)\langle j_x(t+i\lambda )j_x\rangle  \label{kru}
\end{equation}
where $\beta =1/T,$ $\varepsilon \rightarrow 0,$
\begin{equation}
\mathbf{j}=-e\sum_{\mathbf{k}\sigma }\mathbf{v}_{\mathbf{k}\sigma }c_{%
\mathbf{k}\sigma }^{\dagger }c_{\mathbf{k}\sigma }
\end{equation}
is the current operator, $\mathbf{v}_{\mathbf{k}\sigma }=\partial E_{\mathbf{%
k}\sigma }/\partial \mathbf{k}$ is the electron velocity.
Representing the total Hamiltonian in the form
$\mathcal{H}=\mathcal{H}_0+\mathcal{H}^{\prime
}$, the correlator in (\ref{kru}) may be expanded in the perturbation $%
\mathcal{H} ^{\prime }$ \cite{nakano:145,mori:399}. In the second
order we obtain for the electrical resistivity
\begin{equation}
\rho _{xx}=\sigma _{xx}^{-1}=\frac T{\langle j_x^2\rangle
^2}\int_0^\infty dt\langle [j_x,\mathcal{H}^{^{\prime
}}(t)][\mathcal{H}^{^{\prime }},j_x]\rangle
\end{equation}
where $\mathcal{H}^{\prime }(t)$ is calculated with the Hamiltonian $%
\mathcal{H}_0$.

In the HFM situation the band states with one spin projection
only, $\sigma = \mathrm{sign}I,$ are present at the Fermi level
\cite{Irkhin:705}. We consider the case $I>0,$ $\sigma =+$ and
omit the spin indices in the electron spectrum to obtain for the
transport relaxation time $\tau $
defined by $\sigma _{xx}=e^2\langle (v^x)^2\rangle \tau $%
\begin{eqnarray}
\frac 1\tau &=&\frac \pi {4T}\sum_{\mathbf{kk}^{\prime }\mathbf{q}}(v_{%
\mathbf{k}}^x-v_{\mathbf{k}^{\prime }}^x)^2(\mathcal{A}_{\mathbf{kq}%
}^{\uparrow }+\mathcal{A}_{\mathbf{k}^{\prime }\mathbf{,q-k}^{\prime }%
\mathbf{+k}}^{\uparrow })^2N_{\mathbf{q}}(1+N_{\mathbf{q-k}^{\prime }+%
\mathbf{k}})  \nonumber \\
&&\ \ \ \ \ \ \ \ \ \times n_{\mathbf{k}}(1-n_{\mathbf{k}^{\prime
}})
\nonumber \\
&&\times \delta (t_{\mathbf{k}^{\prime }}-t_{\mathbf{k}}-\omega _{\mathbf{q}%
}+\omega _{\mathbf{q-k}^{\prime }+\mathbf{k}})\left/ \sum_{\mathbf{k}}(v_{%
\mathbf{k}}^x)^2\delta (t_{\mathbf{k}})\right.  \label{tau1}
\end{eqnarray}
Averaging over the angles of the vector $\mathbf{k}$ leads to the result $%
1/\tau \propto I^2\Phi $ with

\begin{equation}
\Phi =\sum_{\mathbf{pq}}f_{\mathbf{pq}}\frac{\beta (\omega _{\mathbf{p}%
}-\omega _{\mathbf{q}})|\mathbf{p-q}|}{\exp \beta \omega
_{\mathbf{p}}-\exp \beta \omega
_{\mathbf{q}}}(1+N_{\mathbf{q}})(1+N_{\mathbf{p}})  \label{lam}
\end{equation}
where $f_{\mathbf{pq}}=1$ for $p,q\gg q^{\ast }$ and

\begin{equation}
f_{\mathbf{pq}}=\frac{[\mathbf{p\times
q]}^{2}}{(\mathbf{p-q)}^{2}(q^{\ast })^{2}}\,\,\,(p,q\ll q^{\ast
}).
\end{equation}
As discussed before, the wavevector $q^{\ast }$ determines the
boundary of the region where $\mathbf{q}$-dependence of the
amplitude become important,
so that $t(\mathbf{k+q})-t(\mathbf{k})\simeq \Delta $ at $q\simeq q^{\ast }$%
. In the case $q<q^{\ast }$ the simple perturbation theory fails
and we have to take into account the spin splitting by careful
collecting the terms of
higher orders in $I$. In the simple one-band model of HMF where $%
E_{F}<\Delta $ one has $q^{\ast }\sim \sqrt{\Delta /W}$ ($W$ is
the conduction bandwidth) \cite{Grigin:65,nagaev:1983}. The
quantity $q^{\ast }$
determines a characteristic temperature and energy scale $T^{\ast }=\mathcal{%
D}(q^{\ast })^{2}\propto \mathcal{D}({\Delta /}W).~$

When estimating temperature dependences of resistivity one has to
bear in mind that each power of $p$ or $q$ gives $T^{1/2}.$ At
very low temperatures $T<T^{\ast }$ small quasimomenta
$p,q<q^{\ast }$ yield main contribution to the integrals and
\begin{equation}
\rho (T)\propto (T/T_{C})^{9/2}
\end{equation}
Such a dependence was obtained in the narrow-band case
(double-exchange
model with large $|I|$), where the scale $T^{\ast }$ is absent \cite{kubo:21}%
, and by the diagram approach in the broad-band case
\cite{lutovinov:707}.
At the same time, at $T>T^{\ast }$ the function $f_{\mathbf{pq}}$ in Eq. (%
\ref{lam}) can be replaced by unity to obtain
\begin{equation}
\rho (T)\propto (T/T_{C})^{7/2}
\end{equation}

Generally speaking, $q^{\ast }$ may be sufficiently small provided
that the energy gap is much smaller than $W$, which is the case
for real HMF systems. We consider the model of HMF spectrum
(\ref{DOSbaredn}) where the majority-spin band is metallic and the
minority-spin is semiconducting. For the temperatures $T\ll T_{C}$
both characteristic $q$ and $p$ are small in comparison with the
inverse lattice constant and we can put
\begin{equation}
\mathcal{A}_{\mathbf{kq}}^{\sigma }=\frac{1}{2S}\left( t_{\mathbf{k}%
\downarrow }-t_{\mathbf{k}\uparrow }\right)
\frac{t_{\mathbf{k+q}\uparrow }-t_{\mathbf{k}\uparrow
}}{t_{\mathbf{k+q}\uparrow }-t_{\mathbf{k}\downarrow }}
\label{amp11}
\end{equation}
The wavevector $q^{\ast }$ determines the boundary of the region where $%
\mathbf{q}$-dependence of the amplitude become important, so that
$q^{\ast }=\Delta /v_{F}$ (the same value as for the spin
polarization problem). The corresponding characteristic
temperature and energy scale is
\begin{equation}
T^{\ast }=\mathcal{D}(q^{\ast })^{2}\propto \mathcal{D}({\Delta
/}W)^{2} \label{T*fmm}
\end{equation}
which coincides with the case of an usual ferromagnetic metal. The
above temperature dependences of resistivity are not changed
\cite{irkhin:481}.

Now we treat the two-dimensional ($2D$) situation which may be
appropriate for layered manganites \cite{nagaev:387,Boer:10758}.
At low temperatures we obtain
\begin{equation}
\rho (T<T^{\ast })\propto (T/T_{C})^{7/2}
\end{equation}
At the same time, for $T>T^{\ast }$ we obtain after replacing the
scattering amplitude by unity a logarithmically divergent integral
which should be cut at $T^{\ast }$. Thus we get
\begin{equation}
\rho (T>T^{\ast })\propto (T/T_{C})^{5/2}\ln (T/T^{\ast })
\end{equation}

To discuss the magnetoresistivity we have to introduce the gap in
the magnon spectrum, $\omega _{\mathbf{q\rightarrow
}0}=\mathcal{D}q^{2}+\omega _{0}.$ Provided that the external
magnetic field $H$ is large in comparison with the anisotropy gap,
$\omega _{0}$ is proportional to $H$. In the $3D$ case the
resistivity at $T<T^{\ast }$ is linear in magnetic field,
\begin{equation}
\rho (T,H)-\rho (T,0)\propto -\omega _{0}T^{7/2}/T_{C}^{9/2}
\end{equation}
The situation at $T>T^{\ast }$ is more interesting since the quantity $%
\partial \Phi /\partial \omega _{0}$ contains a divergence which is cut at $%
\omega _{0}$ or $T^{\ast }$. We have at $T>\omega _{0},T^{\ast }$

\begin{equation}
\delta \rho (T,H)\propto -\frac{T^3\omega _0}{[\max
(\omega_0,T^{*})]^{1/2}}
\end{equation}
(of course, at $T<\omega _0$ the resistivity is exponentially
small). A
negative $H$-linear magnetoresistance was observed recently in CrO$_2$ \cite%
{rabe:7}.

The discovery of giant magnetoresistance (GMR) lead to tremendous
activity to understand and develop technology based on
high-density magnetic recording \cite{zutic:323}. Giant
magnetoresistance for metallic multilayers (superlattices)
containing HMF was first predicted in \cite{Irkhin:705}.
NiMnSb-based spin-valve structures using Mo spacer layers
NiMnSb/Mo/NiMnSb/SmCo$_{2}$ were successfully produced
\cite{Hordequin:225}. The associated GMR exhibits a clear
spin-valve contribution about $\Delta R/R\approx 1\%$
\cite{Hordequin:225}. One of the limiting factor for such a small
value is the large resistivity of the Mo layer which determines
limited flow of active electrons exchanged between the two
ferromagnetic layers without being scattered. The giant
\textit{tunneling} magnetoresistance differs by use of dielectric
spacer instead of metallic one. The GMR in tunnel junctions based on
HMF was considered theoretically in Ref. \cite{Rob83,Tkachov:024519} and 
recently this issue became a subject of intensive experimental 
investigations \cite{sakuraba:052508,sa.ha.06,sa.ha.07,gercsi:082512,rybchenko:132509}.

It is important that NQP states do \textit{not} contribute to the
temperature dependence of the resistivity for pure HMF. An
opposite conclusion was made by Furukawa \cite{Furukawa:1954}. He
attempted to calculate low-temperature resistivity of half-metals
with account of the non-rigid-band behavior of the minority band
due to spin fluctuations at finite temperatures and derived that
the unconventional one-magnon scattering process give
$T^{3}$-dependence in resistivity. However, this calculation was
not based on a consistent use of the Kubo formula and, in our
opinion, can be hardly justified.

On the contrary, \textit{impurity} contributions to transport
properties in the presence of potential scattering are determined
mainly by the NQP states
\cite{Irkhin:1733,Irkhin:397,Irkhin:705}). To second order in the
impurity potential $\mathcal{U}$ we derive after neglecting vertex
corrections and averaging over impurities we obtain for the
transport relaxation time
\begin{equation}
\delta \tau _{\mathrm{imp}}^{-1}(E)=-2\mathcal{U}^2\mathrm{Im}\sum_{\mathbf{p%
}}G_{\mathbf{p}\sigma }^{(0)}(E)
\end{equation}
where is the exact Green's function for the ideal crystal. Thus
the contributions under consideration are determined by the energy
dependence of the density of states $N(E)$ for the interacting
system near the Fermi level. The most nontrivial dependence comes
from the non-quasiparticle (incoherent) states with the spin
projection $-\sigma =- \mathrm{sign}I,$ which are present near
$E_F$. Near the Fermi level the NQP contribution is determined by
the magnon density of states $g(\omega )$ and follows a power law,
\begin{equation}
\delta N_{\mathrm{incoh}}(E)\propto \int_0^{\sigma E}d\omega
g(\omega )\propto |E|^\alpha \theta (\sigma E)\,\,(|E|\ll
\overline{\omega }).
\end{equation}
Here $\theta (x)$ is the step function, $E$ is referred to $E_F$; we have $%
\alpha =3/2$ and $\alpha =1$ for $3D$ and $2D$ cases,
respectively. The corresponding correction to resistivity reads
\begin{eqnarray}
\frac{\delta \rho _{\mathrm{imp}}(T)}{\rho ^2} &=&-\delta \sigma _{\mathrm{%
imp}}(T)  \label{rimp} \\
\ &\propto &-\mathcal{U}^2\int dE\left( -\frac{\partial f(E)}{\partial E}%
\right) \delta N_{\mathrm{incoh}}(E)\propto T^\alpha  \nonumber
\end{eqnarray}
The contribution of the order of $T^\alpha $ with $\alpha \simeq
1.65$ (which is not too far from 3/2) was observed in the
temperature dependence of the resistivity for NiMnSb
\cite{Borca:052409} above 100K.  The
half-metallic properties above 100K are being challanged, however.
The incoherent contribution to
magnetoresistivity is given by
\begin{equation}
\delta \rho _{\mathrm{imp}}(T,H)\propto \omega _0\partial \delta N_{\mathrm{%
incoh}}(\sigma T)/\partial T\propto \omega _0T^{\alpha -1},
\end{equation}
so that we obtain a temperature-independent term in the $2D$ case.

The correction to thermoelectric power, which is similar to
(\ref{rimp}), reads (cf. Ref.\cite{Irkhin:1733,Irkhin:397}):
\begin{equation}
\delta \mathcal{Q}(T)\propto \frac{1}{T}\int dE(-\partial
f(E)/\partial E)E\delta N(E)  \label{qe}
\end{equation}
Besides that, an account of higher orders in impurity scattering
leads to the replacement of the impurity potential $V$ by the
$T$-matrix. For the point-like scattering the latter quantity is
given by
\begin{equation}
T(E)=\frac{\mathcal{U}}{1-\mathcal{U}\mathcal{R}(E)},\mathcal{R}(E)=\sum_{%
\mathbf{k}}G_{\mathbf{k}\sigma }(E).  \label{te}
\end{equation}
Expanding (\ref{te}) yields also the term
\begin{equation}
\delta \mathcal{Q}(T)\propto \frac{1}{T}\int dE(-\partial f(E)/\partial E)E%
\mathrm{Re}\delta \mathcal{R}(E)  \label{qte}
\end{equation}
with $\delta \mathcal{R}(E)$ being obtained by analytical continuation from $%
\delta N(E).$ Thus we have $\delta \mathcal{Q}(T)\propto T^{3/2}$.

\subsection{X-ray absorption and emission spectra. Resonant x-ray scattering}

\label{sec:xray}

\bigskip The NQP contributions in the presence of the potential $U,$ that is
induced by the impurity at a lattice site, can be treated in the
$s-d$ exchange model in the representation (\ref{eigen}). The
impurity potential
results in the NQP contribution to this quantity being enhanced for $%
\mathcal{U}<0$ and suppressed for $\mathcal{U}>0$. These results
can be used to consider the manifestations of NQP states in the
core level spectroscopy \cite{Irkhin:479}.

Various spectroscopy techniques such as x-ray absorption, x-ray
emission, and photoelectron spectroscopies (xas, xes, and xps,
correspondingly) give an important information about the
electronic structure of HMF and related compounds, i.e.
ferromagnetic semiconductors and colossal magnetoresistance
materials (see, e.g., Refs. \cite%
{Yarmoshenko:1,Yablonskikh:235117,kurmaev:155105,wessely:235109}).
It is well known \cite{mahan:1990} that many-body effects (e.g.,
dynamical core hole screening) can be important for the core level
spectroscopy even when the system is not strongly correlated in
the initial state. Therefore it is very interesting to study the
interplay of these effects and NQP states which are of essentially
many-body origin themselves.

To consider the core level problem in HMF we use the Hamiltonian
of the $s-d$ exchange model in the presence of the external
potential $\mathcal{U}$ induced by the core hole,

\begin{equation}
\mathcal{H}^{\prime }=\varepsilon _{0}f^{\dagger }f+\mathcal{U}\sum_{\mathbf{%
kk}^{\prime }\sigma }c_{\mathbf{k}\sigma }^{\dagger
}c_{\mathbf{k}^{\prime }\sigma }f^{\dagger }f  \label{Hi}
\end{equation}
where $f^{\dagger },f$ are core hole operators, $\mathcal{U}<0$.
X-ray
absorption and emission spectra are determined by the Green's function \cite%
{mahan:1990}
\begin{equation}
G_{\mathbf{kk}^{\prime }}^{\sigma }(E)=\langle \langle
c_{\mathbf{k}\sigma }f|f^{\dagger }c_{\mathbf{k}^{\prime }\sigma
}^{\dagger }\rangle \rangle _{E}
\end{equation}
As follows the investigation of the sequence of equations of motion \cite%
{Irkhin:479}, in the ladder approximation the spectral density for
two-particle Green's function $G_{\mathbf{kk}^{\prime }}^{\sigma
}(E)~$is equivalent to the one-particle spectral density in the
presence of the core hole potential $\mathcal{U}$ (note that the
ladder approximation is inadequate to describe the xas edge
singularity in a close vicinity of the Fermi level
\cite{mahan:1990}). Thus the core hole problem is intimately
connected with the impurity problem.

\begin{figure}[tbp]
\includegraphics[clip]{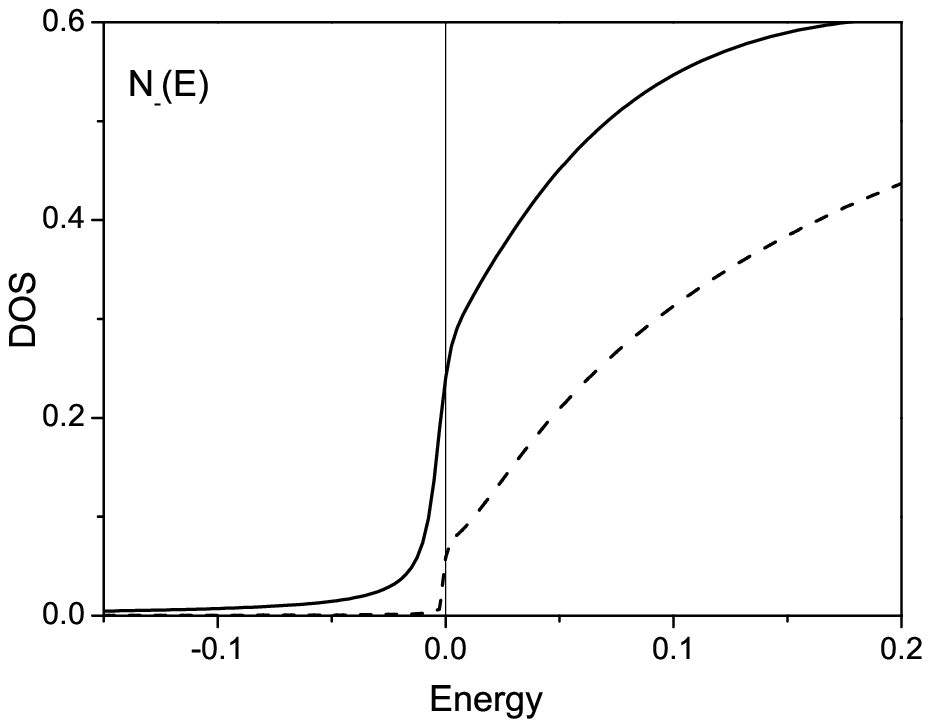}
%
\caption{The local density of states $N_{\mathrm{loc}}^{\downarrow
}(E)$
(solid line) for a half-metallic ferromagnet with $S=1/2,I=0.3,\protect%
\delta =0.01 $ in the presence of the core hole potential
$\mathcal{U}=-0.2.$ The dashed line shows the DOS $N_{\downarrow
}(E)$ for the ideal crystal. The value of $E_F$ calculated from
the band bottom is 0.15.} \label{fig:x3}
\end{figure}

\begin{figure}[tbp]
\includegraphics[clip]{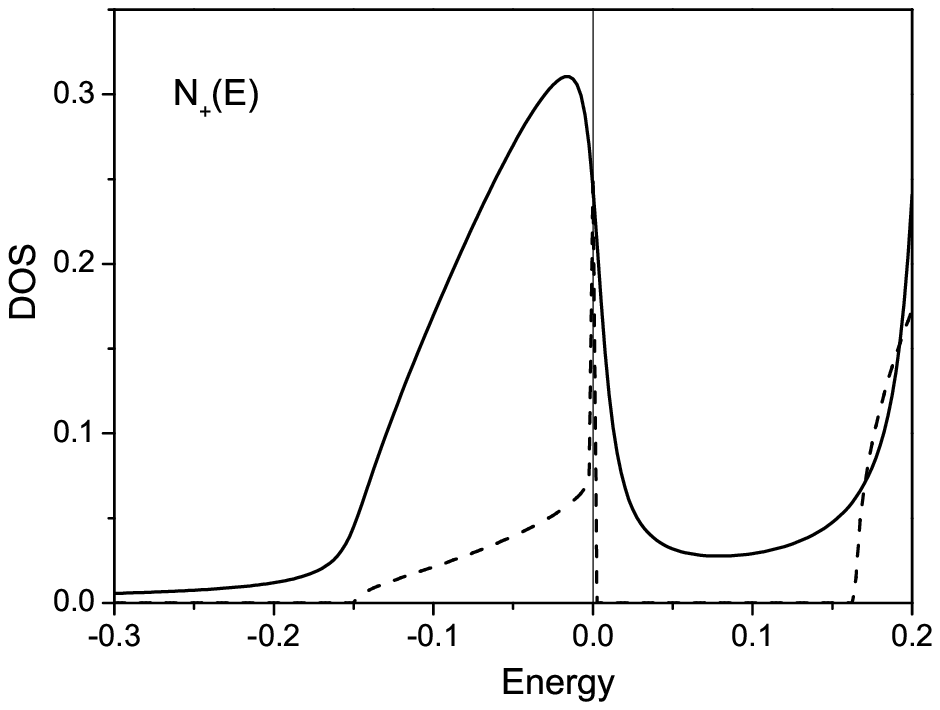}
\caption{The local density of states $N_{\mathrm{loc}}^{\uparrow
}(E)$
(solid line) for a half-metallic ferromagnet with $S=1/2,I=-0.3,\protect%
\delta =0.01 $ in the presence of the core hole potential
$\mathcal{U}=-0.3.$ The dashed line shows the DOS $N_{\uparrow
}(E)$ for the ideal crystal. The value of $E_F$ calculated from
the band bottom is 0.15.} \label{fig:x4}
\end{figure}

Since xas probes empty states and xes occupied states, the local
DOS
\begin{equation}
N_{\mathrm{loc}}^{\sigma }(E)=-\frac{1}{\pi
}\mathrm{Im}G_{00}^{\sigma }(E) \label{gsp}
\end{equation}
describes the absorption spectrum for $E>E_{F}$ and emission spectrum for $%
E<E_{F}.$ To take into account the core level broadening a finite damping $%
\delta $ should be introduced \cite{Irkhin:479}. For small band
filling the \textquotedblleft exciton effects\textquotedblright\
(strong interaction with the core hole) result in a considerable
enhancement of NQP contributions to the spectra in comparison with
those to DOS. The results for semielliptic bare band are shown in
Figs.\ref{fig:x3},\ref{fig:x4}.

To probe the ``spin-polaron'' nature of the NQP states more
explicitly, it would be desirable to use spin-resolved
spectroscopical methods such as
x-ray magnetic circular dichroism (XMCD, for a review see Ref. \cite%
{ebert:1665}). Owing to interference of electron-magnon scattering
and ``exciton'' effects (interaction of electrons with the core
hole), the NQP contributions to x-ray spectra can be considerably
enhanced in comparison with those to DOS of the ideal crystal.
Thus the core level (x-ray absorption, emission and photoelectron)
spectroscopy might be an efficient tool to investigate the NQP
states in electron energy spectrum.

Now we consider NQP effects in resonant x-ray scattering
processes. It was observed recently \cite{kurmaev:155105} that the
elastic peak of the x-ray scattering in CrO$_2$ is observed which
is more pronounced than in usual Cr compounds, e.g., in elemental
chromium. The authors of this work have put forward some
qualitative arguments that the NQP states may give larger
contributions to resonant x-ray scattering than usual itinerant
electron states. Here we can treat this question quantitatively
and estimate explicitly the corresponding enhancement. The
intensity of resonant x-ray emission induced by the photon with
the energy $\omega $ and polarization $q$ is given by the
Kramers-Heisenberg formula
\begin{equation}
I_{q^{\prime }q}(\omega ^{\prime },\omega )\propto \sum_n\left| \sum_l\frac{%
\langle n|C_{q^{\prime }}|l\rangle \langle l|C_q|0\rangle
}{E_0+\omega ^{\prime }-E_l-i\Gamma _l}\right| ^2\delta
(E_n+\omega ^{\prime }-E_0-\omega )  \label{xray}
\end{equation}
Here $q^{\prime }$ and $\omega ^{\prime }$ are the polarization
and energy of the emitted photon, $|n\rangle ,|0\rangle $ and
$|l\rangle $ are the final, initial and intermediate states of the
scattering system, respectively, $E_i$ are the corresponding
energies, $C_q$ is the operator of
the dipole moment for the transition, which is proportional to $%
fc+c^{\dagger }f^{\dagger }$. Assuming for simplicity that $\Gamma
_l$ does not depend on the intermediate state, $\Gamma _l=\Gamma
$, and taking into account only the main x-ray scattering channel
(where the hole is filled from the conduction band) one obtains
\cite{Sokolov:383}

\begin{equation}
I_{\omega ^{\prime }}\propto \left| \sum_\sigma G_{00}^\sigma
(z)\right| ^2 \label{zz}
\end{equation}
where $z=\omega ^{\prime }-E_0+i\Gamma .$ Owing to a jump in the
DOS at the Fermi level, the NQP part of the Green's function
contains a large logarithm $\ln (W/z)$ at small $z$. It means that
the corresponding contribution to the elastic x-ray scattering
intensity ($\omega ^{\prime }=E_0$) is enhanced by a factor of
$\ln ^2(W/\Gamma ),$ which makes a quantitative estimation for the
qualitative effect discussed in Ref. \cite{kurmaev:155105}. Of
course, the smearing of the jump in the density of NQP states by
spin
dynamics is irrelevant provided that $\Gamma \gg \overline{\omega }$ ($%
\overline{\omega }$ is a characteristic magnon frequency).

\section{Modern first-principle calculations}

\subsection{Different functional schemes}

\label{sec:diffunc}

In this section we review contemporary approaches to the
electronic structure calculations with taking into account
correlation effects. Model considerations discussed above
demonstrate relevance of the correlation effects (such as
electron-magnon interactions) for physics of half-metallic
ferromagnets. In order to calculate the electronic structure of
\textit{real} materials we have to solve a complicated many-body
problem for a crystal, corresponding to inhomogeneous gas of
interacting electrons in an external periodic potential:
\begin{eqnarray}
\mathcal{H} &=&\mathcal{H}_{0}+\mathcal{H}_{\mathrm{int}},  \nonumber \\
\mathcal{H}_{0} &=&\sum_{\sigma }\int d\mathbf{r}\psi _{\sigma }^{+}\mathbf{(%
{r})[}-\frac{1}{2}\nabla
^{2}\mathbf{+}V_{\mathrm{ext}}\mathbf{({r})]}\psi
_{\sigma }\mathbf{({r}),}  \nonumber \\
\mathcal{H}_{\mathrm{int}} &=&\frac{1}{2}\sum_{\sigma \sigma
^{\prime }}\int \int d\mathbf{r}d\mathbf{r^{\prime }}\psi _{\sigma
}^{+}\mathbf{({r})}\psi _{\sigma ^{\prime
}}^{+}\mathbf{({r}^{\prime })}V(\mathbf{r-r}^{\prime
})\psi _{\sigma ^{\prime }}\mathbf{({r}^{\prime })}\psi _{\sigma }\mathbf{({r%
}).}
\end{eqnarray}
In this section we use the atomic units $(\hbar =m=e=1)$, $\psi _{\sigma }%
\mathbf{({r})}$ is a field operators for electrons, $V_{\mathrm{ext}}\mathbf{%
({r})}$ describes interaction of electrons with static nuclei
which are supposed to form the periodic crystal lattice and also
may include other
external potentials (defects, electric fields, etc.) and $V\mathbf{({r}-{r}%
^{\prime })=}1/|\mathbf{{r}-{r}^{\prime }|}$ is the Coulumb
interaction between electrons.

The modern view on various practical schemes for solution of this
general many-electron problem is based on its functional
formulation in a framework
of so-called effective action approach \cite%
{Fukuda:865,georges:13,kotliar:865}. The partition function of
electronic system within imaginary-time functional integral
formalism can be expressed as\ an integral over electronic
Grassmann variables:
\begin{eqnarray}
\mathcal{Z} &=&\sum_{\sigma }\int D[\psi _{\sigma }^{+}\psi
_{\sigma
}]e^{-S},  \nonumber \\
S &=&\sum_{\sigma }\int d\mathbf{r}\int_{0}^{\beta }d{\tau }\psi
_{\sigma }^{+}\mathbf{({r}},{\tau }\mathbf{)}\frac{\partial
}{\partial {\tau }}\psi
_{\sigma }\mathbf{({r}},{\tau }\mathbf{)+}\int_{0}^{\beta }d{\tau }\mathcal{H%
}{({\tau }).}
\end{eqnarray}
The free energy of many-electron system $F=-T\ln \mathcal{Z}$ can
be expressed as a function of optimally chosen physical variables
for a given problem. The most accurate scheme corresponds to
Baym-Kadanoff, or Luttinger-Ward functional
\cite{baym:287,luttinger:1417} of the one-electron Green's
function:

\begin{equation}
G_{\sigma }(\mathbf{r-r}^{\prime },{\tau -\tau }^{\prime
}{)=-}\left\langle
T_{{\tau }}\psi _{\sigma }\mathbf{({r}},{\tau }\mathbf{)}\psi _{\sigma }^{+}%
\mathbf{(\mathbf{r}}^{\prime }\mathbf{,}\tau ^{\prime }\mathbf{)}%
\right\rangle
\end{equation}
In this formulation one has to add constraint fields of dual variable $%
\Sigma $ to the action

\begin{equation}
S[\Sigma ]=S+\sum_{\sigma }\int d\mathbf{r}\int
d\mathbf{r}^{\prime }\int_{0}^{\beta }d{\tau }\int_{0}^{\beta
}d{\tau }^{\prime }\Sigma _{\sigma
}(\mathbf{r,r}^{\prime },{\tau ,\tau }^{\prime }{)}G_{\sigma }(\mathbf{r}%
^{\prime },\mathbf{r},{\tau }^{\prime },{\tau )}
\end{equation}
and find the partition function in the presence of the auxiliary
source field:

\begin{equation}
\mathcal{Z}[\Sigma ]=e^{-F[\Sigma ]}=\int D[\psi ^{+}\psi
]e^{-S[\Sigma ]}
\end{equation}
The corresponding Baym-Kadanoff functional is defined as the
Legendre transformation of $F[\Sigma ]$ to the Green's function
variable:

\begin{equation}
F[G]=F[\Sigma ]-\mathrm{Tr}(\Sigma G)
\end{equation}
with further use of the functional derivative $G=\delta F/\delta
\Sigma $ to eliminate the constraint fields. Using free-electron
Green's function corresponding to $\mathcal{H}_{0}$ part of the
Hamiltonian, the final form of the functional with
\textquotedblleft Kohn-Sham\textquotedblright\ decomposition can
be written in the following form:

\begin{equation}
F[G]=-Tr\ln (G_{0}^{-1}-\Sigma )-\mathrm{Tr}(\Sigma G)+\Phi
\lbrack G];
\end{equation}
here $\Phi \lbrack G]$ is the Luttinger generating functional
which can be represented as a sum of all irreducible diagrams
without legs constructed from the exact Green's function $G$ and
bare electron-electron interaction line (bare four-leg vertex)
$V$. The Baym-Kadanoff functional is stationary in both $G$ and
$\Sigma $ and its variation with respect to $\Sigma $ leads to the
Dyson equation
\begin{equation}
G^{-1}=G_{0}^{-1}-\Sigma  \label{DYSON}
\end{equation}
and the $G$-extremum gives the variational identity, $\Sigma
=\delta \Phi /\delta G.$

The Baym-Kadanoff functional allows us in principle to calculate
not only free energy and thus thermodynamic properties of the
system, but also the Green's function and thus the corresponding
excitation spectrum. The main point which makes this scheme rather
useful for model many-body analysis and preserves its broad
practical use in the electronic structure calculation is related
with difficulties to find an exact representation of $F[G]$ even
for simple systems.

In this situations the density-functional scheme of Kohn,
Hohenberg, and Sham \cite{Hohenberg:B864,kohn:A1133} turns out to
be the most successful scheme for the electronic structure
calculations of an electronic systems with not too strong
correlations. For this purpose, the functional of the static
electronic density

\begin{equation}
\rho _{\sigma }(\mathbf{r})=-\frac{1}{\pi }\mathrm{Im}G_{\sigma }(\mathbf{r},%
{\tau =0).}
\end{equation}
is constructed. The corresponding constraint fields in the
effective action are related to the Kohn-Sham interaction
potential which is represented as a sum of Hartree and
exchange-correlation (\textit{xc}) parts:

\begin{equation}
V_{\mathrm{int}}=V_{H}+V_{xc}.
\end{equation}
Finally, the Kohn-Sham free-energy functional can be written in
the following form:

\begin{equation}
F[\rho ]=-Tr\ln (G_{0}^{-1}-V_{\mathrm{ext}}-V_{\mathrm{int}})-Tr(V_{\mathrm{%
int}}\rho )+F_{H}[\rho ]+F_{xc}[\rho ].
\end{equation}
where $V_{\mathrm{ext}}$ is an external potential and $F_{H}$ is
the Hartree potential. Again, there is a similar problem: an exact
form of the exchange-correlation functional $F_{xc}[\rho ]$ is,
generally speaking, unknown, and only a formal expression in terms
of the integral over the coupling constant exists
\cite{Harris:1170}. The practical use of the density-functional
theory (DFT) is related with the local-density approximation
(LDA):

\begin{equation}
F_{xc}[\rho ]\approx \int d\mathbf{r}\rho (\mathbf{r})\epsilon _{xc}[\rho (%
\mathbf{r})]
\end{equation}
where $\epsilon _{xc}[\rho ]$ is the exchange-correlation energy
per particle of \textit{homogeneous} electron gas with a given
density. It can be carefully parametrized from the numerically
exact Monte-Carlo calculations \cite{ceperley:566}. Taking into
account spin-dependence of the DFT through $\rho _{\sigma
}(\mathbf{r})$ one can study magnetic properties of complex
materials. This was the method used in the most of electronic
structure calculations referred above; in particular, the concept
itself of the half-metallic ferromagnetism was introduced based on
this kind of calculations \cite{deGroot:2024}. In practice, the
LDA scheme results sometimes in well-known difficulties; in
particular, it underestimates usually energy gaps in
semiconductors. For this reason it can fail to describe properly
the half-metallic state, e.g., in the case of colossal
magnetoresistance manganites \cite{Pickett:1146}. The DFT scheme
is formally exact (assuming that an exact $E_{xc}$ is known) to
find the energy and electronic density of the many-body systems by
minimization of density functional. However, the excitation
spectrum, rigorously speaking, cannot be expressed in terms of the
Kohn-Sham eigenenergies $\varepsilon _{i}$ defined by

\begin{equation}
(-\frac{1}{2}\nabla ^{2}+V_{\mathrm{ext}}+V_{\mathrm{int}})\psi
_{i}=\varepsilon _{i}\psi _{i}.
\end{equation}
(see, e.g., the discussion of NQP contributions to thermodynamic
properties in Sect.\ref{sec:thermo}).

A reasonable scheme which can overcome the difficulties of the DFT
scheme
for the gap problem uses so-called GW approximation proposed by Hedin \cite%
{Hedin:A796}. The functional approach to the GW scheme has been
developed recently \cite{Almbladh:535,chitra:115110,chitra:12715}
and is related with
the free energy functional of both total Green's function $G$ and \textit{%
screened} Coulomb interactions $W=(V^{-1}-\Pi )^{-1}$:

\begin{equation}
F[G,W]=-\mathrm{Tr}\ln (G_{0}^{-1}-\Sigma )-Tr(\Sigma G)+\frac{1}{2}\mathrm{%
Tr}\ln (V^{-1}-\Pi )+\frac{1}{2}\mathrm{Tr}(\Pi W)+F_{H}[\rho
]+\Phi \lbrack G,W]
\end{equation}
where $\Pi $ is the polarization operator. Earlier a similar
approach was used in the theory of phonon-induced
superconductivity of disordered systems \cite{Anokhin:2468}. In
the GW approximation only the lowest-order diagram in the screened
interactions is included into the generating functional:

\begin{equation}
\Phi \lbrack G,W]=\frac{1}{2}\mathrm{Tr}(GWG)
\end{equation}
In this case the polarization operator\ $\Pi $ which serves as a
constraint field for the screened Coulomb interactions $W$ has the
simplest form:

\begin{equation}
\Pi =-2\frac{\delta \Phi \lbrack G,W]}{\delta W}=-GG
\end{equation}
and the corresponding electron self-energy reads

\begin{equation}
\Sigma =\frac{\delta \Phi \lbrack G,W]}{\delta G}=GW
\end{equation}
The GW scheme gives an accurate estimation of the screened Coulomb
interactions in solids and can be used to define the
first-principle values of local Hubbard-like multi-orbital
energy-dependent interactions for correlated local orbitals $\phi
_{i}(\mathbf{r})$ which describe $d$ states of transition metal
ions \cite{aryasetiawan:195104}:

\begin{equation}
U_{ijkl}(\omega )=\left\langle \phi _{i}\phi
_{j}|\widetilde{W}(\omega )|\phi _{k}\phi _{l}\right\rangle
\label{frequency}
\end{equation}
where $\widetilde{W}$ does not take into account the effects of
$d-d$ screening, the latter being explicitly taken into account
further within an effective low-energy Hubbard-like model. The
numerical estimation of $U$ for metallic nickel
\cite{aryasetiawan:195104} shows relatively weak energy dependence
within the $d$-band energy width and the static values of the
order of 2-4~eV, in a good agreement with the experimental values
of the Hubbard parameters \cite{Sawatzky:10674}.

The success of GW approximation is closely related to the fact
that the bare Coulomb interaction $V$ is strongly screened in
solids and thus one can use the lowest-order approximation for
$\Phi \lbrack G,W]$. On the other hand, the spin dependence of
self-energy in GW scheme comes only from the spin dependence of
the Green's function $G_{\sigma }(\mathbf{r},{\tau )}$ and not
from the effective interactions $W$. In the Baym-Kadanoff
formalism this means that only the density-density channel was
taken into account in the screening of the Coulomb interactions.
It is well known that the Hund's
intraatomic exchange interactions are weakly screened in crystals \cite%
{Sawatzky:10674}, and strong spin-flip excitation processes will
modify electronic self-energy in itinerant electron magnets. In
particular, these processes are responsible for electron-magnon
interaction and lead to appearance of the NQP states in the gap
region for half-metallic ferromagnets.

An accurate treatment of the effects of local screened Coulomb $U$
and exchange $J$ interactions beyond GW or DFT methods can be
carried out within the dynamical mean-field theory (DMFT) combined
with the GW or LDA/GGA functionals. The DMFT scheme defines the
best local approximation for the self-energy, which uses the
mapping of the original many-body system with Hubbard-like
interactions onto multi-orbital quantum impurity model in the
effective electronic bath under the self-consistency condition \cite%
{georges:13}. The corresponding GW+DMFT scheme
\cite{biermann:086402} or spectral-density functional theory
\cite{savrasov:245101} is probably the best known way to treat
correlation effects in the electronic structure of real materials.
However, it is still very cumbersome and computationally
expensive; also, methods of work with the frequency-dependent
effective interaction (\ref{frequency}) are not developed enough
yet (for examples of first attempts, see
\cite{rubtsov:035122,savkin:026402}). The only way to consider
effects of spin-flip processes on the electronic structure of real
materials is a simplified version of \ a general spectral-density
functional known as the LDA+DMFT approach
\cite{anisimov:7359,lichtenstein:6884}. In a sense, one can
consider the LDA+DMFT and LDA as complementary approaches. In both
the cases we split a complicated many-body problem for a crystal
into a \textit{one}-body problem for the crystal and many-body
problem for some appropriate auxiliary system where we can hope to
calculate the correlation effects more or less accurately. For LDA
we choose the homogeneous electron gas as this auxiliary system.
For the LDA+DMFT it is an atom in some effective medium. The
latter choice is optimal to consider atomic-like features of $d$
or $f$ electrons in solids. The local Green's function in magnetic
solids is obtained from the effective impurity action with the
static (frequency-independent) Hubbard-like multi-orbital
interactions,

\begin{equation}
S_{\mathrm{imp}}=-\sum_{ij}\int_{0}^{\beta }d\tau \int_{0}^{\beta
}d\tau ^{\prime }c_{i}^{+}(\tau )\mathcal{G}_{ij}(\tau -\tau
^{\prime })c_{j}(\tau ^{\prime
})+\frac{1}{2}\sum_{ijkl}\int_{0}^{\beta }d\tau c_{i}^{+}(\tau
)c_{j}^{+}(\tau )U_{ijkl}c_{k}(\tau )c_{l}(\tau ),
\end{equation}
where $c_{i}(\tau )$ are the fermionic Grassmann variables for
localized correlated $d$-orbitals $\phi _{i}(\mathbf{r})$ and
$\mathcal{G}_{ij}$ is so-called bath Green's function which is
defined self-consistently within the single-particle lattice
model. The corresponding interacting local Green's function

\begin{equation}
G_{ij}(\tau -\tau ^{\prime })=-\langle T_{{\tau }}c_{i}(\tau
)c_{j}^{+}(\tau ^{\prime }\mathbf{\mathbf{)\rangle
}}_{S_{\mathrm{imp}}}
\end{equation}
can be found, within numerically exact Quantum Monte-Carlo scheme \cite%
{hirsch:4059}, or some perturbative approach which treats
accurately
spin-flip excitation processes in the particle-hole channel \cite%
{katsnelson:1037}. The corresponding self-energy matrix of the
impurity model

\begin{equation}  \label{Sig}
\Sigma =\mathcal{G}^{-1}-G^{-1}
\end{equation}
can be used in the spectral-density functional:

\begin{equation}
F[\mathcal{G}]=-Tr\ln (G_{0}^{-1}-\Sigma )-Tr(\Sigma
\mathcal{G})+\Phi \lbrack \mathcal{G}]
\end{equation}
and satisfies self-consistent equation for the bath Green's
function:

\begin{equation}
\mathcal{G}(\omega
)=\sum_{\mathbf{k}}[G_{0}^{-1}(\mathbf{k},\omega )-\Sigma (\omega
)]^{-1}+\Sigma (\omega )  \label{Gk}
\end{equation}

\subsection{LDA+DMFT: the Quantum Monte Carlo solution of the impurity
problem}

\label{sec:lda+dmft}

Now we describe the most rigorous way to solve an effective
impurity problem
using the multi-band Quantum Monte Carlo (QMC) method \cite{Rozenberg:R4855}%
. In the framework of LDA+DMFT this approach was used first in Ref.\cite%
{Katsnelson:8906} for the case of ferromagnetic iron.

We start from the many-body Hamiltonian in the LDA$+U$ form \cite%
{Anisimov:767}:

\begin{eqnarray}
\mathcal{H}
&=&\mathcal{H}_{\mathrm{LDA}}^{dc}+\frac{1}{2}\sum_{i\{\sigma
m\}}U_{m_{1}m_{2}m_{1}^{\prime }m_{2}^{\prime
}}^{i}c_{im_{1}\sigma }^{+}c_{im_{2}\sigma ^{\prime
}}^{+}c_{im_{2}^{\prime }\sigma ^{\prime
}}c_{im_{1}^{\prime }\sigma }  \nonumber  \label{HAM} \\
\mathcal{H}_{\mathrm{LDA}}^{dc} &=&\sum_{ij\sigma
\{m\}}h_{m_{1}m_{2}}^{ij}c_{im_{1}\sigma }^{+}c_{jm_{2}\sigma
}-E^{dc}
\end{eqnarray}
where $(ij)$ represents different crystal sites, $\{m\}$ label
different orbitals, $\{\sigma \}$ are spin indices, and
$t_{m_{1}m_{2}}^{ij}$ are the hopping parameters. The Coulomb
matrix elements are defined by
\begin{equation}
U_{m_{1}m_{2}m_{1}^{\prime }m_{2}^{\prime }}^{i}=\int \int d\mathbf{r}d%
\mathbf{r}^{\prime }\Psi _{im_{1}}^{*}(\mathbf{r})\Psi _{im_{2}}^{*}(\mathbf{%
r}^{\prime })V_{ee}(\mathbf{r}-\mathbf{r}^{\prime })\Psi _{im_{1}^{\prime }}(%
\mathbf{r})\Psi _{im_{2}^{\prime }}(\mathbf{r}^{\prime })
\end{equation}
where $V_{ee}(\mathbf{r}-\mathbf{r}^{\prime })$ is the screened
Coulomb interaction which remains to be determined. We follow
again the spirit of the LDA$+U$ approach by assuming that within
the atomic spheres these interactions retain to a large measure
their atomic nature. Moreover, the values of screened Coulomb
($U$) and exchange ($J$) interactions can be calculated within the
supercell LSDA approach \cite{anisimov:7570}: the elements of the
density matrix $n_{mm^{\prime }}^{\sigma }$ are to be constrained
locally, and the second derivative of the LSDA energy with respect
to the variation of the density matrix yields the wanted
interactions. In a spherical approximation, the matrix elements if
$V_{ee}$
can be expressed in terms effective Slater integrals $F^{(k)}$ \cite%
{JUDD:1963} as
\begin{equation}
\langle m,m^{\prime \prime }|V_{ee}|m^{\prime },m^{\prime \prime
\prime }\rangle =\sum_{k}a_{k}(m,m^{\prime },m^{\prime \prime
},m^{\prime \prime \prime })F^{(k)},  \label{slater}
\end{equation}
where $0\leq k\leq 2l$ and
\[
a_{k}(m,m^{\prime },m^{\prime \prime },m^{\prime \prime \prime
})=\frac{4\pi }{2k+1}\sum_{q=-k}^{k}\langle lm|Y_{kq}|lm^{\prime
}\rangle \langle lm^{\prime \prime }|Y_{kq}^{\ast }|lm^{\prime
\prime \prime }\rangle
\]
For \textit{d }electrons one needs $F^{(0)},F^{(2)}$ and
$F^{(4)}$; they are connected with the Coulomb- and Stoner
parameters $U$ and $J$ by $U=F^{(0)}$ and
$J=(F^{(2)}+F^{(4)})/14$, while the ratio $F^{(2)}/F^{(4)}$ is to
a good
accuracy a constant, about $0.625$ for the 3$d$ elements \cite%
{deGroot:5459,Anisimov:16929}. $\mathcal{H}_{\mathrm{LDA}}^{dc}$
represents the LDA Hamiltonian corrected by double counting of
average static Coulomb
interaction that is already presented in LDA \cite{Anisimov:767}. The index $%
i$ for the $U^{i}$ has a meaning only for the same correlated
sites as the orbital indices $\{m\},$ unlike the LDA term
$h_{m_{1}m_{2}}^{ij}$ (one-particle Hamiltonian parameters) where
we have the contribution of all the sites and orbitals in the unit
cell.

The one-particle spin-polarized LDA+DMFT Green's function $G_{\sigma }(%
\mathbf{k},\omega )$ is related to the LDA Green's function and
the local self-energy $\Sigma _{\sigma }(\omega )$ via the Dyson
equation
\begin{equation}
G_{\sigma }^{-1}(\mathbf{{k},}\omega \mathbf{)}=\omega +\mu -\mathcal{H}_{%
\mathrm{LDA,\sigma }}^{dc}\mathbf{({k})}-\Sigma _{\sigma }(\omega
) \label{eq:dyson}
\end{equation}
where $\mathcal{H}_{\mathrm{LDA,\sigma }}^{dc}(\mathbf{{k})}$ is
the LDA
Hamiltonian in local orthogonal basis set depending on the Bloch vector $%
\mathbf{k}$, and $\mu $ is the chemical potential. In order to
avoid \textquotedblleft double counting\textquotedblright , we can
just subtract
the static part of the self-energy, $E^{dc}=\mathrm{Tr}\Sigma _{\sigma }(0)$%
. It has been proven that this type of \textquotedblleft
metallic\textquotedblright\ double-counting is suitable for
moderately correlated $d$ electron systems
\cite{lichtenstein:067205}.

The standard QMC scheme for local Coulomb interactions takes into
account only density-density like interactions, although the new
continuous-time QMC \cite{rubtsov:035122,savkin:026402} can
overcome this problem and include all the elements of interaction
vertex. We use the functional integral formalism and describe the
discrete Hubbard-Stratonovich transformations for calculating the
partition functions and corresponding Green's function. In this
method the local Green's function is calculated for the imaginary
time
interval $\left[ 0,\beta \right] $ with the mesh $\tau _{l}=l\Delta \tau $, $%
l=0,...,L-1$ ($\Delta \tau =\beta /L,~\beta =1/T$) by using the
path-integral formalism \cite{georges:13}. The multi-orbital DMFT
problem with density-density interactions is described by the
following effective impurity action

\begin{equation}
S=-\int_{0}^{\beta }d\tau \int_{0}^{\beta }d\tau ^{\prime
}\sum_{i,j}c_{i}^{+}(\tau )\mathcal{G}_{ij}(\tau -\tau ^{\prime
})c_{j}(\tau ^{\prime })+\frac{1}{2}\int_{0}^{\beta }d\tau
\sum_{i,j}n_{i}(\tau )U_{ij}n_{j}(\tau )  \label{path}
\end{equation}
where $i=\{m,\sigma \}$ labels both orbital and spin indices (we
remind that we have no site indices since we are now solving the
one-site effective impurity problem). Thus we truncate the
original four-index rotationally invariant vertex and use only
two-index approximation for it. This is a price we should pay for
more exact way to solve the effective impurity
problem. Without spin-orbital coupling we have $\mathcal{G}_{ij}=\mathcal{G}%
_{m,m^{\prime }}^{\sigma }\delta _{\sigma \sigma ^{\prime }}$.

In the auxiliary fields Green-function QMC scheme one can use the
discrete Hubbard-Stratonovich transformation introduced by Hirsch
\cite{hirsch:4059}
\begin{equation}
\exp \left[ -\Delta \tau U_{ij}\left( n_{i}n_{j}-\frac{1}{2}%
(n_{i}+n_{j})\right) \right] =\frac{1}{2}\sum_{s_{ij}=\pm 1}\exp
\left[ \lambda _{ij}s_{ij}(n_{i}-n_{j})\right]  \label{Hirsch1}
\end{equation}
where $S_{ij}(\tau )$ are the auxiliary Ising fields for each pair
of spins, orbitals and time slices with the strength:
\begin{equation}
\lambda _{ij}=\mathrm{arccosh}\left[ \exp \left( \frac{\Delta \tau }{2}%
U_{ij}\right) \right]  \label{Hirsch2}
\end{equation}
Using Hirsch's transformation (\ref{Hirsch1}), (\ref{Hirsch2}) we
can transform the non-linear action to a normal Gaussian one (for
a given configuration of the auxiliary Ising fields $s_{ij}$) and
integrate out exactly fermionic fields in the functional integral
( \ref{path}). As a result, the partition function and Green's
function matrix have the form \cite{georges:13}
\begin{eqnarray}
Z &=&\frac{1}{2^{N_{f}L}}\sum_{s_{ij}(\tau )}\det
[\widehat{G}^{-1}(s_{ij})]
\nonumber \\
\widehat{G} &=&\frac{1}{Z}\frac{1}{2^{N_{f}L}}\sum_{s_{ij}(\tau )}\widehat{G}%
(s_{ij})\det [\widehat{G}^{-1}(s_{ij})]
\end{eqnarray}
where $N_{f}$ is the number of Ising fields, $L$ is the number of
time
slices, and $\widehat{G}(s_{ij})$ is the Green's function of \textit{%
non-interacting} fermions for a given configuration of the
external Ising fields:
\begin{eqnarray}
G_{ij}^{-1}(s) &=&\mathcal{G}_{ij}^{-1}+\Delta _{i}\delta
_{ij}\delta _{\tau
\tau ^{\prime }}  \nonumber \\
\Delta _{i} &=&(e^{V_{i}}-1)  \nonumber \\
V_{i}(\tau ) &=&\sum_{j(\neq i)}\lambda _{ij}s_{ij}(\tau )\sigma
_{ij}.
\end{eqnarray}
Here we introduce the generalized Pauli matrix
\begin{equation}
\sigma _{ij}=\left\{
\begin{array}{c}
+1,~i<j \\
-1,~i>j%
\end{array}
\right. .
\end{equation}
To calculate the Green's function $G_{ij}(s)$ for an arbitrary
configuration of the Ising fields one can use the Dyson equation
\cite{hirsch:4059}:
\begin{equation}
G^{\prime }=[1+(1-G)(e^{V^{\prime}-V}-1)]^{-1}G
\end{equation}
where $V$ and $G$ are the potential and Green's function before
the Ising spin flip, and $V^{\prime }$ and $G^{\prime }$ after the
flip. The QMC important sampling scheme allows one to integrate
over the Ising fields with
$|\det [\widehat{G}^{-1}(S_{ij})]|$ being a stochastic weight \cite%
{hirsch:4059,georges:13}. Using the output local Green's function
from QMC
and input bath Green's functions the new self-energy is obtained via Eq.(\ref%
{Sig}), the self-consistent loop being closed through
Eq.(\ref{Gk}). The main problem of the multi-band QMC formalism is
the large number of the auxiliary fields $S_{mm^{\prime }}^{l}$.
For each time slice $l$ it is equal to $M(2M-1)$ where $M$ is the
total number of the orbitals, which gives 45 Ising fields for
\textit{\ d} states case and 91 fields for \textit{f} states.
Analytical continuation of the QMC Green's functions from the
imaginary time to the real energy axis can be performed by the
maximum entropy method \cite{Jarrell:134}. It is important to
stress that for the diagonal Green's function $G_{ij}=G_{i}\delta
_{ij}$ the determinant ratio is always positive. This means that
the sign problem, which is the main
obstacle for the application of the QMC method to fermionic problems \cite%
{troyer:170201}, does not arise in this case. Real computational
experience shows that even for generic multiband case the sign
problem for the effective impurity calculations is not serious.

\begin{figure}[h]
\centerline{\psfig{file=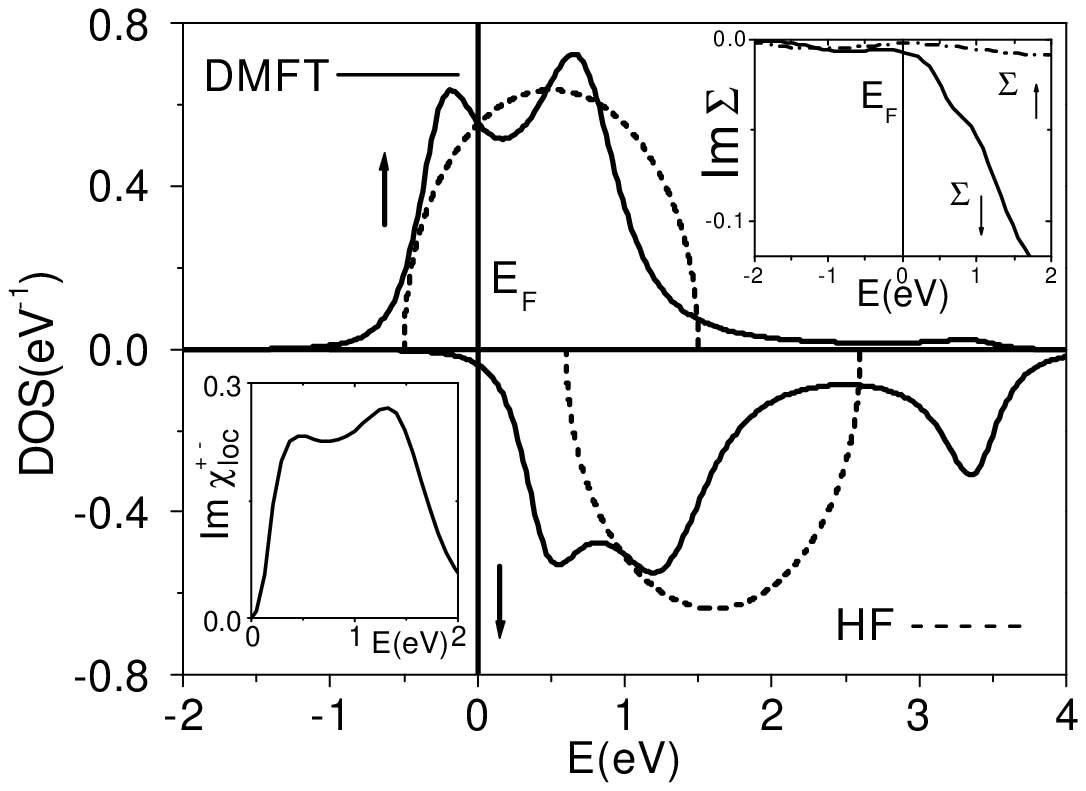,height=3.0in,angle=0}}
\caption{Density of states for HMF in the Hartree-Fock (HF)
approximation (dashed line) and in QMC solution of DMFT problem
for semi-circular model
(solid line) with the band-width $W$=2~eV,Coulomb interaction $U$=2~eV, $%
\Delta $=0.5~eV, chemical potential $\protect\mu $=-1.5~eV and temperature $%
T $=0.25~eV. Insets: imaginary part of the the spin-flip
susceptibility (left) and imaginary part of selfenergy (right).\cite{chioncel:144425}}
\label{model_bethe}
\end{figure}

It is worthwhile to illustrate QMC scheme for correlation effects
in the electronic structures of HMF by using a simple example.
Since solving the full one-band Hubbard model
\begin{equation}
\mathcal{H}=-\sum_{i,j,\sigma }t_{ij}(c_{i\sigma }^{\dagger
}c_{j\sigma }+c_{j\sigma }^{\dagger }c_{i\sigma
})+U\sum_{i}n_{i\uparrow }n_{i\downarrow }  \label{ham}
\end{equation}
is difficult (see discussion in Section \ref{sec:Hubbard}), we
treat the Dynamical Mean Field Theory (DMFT) \cite{georges:13}
which is formally exact in the limit of infinite dimensionality.
Following this approach we will consider the Bethe lattice with
coordination $z\rightarrow \infty $ and nearest neighbor hoping
$t_{ij}=t/\sqrt{z}$. In this case a semielliptic
density of states is obtained as a function of the effective hopping $t$, $%
N(\epsilon )=(2\pi t^{2})^{-1}\sqrt{4t^{2}-\epsilon ^{2}}$. In
order to stabilize the \textquotedblleft toy\textquotedblright\
model in the HMF state, we add an external magnetic spin splitting
term $\Delta $ which mimics the local Hund polarization. This HMF
state corresponds to a mean-field (HF) solution denoted in
Fig.~\ref{model_bethe} as a dashed line.

The effective medium Green's function ${\mathcal{G}}_{\sigma }$ is
connected with the local Green's function $G_{\sigma }$ on the
Bethe lattice through the self-consistency condition
\cite{georges:13}
\begin{equation}
{\mathcal{G}}_{\sigma }^{-1}=i\omega _{n}+\mu -t^{2}G_{\sigma
}-1/2\sigma \Delta  \label{slfc}
\end{equation}
where $\omega _{n}=(2n+1)\pi T~(n=0,\pm 1,\pm 2...)$ are the
Matsubara frequencies. The Green's function corresponding to the
DMFT effective action, Eq.(\ref{path}), $G_{\sigma }(\tau -\tau
^{\prime })=-\langle T_{\tau }c_{\sigma }(\tau )c_{\sigma
}^{\dagger }(\tau ^{\prime })\rangle _{S_{eff}}$, has been
calculated using the Quantum Monte Carlo scheme within the
so-called exact enumeration technique \cite{georges:13}, by using
the
time discretization parameter $L=25$ (see for details \cite{chioncel:144425}%
). We emphasize that due to the symmetry of the ferromagnetic
state the local $G_{\sigma }$ and the effective medium Green's
functions are diagonal in spin space, even in the presence of the
interactions which enable the spin-flip scattering process. The
magnon excitation can be studied through the two-particle
correlation function
\begin{equation}
\chi _{\mathrm{loc}}^{+-}(\tau -\tau ^{\prime })=\langle
S^{+}(\tau )S^{-}(\tau ^{\prime })\rangle =\langle T_{\tau
}c_{\uparrow }^{\dagger }(\tau )c_{\downarrow }(\tau
)c_{\downarrow }^{\dagger }(\tau ^{\prime })c_{\uparrow }(\tau
^{\prime })\rangle _{S_{eff}}  \label{chipm}
\end{equation}
which is obtained by using the QMC procedure \cite{Jarrell:168}.
Being local, this function is insufficient to find the
$\mathbf{q}$-dependence of the magnon spectrum, but yields only a
general shape\ of the magnon density of states.

The DMFT results are presented in Fig. \ref{model_bethe}. In
comparison with a simple Hartree-Fock solution one can see an
additional well-pronounced feature appearing in the spin-down gap
region, just above the Fermi level: the non-quasiparticle states
which are visible in both spin channels of DOS around 0.5~eV. In
addition, a many-body satellite appears at 3.5~eV.

The left inset of Fig.~\ref{model_bethe}, represents the imaginary
part of
the local spin-flip susceptibility. One can see a well pronounced shoulder ($%
\simeq 0.5$~eV), which is related to a characteristic magnon excitation \cite%
{Irkhin:4173,Irkhin:705,irkhin:7151}. There is a broad maximum at
about $1$ eV, which corresponds to the Stoner excitation energy.
The right inset of Fig. \ref{model_bethe}, represents the
imaginary part of self-energy. The spin-up channel can be
described by a Fermi-liquid type behavior with a parabolic energy
dependence $-\mathrm{Im}\Sigma ^{\uparrow }\propto (E-E_{F})^{2}$,
whereas in the spin down channel the NQP shoulder at 0.5~eV is
visible. Due to the relatively high temperature ($T=0.25$~eV) in
the QMC calculation the NQP tail extends below the Fermi level
(remember that at
zero temperature the tail should vanish exactly at the Fermi level, Sect.\ref%
{nqp}).

\subsection{Spin-polarized \textit{T}-matrix fluctuating exchange
approximation}

\label{sec:Tflex}

Most of HMF materials are moderately or weakly correlated systems.
This means that one can use some perturbative approaches which
make computations much less expensive and allow one to work with
the complete four-index Coulomb interaction matrix. A very
efficient perturbative scheme has been proposed by Bickers and
Scalapino \cite{bickers:206} and was called fluctuating exchange
approximation (FLEX). This was generalized to the
multiband case and used in the context of the DMFT in \cite%
{lichtenstein:6884,katsnelson:1037}. The latter step means that
this approach is used not directly for the whole crystal, but for
the effective impurity problem, so that the momentum dependence of
the Green's functions is neglected. On the other hand,
self-consistency of the DMFT procedure makes the description of
the \textit{local }effects in perturbative schemes more accurate.
For example, in the case of one-band half-filled Hubbard model the
second-order perturbation expression for the self-energy in the
context of the DMFT provides correct atomic limit and, actually,
very accurate description of the metal-insulator transition
\cite{kajueter:131}. A starting point in the FLEX approximation is
the separation of different interaction channels. The
symmetrization of \textit{bare U } matrix is done over
particle-hole and particle-particle channels:
\begin{eqnarray}
U_{m_{1}m_{1}^{\prime }m_{2}m_{2}^{\prime }}^{d}
&=&2U_{m_{1}m_{2}m_{1}^{\prime }m_{2}^{\prime
}}^{i}-U_{m_{1}m_{2}m_{2}^{\prime }m_{1}^{\prime }}^{i}  \nonumber \\
U_{m_{1}m_{1}^{\prime }m_{2}m_{2}^{\prime }}^{m}
&=&-U_{m_{1}m_{2}m_{2}^{\prime }m_{1}^{\prime }}^{i}  \nonumber \\
U_{m_{1}m_{1}^{\prime }m_{2}m_{2}^{\prime }}^{s} &=&\frac{1}{2}%
(U_{m_{1}m_{1}^{\prime }m_{2}m_{2}^{\prime
}}^{i}+U_{m_{1}m_{1}^{\prime
}m_{2}^{\prime }m_{2}}^{i})  \nonumber \\
U_{m_{1}m_{1}^{\prime }m_{2}^{\prime }m_{2}}^{t} &=&\frac{1}{2}%
(U_{m_{1}m_{1}^{\prime }m_{2}^{\prime
}m_{2}}^{i}-U_{m_{1}m_{1}^{\prime }m_{2}m_{2}^{\prime }}^{i})
\end{eqnarray}
The above expressions are the matrix elements of bare interaction
which can be obtained with the help of the pair operators
corresponding to different channels:

\begin{itemize}
\item particle-hole density
\begin{equation}
d_{12} = \frac 1 {\sqrt2 } ( c_{1 \uparrow}^+ c_{2 \uparrow} +
c_{1 \downarrow} ^+ c_{2 \downarrow} )
\end{equation}

\item particle-hole magnetic
\begin{eqnarray}
m_{12}^0 &=& \frac 1 {\sqrt2 } ( c_{1 \uparrow}^+ c_{2 \uparrow} -
c_{1
\downarrow} ^+ c_{2 \downarrow} )  \nonumber \\
m_{12}^+ &=& c_{1 \uparrow} ^+ c_{2 \downarrow} \\
m_{12}^+ &=& c_{1 \downarrow} ^+ c_{2 \uparrow}  \nonumber
\end{eqnarray}

\item particle-particle singlet
\begin{eqnarray}
s_{12} &=&\frac{1}{\sqrt{2}}(c_{1\downarrow }c_{2\uparrow
}-c_{1\uparrow
}c_{2\downarrow })  \nonumber \\
s_{12}^{+} &=&\frac{1}{\sqrt{2}}(c_{1\downarrow }^{+}c_{2\uparrow
}^{+}-c_{1\uparrow }^{+}c_{2\downarrow }^{+})
\end{eqnarray}

\item particle-particle triplet
\begin{eqnarray}
t_{12}^{0} &=&\frac{1}{\sqrt{2}}(c_{1\downarrow }c_{2\uparrow
}-c_{1\uparrow
}c_{2\downarrow })  \nonumber \\
t_{12}^{+0} &=&\frac{1}{\sqrt{2}}(c_{1\downarrow }^{+}c_{2\uparrow
}^{+}-c_{1\uparrow }^{+}c_{2\downarrow }^{+})  \nonumber \\
t_{12}^{\pm } &=&c_{1\uparrow ,\downarrow }c_{2\downarrow
,\uparrow }
\nonumber \\
t_{12}^{\pm } &=&c_{1\uparrow ,\downarrow }^{+}c_{2\downarrow
,\uparrow }^{+}
\end{eqnarray}

These operators describe the correlated movements of the electrons
and holes below and above the Fermi level and therefore play an
important role in
defining the spin dependent effective potentials $W_{m_{1}m_{2}m_{3}m_{4}}^{%
\sigma \sigma ^{\prime }}$.
\end{itemize}

In the spin polarized $\mathit{T}$-matrix fluctuating exchange
approximation (SPTF) scheme \cite{katsnelson:9} the
particle-particle interactions are described in the $T$-matrix
approach \cite{galitski:1011,kanamori:276} where for the effective
interaction the sum over the ladder graphs are carried out with
the aid of the so called $T-$matrix which obeys the
Bethe--Salpether-like integral equation:
\begin{eqnarray}
\langle 13|T^{\sigma \sigma ^{\prime }}(i\Omega )|24\rangle
&=&\langle
13|v|24\rangle  \nonumber \\
&-&\frac{1}{\beta }\sum_{\omega }\sum_{5678}G_{56}^{i\sigma
}(i\omega )G_{78}^{i\sigma ^{\prime }}(i\omega )G(i\Omega -i\omega
)\langle 68|T^{\sigma \sigma ^{\prime }}(i\Omega )|24\rangle
\end{eqnarray}
The corresponding contributions to the self-energy is described by
the Hartree and Fock diagrams with the formal replacement of the
bare interaction by the $T-$matrix:
\begin{eqnarray}
\Sigma _{12,\sigma }^{(TH)}(i\omega ) &=&\frac{1}{\beta
}\sum_{\Omega }\sum_{34\sigma ^{\prime }}\langle 13|T^{\sigma
\sigma ^{\prime }}(i\Omega
)|24\rangle G_{43}^{\sigma ^{\prime }}(i\Omega -i\omega ) \\
\Sigma _{12,\sigma }^{(TF)}(i\omega ) &=&-\frac{1}{\beta
}\sum_{\Omega }\sum_{34\sigma ^{\prime }}\langle 14|T^{\sigma
\sigma }(i\Omega )|32\rangle G_{34}^{i\sigma }(i\Omega -i\omega )
\end{eqnarray}
The \textquotedblleft Hartree\textquotedblright\ contribution
dominates for small concentration of electrons and holes; these
two contributions together contain all second-order in $V$ terms
in the self-energy.

Combining the density and magnetic parts of the particle-hole
channel we can
write down the expression for the interaction part of the Hamiltonian \cite%
{lichtenstein:6884,katsnelson:1037}.
\[
\mathcal{H}_{U}=D\ast H^{U}\ast D^{+}
\]
\begin{equation}
\mathcal{H}_{U}=\frac{1}{2}\mathrm{Tr}(D^{+}\ast V^{\Vert }\ast
D+m^{+}\ast V_{m}^{\bot }\ast m^{-}+m^{-}\ast V_{m}^{\bot }\ast
m^{+})
\end{equation}
where $D$ is a row vector with elements $(d,m^{0})$, and $D^{+}$
is a column vector with elements $(d^{+}m_{0}^{+})$, $\ast $
stands for the matrix multiplication with respect to the pairs of
orbital indices. It follows from the model consideration presented
above that for proper describing the effects of electron-magnon
interactions it is important to replace the bare
spin-flip potential by a static limit of the \textit{T}-matrix (see Eqs.(\ref%
{Edwards:L327}) and (\ref{sss})). With this replacement, the
expression of the effective potential is:
\begin{eqnarray}
V^{\Vert }(i\omega ) &=&\frac{1}{2}\left(
\begin{array}{cc}
V^{dd} & V^{dm} \\
V^{md} & V^{mm}%
\end{array}
\right)  \nonumber \\
(V_{m}^{\bot })_{1234} &=&\langle 13|T^{\uparrow \downarrow
}|42\rangle
\end{eqnarray}
The matrix elements of the effective interaction for $z$
(longitudinal) spin fluctuations are:
\begin{eqnarray}
V_{dd} &=&\frac{1}{2}\sum_{\sigma }(\sum_{\sigma ^{\prime
}}\langle 13|T^{\sigma \sigma ^{\prime }}|42\rangle -\langle
13|T^{\sigma ^{\prime
}\sigma ^{\prime }}|42\rangle )  \nonumber \\
V_{dm}=V_{md} &=&\frac{1}{2}\sum_{\sigma \sigma ^{\prime }}\sigma
(\langle 13|T^{\sigma \sigma }|42\rangle -\langle 13|T^{\sigma
\sigma }|24\rangle
+\langle 13|T^{\sigma ^{\prime }\sigma }|42\rangle )  \nonumber \\
V_{mm} &=&\frac{1}{2}\sum_{\sigma }(\sum_{\sigma ^{\prime }}\sigma
\sigma ^{\prime }\langle 13|T^{\sigma \sigma ^{\prime }}|42\rangle
-\langle 13|T^{\sigma ^{\prime }\sigma ^{\prime }}|42\rangle )
\end{eqnarray}
Further we introduce the expressions for the generalized
longitudinal ($\chi ^{\Vert }$) and transverse ($\chi ^{\bot }$)
susceptibilities
\begin{eqnarray}
\chi ^{\bot }(i\omega ) &=&[1+V_{m}^{\bot }\Gamma ^{\uparrow
\downarrow
}(i\omega )]^{-1}\ast \Gamma ^{\uparrow \downarrow }(i\omega ) \\
\chi ^{\Vert }(i\omega ) &=&[1+V^{\Vert }\ast \chi _{0}^{\Vert
}(i\omega )]^{-1}\ast \chi _{0}^{\Vert }(i\omega )
\end{eqnarray}
where $\Gamma (i\omega )$ is the Fourier transform of the empty
loop:
\begin{equation}
\Gamma _{m_{1}m_{2}m_{3}m_{4}}^{\sigma \sigma ^{\prime }}(\tau
)=-G_{m_{2}m_{3}}^{i\sigma }(\tau )G_{m_{4}m_{1}}^{i\sigma }(-\tau
)
\end{equation}
and the bare matrix longitudinal susceptibility is:
\begin{equation}
\chi _{0}^{\Vert }(i\omega )=\frac{1}{2}\left(
\begin{array}{cc}
{\Gamma }^{\uparrow \uparrow }+{\Gamma }^{\downarrow \downarrow } & {\Gamma }%
^{\uparrow \uparrow }-{\Gamma }^{\downarrow \downarrow } \\
{\Gamma }^{\uparrow \uparrow }-{\Gamma }^{\downarrow \downarrow } & {\Gamma }%
^{\uparrow \uparrow }+{\Gamma }^{\downarrow \downarrow }%
\end{array}
\right)  \label{chilo}
\end{equation}

The four matrix elements correspond to the density-density $(dd)$,
density-magnetic $(dm^{0})$, magnetic-density $(m^{0}d)$ and
magnetic-magnetic channels $(m^{0}m^{0})$ and couple longitudinal
magnetic fluctuation with density magnetic fluctuation. In this
case the particle hole contribution to the self-energy is:
\begin{equation}
\Sigma _{12\sigma }^{(ph)}(\tau )=\sum_{34\sigma ^{\prime
}}W_{1342}^{\sigma \sigma ^{\prime }}(\tau )G_{34}^{\sigma
^{\prime }}  \label{selfph}
\end{equation}
with the particle-hole fluctuation potential matrix
\begin{equation}
W^{\sigma \sigma ^{\prime }}(i\omega )=\left(
\begin{array}{cc}
{W}_{\uparrow \uparrow } & {W}_{\uparrow \downarrow } \\
{W}_{\downarrow \uparrow } & {W}_{\downarrow \downarrow }%
\end{array}
\right)
\end{equation}
and the spin-dependent effective potentials are defined by
\begin{eqnarray}
W_{\uparrow \uparrow } &=&\frac{1}{2}V^{\Vert }\ast (\chi ^{\Vert
}-\chi
_{0}^{\Vert })\ast V^{\Vert }  \nonumber \\
W_{\downarrow \downarrow } &=&\frac{1}{2}V^{\Vert }\ast
(\tilde{\chi}^{\Vert
}-\tilde{\chi}_{0}^{\Vert })\ast V^{\Vert }  \nonumber \\
W_{\uparrow \downarrow } &=&\frac{1}{2}V_{m}^{\bot }\ast (\chi
^{+-}-\chi
_{0}^{+-})\ast V_{m}^{\bot }  \nonumber \\
W_{\downarrow \uparrow } &=&\frac{1}{2}V_{m}^{\bot }\ast (\chi
^{-+}-\chi _{0}^{-+})\ast V_{m}^{\bot }
\end{eqnarray}
where $\tilde{\chi}^{\Vert },\tilde{\chi}_{0}^{\Vert }$ differ
from the values of $\chi ^{\Vert },\chi _{0}^{\Vert }$ by the
replacement $\Gamma ^{\uparrow \uparrow }\leftrightarrow \Gamma
^{\downarrow \downarrow }$ in equation (\ref{chilo}). The final
expression for the SPTF self-energy is given by:
\begin{equation}
\Sigma =\Sigma ^{(TH)}+\Sigma ^{(TF)}+\Sigma ^{(ph)}
\end{equation}

Due to off-diagonal spin structure of the self-energy $\Sigma
^{(ph)}$ this method can be used to consider the non-quasiparticle
states in HMF. A more
detailed justification and description of the method is presented in \cite%
{katsnelson:9}. This approach was also generalized to the case of
strong
spin-orbit coupling and used for actinide magnets \cite%
{pourovskii:060506,pourovskii:115106}. In that case separation
into density and magnetic channels is not possible and one should
work with the $4\times 4 $ supermatrices for the effective
interaction.

Recently several LDA+DMFT calculations of different HMF materials
have been carried out. Now we review briefly the results of these
calculations focusing mainly on non-quasiparticle states resulting
from the electron-magnon interactions. LDA+DMFT is the only
contemporary practical way to consider them in the electronic
structure calculations.

\section{Electronic structure of specific half-metallic compounds}
\label{sec:elstr_hmf}

\subsection{Heusler alloys.}

\subsubsection{NiMnSb: electronic structure and correlations}
\label{sec:NiMnSb1}

The intermetallic compound NiMnSb crystallizes in the cubic
structure of
MgAgAs type (C1$_{b}$) with the $fcc$ Bravais lattice (space group $F%
\overline{4}3m=T_{d}^{2}$). The crystal structure is shown in Fig.~\ref%
{NiMnSbstruc}. This structure can be described as three
interpenetrating \textit{fcc} lattices of Ni, Mn and Sb. A
detailed description of the band structure of semi-Heusler alloys
was given by using electronic structure calculations analysis
~\cite{deGroot:2024},
\cite{ogut:10443,galanakis:134428,nanda:7307,kulatov:343,Galanakis:315213}; we
briefly summarize the results.

\begin{figure}[t]
\includegraphics[width=\columnwidth]{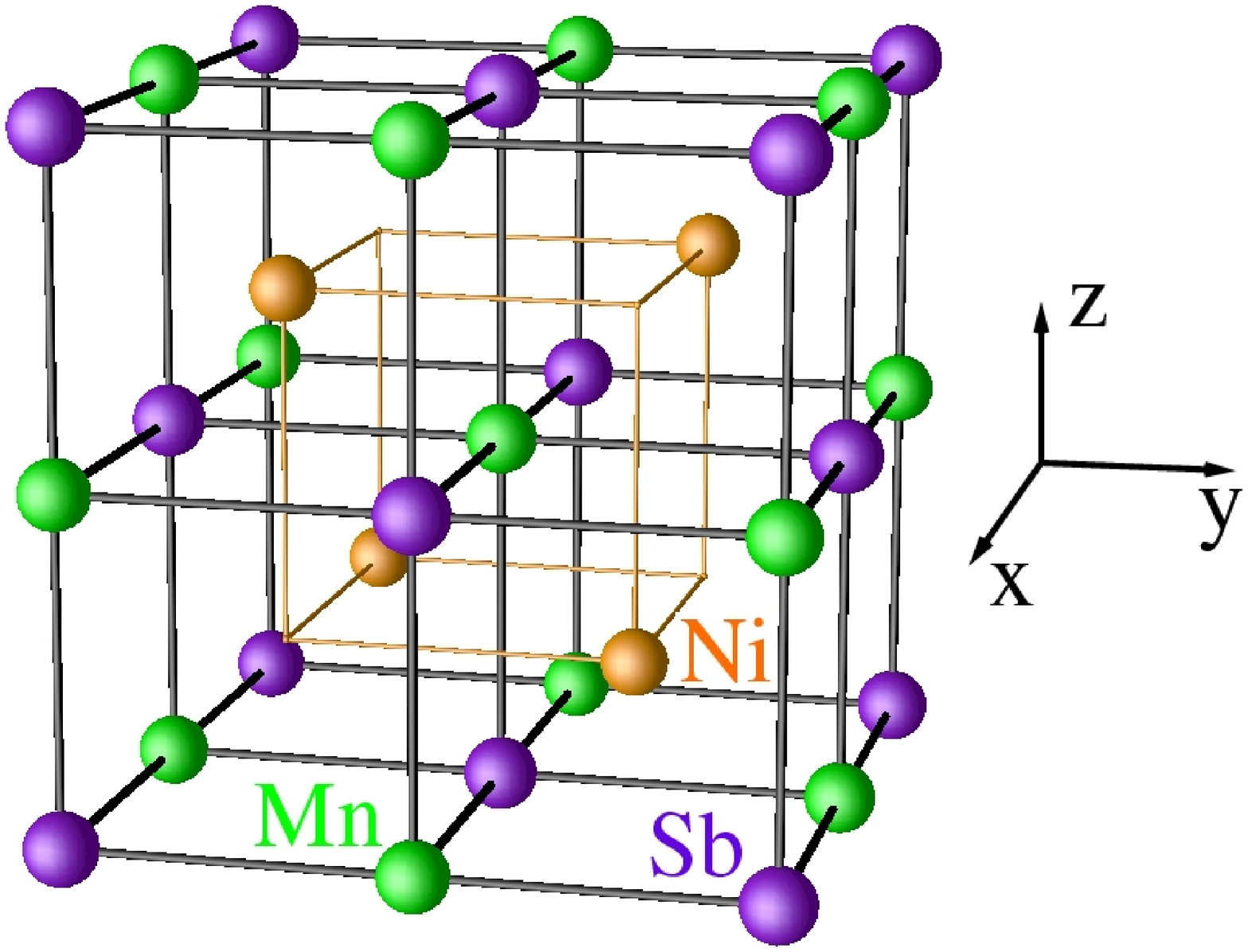}
\caption{(Color online) C1$_b$ structure with the \textit{fcc}
Bravais lattice (space group $F\overline{4}3m$). Mn (green) and Sb
(purple) atoms are located at (0, 0, 0) and ($\frac{1}{2}$,
$\frac{1}{2}$, $\frac{1}{2}$) forming the rock salt structure
arrangement. Ni (orange) atom is located in
the octahedrally coordinated pocket, at one of the cube center positions ($%
\frac{3}{4}$, $\frac{3}{4}$, $\frac{3}{4}$) leaving the other ($\frac{1}{4}$%
, $\frac{1}{4}$, $\frac{1}{4}$) empty. This creates voids in the
structure. \cite{yamasaki:024419} } \label{NiMnSbstruc}
\end{figure}


To obtain the
minority spin gap, not only the Mn$-d $--Sb$-p$ interactions, but
also Mn$-d$--Ni$-d$ interactions are to be taken into account.
Moreover, the loss of inversion symmetry produced by C1$_{b}$
structure (the symmetry lowering from $O_{h}$ in the L2$_{1}$ structure to $%
T_{d}$ in the C1$_{b}$ structure at Mn site) is an essential
additional ingredient. All the above interactions combined with
the $T_{d}$ symmetry lead to a nonzero anticrossing of bands and
to gap opening. 

\begin{figure*}[t]
\includegraphics[width=\textwidth]{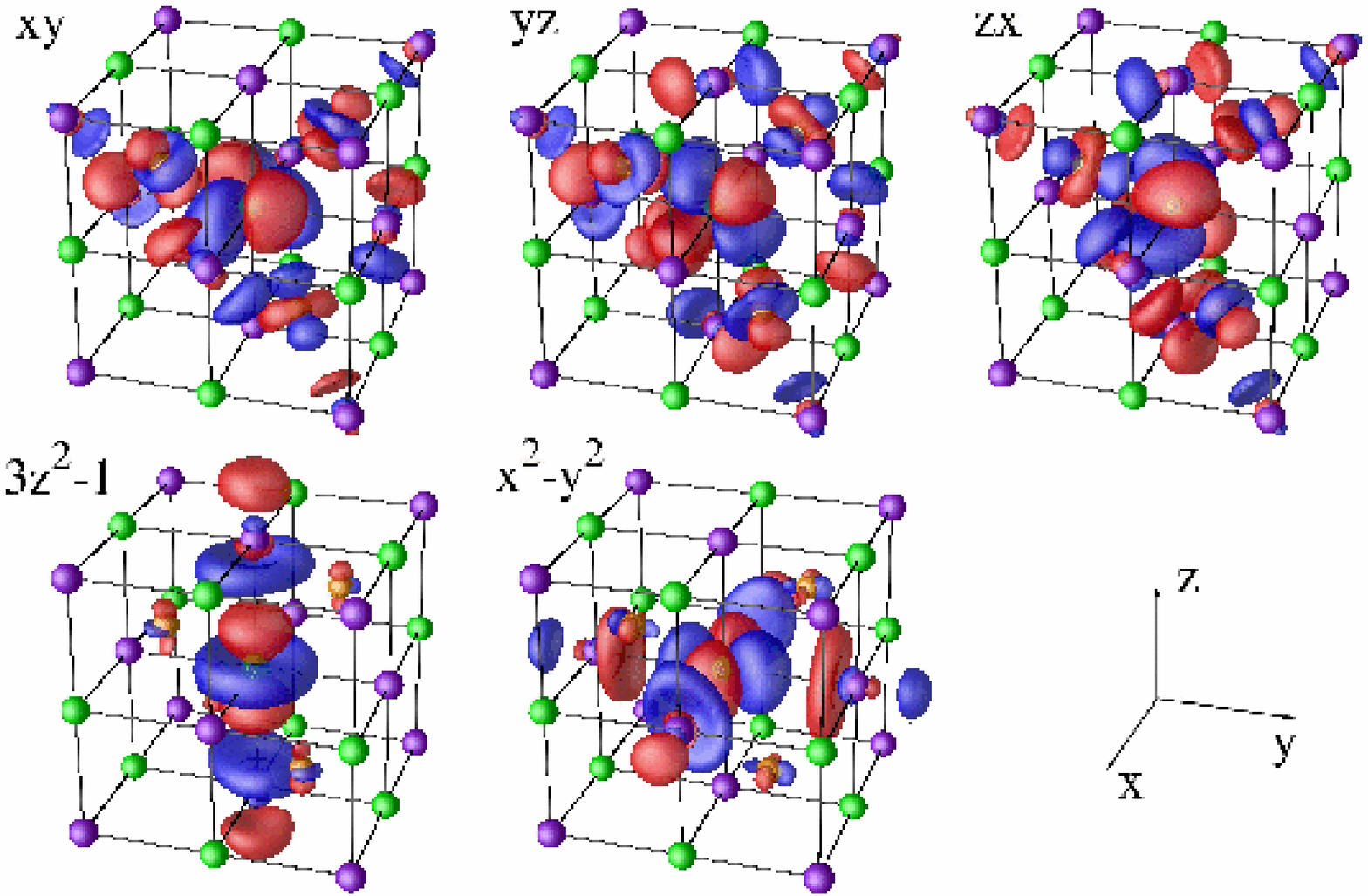}
\caption{(Color online) NMTO Mn-$d$ Wannier orbitals of NiMnSb. Ni
is orange, Mn is green, and Sb is purple. Red/blue indicates a
positive/negative sign. \textit{Upper panel}: $t_{2g}$ orbitals;
$d_{xy}$ (left), $d_{yz}$ (middle), $d_{zx}$ (right). The
triply-degenerate $t_{2g}$
orbitals can be obtained by the permutation of axes. \textit{Lower panel}: $%
e_g$ orbitals; $d_{3z^2-1}$ (left), $d_{x^2-y^2}$ (middle). These
$e_g$ orbitals are doubly degenerated.\cite{yamasaki:024419}} \label{Mndorbit}
\end{figure*}


The large exchange
splitting of Mn atom (producing the local Mn magnetic moment of about$~$3.7 $%
\mu_B$) is crucial to induce a half-metallic structure. In the
spin-polarized calculation the position of $t_{2g}$ and $e_{g}$ Ni
states is changed slightly, so that the exchange splitting on Ni
is not large. The local Ni magnetic moment calculation gives a
value around 0.3$\mu_B$.

\begin{figure}[t]
\rotatebox{270}{\includegraphics[height=\columnwidth]{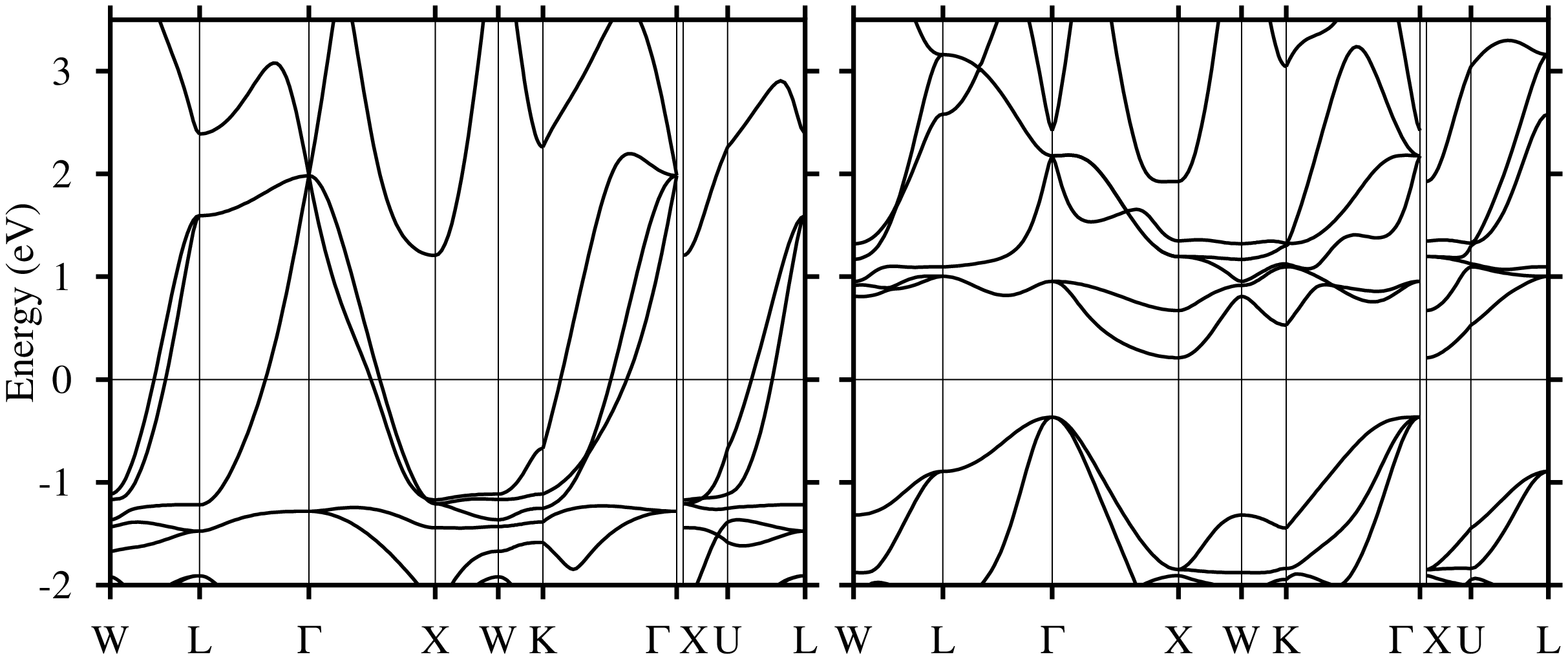}}
\caption{Full basis set spin-polarized (ferromagnetic) bands for
NiMnSb; majority spin (left) and minority spin (right). The
high-symmetry points are; $W(\frac{1}{2},1,0)$,
$L(\frac{1}{2},\frac{1}{2},\frac{1}{2})$, $\Gamma
(0,0,0)$, $X(0,1,0)$, $K(\frac{3}{4},\frac{3}{4},0)$, in $W$--$L$--$\Gamma $%
--$X$--$W$--$K$--$\Gamma $ line, and $X(0,0,1)$, $U(\frac{1}{4},\frac{1}{4}%
,1)$, $L(\frac{1}{2},\frac{1}{2},\frac{1}{2})$, in $X$--$U$--$L$
line.\cite{yamasaki:024419}} \label{bnds_sp}
\end{figure}

The non-spin-polarized result has a striking resemblance to the
majority spin-polarized calculations, presented in Fig.
\ref{bnds_sp}. Kulatov \textit{et al.} explained half-metallicity
of NiMnSb by the extended Stoner factor calculations in the
rigid-band approximation ~\cite{kulatov:343}: the minority
spin-band gap opens due to the exchange splitting which shifts
minority bands, so that they become empty. 

There is an excellent agreement between the first band-structure calculation ~\cite%
{deGroot:2024} and recent N-th order muffin-tin orbitals (NMTO)
investigations~\cite{yamasaki:024419,Andersen:16219,zurek:1934}.
NMTO Wannier Mn $d$ orbitals are shown in Fig.~\ref{Mndorbit}. The
triple-degenerate manganese $t_{2g}$ orbitals are very complicated
due to the hybridization with Ni $d$ and Sb $p~$states. The
$d_{xy}$ orbital at Mn
site is deformed by antibonding with Ni $d~$state directed tetrahedrally to [%
$11\bar{1}$], [$\bar{1}\bar{1}\bar{1}$], [$1\bar{1}1$], and
[$\bar{1}11$].
The same Ni $d~$orbitals couple with Sb $p$ states. The direct Mn-$d_{xy}$%
--Sb-$p$ $\pi $ coupling is not seen since the distance is $d$(Mn--Sb):$d$%
(Ni--Sb) $=1:\sqrt{3}/2$. Therefore the Ni-$d$--Sb-$p$
interactions are more favorable. The dispersion of the Mn
$t_{2g}~$bands is mainly due to hopping via the tails of Sb $p$
and Ni $d~$orbitals. On the other hand, the next-nearest neighbor
(n.n.n.) $d-d$ hopping of $t_{2g}$ orbital is small. The $e_{g}$
orbitals at Mn site are much easier to understand: they point
towards Sb atoms, and a strong $pd\sigma $ coupling between Sb $p$ and Mn e$%
_{g}~$states is seen. This induces large n.n.n. $d$-$d$ hoppings.

\begin{figure}[t]
\rotatebox{0}{\includegraphics[height=\columnwidth]{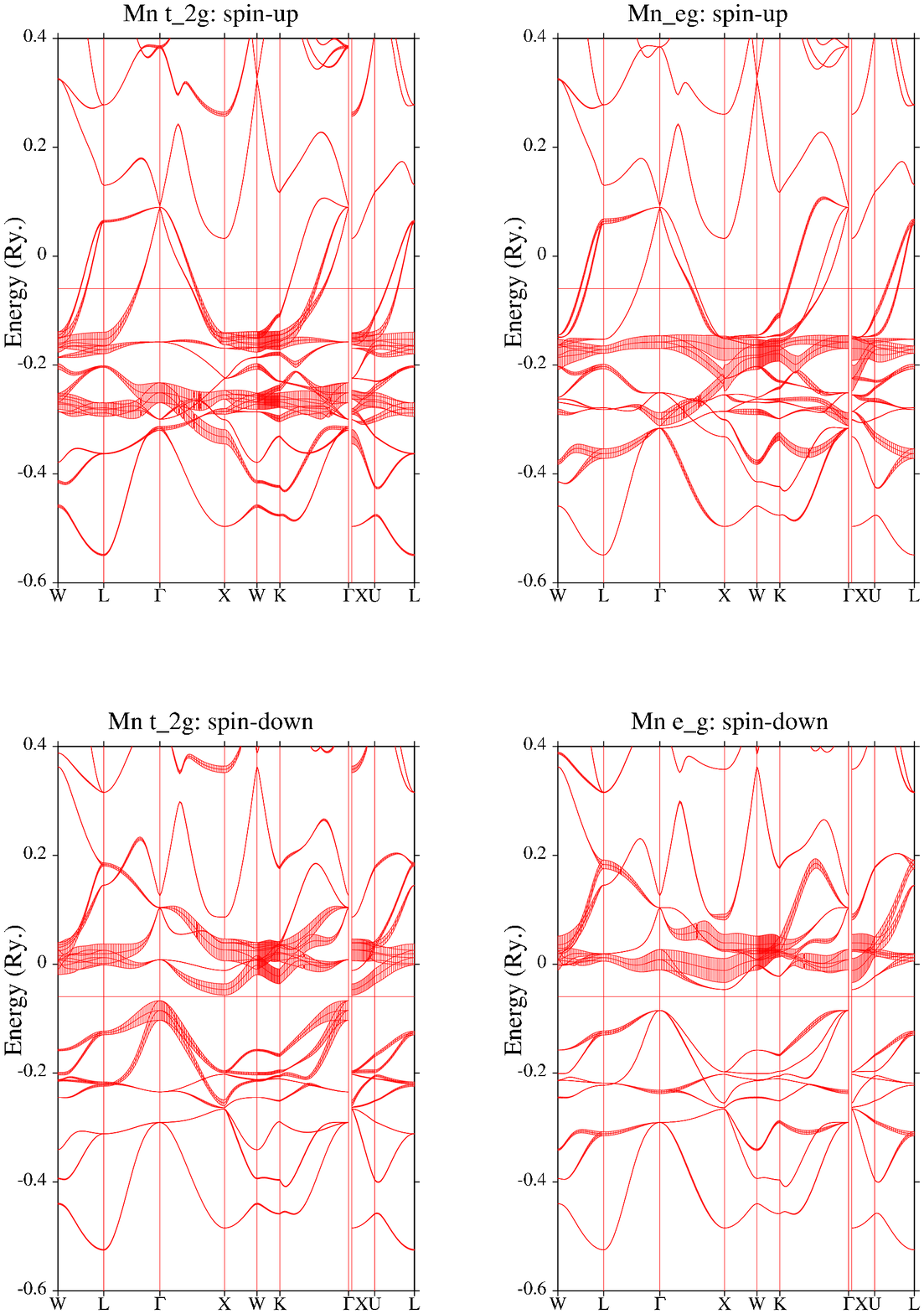}}
\caption{Decorated (fat) bands for the spin-polarized
(ferromagnetic) NiMnSb; Mn majority spin (up) and minority spin
(down). The high-symmetry
points are; $W(\frac{1}{2},1,0)$, $L(\frac{1}{2},\frac{1}{2},\frac{1}{2})$, $%
\Gamma (0,0,0)$, $X(0,1,0)$, $K(\frac{3}{4},\frac{3}{4},0)$, in $W$--$L$--$%
\Gamma $--$X$--$W$--$K$--$\Gamma $ line, and $X(0,0,1)$, $U(\frac{1}{4},%
\frac{1}{4},1)$, $L(\frac{1}{2},\frac{1}{2},\frac{1}{2})$, in
$X$--$U$--$L$ line \cite{Yamasaki:private}.} \label{bnds_sp_fatMn}
\end{figure}

The Wannier orbital analysis for NiMnSb \cite{yamasaki:024419}
confirms the previous conclusions of Ref.\cite{deGroot:2024} about
the role of $p$-$d$
hybridization as well as the role of $d$-$d$ hybridization\cite%
{galanakis:134428}. The LDA partial density of states for
half-metallic NiMnSb and the fat-band structure which marks the
main orbital character of
valence states are presented in the Fig. \ref{bnds_sp_fatMn},\ref%
{bnds_sp_fatNi}, \ref{bnds_sp_fatSb}. One can see that the spin-up
Mn $d$ and Ni $d$ states are located at the same energy region,
while spin-down bands are separate due to the significant Mn
exchange splitting. The top of the valence spin-down bands form by
the Mn $t_{2g}$, Sb $p$ and Ni $t_{2g}$ orbitals, while bottom of
the conduction bands are due to the Mn $e_{g}$ and $t_{2g}$
states. This means that large Mn spin-splitting and Sb mediated 
indirect Mn-Ni interactions are responsible for formation of  
the half-metallic gap. One can see this complicated Mn $t_{2g}$
valence orbitals
with large contributions of Sb $p$ and Ni $t_{2g}$ states in the Fig. (\ref%
{Mndorbit}). The physical picture of the downfolding analysis
should not change very much if one explicitly includes Sb $p$ and
Ni $d$ orbitals.

\begin{figure}[t]
\rotatebox{0}{\includegraphics[height=\columnwidth]{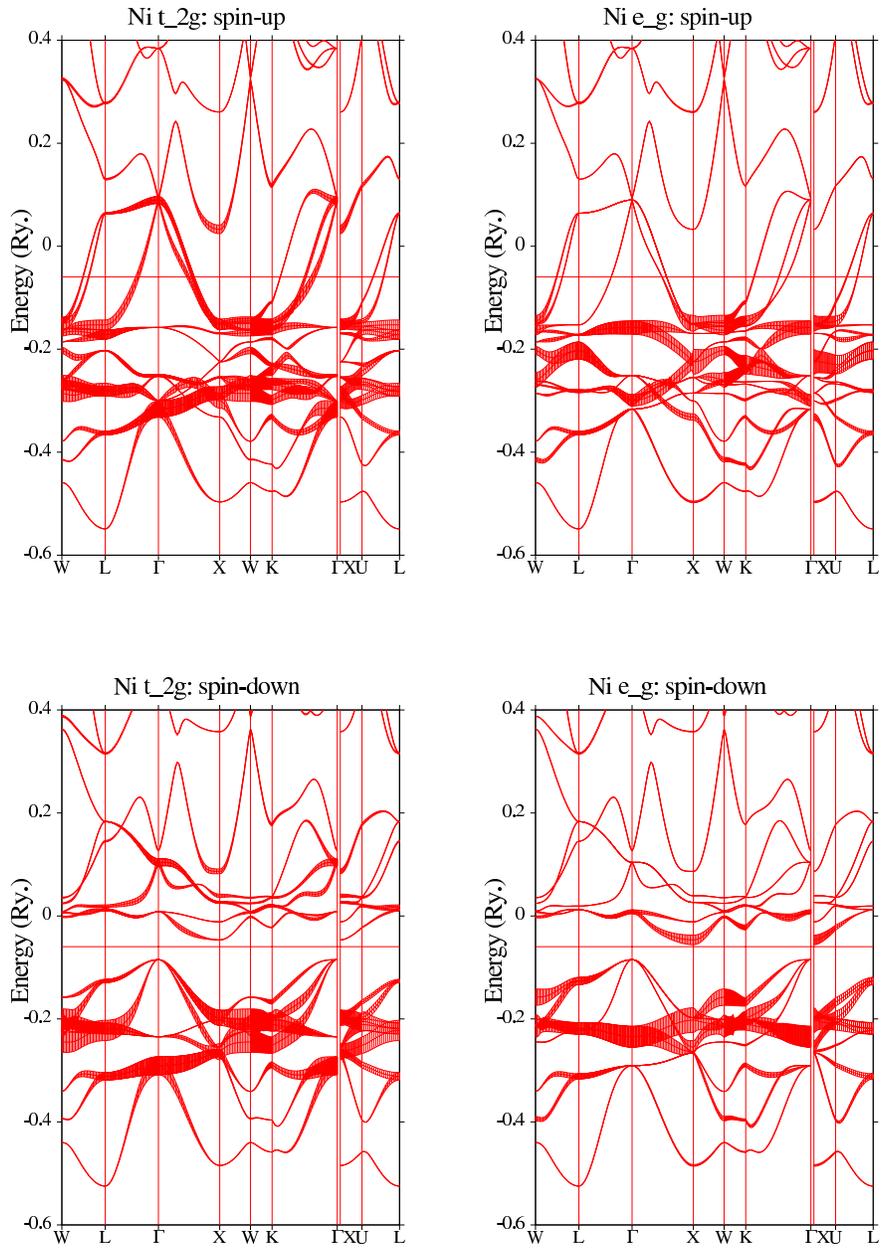}}
\caption{Decorated (fat) bands for the spin-polarized
(ferromagnetic) NiMnSb; Ni majority spin (up) and minority spin
(down) \cite{Yamasaki:private}.} \label{bnds_sp_fatNi}
\end{figure}

\begin{figure}[t]
\rotatebox{0}{\includegraphics[height=\columnwidth]{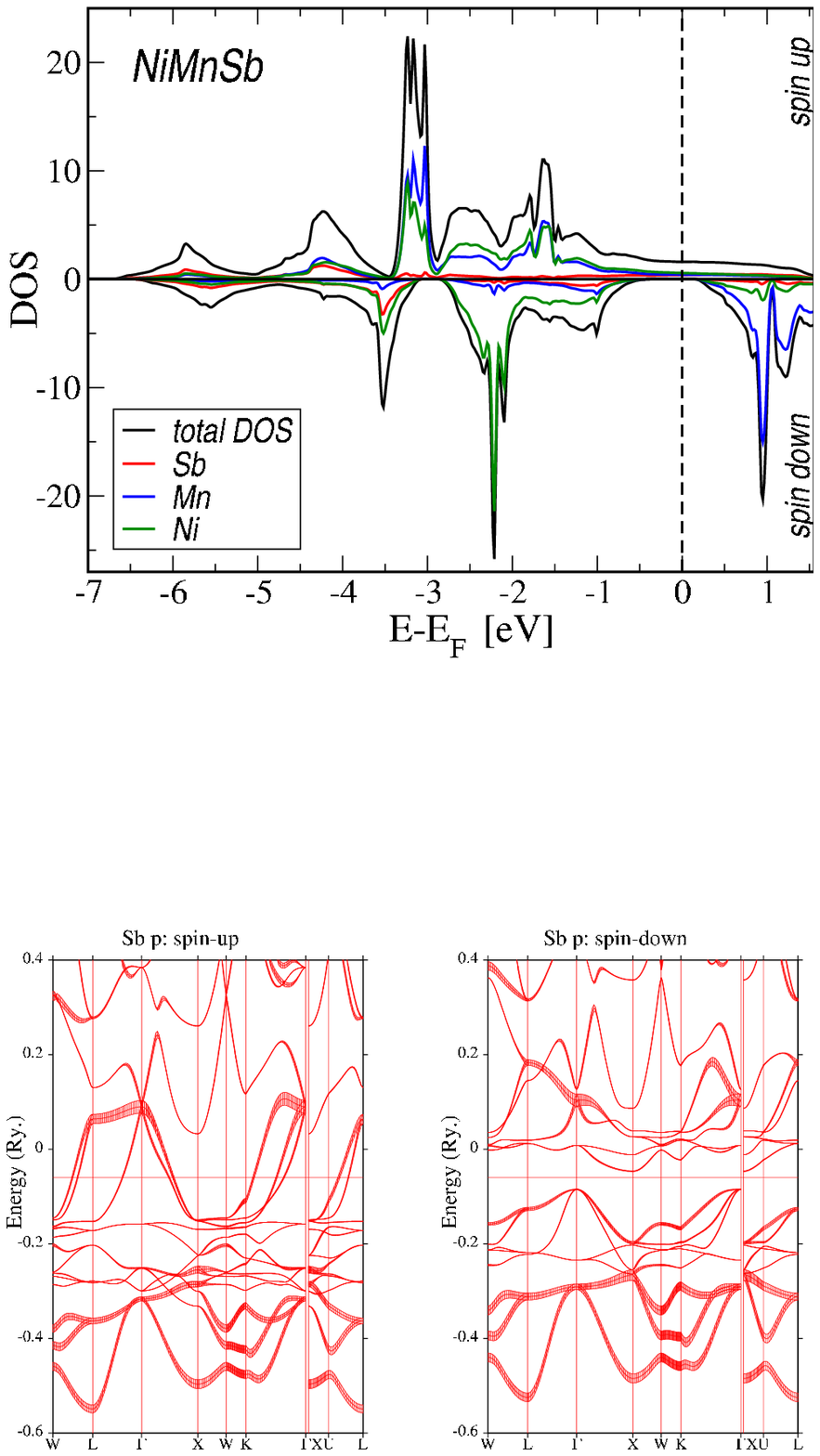}}
\caption{Upper figure: FLAPW calculation of spin-polarized NiMnSb
\cite{Lezaic:private}. Lower figures: Decorated (fat) bands for
the spin-polarized (ferromagnetic) NiMnSb; Sb majority spin (up)
and minority spin (down) \cite{Yamasaki:private}.}
\label{bnds_sp_fatSb}
\end{figure}

We discuss now in detail the prototype half-metallic ferromagnet
NiMnSb where the gap is situated in the spin-down (minority)
channel. The temperature dependence of the HMF electronic
structure and stability of half-metallicity against different
spin-excitations are crucial for practical applications in
spintronics. A simple attempt to incorporate finite-temperature
effects \cite{skomski:544,dowben:7948}, leading to static
non-collinear spin-configurations, shows a mixture of spin up and
spin down
density of states that destroy the half-metallic behavior. In Ref. \cite%
{chioncel:144425} a different more natural approach was used to
investigate the proper effect of dynamical spin fluctuations on
the electronic structure at $T<T_{C}$, within the half-metallic
ferromagnetic state.

The LDA+DMFT calculation for NiMnSb \cite{chioncel:144425} was the
first application of the combined electronic structure and
many-body technique to HMF. Aryasetiawan \textit{et al} pointed
out recently \cite{aryasetiawan:195104} that a rigorous way to
define the screened frequency-dependent on-site Coulomb
interactions matrix elements for correlated states related with
generalized GW scheme where $d-d$ screening is suppressed to
preserve the double counting of in the model approach. However, in
practice for realistic materials, the elimination of degrees of
freedom is a very difficult procedure. To find the average Coulomb
interaction on the $d$ atoms $U$ and corresponding exchange
interactions $J$, a more simple approach, the
constrained LDA scheme~ \cite%
{dederichs:2512,norman:8896,mcmahan:6650,gunnarsson:1708,hybertsen:9028,anisimov:7570}%
, can be used. In this approach the Hubbard $U$ is calculated from
the variation of the total energy with respect to the occupation
number of the localized orbitals. In such a scheme, the metallic
screening is rather inefficient for 3$d$ transition metals, and
effective $U$ is of the order of 6 eV~\cite{anisimov:7570}. The
perfect metallic screening will lead to a smaller value of $U$.
Unfortunately, there are no reliable scheme to calculate $U$
within constrained LDA for metals \cite{solovyev:045103}, and
in the works~ \cite%
{chioncel:144425,chioncel:085111,chioncel:137203,chioncel:197203}
some intermediate values were chosen, $U=2\div 4$~eV and
$J=0.9$~eV. The recent analysis of angle-resolved photoemission
and the LDA theory indicates a shift of spectral function of the
order of $0.5\div 1$ eV, which can be attributed to correlation
effects beyond LDA \cite{correa:125316}. First angle-resolved
photoemission results \cite{Kisker:21} in generally agree with the
LDA band structure, but demonstrate the same shift of
quasiparticle
dispersion of the order of $0.5$ eV below the Fermi level for spin-down Mn $%
t_{2g}$ bands. This can easily lead to the effective Hubbard
interactions in the static mean-field approximation of the order
of $U^{\ast }=1$ eV, although one should carefully investigate the
effects of spin-orbital splitting.  Since the spherically-average
effective Hubbard interactions is equal to $U^{\ast }=U-J$ and the
value of intra-atomic exchange interactions is not screened much in
solids being of the order of $J=1$~eV, we can conclude from the
photoemission experiments \cite{Kisker:21} and resonant x-ray
scattering \cite{Yablonskikh:235117} that the Hubbard interactions
$U=2$ eV is quite reasonable. An account of the correlation
effects in the framework of the LDA+DMFT method also improves the
description of magnetooptical properties of NiMnSb
\cite{chadov:140411}.

\begin{figure}[h]
\centerline{\psfig{file=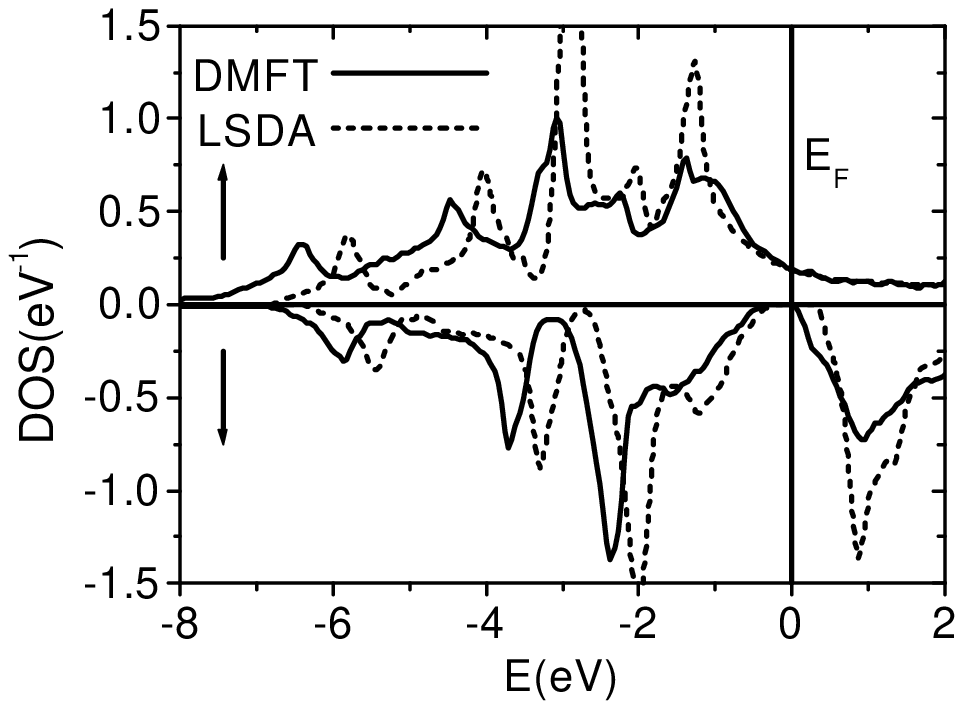,height=2.0in}
\psfig{file=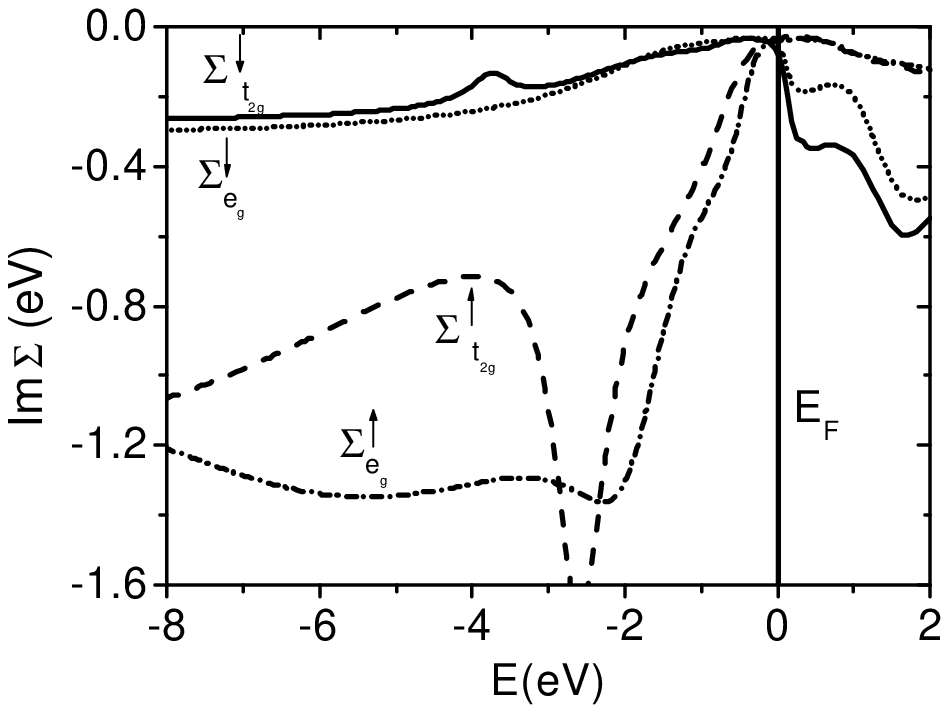,height=2.0in}} \caption{Density of
states for HMF NiMnSb in LSDA scheme (dashed line) and in LDA+DMFT
scheme (solid line) with effective Coulomb interaction $U$=3 eV,
exchange parameter $J$=0.9 eV and temperature $T$=300 K. The
non-quasiparticle state is evidenced just above the Fermi level.
The imaginary part of self-energies ${\rm Im}\Sigma
_{d}^{\downarrow }$ for $t_{2g}$
(solid line) and $e_{g}$ (dotted line), ${\rm Im}\Sigma _{d}^{\uparrow }$ for $%
t_{2g}$ (dashed line) and $e_{g}$ (dashed dotted line)
respectively.} \label{nimnsb}
\end{figure}

The typical results for density of states using LDA and LDA+DMFT
is presented in Fig. \ref{nimnsb}. The LDA+DMFT density of states
shows the existence of non-quasiparticle states in the LDA gap of
the spin down channel just above the Fermi level.
It is important to mention that the magnetic moment per formula
unit is not sensitive to the $U$ values. For a temperature
$T=300$~K the calculated magnetic moment, $\mu =3.96\mu _{B}$, is
close to the zero-temperature LDA-value which is integer, $\mu =4$
$\mu _{B}$. This means that the half-metallic state is stable with
respect to switching on the correlation effects. The DMFT gap in
the spin down channel, defined as the distance between the
occupied part and starting point of NQP tail, is also not very
sensitive to $U$. The total DOS is also weakly $U$-dependent due to the $T$%
-matrix renormalization effects.

\begin{figure}[h]
\centerline{\psfig{file=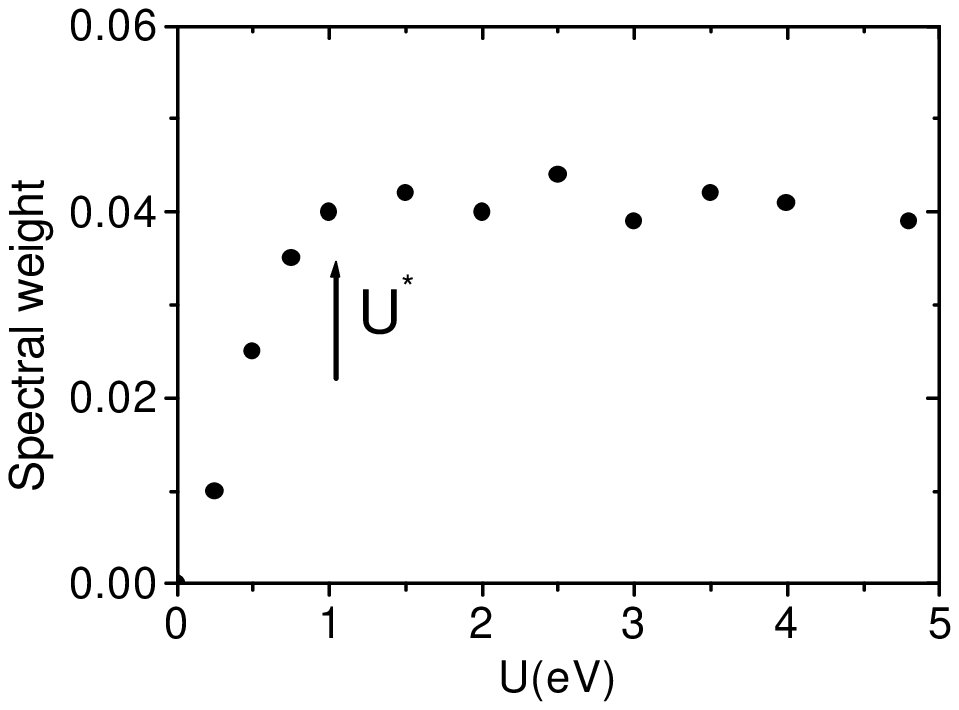,height=3.0in}}
\caption{Spectral weight of the non-quasiparticle state,
calculated as a function of average on-site Coulomb repulsion $U$
at temperature $T$=300 K.\cite{chioncel:144425}} \label{spec}
\end{figure}

In comparison with the LDA result, a strong spectral weight
transfer is present for the unoccupied part of the band structure
due to appearance of
the non-quasiparticle states in the energy gap above the Fermi energy (Sect.%
\ref{nqp}). Their spectral weight is not too small and has a
relatively weak dependence on the $U$ value (Fig. \ref{spec}),
which is also a consequence of the $T$-matrix renormalization
\cite{katsnelson:9}.

For spin-up states we have a normal Fermi-liquid behavior $-\mathrm{Im}%
\Sigma _{d}^{\uparrow }(E)\propto (E-E_{F})^{2}$ with a typical
energy scale of the order of several eV. The spin-down self energy
behaves in a similar way below the Fermi energy, with a bit
smaller energy scale. At the same time, a significant increase in
$\mathrm{Im}\Sigma _{d}^{\downarrow }(E)$ with a much smaller
energy scale (tenths of eV) is observed right above the
Fermi level, which is more pronounced for $t_{2g}$ states (Fig. \ref{nimnsb}%
). The NQP states are visible in the spin-down DOS, Fig.
\ref{nimnsb}, at the same energy scale as the imaginary part of
$\Sigma ^{\downarrow }$. A
similar behavior is evidenced in the model calculation of Fig. \ref%
{model_bethe}. The NQP spectral weight in the density of states (Fig. \ref%
{nimnsb}) is proportional to the imaginary part of the
self-energy.

From the general many-body theory, the DMFT approach is
neglecting the momentum dependence of the electron self-energy. In
many cases, such as the Kondo effect and the Mott metal-insulator
transition the energy dependence of the self-energy is obviously
more important than the momentum dependence
and, therefore, the DMFT scheme is adequate to consider these problems \cite%
{georges:13}. In the case of itinerant electron ferromagnetism,
the situation is much less clear. However, the LDA+DMFT treatment
of finite temperature magnetism and electronic structure in Fe and
Ni appeared to be quite successful \cite{lichtenstein:067205}.
Experimentally, even in itinerant electron paramagnets which are
close to ferromagnetic instability, such as Pd, the momentum
dependence of the self-energy does not seem to be essential
\cite{Joss:5637}. One can expect that in magnets with well defined
local moments such as half-metallic ferromagnets the local DMFT
approximation for the self-energy should be even more accurate. In
particular, as discussed above, it can be used for the
calculations of spin-polaronic or non-quasiparticle effects in
these materials.

\subsubsection{Impurities in HMF: lanthanides in NiMnSb}

\label{sec:la_nimnsb}

Here we discuss the introduction of non-intrinsic deffects in 
NiMnSb, which preserves the half-metallic properties on one hand
while optimizing the magnetic disorder on the other hand \cite{attema:S5517}.
Good candidates are the rare-earth impurities in NiMnSb \cite%
{attema:S5517,Chioncel:thesis}. The motivation of this choice is connected with
the substantial spin-orbit interaction in the rare-earth localized
$f$ shell. Besides that, the origin of the band gap in NiMnSb is
closely related to the band gap in III-V semiconductors: it is
expected that a substitution of some of the tetravalent element in
NiMnSb by a lanthanide preserves the essential feature of the
half-metal: the band gap for one spin direction.

These systems can be really synthesized, compounds RNiSb (for
heavy rare earths R) existing with exactly the same crystal
structure as NiMnSb. The rare-earth atoms which show both a large
spin and orbital moment can be expected to introduce a large
spin-orbit interaction, in other words around 1/4 (Nd) and 3/4
(Er) in the lanthanides series. Total energy calculations allow
one to evaluate the coupling between the rare-earth ($4f$)
impurity spins and the manganese ($3d$) conduction electron spins.
For a substantial coupling the fluctuations of the Mn ($3d$)
conduction electron spins, i.e. the spin wave, might be blocked,
thus the magnon branch is qualitatively changed. In contrast with
the clean limit (pure NiMnSb), the magnon spectra of
NiMn$_{1-x}$R$_{x}$Sb present a fragmented structure with several
gaps in the Brillouin zone. This fragmentation implies that the
finite temperature effects are diminished for a suitably chosen
rare-earth.

The ab-initio electronic structure calculations were carried out
using the scalar relativistic linear muffin-tin orbital (LMTO)
method within the
atomic-sphere approximation in two flavors: LDA and LDA$+U$ \cite%
{Anisimov:767,Andersen:3060,Andersen:2571}.
To evaluate the coupling
between the rare-earth and the Mn sublattices, ferro- and
ferrimagnetic structures were taken as the initial state of the
calculation and they were preserved after the self-consistent
calculation.

A simplified model was used that captures the complex interplay of the Mn ($%
3d$) itinerant conduction band electrons and the localized $4f$
electrons, the latter carrying a strong magnetic moment. This
model deals with a mean-field decoupling, in which the Mn $3d$ and
the R $f$ states are described by the LDA$+U$ whereas the $3d-4f$
interaction is treated as perturbation. The corresponding
mean-field Hamiltonian can be written in the form:
\begin{equation}
\mathcal{H}\simeq \mathcal{H}_{\mathrm{LDA+U}}-J\sum_{i}\sigma
_{i}^{3d}S_{i+\delta }^{f}  \label{3d-4f}
\end{equation}
where the spin of the conduction electron at site \textbf{$R_{i}$}
is denoted by $\sigma _{i}^{3d}$ and $S_{i+\delta }^{f}$
represents the spin of the 4f shell at the \textbf{$R_{i+\delta
}$} site. The Mn $d$ local moment fluctuations could be quenched
by a strong $f-d$ coupling, which affects the magnon excitations.
The strength of such a coupling was evaluated by
calculating in an ab-initio fashion the total energy of NiMn$_{1-x}$R$_{x}$%
Sb compounds for a parallel/antiparallel $f-d$ coupling \cite%
{Chioncel:thesis,attema:S5517}. Given the geometry of the cell the
lanthanide substitution is realized in the $fcc$-Mn sublattice, so
12 pairs
of R($4f$)-Mn($3d$) are formed. The $f-d$ coupling is calculated as the $%
E_{\uparrow \uparrow }-E_{\downarrow \uparrow }$ energy
corresponding to a pair. 
Adopting a two sublattices model described by the Hamiltonian
(\ref{3d-4f}), i.e. the sublattice of lathanides R($4f$) being
antiferro/ferromagnetically oriented with respect to the Mn($3d$) 
sublattice, the $J$ values correspond to the intersublattice couplings. In
pure NiMnSb the ferromagnetic Curie temperature $T_{C}=740$K is
determined by the strength of the Mn-Mn ($3d-3d$) sublattice
interaction. For a small lanthanide content one can expect that this 
interaction in NiMn$_{1-x}$R$%
_{x} $Sb compounds is on the same scale as in pure NiMnSb.
Therefore the substitution introduces the competition between the
intra- and intersublatice interactions which are crucial
parameters for any practical
applications. In the case of Nd  there is a large $%
3d-4f$ \ coupling which should dominate over the $3d-3d$ coupling.
For
temperatures lower than the Curie temperature corresponding to the Ni$_{8}$Mn%
$_{7}$NdSb$_{8}$ compound, $T_{C}^{\mathrm{Nd}}$, the $3d-4f$ \
coupling interaction will lock the Mn ($3d$) magnetic moment
fluctuations decreasing the available number of magnon states.
Above $T_{C}^{\mathrm{Nd}}$ the thermal fluctuations already took
away the long range order of the Mn ($3d$) sublattice, and there
are no available magnon states at all. As a consequence, the Nd
substitution can be attractive for the high-temperature
applications where the $3d-4f$ coupling might play an important
role. In the cubic structure of NiMnSb the uniaxial anisotropy is
completely missing. Nevertheless, the $4f$ impurity spin-orbit
coupling lifts the degeneracy in the $\Gamma $ point. The lowest
$3d-4f$ coupling is realized in the case of Ho.

\begin{figure}[h]
\centerline{\psfig{file=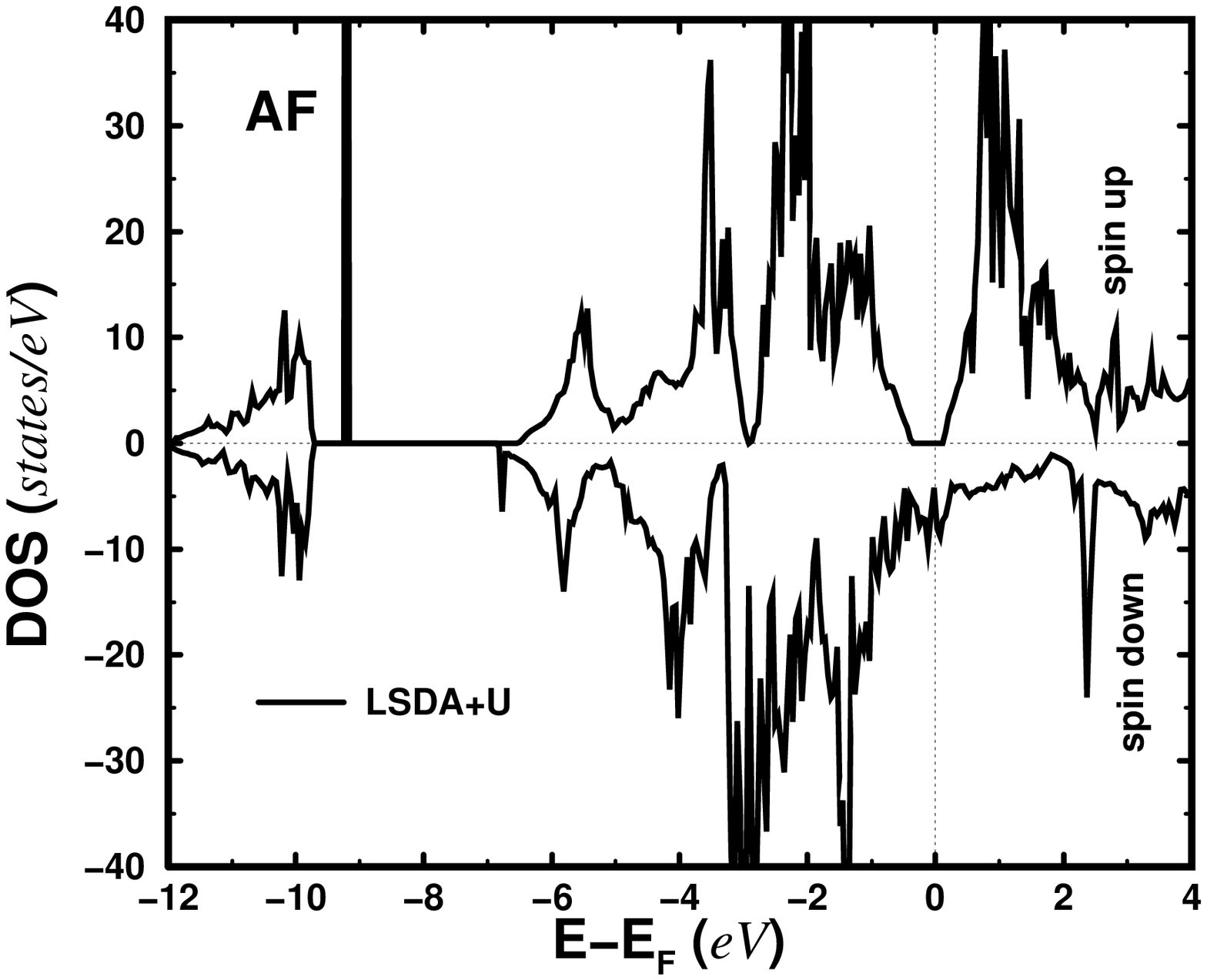,height=2.25in}
\psfig{file=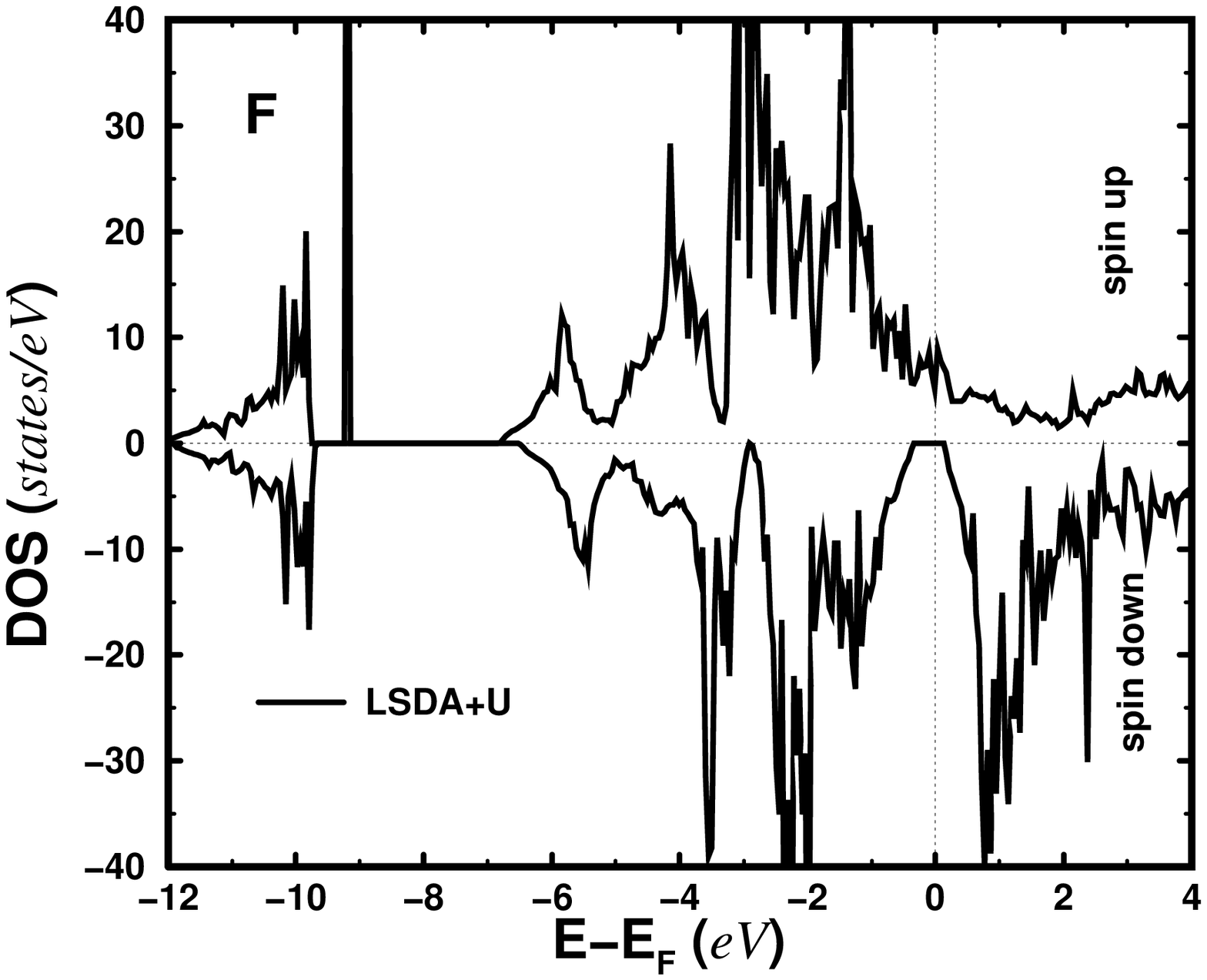,height=2.25in}}
\caption{Density of states for the half-metallic ferromagnet Ni$_8$Mn$_7$HoSb%
$_8$ in the case of anti-parallel $3d-4f$ coupling left, and
parallel coupling right.} \label{ho_nimnsb}
\end{figure}

The LDA$+U$ density of states for HoNi$_{8}$Mn$_{7}$Sb$_{8}$ for
$3d-4f$ antiparallel and parallel couplings and $U_{f}=9$~eV value
of the on site Coulomb interaction is presented in Fig.
(\ref{ho_nimnsb}). On one hand for
the antiparallel coupling the magnetic moments are $\mu _{\mathrm{Ho}%
}^{AF}=-4.09\mu _{B},\mu _{\mathrm{Mn1}}^{AF}=3.73\mu _{B},\mu _{\mathrm{Mn2}%
}^{AF}=3.80\mu _{B}$ and the gap of $0.55$~eV is situated in the
majority spin channel. On the other hand for parallel coupling the
magnetic moments
have almost the same magnitude $\mu _{\mathrm{Ho}}^{F}=3.96\mu _{B},\mu _{%
\mathrm{Mn1}}^{F}=3.72\mu _{B},\mu _{\mathrm{Mn2}}^{F}=3.80\mu
_{B}$ with a similar gap situated in the minority spin channel. It
is important to mention that in the case of
$\mathrm{Ni}_{8}\mathrm{Mn}_{7}\mathrm{HoSb}_{8}$ the Ho($4f$)
orbitals do not hybridize with the Mn($3d$) orbitals near the
Fermi level. As one can see in Fig. (\ref{ho_nimnsb}), the
behavior of DOS
near the Fermi level is very similar, so that the nature of carriers around $%
E_{F}$ is not changed.
\begin{figure}[h]
\mbox{\epsfig{file=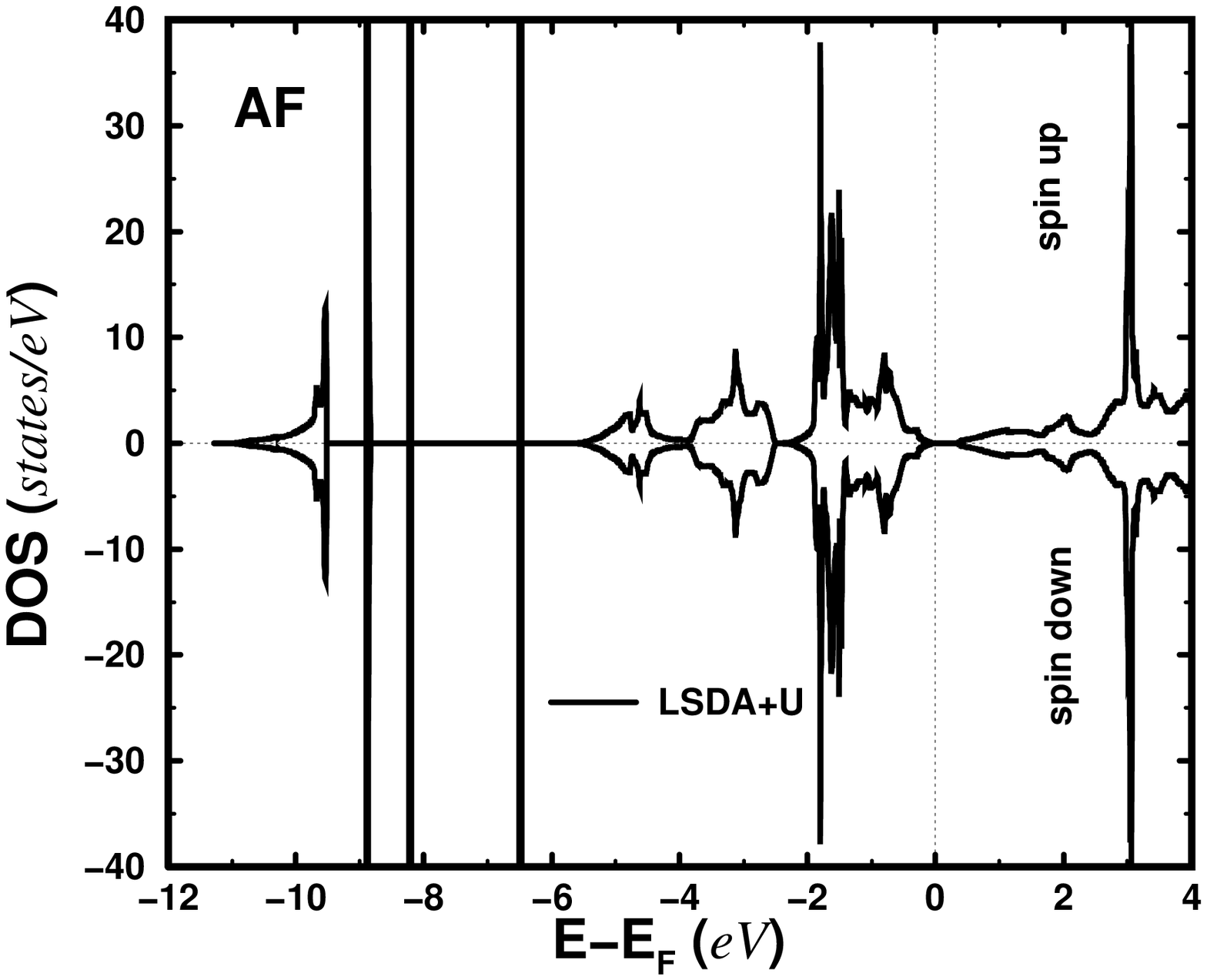,height=2.00in,width=2.5in}}
\caption{Density of states of semiconducting HoNiSb.\cite{Chioncel:thesis}}
\label{honisb}
\end{figure}
The electronic structure calculation for the antiferromagnetic
HoNiSb compound was performed in
\cite{Chioncel:thesis,attema:S5517}. This is known to be a
semiconducting material with interesting transport properties
(giant magnetoresistance effect
\cite{Karla:42,Karla:215,Karla:294,Pierre:74}). The
inverse susceptibility curves show a Curie-Weiss behavior down to $10$K \cite%
{Karla:42,Karla:215,Karla:294,Pierre:74}. The onset of the
antiferromagnetic
ordering was estimated from susceptibility measurements between $1.5$ and $%
2.5~$K, and the neutron diffraction data indicated an
antiferromagnetic
propagation vector $(1/2,1/2,1/2)$ \cite%
{Karla:42,Karla:215,Karla:294,Pierre:74}. In order to describe the
antiferromagnetic ground state, a rhombohedral description of the Ho$_{2}$Ni$%
_{2}$Sb$_{2}$ unit cell was used. The atoms $\mathrm{Ho}_{1}(0,0,0)$ and $%
\mathrm{Ho}_{2}(1,1,1)$ acquire magnetic moments of $\mu _{\mathrm{Ho}%
_{1}}=4.0\mu _{B}$ and $\mu _{\mathrm{Ho}_{2}}^{F}=-4.0\mu _{B},$
respectively. The antiferromagnetic insulating state shows a gap of $0.29$%
~eV in agreement with the experimental results. There are some
qualitative features which are similar for the antiferromagnetic
Ho$_{2}$Ni$_{2}$Sb$_{2}$ and the HoNi$_{8}$Mn$_{7}$Sb$_{8}$
compounds: the almost identical magnetic
moment $\mu _{\mathrm{Ho}}$ and positions of occupied and unoccupied Ho($4f$%
) peaks in DOS. Perhaps the most important observation is that the
spin-down channel in Ho$_{2}$Ni$_{2}$Sb$_{2}$ is isoelectronic to
that in NiMnSb, so
that the Ho substitution would preserve the half-metallicity of HoNi$_{8}$Mn$%
_{7}$Sb$_{8}$ in the minority spin channel.

\subsubsection{FeMnSb: a ferrimagnetic half-metal}

\label{sec:femnsb}

Early theoretical studies demonstrated that the gap in the
minority spin channel is stable with respect to change of the
\textit{3d} atom X = Fe, Co, Ni in the XMnSb compounds
\cite{deGroot:330,Kubler:257}. Noticeable difference between Ni-
and Fe-based Heusler alloys is that NiMnSb is a
half-metallic ferromagnet with a very small value of Ni magnetic moment ($%
0.2\mu _{B}$), whereas in FeMnSb the antiferromagnetic coupling between Fe ($%
-1\mu _{B}$) and Mn ($3\mu _{B}$) moments stabilizes the gap and
the half-metallic ferrimagnetic electronic structure
\cite{deGroot:330}. Unfortunately, the ternary compound FeMnSb
does not exist, but indications of its magnetic and
crystallographic properties were obtained by extrapolating the
series of Ni$_{1-x}$Fe$_{x}$MnSb \cite{deGroot:330} to high Fe
concentration.

\begin{figure}[h]
\includegraphics[angle=0,width=\linewidth]{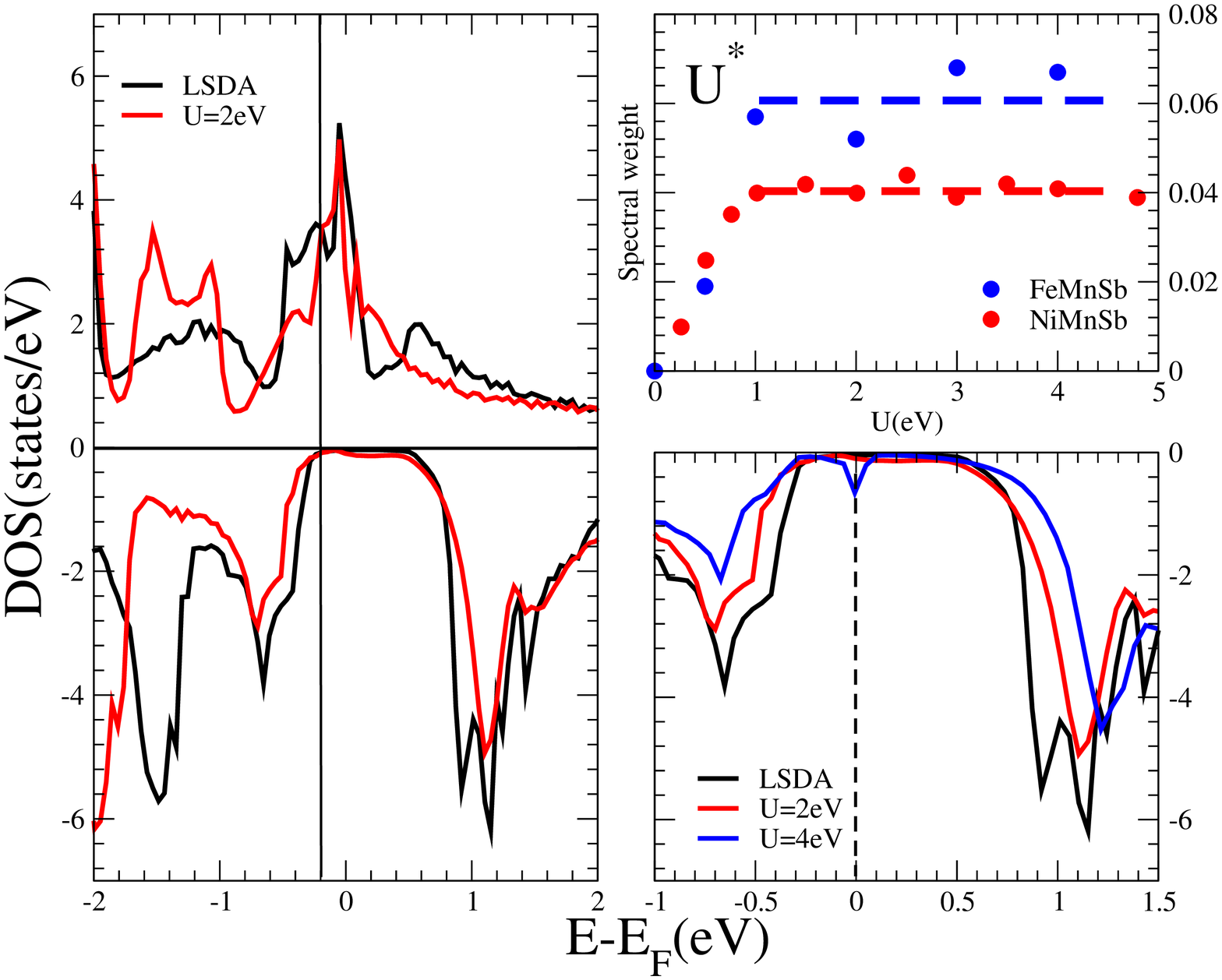}
\caption{(Color online) Left: Density of states of half-metallic
FeMnSb, LSDA (black line) and LSDA+DMFT (red line), for the
efective Coulomb
interaction $U$=2~eV exchange parameter $J$=0.9~eV and temperature $T$%
=300~K. Lower right panel: zoom around $E_{F}$ for different
values of $U$. Upper right panel: Spectral weight of the NQP
states calculated as a function of $U$. The values obtained for
NiMnSb \protect\cite{chioncel:144425} are plotted for comparison.\cite{chioncel:137203}
} \label{fig:DOSFeMnSb}
\end{figure}

In the non-relativistic approximation there are two essentially
different sources for states in the gap at finite temperatures.
First, there is the simple effect of gap filling due to disorder,
i.e. due to scattering on static (classical) spin fluctuation or
thermal magnons; this is symmetric with respect to the Fermi
energy. On the contrary, the correlation effects result in an
asymmetry in the gap filling, spin-down non-quasiparticle states
appearing above the Fermi level. One has to take into account also
spin-orbit coupling effects mixing the spin-up and spin-down
states due to non-zero elements of spin-orbital interactions
$V_{SO}^{\sigma ,\sigma ^{\prime }}$ (see the discussion in
Sect.\ref{nqp}). To illustrate the differences between the static
and dynamic effects we plot the DOS of the LDA+DMFT calculations
which should be compared with recent results including SO coupling
\cite{deGroot:330,Mavropoulos:S5759}.

A relatively weak dependence of the NQP spectral weight on $U$ (Fig. \ref%
{fig:DOSFeMnSb}) is evidenced for both NiMnSb and FeMnSb
compounds. A \textquotedblleft saturation\textquotedblright\ of
the spectral weight for FeMnSb takes place for almost the same
value, $U^{\ast }\simeq 1$~eV, as in the case of NiMnSb, which is
in agreement with experimental observation of
relatively weak correlation effects in Heusler HMF like PtMnSb \cite%
{Kisker:21}. The small value of effective Hubbard parameter can be
understood in terms of the large $T$-matrix renormalization of the
Coulomb interactions \cite{katsnelson:1037,katsnelson:9}. The
spectral weight values
for FeMnSb are larger in comparison with those obtained for NiMnSb \cite%
{chioncel:144425}, which can be attributed to a larger majority
spin DOS at the Fermi level.

The spin-orbit coupling produces a peak in the minority-spin
channel close to the Fermi level \cite{Mavropoulos:S5759}, which
is an order of magnitude smaller than the spectral weight of the
NQP states. According to the SO results, the polarization at the
Fermi level for NiMnSb and FeMnSb are almost the same. In
contrast, the calculation \cite{chioncel:137203} shows that the
spectral weight of the NQP states in FeMnSb is almost twice as
large as the value for NiMnSb.

\begin{figure}[h]
\includegraphics[angle=0,width=0.95\linewidth]{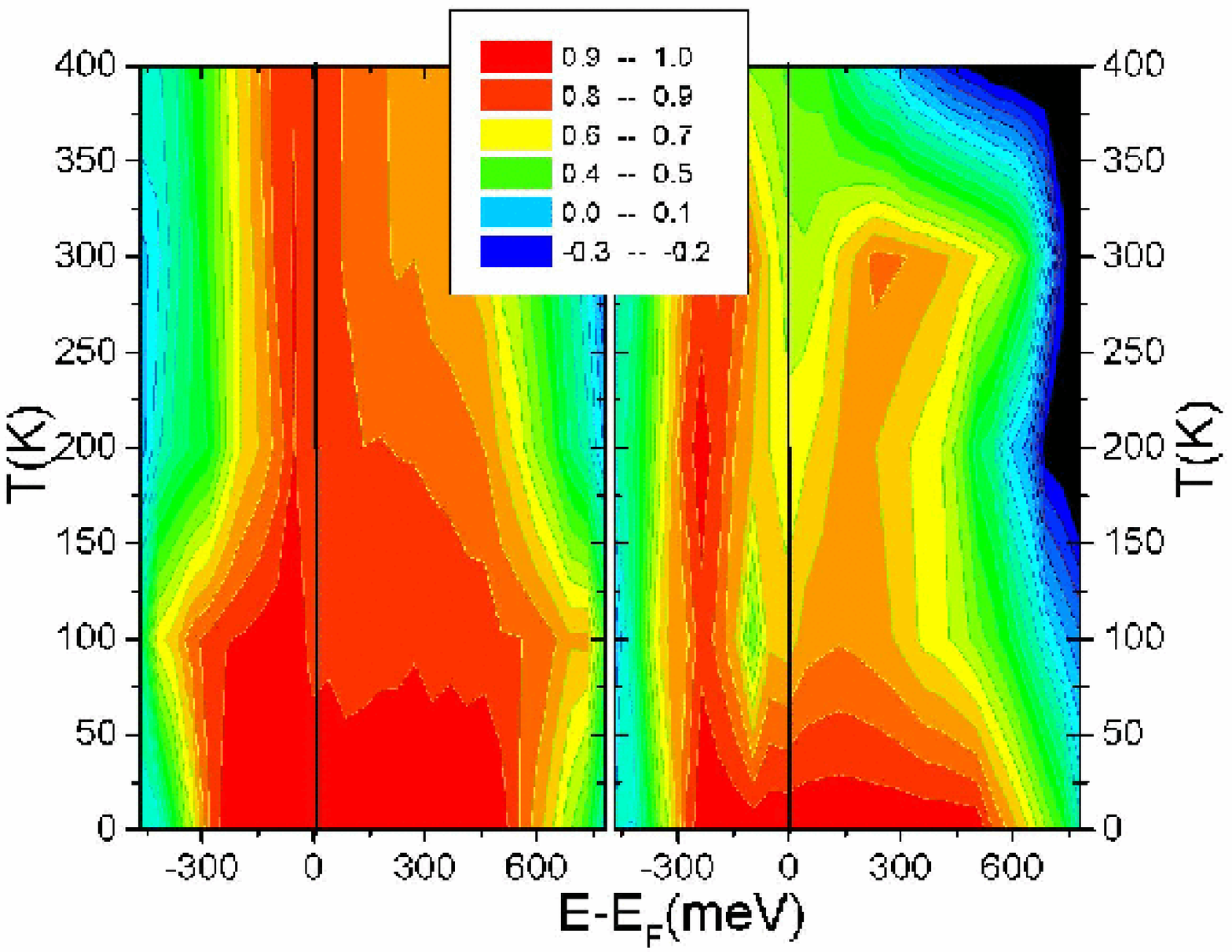}
\caption{(Color online) Contour plots of polarization as a function
of energy 
and temperature for different values of local Coulomb interaction U. Left $U$%
=2~eV, right $U$=4~eV. The LSDA polarization is plotted as the
$T$=0~K
temperature result. The asymmetry of the NQP states, is clearly visible for $%
U$=4~eV. \cite{chioncel:137203}} \label{pol_t} 
\end{figure}

To discuss the influence of temperature and local Coulomb
interactions on the polarization in FeMnSb compound, we present
results of LDA+DMFT calculations for $T\leq 400~\mathrm{K}$ and
different $U$'s. Fig.\ref{pol_t} presents the contour plot of spin
polarization $P(E,T)$ as a function of energy $E$ and temperature
$T$ for $U=2$ and 4 eV. The LDA value, plotted
for convenience as the $T=0~$K result, shows a gap with magnitude $0.8~%
\mathrm{eV,}$ in agreement with previous calculations \cite%
{Mavropoulos:S5759}.

One can see a peculiar temperature dependence of the spin
polarization. The
NQP features appear for $E-E_{F}\geq 0$ and are visible in Fig. \ref%
{fig:DOSFeMnSb} for $U=2$~eV and $T=300$K. Increasing the value of
$U$ from 2~eV to 4~eV, the NQP contribution in depolarization
becomes more significant. When the tail of the NQP states crosses
the Fermi level, a drastic depolarization at Fermi level takes
place, for $U=4$~eV the NQP contribution being pinned to the Fermi
level.

\begin{figure}[h]
\includegraphics[angle=0,width=0.95\linewidth]{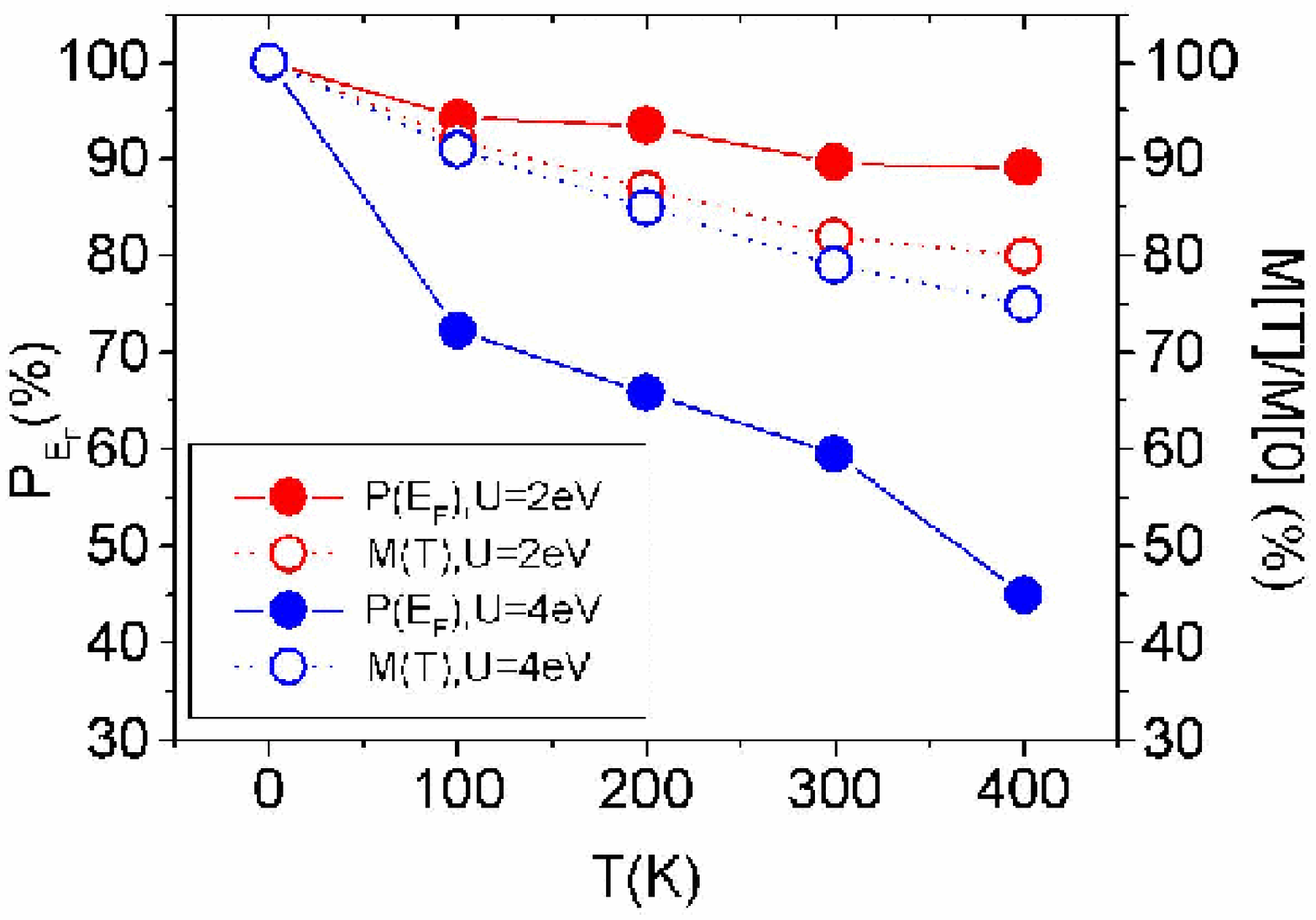}
\caption{(Color online) Temperature dependent polarization at the
Fermi level, $P(E=E_F, T)$ (solid line) and magnetization (dashed
line) for different values of local Coulomb interaction $U$.\cite{chioncel:137203} }
\label{pm_t}
\end{figure}

One can see a clear distinction between the finite-temperature
behavior of the polarization and magnetization, shown in Fig.
\ref{pm_t} for different values of $U$. 
For $U=4eV$, already at 100K there  
is a strong depolarization about 25$\%$. 
On contrary, it is interesting to note that the reduced magnetization $%
M(T)/M(0)$ decreases slowly in the temperature range shown in Fig. \ref%
{pol_t}. This reduction is a consequence of the finite temperature
excitations, i.e. spin-flip processes, affecting both spin
channels. In the minority spin channel, NQP states are formed, and
in the majority channel a
spectral weight redistribution around the Fermi level (Fig.\ref%
{fig:DOSFeMnSb}) contributes to the depolarization. The
corresponding depolarization increases with the strength of
correlations. The density of NQP states displays a rather strong
temperature dependence (Sect.\ref{pol})
and results in an asymmetry that is visible in Figs.~\ref{fig:DOSFeMnSb} and %
\ref{pol_t}.

NQP states dominate in the depolarization of this class of Heusler
compounds, while spin-orbit contributions are much smaller. In
addition, many-body effects are more pronounced in FeMnSb than in
NiMnSb. This is connected with the larger DOS in the majority spin
channel in the former material. Therefore, doping of NiMnSb by Fe
could be an interesting issue to investigate the interplay between
alloying and many-body effects. The LDA+DMFT calculations for
NiMnSb supercell containing 25 \% Fe impurities show a
half-metallic character at the LDA level, with the same strong
correlation-induced depolarization effects as in pure FeMnSb.
Therefore many-body effects for this material are of primary
importance even in the presence of disorder. Correlation effects
on surfaces of half-metals were discussed recently and it was
shown that these states can be probed both directly and via their
effect on surface states \cite{irkhin:104429}.

\subsubsection{Co$_2$MnSi: a full-Heusler ferromagnet}

The origin of  minority band gap in full Heuslers was discussed by
Galanakis et al. \cite{Galanakis:765,Galanakis:315213}. Basing on the analysis of
the band structure calculations, it was shown that the $3d$
orbitals of Co atoms from the two different sublattices, Co$^1
(0,0,0)$ and Co$^2 (1/2,1/2,1/2)$, couple and form bonding hybrids
Co$^1$($t_{2g}$/$e_{g}$) - Co$^2$($t_{2g}$/$e_g$). In other words,
the $t_{2g}$/$e_{g}$ orbitals of one of the Co atom can couple
only with the $t_{2g}$/$e_{g}$ orbitals of  other Co atom. Further
on, the Co-Co hybrid bonding orbitals hybridize with
Mn(d)-$t_{2g}$,$e_g$ manifold, while the Co-Co hybrid antibonding
orbitals remain uncoupled owing to their symmetry. The Co-Co
hybrid antibonding  $t_{2g}$ is situated below the Fermi energy
$E_F$ and the Co-Co hybrid antibonding $e_{g}$ is unoccupied and
lies just above the Fermi level. Thus, due to the missing
Mn(d)-$t_{2g}$,$e_g$ and Co-Co hybrid antibonding hybridization,
the Fermi energy is situated within the minority gap formed by the
triply degenerate Co-Co antibonding $t_{2g}$ and the double
degenerate Co-Co antibonding $e_g$.

\begin{figure}[tbh]
\begin{center}
\includegraphics[width=0.75\columnwidth, angle=0]{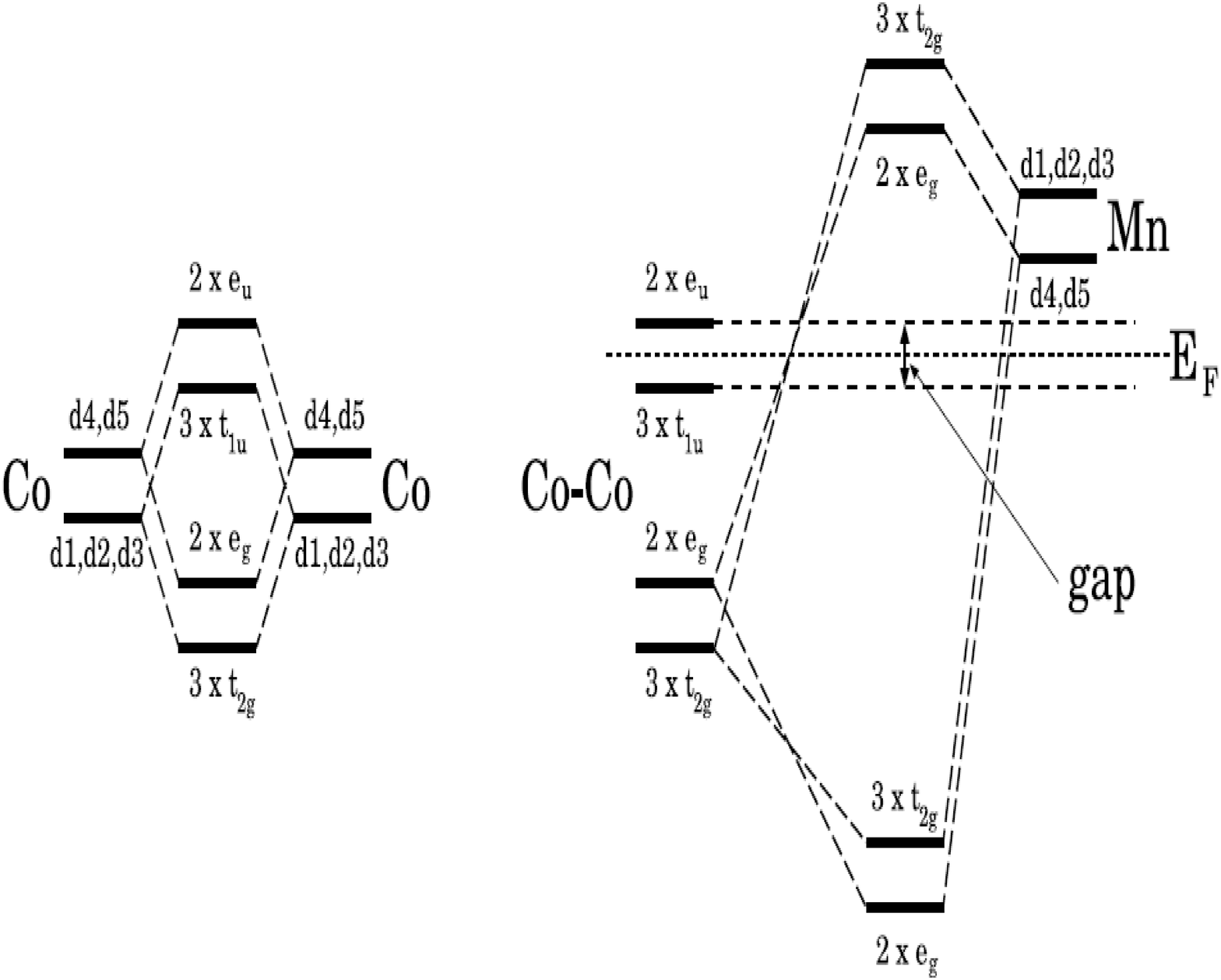}
\end{center}
\caption{Schematic illustration of the gap formation in
Co$_{2}$MnZ compounds with Z=Al, Si, Ge,Sn \protect\cite{Galanakis:765}.}
\label{gap_L21}
\end{figure}

\begin{figure}[tbh]
\begin{center}
\includegraphics[width=0.75\columnwidth, angle=0]{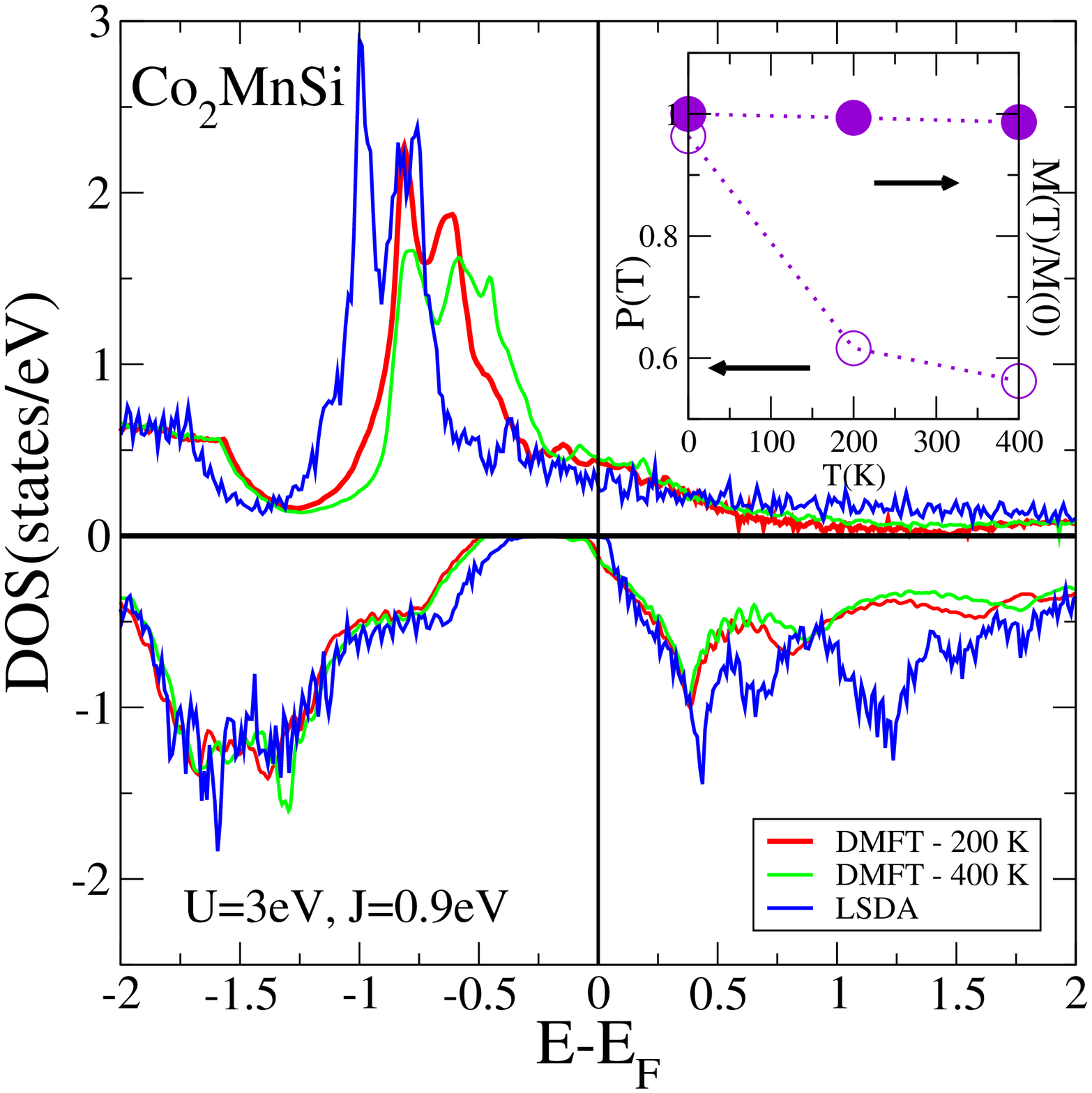}
\end{center}
\caption{Total density of states of Co$_2$MnSi full-Heusler alloy.
Notice that the gap is formed between the occupied Co-Co
antibonding $t_{2g}$ orbitals and empty Co-Co antibonding $e_g$
orbitals. The LDA+DMFT results are presented as well for $U=3$eV,
and $J=0.9$eV, and different temperatures. The inset shows the
finite temperature spin-polarization.} \label{fig:dos-Co2MnSi}
\end{figure}

Fig. \ref{fig:dos-Co2MnSi} shows the results of DOS calculations
using the LDA and LDA+DMFT schemes. The inset presents the spin
polarization $P(E_F)=(N_{\uparrow} (E_F)-N_\downarrow (E_F))/
(N_\uparrow (E_F)+N_\downarrow (E_F))$. One can see that in the
minority spin channel asymmetric NQP states are formed, while in
the majority a spectral weight redistribution takes place, which
contributes to the depolarization. Contrary to the FeMnSb
\cite{chioncel:137203}, where the density of NQP states shows a
rather strong temperature dependence, in the full-Heusler
Co$_2$MnSi the temperature dependence is not so significant,
similar to the result obtained for NiMnSb \cite{chioncel:144425}.

\subsection{Half-metallic materials with zinc-blende structure}

\label{sec:ZB}One of the strongest motivations to investigate
magnetic semiconductors and half-metallic ferromagnets is the
possibility to design and produce novel stable structures on
semiconducting substrates with new interesting properties. From
this point of view first-principle studies are an excellent
starting point to predict new systems having the desired
properties. Using full-potential density-functional method, all
the 3d transition metal pnictides and chalcogenides with wurtzite
structure were investigated systematically in
Ref.\cite{xie:134407} in order to find half-metallic ferromagnets.
These can be fabricated as thin films with thickness large for
real spintronic applications. Nine of the wurtzite phases, MnSb,
CrAs, CrSb, VAs, VSb, CrSe, CrTe, VSe, and VTe, were found to be
robust half-metallic ferromagnets with large half-metallic gaps
(0.2$\div $1~eV). Being compatible with the III-V and II-VI
semiconductors, these half-metallic ferromagnetic phases, when
realized experimentally, would be useful in spintronic and other
applications. At the same time, zinc-blende (ZB) phases of MnAs,
CrAs and CrSb have been fabricated successfully in the form of
nanodots, ultrathin films and multilayers, respectively. A study
within density-functional theory \cite{liu:172411} predicted for
the ZB CrSb half-metallic ferromagnetism with a magnetic moment of
3.0 $\mu _{B}$ per formula.

\subsubsection{CrAs: tunable spin transport}

Recently Akinaga \textit{et al.} \cite{akinaga:L1118} found the
possibility to fabricate ZB-type CrAs half-metallic ferromagnetic
material. Experimental data confirmed that this material is
ferromagnetic with the magnetic moment
of $3\mu _{B}$, in agreement with theoretical predictions \cite%
{akinaga:L1118}. According to this calculation, this half-metallic
material has a gap of about 1.8~eV in the minority spin channel,
which has attracted much attention to this potential candidate for
spintronic applications, keeping in mind also its high Curie
temperature $T_{C}$ about $400$K. Recent experiments on CrAs
epilayers grown on GaAs$(001)$ evidenced an orthorhombic
structure, different from the ZB one, so the structure is rather
sensitive to the preparation process \cite{etgens:2004}. However,
it is highly desirable to explore the possibility of existence of
half-metallic ferromagnetism in materials which are compatible
with practically important III-V and II-IV semiconductors. For
this purpose efforts have been made to
investigate metastable ZB structures, such as CrAs \cite%
{akinaga:L1118,mizuguchi:7917}.

It is interesting to explore the mechanisms of half-metallic
ferromagnetism at finite temperature from a realistic electronic
structure point of view. Theoretical studies \cite{shirai:6844} of
the $3d$ transition metal monoarsenides have shown that the
ferromagnetic phase of ZB structure CrAs compound should be more
stable than the antiferromagnetic one. The many-body effects are
very sensitive to structural properties of the artificially
produced CrAs compound \cite{chioncel:085111}. Similar electronic
structure calculations concerning the stability of the
half-metallic ferromagnetic state in the ZB structure have been
carried out \cite{xie:2003}. The LDA+DMFT calculations were
carried out for three lattice constants: the GaAs ($5.65$\AA ),
InAs ($6.06$\AA ) and the \textquotedblleft
equilibrium\textquotedblright\ value ($a_{\mathrm{eq}}=5.8$\AA )
obtained by density functional calculations \cite{shirai:6844}.
The corresponding LDA
computational results agree with previous ones \cite%
{akinaga:L1118,shirai:6844,galanakis:104417}. In order to evaluate
the average Coulomb interaction on the Cr atoms and the
corresponding exchange interactions, the constrained LDA method
\cite{anisimov:7570} was used in Ref \cite{chioncel:085111}, which
yielded $U=6.5~$eV and $J=0.9$~eV. It is important to note that
the values of the average Coulomb interaction
parameter slightly decrease going from the GaAs ($U=6.6$~eV) to InAs ($%
U=6.25 $~eV) lattice constants, \cite{chioncel:085111}
which is in agreement with a naive picture of a less effective
screening due to increasing of the distances between the atoms.

The typical insulating screening used in the constraint calculation \cite%
{anisimov:7570} should be replaced by a metallic kind of
screening. The metallic screening will lead to a smaller value of
$U$. Since there are no reliable schemes to calculate $U$ in
metals, some intermediate value were chosen, $U=2$~eV and
$J=0.9$~eV. It is important to realize that there are no
significant changes in the values of the average Coulomb
interaction for the lattice structures studied, the exchange
interaction being practically constant. Note that the physical
results are not very sensitive to the value of $U$, as it was
demonstrated for NiMnSb \cite{chioncel:144425}.

The LSDA and LSDA+DMFT calculation for the density of states, DOS,
is presented in Fig. \ref{200_cras_DOS}. Depending on the lattice
constant the Cr and As atoms loose electrons and this charge is
gained by the vacant sites. As a result, the Fermi level is moved
from the right edge of the gap (for the GaAs substrate) towards
the middle of the gap (for the InAs substrate). The Cr moments are
well localized due to a mechanism similar to
localization of the magnetic moment on the Mn atom in NiMnSb \cite%
{deGroot:2024}. The local Cr spin moment is more than $3\mu _{B}$,
and the
As atom possesses a small induced magnetic moment (of order of $-0.3\mu _{B}$%
) coupled antiparallel to the Cr one. The results are presented in table \ref%
{tab2_cras}. Cr magnetic moments calculated in DMFT increase in
comparison with the LDA results due to the localization tendency
of the Cr$~3d$ states, which is a consequence of correlation
effects.

According to the calculations of Ref.\cite{chioncel:085111}, the
system remains half-metallic with a rather large band gap (about
1.8~eV) for all the lattice constants compared with the band gap
of the NiMnSb which is only 0.75~eV \cite{deGroot:2024}. In Fig.
\ref{200_cras_DOS} the non-quasiparticle states are visible for
lattice parameter higher than the equilibrium one, with a
considerable spectral weight in the case of the InAs substrate.
This situation is very favorable for the experimental
investigation of the NQP states.

Comparing the electronic structure of CrAs growing on InAs or GaAs
substrates we can conclude that the most significant change in the
electronic structure is related with the As $p$ states. Having a
larger lattice constant in the case of InAs substrates, the Cr
atom acquires a slightly larger magnetic moment. Nevertheless, in
the LDA calculations the magnetic moment per unit cell is integer,
$3\mu _{B}$. Expanding the lattice constant from GaAs to InAs
lattice the Cr states become more \textquotedblleft
atomic\textquotedblright , and therefore the spin magnetic moment
increases. This is reflected equally in the charge transfer which
is smaller for the InAs lattice parameters. A larger Cr moment
induces a large spin polarization of the As $p$ states,
compensating the smaller $p-d$ hybridization, the total moment
retaining its integer value of $3\mu _{B}$
\cite{galanakis:104417}.

The essential difference of the many-body electronic structure for
the lattice constants of GaAs and InAs is completely due to the
difference in the position of the Fermi energy with respect to the
minority-spin band gap, whereas the self-energy characterizing the
correlation effects is not changed too much (Fig. \ref{sigma}).
The total density of states $N\left(
E\right) $ is rather sensitive to the difference between the band edge $%
E_{c} $ and the Fermi energy $E_{F}$. If this difference is very
small
(i.e., the system is close to the electronic topological transition $%
E_{c}\rightarrow E_{F}$) one can use a simple expression for the
singular contribution to the bare density of states, $\delta
N_{0}\left( E\right) \propto \sqrt{E-E_{c}}$ $\left(
E>E_{c}\right) $. The appearance of the complex self-energy
$\Sigma \left( E\right) =\Sigma _{1}\left( E\right) -i\Sigma
_{2}\left( E\right) $ \ changes the singular contribution as
\begin{equation}
\delta N\left( E\right) \propto \lbrack \sqrt{Z_{1}^{2}\left(
E\right) +\Sigma _{2}^{2}\left( E\right) }+Z_{1}\left( E\right)
]^{1/2}
\end{equation}
where $Z_{1}\left( E\right) =E-E_{c}-\Sigma _{1}\left( E\right) $ (cf. Ref. %
\onlinecite{katsnelson:63}). Assuming that the self-energy is
small in comparison with $E-E_{c}$ one can find for the states in
the gap $\delta N\left( E\right) \propto \Sigma _{2}\left(
E\right) /\sqrt{E_{c}-E}$ $\left( E<E_{c}\right) .$ One can see
that the shift of the gap edge changes drastically the density of
states for the same $\Sigma _{2}\left( E\right) .$

A practical use of tunable properties of NQP states in CrAs grown
on different substrates is possible. For most of the applications
room-temperature and the stability of the ferromagnetic state are
the most important prerequisites. The ferromagnetic CrAs might be
grown on III-V semiconductors similar to the zinc-blende CrSb
\cite{zhao:2776}. The presence of the NQP states was obtained in
Ref. \cite{chioncel:085111} for CrAs lattice parameters larger
than $5.8$\AA . It was found experimentally that at 300K, around
this value of the lattice parameter, a stable solid solution of
Ga$_{0.65}$In$_{0.35}$As is formed \cite{harland:2715}. Thus,
from the practical point of view, a $65\%$ of gallium in a Ga$_{x}$In$_{1-x}$%
As compound would constitute the ideal substrate for the CrAs
half-metal. This could be a part of the epitaxial III-V structure
providing an easy way to integrate with the existing semiconductor
technology.

To determine possible substrates for growth of layered
half-metallic materials, electronic structure calculations were
carried out for lattice parameters in the range 5.60$\div $
6.03\AA\ \cite{fong:239}. According to these calculations, growth
with minimal strain might be accomplished in a half-metallic
multilayer system grown on InAs substrate which would be the best
choice to evidence the NQP states, since the Fermi energy is
situated in this case far enough from the bottom of the conduction
band.

\begin{figure}[h]
\centerline{\psfig{figure=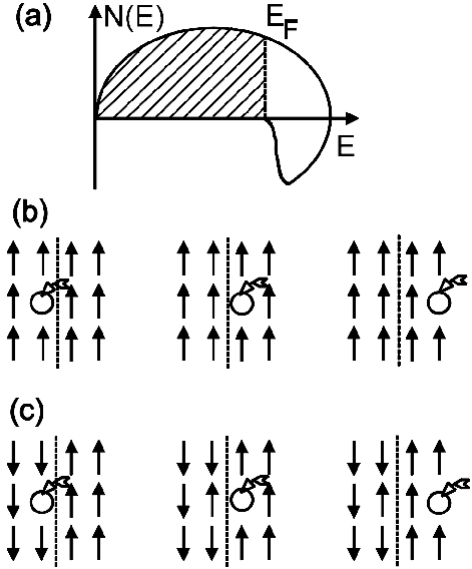,height=3.15in}}
\caption{The tunneling transport between strongly correlated
ferromagnets. The density of states in the lower Hubbard band (a)
is provided by standard current states for majority-spin electrons
(above) and by non-quasiparticle states for minority-spin
electrons (below), the latter contribution being non-zero only
above the Fermi energy (occupied states are shadowed). However,
the tunneling is possible both for parallel (b) and antiparallel
(c) magnetization directions. \cite{chioncel:085111}} \label{model_cras}
\end{figure}

\begin{figure}[h]
\centerline{\psfig{figure=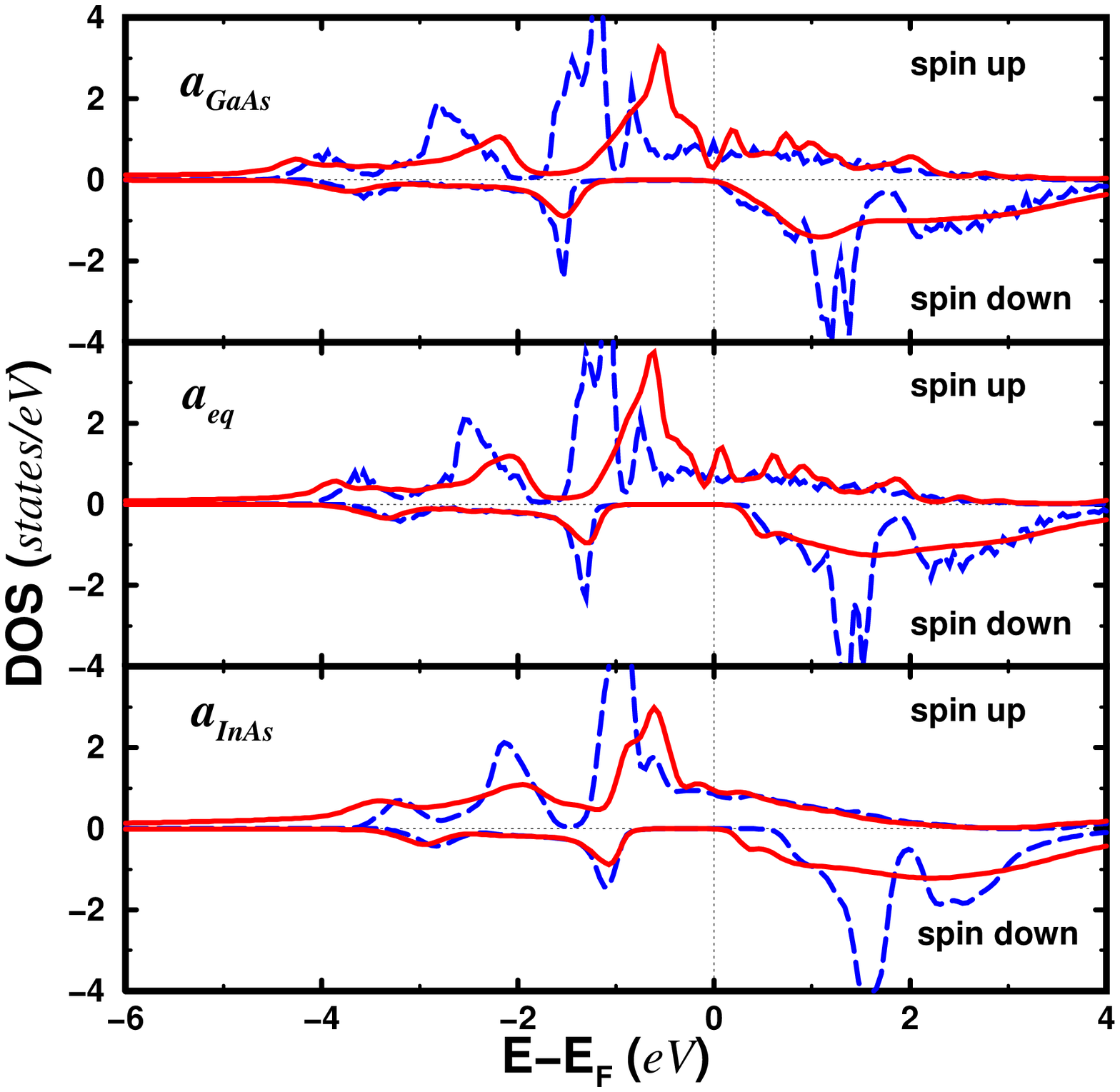,height=3.0in}}
\caption{(Color online) Cr density of states calculated in LSDA
(dashed
line) and LSDA+DMFT (solid line) methods corresponding to a temperature of $%
T $=200K, average Coulomb interaction parameter $U$=2~eV and exchange $J$%
=0.9~eV. The non-quasiparticle states are clearly visible for
lattice parameters larger than $a_{eq}=5.8\mathring{A}$, in the
unoccupied part for minority spin channel just above the Fermi
level, around 0.5~eV. \cite{chioncel:085111}} \label{200_cras_DOS}
\end{figure}

\begin{figure}[h]
\centerline{\psfig{figure=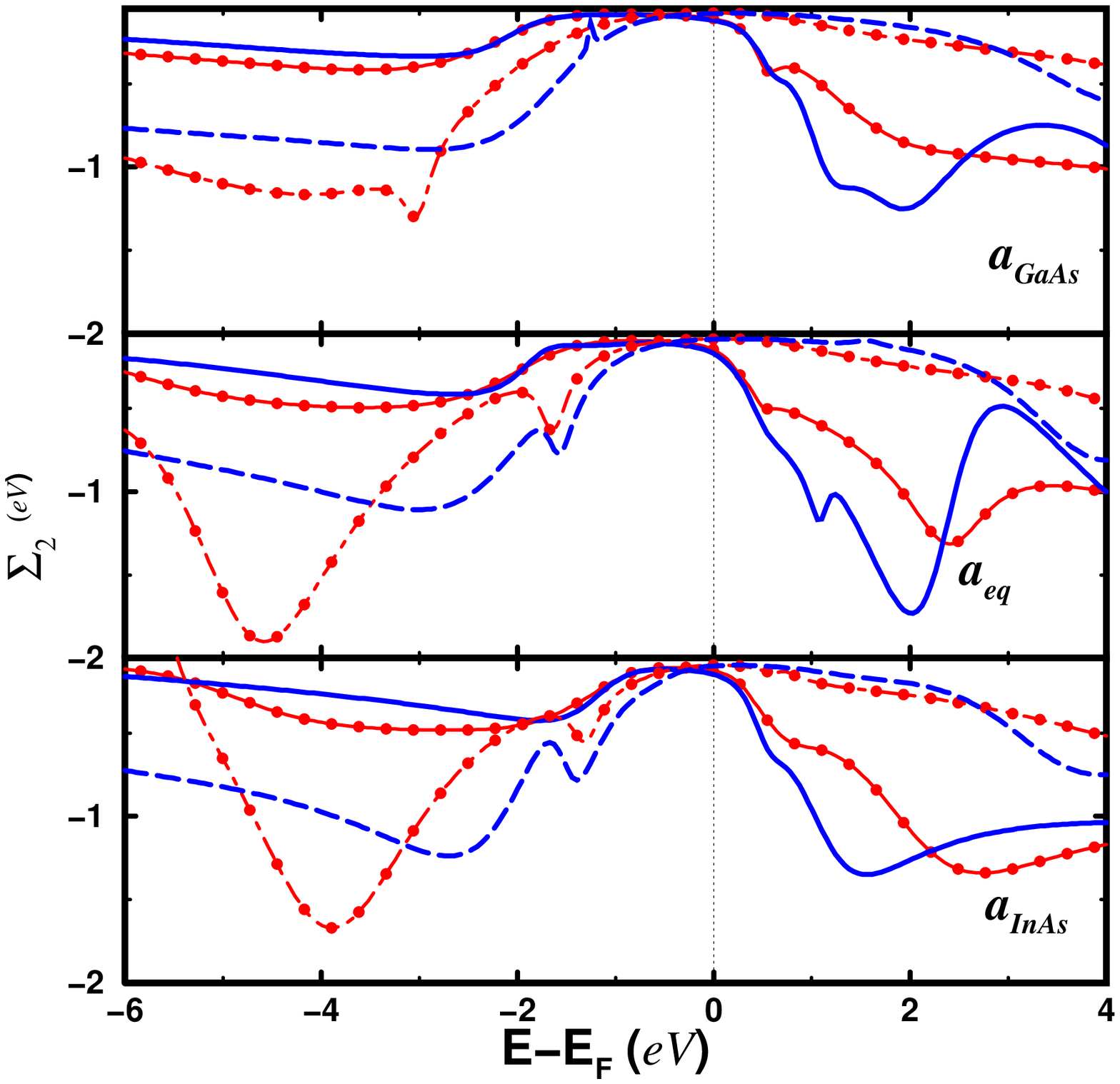,height=3.0in}}
\caption{(Color online) Energy dependences of imaginary parts of
the electron self-energy $\Sigma_2(E)$, for lattice constants of
GaAs (a), equilibrium one (b), and of InAs (c): $e_g$ down solid
line, $t_{2g}$ down decorated solid line, $e_g$ up dashed line,
$t_{2g}$ up decorated dashed line. \cite{chioncel:085111}} \label{sigma}
\end{figure}

A very high sensitivity of the minority-electron DOS near $E_{F}$
to the lattice constant opens a new interesting opportunity.
Suppose we have an antiparallel orientation of the magnetizations
in the CrAs-based tunnel junction (such as shown in Fig.
\ref{model_cras}c), then the $\mathcal{I}-V$ characteristic is
determined by the density of the NQP states. Thus if we will
influence the lattice constant (e.g., using a piezoelectric
material) one can modify the differential conductivity. This makes
CrAs a very promising material with tunable characteristics which
opens new ways for applications in spintronics.

\begin{table}[tbp]
\begin{tabular}{c|ccccc|ccc}
& Cr & As & E & E1 & Total & $T$ & $U$ & $J$ \\
& $(\mu_B)$ & $(\mu_B)$ & $(\mu_B)$ & $(\mu_B)$ & $(\mu_B)$ &
$(K)$ & (eV) & (eV) \\ \hline\hline $\mu_{\mathrm{LDA}}^{\rm GaAs}
$ & 3.191 & -0.270 & -0.009 & 0.089 & 3.00 & - & -
& - \\
$\mu_{\mathrm{DMFT}}^{\rm GaAs}$ & 3.224 & -0.267 & -0.023 & 0.067
& 3.00 & 200 & 2 & 0.9 \\ \hline $\mu_{\mathrm{LDA}}^{\rm eq.} $ &
3.284 & -0.341 & -0.018 & 0.076 & 3.00 & - & -
& - \\
$\mu_{\mathrm{DMFT}}^{\rm eq.}$ & 3.290 & -0.327 & -0.024 & 0.068
& 3.00 & 200 & 2 & 0.9 \\ \hline $\mu_{\mathrm{LDA}}^{\rm InAs} $
& 3.376 & -0.416 & -0.025 & 0.066 & 3.00 & - & -
& - \\
$\mu_{\mathrm{DMFT}}^{\rm InAs}$ & 3.430 & -0.433 & -0.033 & 0.043
& 3.00 & 200
& 2 & 0.9%
\end{tabular}
\vspace{0.25cm} \caption{Summary of the results of the
calculations  \cite{chioncel:085111}. CrAs magnetic moments
corresponding to the GaAs, InAs and the equilibrium lattice
constant $a_{eq}$. For the latter one the value $a_{eq}=5.8$~\AA
was used. Parameters of the DMFT calculations are  presented in
the last three columns of the table. \cite{chioncel:085111}} \label{tab2_cras}
\end{table}

\subsubsection{VAs: correlation-induced half-metallic ferromagnetism?}

\label{sec:vas}

Interesting materials for spintronics applications are
ferromagnetic semiconductors \cite{nagaev:1983,ohno:951,ohno:110}.
Candidate systems are ordered compounds such as europium
chalcogenides (e.g., EuO) and chromium spinels (e.g.,
CdCr$_{2}$Se$_{4}$) \cite{nagaev:1983}, as well as diluted
magnetic semiconductors (e.g., Ga$_{1-x}$Mn$_{x}$As) \cite{ohno:951,ohno:110}%
. Unfortunately, all of them have Curie temperatures much lower
than room temperature. On the other hand, VAs in the zinc-blende
structure is, according to density-functional
calculations~\cite{galanakis:104417}, a
ferromagnetic semiconductor with a high Curie temperature. Unlike CrAs~\cite%
{akinaga:L1118}, CrSb~\cite{zhao:2776}, and
MnAs~\cite{okabayashi:233305}, VAs has not yet been experimentally
fabricated in the zinc-blende structure, but the increasing
experimental activity in the field of the (structurally
metastable) zinc-blende ferromagnetic compounds is promising in
this respect.

The main result including many-body correlation effects is displayed in Fig.~%
\ref{200_vas_DOS}. While this material is expected to be a
ferromagnetic semiconductor from density-functional theory
(LSDA/GGA) or static LSDA+$U$ calculations, the inclusion of
dynamic Coulomb correlations within the LSDA+DMFT approach
predicts a majority-spin band metallic behavior due to the closure
of the gap ($\approx $50 meV). However, since the minority-spin
band gap ($\simeq 1$eV) remains finite, the material is found to
be a half-metallic ferromagnet. To our knowledge, this is a first
example where
dynamic correlations transform a semiconductor into a half-metal \cite%
{chioncel:197203}. This remarkable result demonstrates the
relevance of many-body effects for spintronic materials.
\begin{figure}[h]
\centerline{%
\psfig{figure=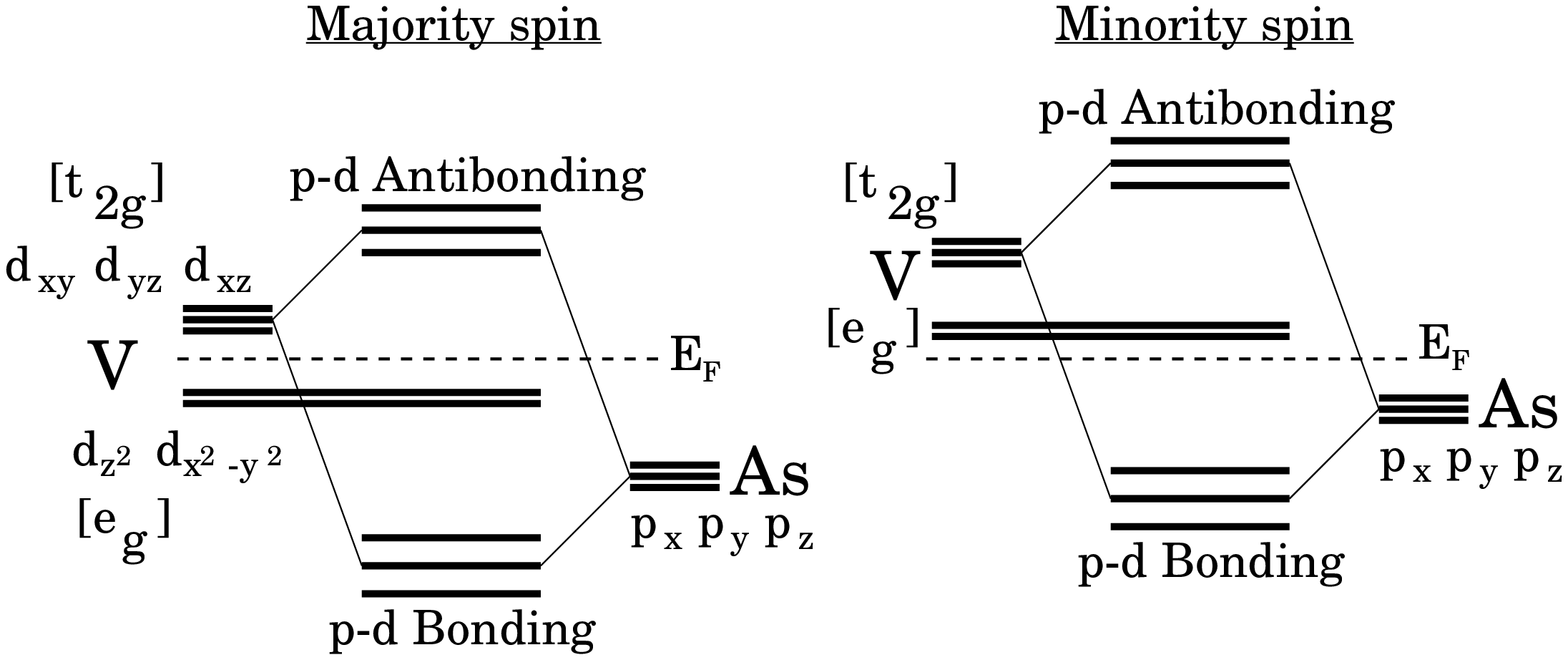,height=3.15in,angle=270}}
\caption{Schematic representation of the $p$-$d$ hybridization and
bonding-antibonding splitting in VAs.\cite{chioncel:197203} } \label{fig:bond_antibond}
\end{figure}

The important features of the electronic structure of VAs \cite%
{galanakis:104417} are shown schematically in
Fig.~\ref{fig:bond_antibond}. The $t_{2g}$ states hybridize with
the neighboring As $p$ states, forming wide bonding and
antibonding hybrid bands. In contrast, the $e_{g}$ states form
mainly non-bonding narrow bands. The Fermi level lies between
$e_{g}$ and antibonding $t_{2g}$ states in the majority-spin
bands, and between bonding $t_{2g}$ and $e_{g}$ in the
minority-spin bands.
The spin moment, concentrated mainly at the V atoms, is an integer
of exactly $M=2$~$\mu _{B}$ per unit formula, which is obvious
from counting the occupied bands for two spin directions.

The exchange constants of VAs were calculated within GGA and
adiabatic spin
dynamics approach, similar to the one used by Halilov \textit{et al.}~\cite%
{halilov:293}. 
Using these exchange parameters in a Monte Carlo simulation of the
corresponding classical Heisenberg Hamiltonian yields a Curie temperature $%
T_{C}=820$~K \cite{chioncel:197203} by the fourth-order cumulant
crossing point. This result agrees with the value of $T_{C}=830$~K
calculated in Ref.~ \cite{sanyal:054417} by using a similar
method. The high Curie point is well above the room temperature,
which makes VAs a very promising candidate for applications in
spintronics.


Static correlations were taken into account within the LDA+$U$
method using similar values of effective interactions parameters
$U=2$~eV and $J=0.9$~eV as in the case of Heusler alloys. It will
give just an estimation of correlations effects in VAs, since for
exact value of $U$ one need to perform a complicated analyse of
screening effects in this compound \cite
{kotliar:865,aryasetiawan:195104}. The theoretically determined
equilibrium lattice parameter, $a=5.69$\AA , and a broadening
$\delta $ of about $15$K, which allows the majority spin gap to be
clearly resolved, were used. For different lattice parameters
(e.g., as for InAs) the LDA results agree with the previous ones
\cite{sanyal:054417,galanakis:104417}. The GGA DOS is shown in
Fig.~\ref{200_vas_DOS}. The main difference in the GGA+$U$
spectrum is that the occupied localized majority $e_{g}$ states
are expected to shift to even lower energy, while the unoccupied
minority $e_{g}$ states to higher
energy. The semiconducting character does not change, since the $e_{g}$ and $%
t_{2g}$ bands remain separated for both spins; the majority-spin
gap slightly increases, but remains small.

\begin{figure}[h]
\centerline{\psfig{figure=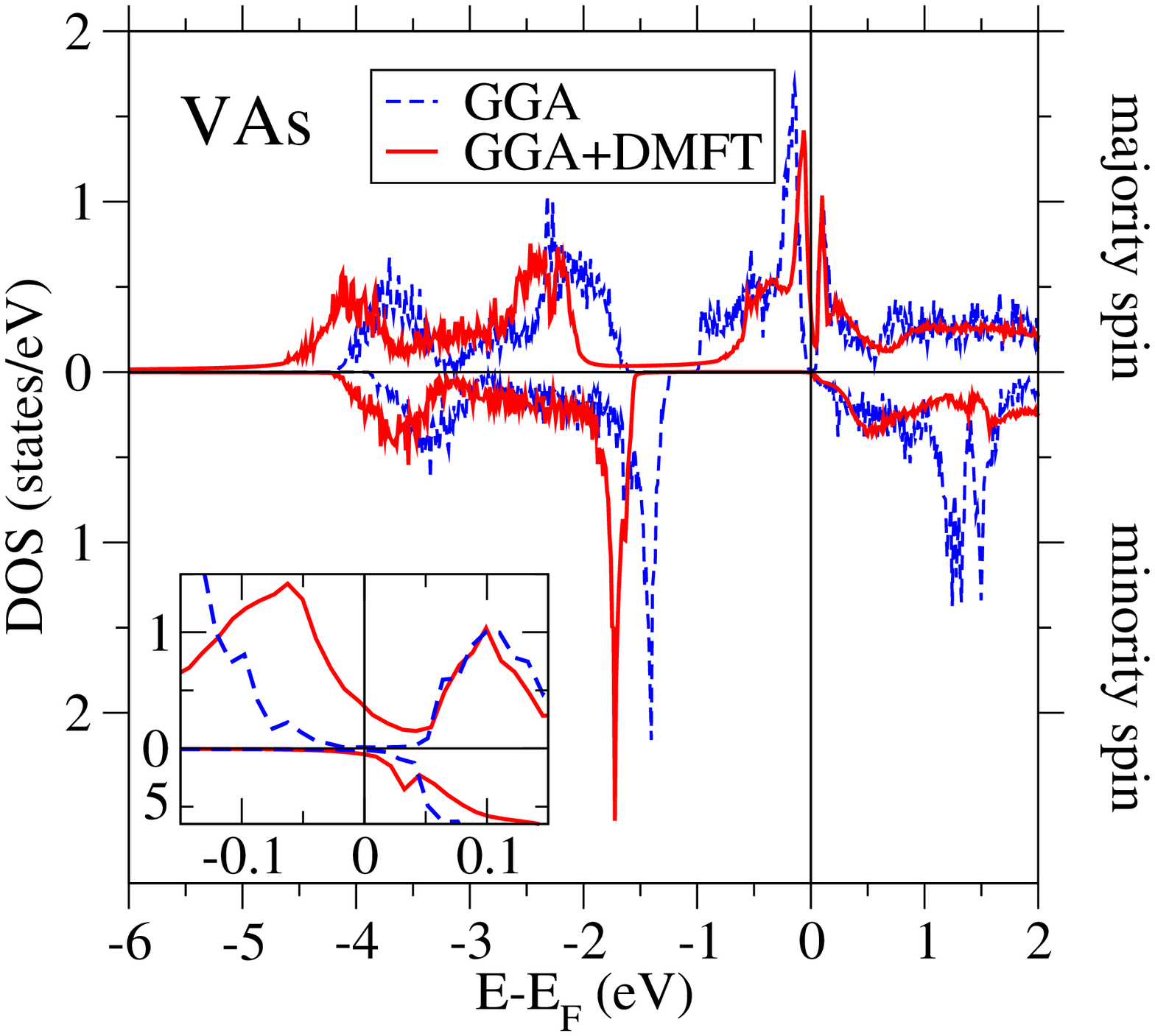,height=3.15in}}
\caption{(color online) DOS of VAs within the GGA (dashed blue
line) and
GGA+DMFT (solid red line) for a temperature of $T$=200~K, $U$=2~eV and $J$%
=0.9~eV. Inset: Focus around $E_F$ showing the semiconducting gap
within the GGA. To illustrate the minority-spin NQP states a ten
times larger scale for the spin down channel is used.\cite{chioncel:197203} }
\label{200_vas_DOS}
\end{figure}

In order to investigate dynamic correlation effects in VAs the
fully
self-consistent in spin, charge and self-energy LSDA+DMFT scheme \cite%
{chioncel:235106,chioncel:144425,chioncel:085111} was used. The
computational results for the DFT in the GGA approximation and
LDA+DMFT densities of states are presented in Fig.
\ref{200_vas_DOS}. The non-quasiparticle states in the minority
spin band are visible just above
the Fermi level (inset), predicted also by previous calculations \cite%
{chioncel:144425,chioncel:085111}. The weak spectral weight of NQP
states is due to that the Fermi level is close to the right edge
of the minority spin
gap, as discussed for CrAs having a similar structure \cite{chioncel:085111}%
. The local spin moments at V atoms do not change significantly
(by less than 5\%). However, in the case of VAs another
correlation effect appears: the small majority-spin gap at $E_{F}$
closes, making the material half-metallic \cite{chioncel:197203}.

In order to investigate the mechanism of the gap closing for the
majority
spin channel, the behavior of the electron self-energy was investigated \cite%
{chioncel:197203}. For the majority-spin, $\mathrm{Im}\Sigma
_{\uparrow
}(E)\sim (E-E_{F})^{2}$ 
which indicates a Fermi-liquid behavior, as opposed to
$\mathrm{Im}\Sigma _{\downarrow }(E)$ which shows a suppression
around $E_{F}$ due to the band gap, as well as a peculiar behavior
for $E>E_{F}$ related to the existence of NQP states.

From the Dyson equation~(\ref{eq:dyson}) one can see that the real part $%
\mathrm{Re}\Sigma _{\sigma }(E)$ causes a shift of the LDA energy
levels. Therefore, due to the non-zero $\Sigma _{\uparrow
}^{e_{g}}$, the $e_{g}$ orbitals in the close vicinity of the
Fermi level are pushed closer to $E_{F}
$. This renormalization is connected with the large absolute value of $%
\mathrm{Re}(\partial \Sigma /\partial E)_{E_{F}}<0$. This causes
occupied levels to be shifted to higher energy and unoccupied
levels to lower energy. Note that this effect is completely
opposite to the LDA$+U$ results discussed above. In addition to
this shift, the $e_{g}$ peak is broadened by
correlations, its tail reaching over the Fermi level (Fig.~\ref{200_vas_DOS}%
, inset). Thus the finite-temperature LDA+DMFT calculations
demonstrate the closure of the narrow gap in the spin-up channel,
which is produced by the correlation-induced Fermi-liquid
renormalization and spectral broadening. At
the same time, NQP states appear for the minority-spin channel just above $%
E_{F}$.

The slope of the majority-spin self-energy is almost a constant as
a function of temperature at low $T$: $\mathrm{Re}(\partial \Sigma
_{\uparrow }/\partial E)_{E_{F}}\simeq -0.4$ between 200~K and
500~K. The quasiparticle weight, which measures the overlap of the
quasiparticle wave function with
the original one-electron one for the same quantum numbers, is $%
Z=(1-\partial \mathrm{Re}\Sigma _{\uparrow }/\partial
E)^{-1}\simeq 0.7.$ As a consequence, the closure of the gap in
the majority channel is a quantum effect originating from the
multi-orbital nature of the local Coulomb interaction (energy
states are shifted towards $E_{F}$) rather than an effect of
temperature. A similar gap closure is obtained for larger values
of $U$, namely, $U=4$ and 6~eV, although the latter values should
be taken with some caution in the FLEX calculation, which is in
principle appropriate
only in weak to intermediate coupling. As a general tendency, increasing $%
U^{\ast }=U-J$ produces a stronger Fermi-liquid renormalization in
the majority spin channel, the same effect being evidenced for
$J=0$.

%
%

On the one hand, density-functional theory calculations within the
GGA (Ref. \cite{chioncel:197203}) predict this material to be a
ferromagnetic semiconductor with a tiny gap of about 50~meV in the
majority-spin states and large gap of the order of 1 eV for
minority-spin states. 
Quantum effects, such as spin and orbital
fluctuations, described by LDA+DMFT destroy the narrow gap and
turn the material into an half-metallic ferromagnet. 

On the other hand, several other mechanisms  could contribute to the band-gap
narrowing  with increasing temperature. 
A well studied example is electron-phonon interaction.
In another semiconductor with the zinc-blende structure, GaAs, 
it amounts to 50meV at 200K \cite{paes.99}.
Also spin-orbit 
coupling can be essential when considering 
closing of the gap.

The LDA/GGA calculation supplemented by a Monte Carlo simulation
also predicts a high Curie temperature of 830~K
\cite{sanyal:054417}, which makes this material of interest for
technological applications. One can expect that $T_{C}$ is not
strongly affected by dynamical correlation, for the same reason as
the
effective exchange interaction parameters (see recent works \cite%
{katsnelson:9}).

The revealed half-metallic (instead of semiconducting) behavior
has important consequences in the potential applications of VAs in
spintronics.
In contrast to all semiconductor-based spin-injection devices~\cite%
{zutic:323} which avoid the resistivity mismatch problem,
half-metals can be applied to obtain giant magnetoresistance or,
provided that interface states are
eliminated~\cite{mavropoulos:174428}, tunneling magnetoresistance
effects. We see that correlation effects play a decisive role in
the prediction of new spintronic materials. The metallic nature of
the majority spin channel would be visible in resistivity
measurements. Therefore, the experimental realization of
zinc-blende VAs would provide a test of this prediction. Further
research should address the issue of the stability of the
half-metallic ferromagnetic state in a zinc-blende structure. Some
work in this direction has been already carried out
\cite{shirai:6844,xie:2003}.

\subsection{Half-metallic transition metal oxides}

\subsubsection{CrO$_{2}$: a rutile structure half-metallic ferromagnet}

\label{sec:cro2}

\begin{figure}[t]
\includegraphics[width=\columnwidth]{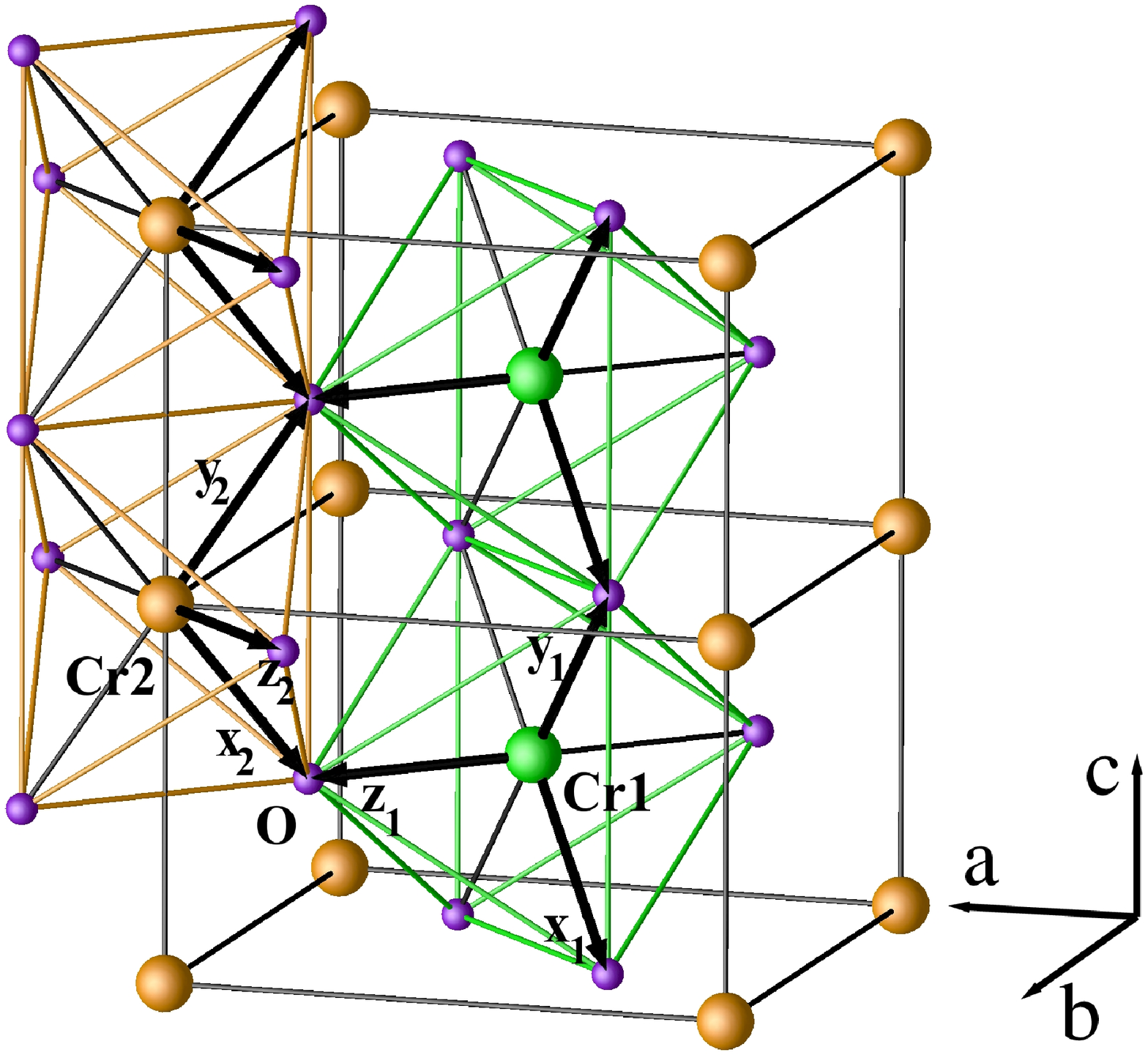}
\caption{(Color online) CrO$_2$ (rutile) structure. Cr1 (green)
and Cr2
(orange) are located at (0, 0, 0) and $(\frac{1}{2}, \frac{1}{2}, \frac{1}{2}%
)$. Cr atoms are octahedrally coordinated by oxygen atoms
(purple). The
local coordinate system is used for each Cr atom; $\hat{\mathbf{x}}_1=-\frac{%
1}{2}\hat{\mathbf{a}} +\frac{1}{2}\hat{\mathbf{b}} -\frac{1}{\protect\sqrt{2}%
}\hat{\mathbf{c}}$, $\hat{\mathbf{y}}_1=-\frac{1}{2}\hat{\mathbf{a}} +\frac{1%
}{2}\hat{\mathbf{b}} +\frac{1}{\protect\sqrt{2}}\hat{\mathbf{c}}$, $\hat{%
\mathbf{z}}_1= \frac{1}{\protect\sqrt{2}}\hat{\mathbf{a}} +\frac{1}{\protect%
\sqrt{2}}\hat{\mathbf{b}}$, and $\hat{\mathbf{x}}_2=-\frac{1}{2}\hat{\mathbf{%
a}} -\frac{1}{2}\hat{\mathbf{b}} -\frac{1}{\protect\sqrt{2}}\hat{\mathbf{c}}$%
, $\hat{\mathbf{y}}_2=-\frac{1}{2}\hat{\mathbf{a}} -\frac{1}{2}\hat{\mathbf{b%
}} +\frac{1}{\protect\sqrt{2}}\hat{\mathbf{c}}$, $\hat{\mathbf{z}}_2=-\frac{1%
}{\protect\sqrt{2}}\hat{\mathbf{a}} +\frac{1}{\protect\sqrt{2}}\hat{\mathbf{b%
}}$. $\hat{\mathbf{x}}_{1,2}$ and $\hat{\mathbf{y}}_{1,2}$ are
approximately point to $O$ atom, and $\hat{\mathbf{z}}_{1,2}$ are
exactly point to $O$ atom. The local axes are transformed into
each other by a  90$^\circ$ rotation around the crystal $c$ axis. \cite{yamasaki:024419}}
\label{CrO2struc}
\end{figure}


Chromium dioxide CrO$_{2}$ has a rutile structure with $a=4.421$ \AA , $%
c=2.916$ \AA\ ($c/a=0.65958$) and internal parameter $u=0.3053$.~\cite%
{porta:157} The Cr atoms form a body-center tetragonal lattice and
are surrounded by a slightly distorted octahedron of oxygen atoms.
The space group of this compound is non-symmorphic
($P4_{2}/mnm=D_{4h}^{14}$). The Cr ions are in the center of
CrO$_{6}$ octahedra, so that the $3d$ orbitals are split into a
$t_{2g}$ triplet and an excited $e_{g}$ doublet. The $e_{g}$
states with only two valence $3d$ electrons are irrelevant, and only the $%
t_{2g}$ orbitals are to be considered. The tetragonal symmetry
distorts the
octahedra, which lifts the degeneracy of the $t_{2g}$ orbitals into a $%
d_{xy} $ ground state and $d_{yz+zx}$ and $d_{yz-zx}$ excited states~\cite%
{levis:10253,korotin:4305} (see Fig. \ref{CrO2struc}, a \textit{local }%
coordinate system is used for every octahedron). A double exchange
mechanism for two electrons per Cr site was
proposed~\cite{schlottmann:174419}.
According to this, the strong Hund's rule together with the distortion of CrO%
$_{6}$ octahedra leads to localization of one electron in the
$d_{xy}$ orbital, while the electrons in the $d_{yz}$ and $d_{xz}$
are itinerant.

Measurements of the magnetic susceptibility in the paramagnetic
phase show a
Curie-Weiss-like behavior indicating the presence of local moments~\cite%
{cham.77}, which suggests a mechanism of ferromagnetism beyond the
standard band or Stoner-like model.

Several recent experimental investigations of photoemission~\cite%
{tsujioka:15509}, soft x-ray absorption~\cite{stagarescu:9233},
resistivity~ \cite{suzuki:11597}, and optics~\cite{singley:4126}
suggest that electron correlations are essential for the
underlying physical picture in CrO$_{2}$. Schwarz
\cite{schwarz:L211} first predicted the half-metallic band
structure
with a spin moment of $2\mu _{B}$ per formula unit for CrO$_{2}$. Lewis~\cite%
{levis:10253} used the plane-wave potential method and
investigated the energy bands and the transport properties,
characterizing CrO$_{2}$ as a \textquotedblleft bad
metal\textquotedblright\ (a terminology applied earlier to high
temperature superconductors and to other transition metal oxides,
even ferromagnets like SrRuO$_{3}$). A decade later the LSDA+$U$
calculation~\cite{korotin:4305} treated the conductivity in the
presence of large on-site Coulomb interactions and described
CrO$_{2}$ as a negative charge-transfer gap material with a
self-doping. Contrary to the on-site strong correlation
description, transport and optical properties obtained within the
FLAPW method (LSDA and GGA)~\cite{mazin:411} suggest that the
electron-magnon scattering is responsible for the renormalization
of the
one-electron bands. More recent model calculations \cite%
{laad:214421,craco:237203} propose even orbital correlations.~

\begin{figure}[t]
\rotatebox{270}{\includegraphics[height=%
\columnwidth]{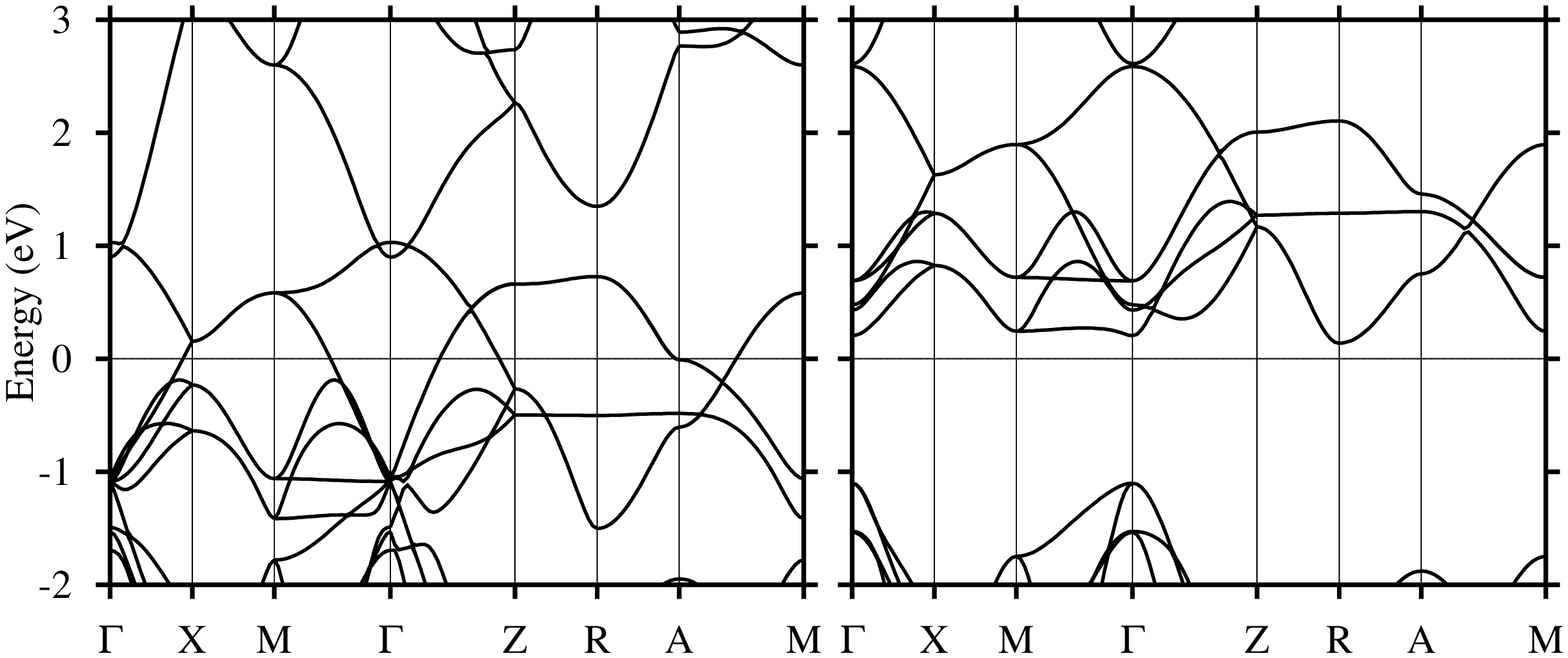}} \caption{Full basis set
spin-polarized (ferromagnetic) bands for CrO$_{2}$; majority spin
(left) and minority spin (right). $E_{F}$ is set to be zero.
The high-symmetry points are $\Gamma (0,0,0)$, $X(0,\frac{1}{2},0)$, $M(%
\frac{1}{2},\frac{1}{2},0)$, $Z(0,0,\frac{1}{2})$, $R(0,\frac{1}{2},\frac{1}{%
2})$, $A(\frac{1}{2},\frac{1}{2},\frac{1}{2})$. \cite{yamasaki:024419} }
\label{bnds_sp_cro2}
\end{figure}


Chemical bonding in rutile-type compounds including CrO$_{2}$ was
analyzed by Sorantin and Schwarz~\cite{sorantin:567}. One can see
that around the Fermi level the bands are primarily chromium $3d$
states of $t_{2g}$ manifold, with $e_{g}$ bands being situated
higher in energy due to the crystal-field splitting. In the spin
polarized case, the exchange splitting
shifts the minority spin $d$ bands above the Fermi level (Fig.~\ref%
{bnds_sp_cro2}). For the majority $t_{2g}$ bands the Fermi level
lies in a pseudogap. Oxygen $p$ $-$ chromium $d$ hybridization
creates both bonding and antibonding hybrid orbitals, with the
bonding orbital appearing in the occupied part and the antibonding
hybrid orbital remaining in the Cr $t_{2g}$ manifold. Half of the
$d_{yz}$ and $d_{zx}$ components of $t_{2g}$ are pushed upward by
antibonding, which explains the $d_{xy}$ dominance in the spin
density. The non-magnetic DOS shows a sharp peak at the Fermi
level, which signals the magnetic instability according to the
usual Stoner argument.

Although there is a significant difference between the $t_{2g}$
and $e_{g}$ orbitals
~\cite{schwarz:L211,sorantin:567,korotin:4305,mazin:411}, the
analysis in the framework of NMTO technique
\cite{Andersen:16219,zurek:1934} shows that their interplay is
important not only for the crystal-field splitting of $t_{2g}$
states, but also for the general bonding in the rutile
structure. The $t_{2g}$ orbitals form the basis set used in Ref. \cite%
{yamasaki:024419} to evaluate the effective hopping Hamiltonian
matrix elements.

In CrO$_{2}$ the bands around the Fermi level are primarily
chromium $3d$ states of $t_{2g}$ manifold, $e_{g}$ bands being
situated higher in energy owing to crystal-field splitting. The
$t_{2g}$ orbitals are further split into single $d_{xy}$ and
nearly degenerate $d_{yz\pm zx}$ bands owing to the orthorhombical
distortion of CrO$_{6}$ octahedra. Despite the differences between
Cr $t_{2g}-e_{g}$ orbitals, their interaction plays an important
role not only in characterizing the crystal-field splitting, but
also in the general picture of bonding in the rutile structure.

Concerning the Coulomb interaction $U$ in CrO$_{2}$, the higher energy $%
e_{g} $ bands, although making no noticeable contribution at the
Fermi
level, could participate in the screening of the $t_{2g}$ orbitals~\cite%
{solovyev:7158,pickett:1201}, thereby giving the values $U=3~$eV and $J=0.87$%
~eV.

Previous LSDA$+U$ \cite{korotin:4305,to.ko.05} and DMFT \cite%
{laad:214421,craco:237203} studies yielded independently a narrow
almost flat band of $d_{xy}$ character, which produces
ferromagnetism in CrO$_{2}$.
In contrast to these results, the fully self-consistent LSDA+DMFT \cite%
{chioncel:cro2} yields, in agreement with a non-local variational
cluster approach \cite{chioncel:cro2}, an itinerant $d_{xy}$
orbital which crosses the Fermi level. Despite the non-localized
character of the orbital, a ferromagnetic phase is still obtained.

\begin{figure}[h]
\includegraphics[width=0.9\linewidth]{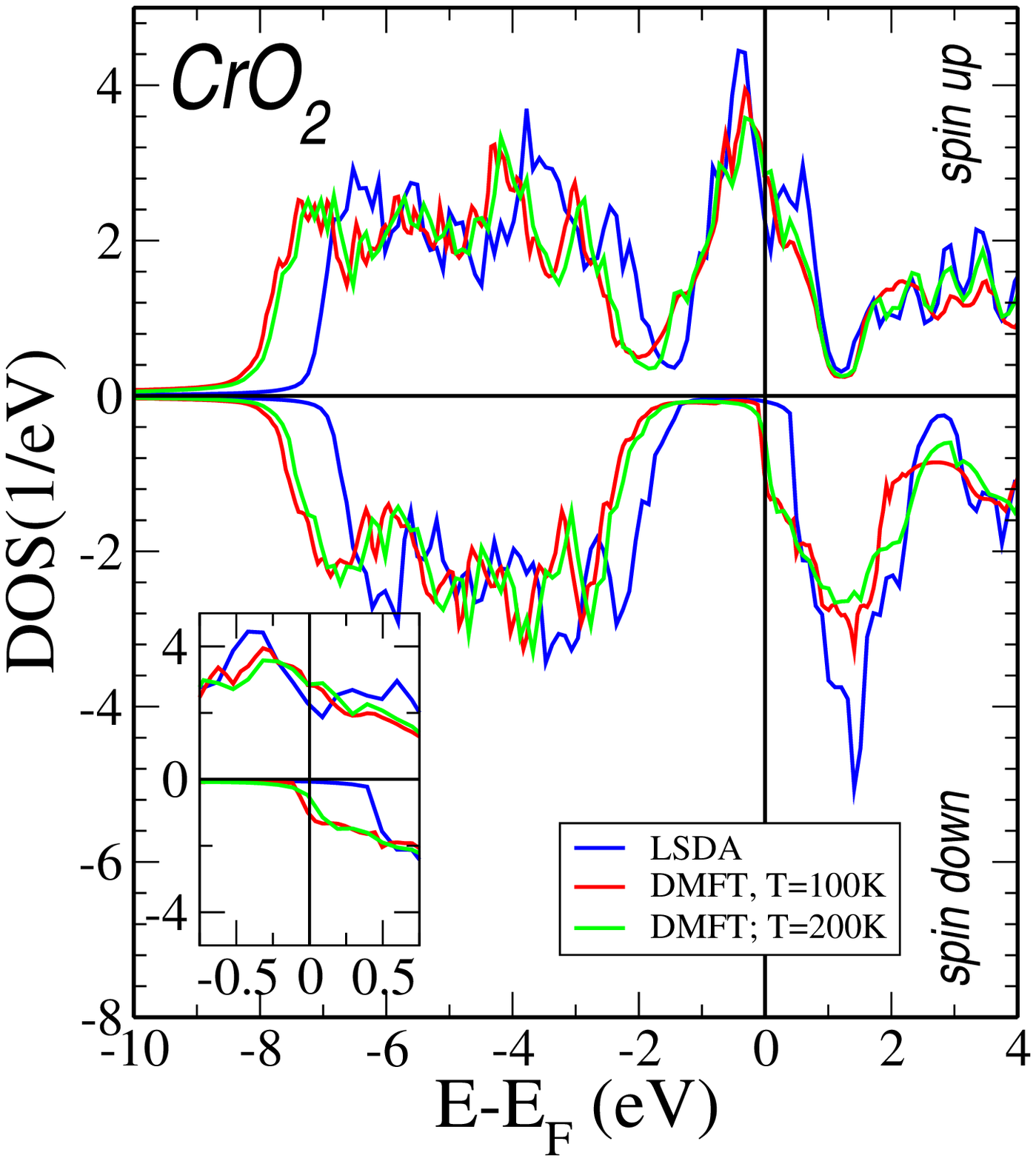}
\caption{(Color online) Density of states obtained within the LSDA
and LSDA+DMFT calculations for different temperatures. The inset
shows the results for a smaller energy window around the Fermi
level. \cite{chioncel:cro2}} \label{dos_cro2}
\end{figure}

Results of the LSDA+DMFT calculation are presented in
Fig.~\ref{dos_cro2} for two different values of $T$ and compared
with the LSDA results. The LSDA Fermi level intersects the
majority-spin bands near a local minimum and lies in the band gap
of the minority spin. Finite temperatures and correlation effects
close this pseudogap around the Fermi level, as can be seen from
the LDA+DMFT results in Fig.\ref{dos_cro2}. No differences can be
observed between the two DMFT results at different temperatures,
except for the smearing of DOS features at larger temperature. For
both spin channels the
DOS is shifted uniformly to lower energies in the energy range $-2\div -6$%
~eV, where predominantly the O($p$) bands are situated. This is
due to the Cr($d$) bands which affect the O($p$) states through
the Cr($d$)$-$O($p$) hybridization, so that O($p$) states
contribute actively to the ferromagnetic ground state formation.

\begin{figure}[h]
\includegraphics[width=0.9\linewidth]{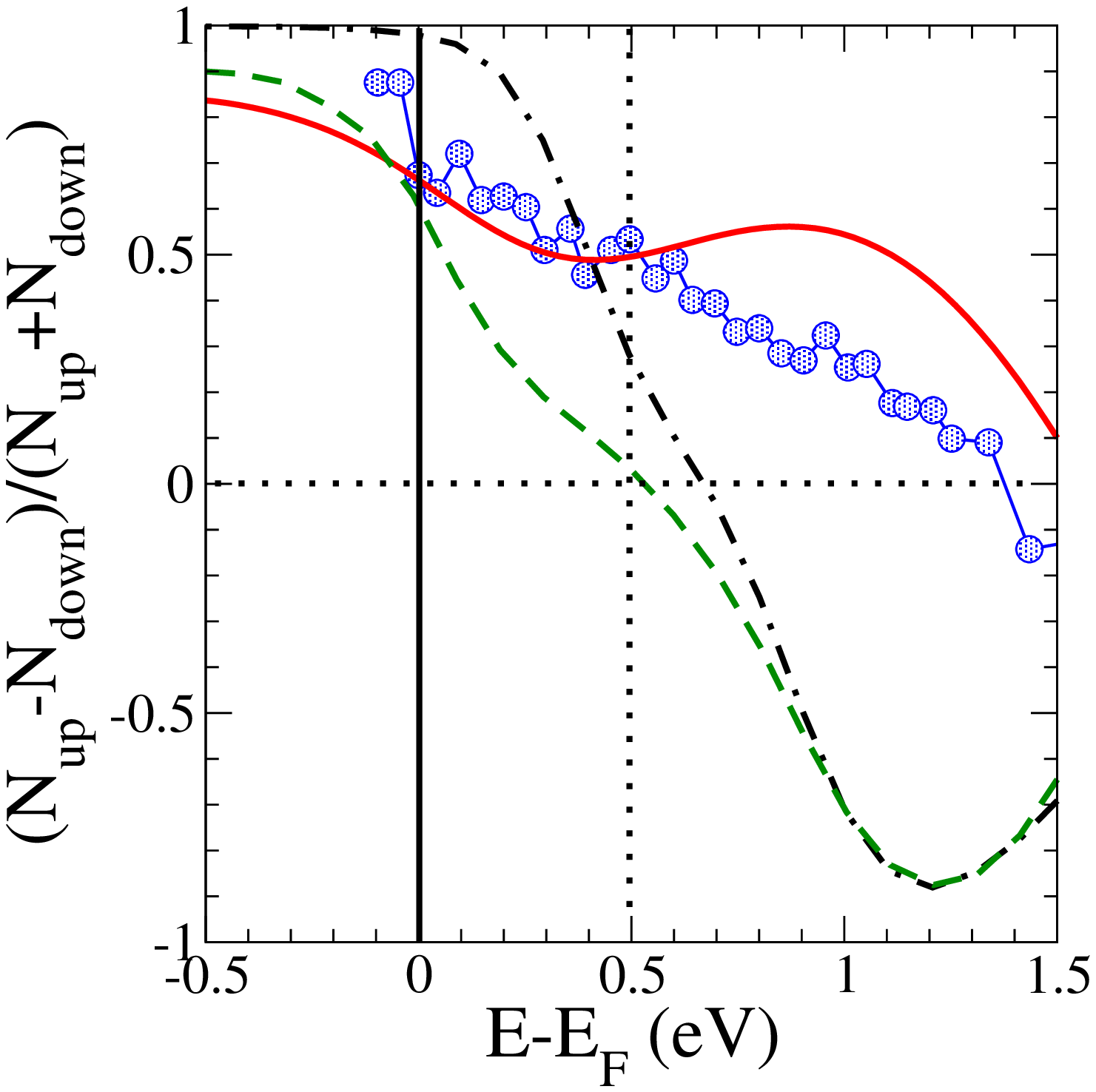}
\caption{(Color online) Energy dependence of spin-polarization of CrO$_2$.
Experimental measurement blue scattered line \cite{huang:214419}, LSDA calculation (black dot-dashed), 
DMFT (green dashed) and non-local variational approach, red continuous line \cite{chioncel:cro2}.}
\label{pol_cro2}
\end{figure}

Fig. \ref{pol_cro2} presents the experimentally measured spin
polarization \cite{huang:214419} in comparison with the
theoretical calculations within the LSDA and finite temperature
LDA+DMFT (T=200~K) \cite{Chioncel:private}. Resonant X-ray
emission spectroscopy \cite{kurmaev:155105} showed the existence
of nearly currentless minority spin states in the vicinity of the
Fermi level, which can be connected to the non-quasiparticle
states \ref{sec:xray}. As described in previous sections
\ref{sec:diffunc} and \ref{sec:elstr_hmf} electronic structure of
several half-metallic ferromagnets reveal the existence of such
NQP states, which are important for the proper description of spin
polarization near the Fermi level. The LSDA+DMFT description,
however, is not sufficient to capture the high energy tail of  the
experimental spin polarization. This can be related with improper
description of unoccupied Cr $e_g$ orbitals in the LSDA, and
probably, also with non-local exchange effects which can be
investigated within a cluster DMFT scheme \cite{kotliar:865}.

The NQP states occur around 0.25~eV in the spin-down channel. At
zero temperatures, the variational cluster results
\cite{chioncel:cro2} yield the states in the minority channel
which are actually far from $E_{F}$, but have a tail vanishing at
$E_{F}$. This is in agreement with low temperature experiments
\cite{soulen:85,ji:5585} which support very high polarization of
CrO$_{2}$.

The effects of local and non-local electronic correlations in
CrO$_{2}$ \cite{chioncel:cro2} change considerably the mean-field
LSDA$+U$ picture, despite the interaction is not too strong. In
particular, in LSDA$+U$ the single occupancy of the Cr $d_{xy}$
orbital is determined by the exchange and crystal-field splitting.
On the other hand, the competition of the latter with correlation
effects, which is taken into account in the DMFT and variational
cluster perturbation theory calculations, induces a ferromagnetic
state with itinerant-type $d_{xy}$ orbitals possessing a large
effective mass rather than with localized moments, in contrast to
previous results
\cite{korotin:4305,to.ko.05,laad:214421,craco:237203}. In the
minority-spin channel, correlations induce NQP states which are
crucial for the occurrence of spin depolarization in CrO$_{2}$.
However, a quantitative analysis of the depolarization would
require the inclusion of additional effects, e.g., disorder or
phonons (sect. \ref{pol}).

\section{Exchange interactions and critical temperatures in half-metallic
compounds}

Owing to the strong interest in the half-metallic ferromagnetism,
the number of theoretical studies of exchange interactions and
calculations of Curie temperatures in Heusler alloys has been
drastically increased last time (see, e.g., recent review \cite{Zhang:315204}). 
The first investigations of the
exchange interactions in half-metals within the DFT formalism was
made by K\"ubler~\cite{ku.wi.83}. The mechanisms of ferromagnetism
in Heusler alloys were discussed on the basis of total-energy
calculations for the ferro- and antiferromagnetic configurations.
Since the antiferromagnetic state of the system is not
half-metallic, such an estimation gives only crude values of
exchange parameters in HMF. Therefore, a more precise evaluation
of the exchange interactions from the first-principle theory is
required. In this section we present the real-space Green's
function and  frozen-magnon techniques to calculate the exchange
parameters, and their applications to full-Heusler, semi-Heusler,
and zinc-blende half-metals.

\subsection{The Green's function formalism}
\label{subsec:GF}

Within the first-principle Green's function approach the exchange
parameters are obtained by mapping the second variation of
electronic band energy to the classical Heisenberg Hamiltonian and
making use of magnetic version of the Andersen's local force
theorem \cite{ma.an.80}.

\begin{equation}\label{exch1}
H_{eff}=- \frac{1}{2}  \sum_{\mu,\nu} \sum_{RR'} J_{\bf R
R'}^{\mu,\nu} {\bf s}_{\bf R}^{\mu} {\bf s}_{\bf R'}^{\nu}
\end{equation}
In Eq. (\ref{exch1}), the indices $\mu$ and $\nu$ mark different
sublattices, ${\bf R}$ and ${\bf R'}$ are the lattice vectors
specifying the atoms within a sublattice, ${\bf s}_{\bf R}^{\mu}$
is the unit vector in the direction of the magnetic moment. By
introducing the generalized notation for site $(i=\mu,{\bf R})$, a
simple and transparent expression for the exchange interaction
parameters is obtained in the following form \cite{li.ka.87}:
\begin{equation}
J_{ij}=\frac 1 {4\pi} \int d\epsilon Im Tr_L \{ \Delta_i
G_{ij}^\uparrow \Delta_j G_{ij}^\downarrow \},
\end{equation}
where $G_{ij}^\sigma $ is the real space Green's function and
$\Delta_i$ is the local exchange splitting for site $i$.

\subsection{The frozen-magnon approach and DFT calculations of
spin spirals} \label{subsec:FM.SS}

The approach is based on the evaluation of the energy of the
frozen-magnon configurations defined by the following atomic polar
and azimuthal angles:
\begin{equation}\label{exch2}
\theta_{\bf R}^{\mu}= \theta; \\
{\phi}_{\bf R}^{\mu}= {\bf q \cdot
R} + {\phi}^{\mu}
\end{equation}
The angle $\theta$ defines the cone of the spin spiral, and the
constant phase ${\phi}^{\mu}$ is normally chosen to be zero. The
magnetic moments of  other sublattices are kept parallel to the
z-axis. Within the classical Heisenberg model (\ref{exch1}) the
energy of the spin-spiral configuration is
\begin{equation}\label{exch3}
E^{\mu \mu}(\theta, {\bf q }) = E^{\mu \mu}_0(\theta) +
J^{\mu \mu}({\bf q }) \cdot sin^2 \theta
\end{equation}
where $ E^{\mu \mu}_0(\theta)$ does not depend on $q$ and the
Fourier transform $J^{\mu \mu}({\bf q })$ is defined by:
\begin{equation}\label{exch4}
J^{\mu \nu}({\bf q }) = \sum_{\bf R}J^{\mu \nu}_{0\bf R}
exp ( i {\bf q \cdot R} )
\end{equation}
In the case where $\nu=\mu$ the sum in Eq. (\ref{exch4}) does not
include the local interaction with $R=0$. Calculating $E^{\mu
\mu}(\theta, {\bf q })$ for a regular $q$-mesh in the Brillouin
zone of the crystal and performing the inverse Fourier
transformation one gets exchange parameters $J^{\mu \mu}_{0\bf R}$
for sublattice $\mu$.

The Curie temperature can be estimated within the mean-field
(MFA),
\begin{equation}\label{mfa_tc}
k_BT_C^{MFA}=\frac 2 3 \sum_{\bf R}J^{\mu \nu}_{0\bf R} =\frac M
{6\mu_B} \frac 1 N \sum_{\bf q} \omega({\bf q} )
\end{equation}
and random phase approximation (RPA)
\begin{equation}\label{rpa_tc}
\frac 1 {k_BT_C^{RPA}}= =\frac {6\mu_B}{M} \frac 1 N \sum_{\bf q}
\frac 1 {\omega({\bf q})}
\end{equation}
with $\omega(q)$ the spin-wave dispersion, $N$ the number of $q$
points in the first Brillouin zone, and $M$ the atomic magnetic
moment. In the MFA the Curie temperature is determined by the
arithmetic average of the magnon energies, while in the RPA $T_C$
is determined by the harmonic average. Therefore the value of
$T_C$ within MFA is  larger than the RPA one, the two values being
equal only provided that the magnon spectrum is dispersionless.

\subsection{First-principle calculations}

\subsubsection{Semi-Heusler C1$_b$ alloys}
\begin{figure}[tbh]
\begin{center}
\includegraphics[width=0.95\columnwidth, angle=0]{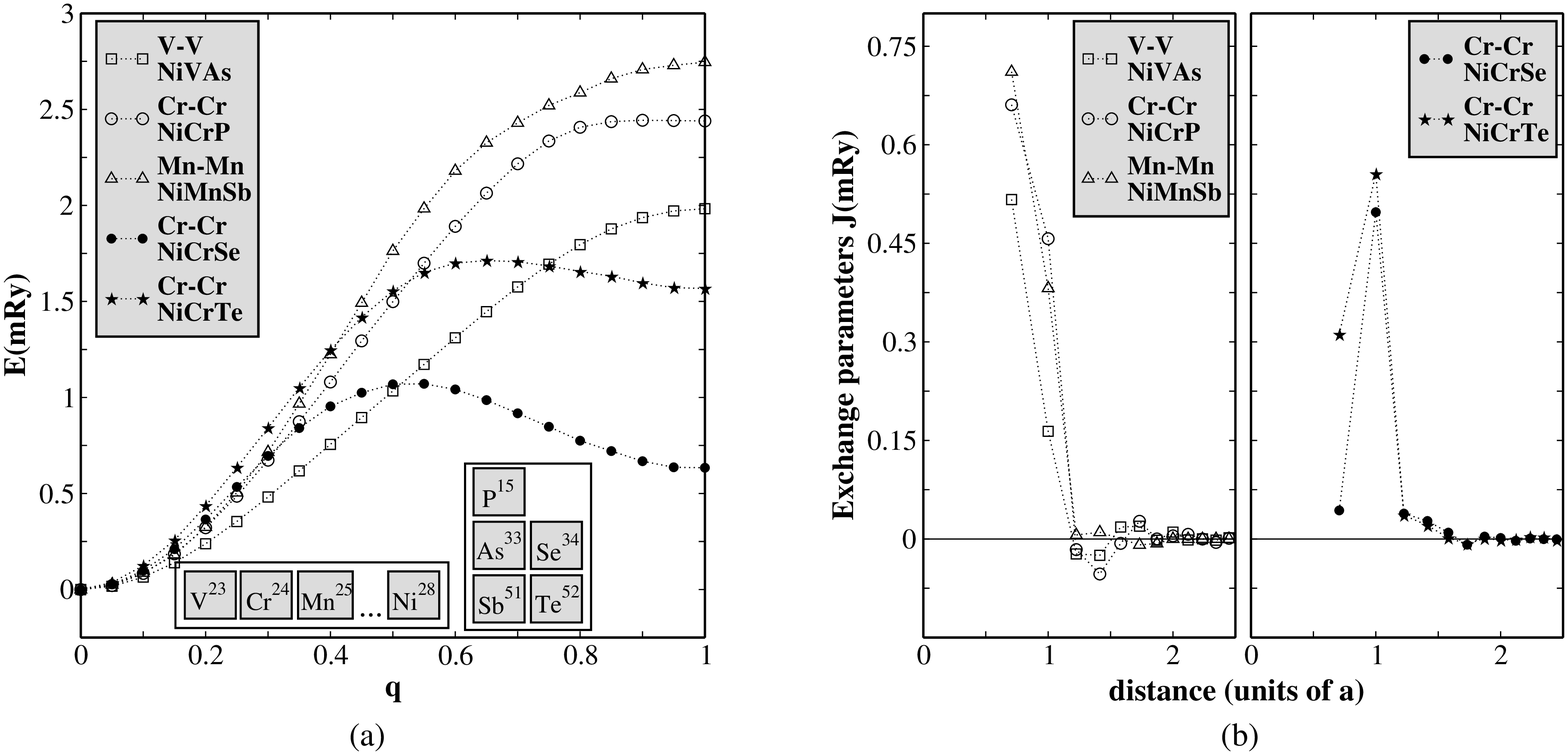}
\end{center}
\caption{Left pannel(a): Frozen-magnon dispersion for NiCrZ (Z=P,Se,Te).
Right pannel(b): Interatomic exchange interactions. \protect\cite{sasioglu:063523}.}
\label{fm_semi}
\end{figure}

First-principle studies of  exchange interactions and  magnetic
phase transitions for NiCrZ (Z=P,Se,Te), NiVAs, NiMnSb, CoMnSb
were carried out by many authors
\cite{sasioglu:063523,rusz:214412} In fig. \ref{fm_semi} the
frozen-magnon dispersion and the exchange interactions are
presented \cite{sasioglu:063523}. The exchange interactions are
Fourier transforms of the frozen-magnon spectrum.

A remarkable feature of the exchange interactions is their
short-range character, the Curie temperature being determined by
the interaction within  first two coordination spheres. The
deviation from the collinear alignment of magnetic moments can be
also characterized by magnon energies  \cite{sand.98,sa.br.03}.
The deviation leads to the mixing of the majority and minority
spin states, which makes half-metallicity less favorable. A
detailed discussion of different depolarization mechanisms is
given in section \ref{pol}.

\subsubsection{Full-Heusler L2$_1$ alloys}
Recently the studies of the interatomic exchange interactions in
several full-Heusler compounds were reported by Kurtulus et al.
\cite{kurtulus:014425} Sasioglu et al. studied the exchange
interactions in non-half-metallic Ni$_2$MnZ (Z=Ga, In, Sn, Sb) and
half-metallic Mn$_2$VZ (Z=Al,Ge). The importance of the
intersublattice exchange interactions has been demonstrated. For
 Mn$_2$VZ (Z=Al,Ge) it was shown that the ferrimagnetic
coupling between the V and Mn moments stabilizes the ferromagnetic
alignment of the Mn moments. 

\begin{figure}[tbh]
\begin{center}
\includegraphics[width=0.75\columnwidth, angle=0]{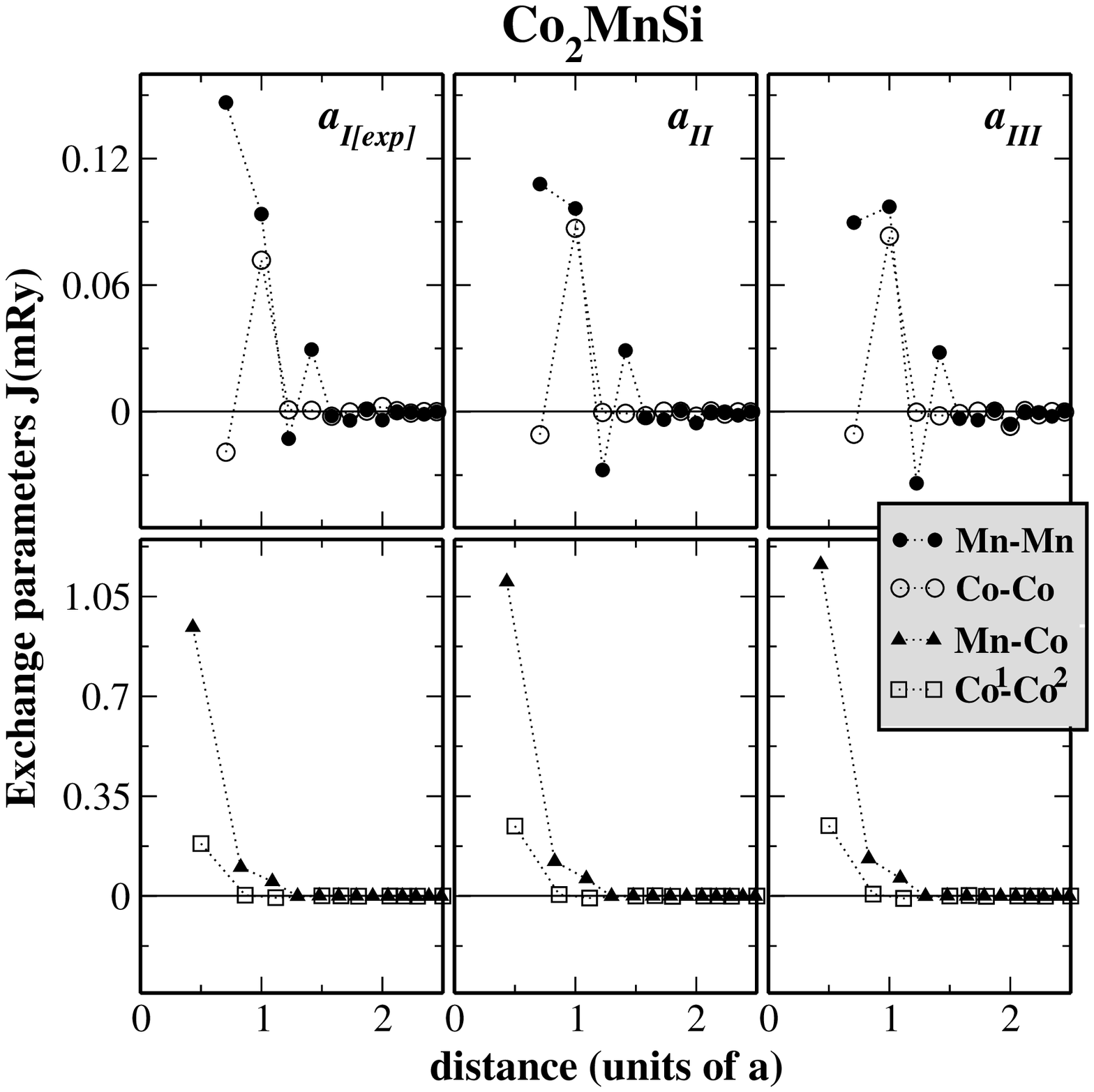}
\end{center}
\caption{Exchange constants for Co$_{2}$MnSi as a function of interatomic
distance. \protect\cite{sasioglu:184415}.}
\label{j_L21}
\end{figure}

In Co$_2$MnZ (Z=Ga, Si, Ge, Sn) the
presence of Co atoms makes the interaction more complicated
\cite{kurtulus:014425}. The interaction between the Co atoms in 
the same sublattice ,Co1(2)-Co1(2),  and between Co atoms at different 
sublattices, Co1(2)-Co2(1) has to be taken into account. This approach 
gives results which goes beyond the initial approach of  K\"ubler 
\cite{ku.wi.83}. From the $J_{ij}$ values it is clear that the 
exchange interactions are relatively short ranged and do not exceed
the first four neighbors in each sublattice. The main exchange parameter 
corresponds to the nearest neighbor Co(1)-Mn interaction. This already
gives 70$\%$ of the total contribution to $J$ and is about 
ten times larger than the Co-Co and Mn-Mn interactions \cite{kurtulus:014425}.
Thus, it was concluded that the Co-Mn interactions are responsible 
for the stability of ferromagnetism (see Fig. \ref{j_L21}). 

The interaction changes
weakly with the contraction of the lattice \cite{sasioglu:184415}
interactions between Co at different sublattices favors
ferromagnetism and is stronger than the ferromagnetic interaction
between the Mn atoms. The intrasublattice Mn-Mn ad Co-Co
interactions have different signs for different distances between
atoms, which means a RKKY-type interaction.

\subsubsection{Zinc-blende half-metals}
Besides the large number of results on the electronic properties
of zinc-blende half-metals, the exchange interactions constitute
an important aspect to understand the stability of
half-metallicity in these structures. Shirai \cite{shirai:6844}
obtained that in VAs, CrAs and MnAs ferromagnetism is
energetically favorable in comparison with the antiferromagnetic
state, unlike FeAs where an opposite effect was demonstrated.
Sakuma \cite{saku.02} predicted
 ferromagnetism in the isoelectronic MnSi, MnGe and MnSn
compounds.  Similarly to CrAs, the ferromagnetism in these systems
is stabilized by short-range interactions (direct Mn-Mn and
indirect through sp atoms), giving a Curie temperature of 1000K
\cite{saku.02}. $T_C$ for the VAs, CrAs and MnAs was calculated
also by K\"ubler \cite{kubl.03} yielding the same range of
magnitudes. Using GaAs and InAs lattice constants, Sanyal et al
\cite{sanyal:054417} calculated $T_C$ for VAs, CrAs and MnAs.
Sasioglu et al \cite{sa.ga.05} calculated the exchange parameters
for a large number of pnictides.

\begin{figure}[tbh]
\begin{center}
\includegraphics[width=0.75\columnwidth, angle=0]{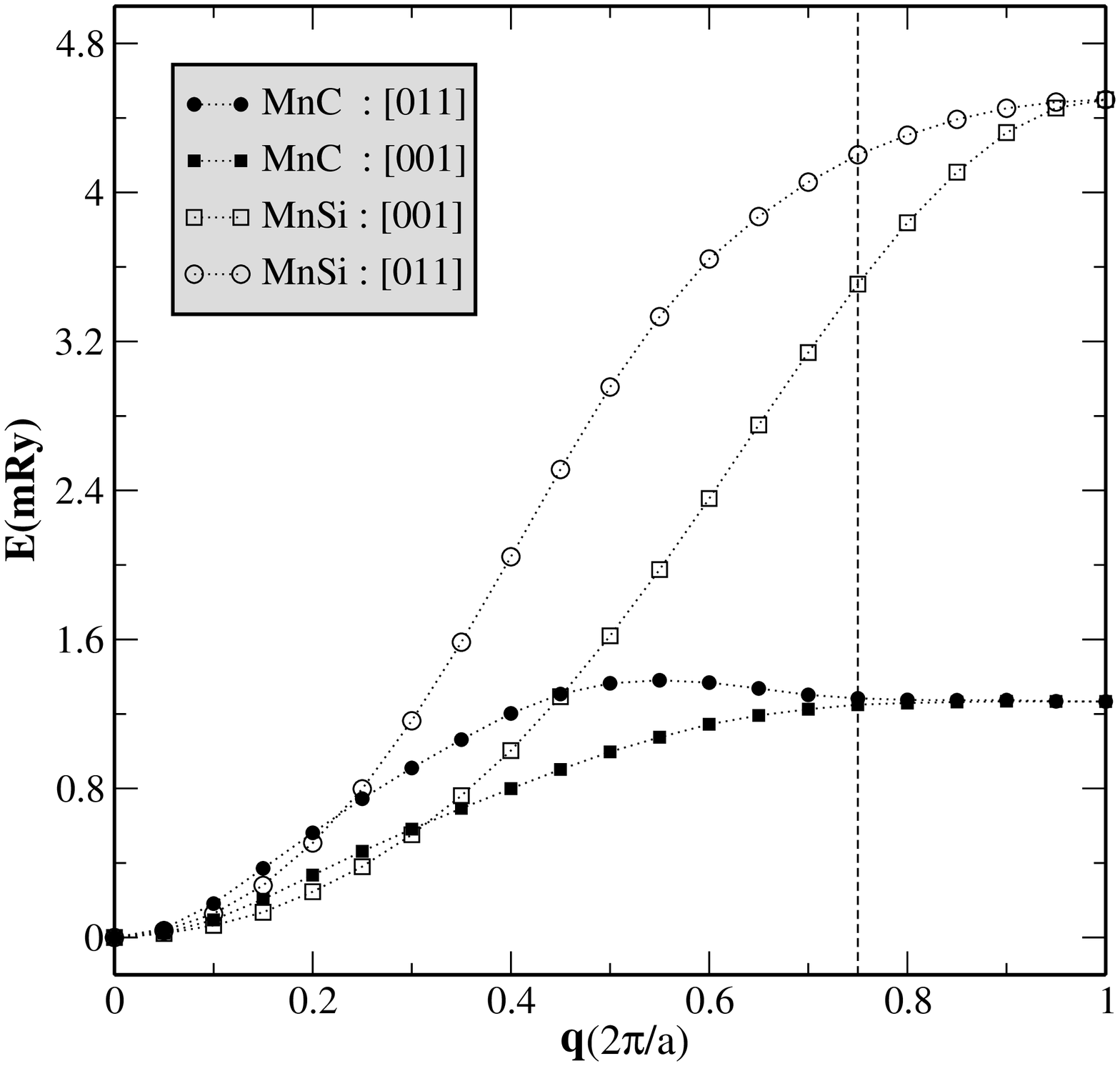}
\includegraphics[width=0.75\columnwidth, angle=0]{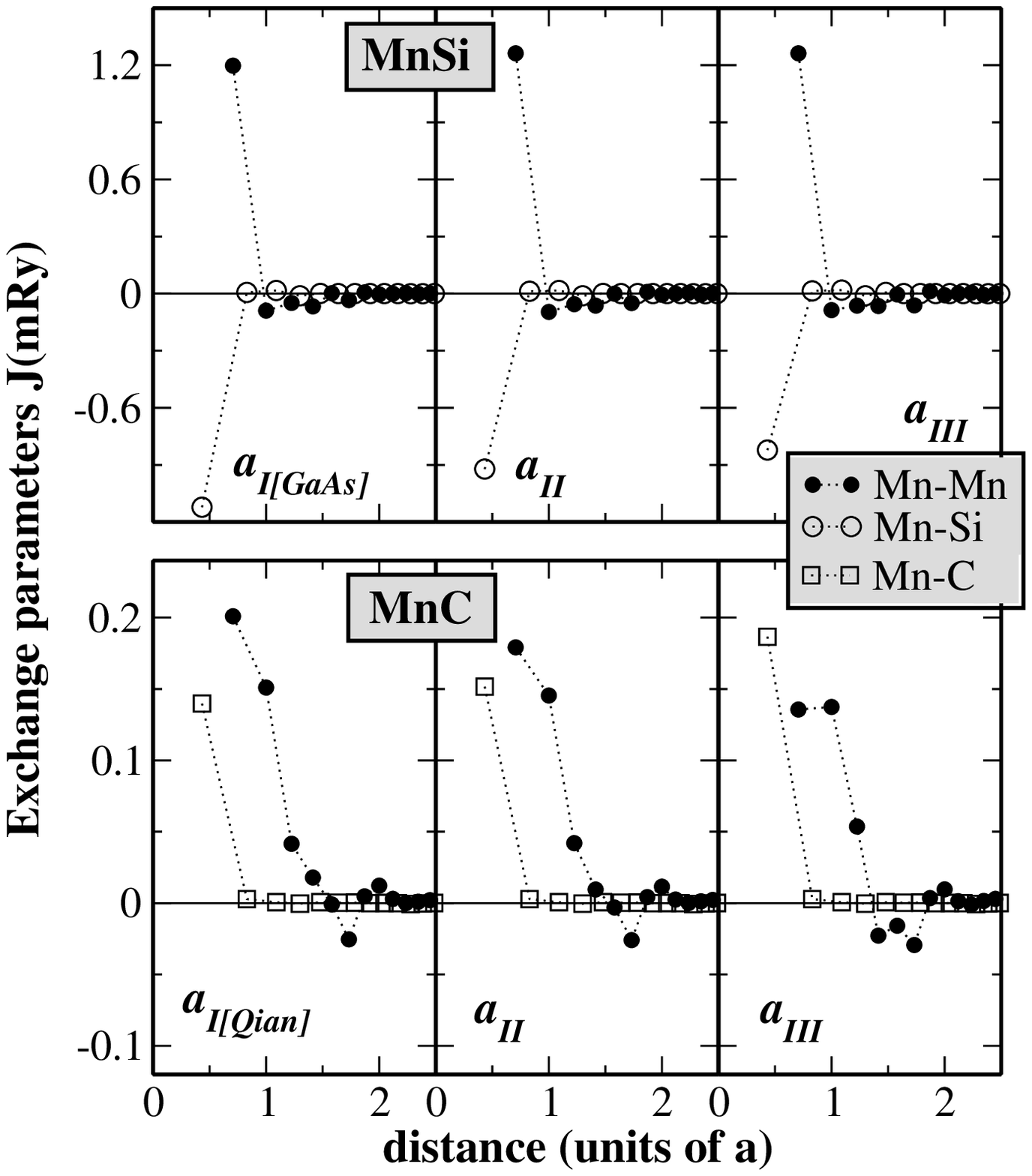}
\end{center}
\caption{Upper pannel: Frozen-magnon energies for MnC and MnSi in the 
$[00q]$ and $[0qq]$ directions. Vertical broken line shows the Brillouin
zone boundary. Lower panel: Exchange constants as a function of interatomic
distance. \protect\cite{sa.ga.05}.}
\label{fm_mnc}
\end{figure}

MnC presents an interesting situation since its half-metallic gap
is situated in majority spin channel. The Curie temperature was
found to be 500K. Fig.\ref{fm_mnc} represents the frozen-magnon
energies for a selected direction in the Brillouin zone and the
exchange constants. Ferromagnetism is stabilized by the direct
Mn--Mn interactions and Mn-C ferromagnetic coupling. A remarkable
feature of  MnC is  small difference between the MFA and RPA
values of $T_C$.\cite{sa.ga.05}.

\section{Conclusions}

The idea of half-metallic ferromagnetism appeared as a result of
band-structure calculations \cite{deGroot:2024}. During a long
time, dominant activity in this field was theoretical.
Conceptually, HMF are of interest from the point of view they
provide an opportunity to probe in a clear form some essentially
many-particle effects \cite{irkhin:7151}.
Whereas for generic metallic system the Landau Fermi-liquid theory \cite%
{Nozieres:1964,Vonsovsky:1989} works, most correlation effects
being hidden in a parameter renormalization (such as effective
mass, magnetic moment etc.), in HMF the spin-polaronic effects
lead to a qualitatively new feature: occurrence of
non-quasiparticle (incoherent) states in the energy gap for one of
spin projections near the Fermi level. On the contrary, similar
effects of electron-magnon interactions in traditional
itinerant-electron ferromagnets are mixed with other kind
renormalizations (e.g., electron-phonon, electron-electron ones
etc.). The NQP states occur only above (below) $E_{F}$ for
minority spin (majority spin)\ gap. Therefore HMF are ideal
objects to investigate effects of electron-magnon interactions
\cite{irkhin:7151}.

Even stronger motivation to study HMF is connected with
perspectives to use them in giant magnetoresistance and tunnel
magnetoresistance \cite{Rob83,Irkhin:705,Prinz:1660} devices. This
initiated great
theoretical activity in the field of heterostructures containing HMF \cite%
{Tkachov:024519,irkhin:481}. At the same time, the interest in the
search and prediction of new HMF was growing on the basis of
band-structure calculations, as well as attempts to understand
better the features of electronic structure and chemical bonding
which are relevant for half-metallicity (see
Sect.\ref{sec:elstr_hmf}).

Recently, numerous attempts have been performed to build
heterostructures with HMF, such as Heusler alloys \cite{sakuraba:052508,sa.ha.06,sa.ha.07,gercsi:082512}, CrO$%
_{2}$ \cite{miao:022511}, Fe$_{3}$O$_{4}$ \cite{zhao:052506,rybchenko:132509}%
. Therefore the half-metallicity predicted by electronic structure
calculations becomes practically applicable. Nevertheless, direct
experimental evidences of half-metallic structure for specific
compounds are still rather poor. 
Perhaps, the unique method of testing genuine, bulk, half-metallic properties 
remains the spin-resolved positron annihilation. This underexposed
technique enables the direct measurement of the spin-polarization in the
bulk. 
Advanced techniques borrowed from semiconductor technologies which
access spatially resolved spin polarization at the Fermi level
would be interesting alternatives for positron annihilation.
Although being extensively used to characterize semiconductors,
they are poorly known for the spintronics community.

Among other experiments expected to advance the field we mention the STM
\cite{irkhin:104429},
spin-polarized photoemission \cite{Park:794}, and Andreev reflection \cite%
{soulen:4589}. Investigations of nuclear magnetic relaxation rate
should be specially mentioned since the absence of the Korringa
relaxation is a clear sign of the half-metallicity
\cite{irkhin:401} and NMR gives is not surface sensitive and gives
direct information on bulk properties. Also data of
core-level spectroscopies, especially XMCD would be very useful (see Sect.%
\ref{sec:xray}). As for the band-structure calculations,
application of state-of-art methods taking into account
correlation effects, such as GW or DMFT (see
Sect.\ref{sec:diffunc}) looks very promising. Another perspective
direction is the use of electron-structure calculations to search
essentially new types of HMF like $sp$ electron (or anionic) magnets \cite%
{Edwards:7209,Attema:16325}.

\section*{Acknowledgments}

We are grateful for enlightening discussions with O. K. Andersen,
E. Arrigoni, V. Antonov, S. Bl{\"u}gel, P. Bruno, C. Carbone, P.
H. Dederichs, P. A.  Dowben, H. Ebert, D. M. Edwards, O. Eriksson,
A. J. Freeman, P. Fulde, A. Georges, G. Kotliar, J.  K{\"u}bler,
Ph. Mavropoulos, I. I. Mazin, J. Minar, W. Nolting, W. E. Pickett,
L. M. Sandratskii, G. Sawatzky, D. J. Singh, L. Vitos, D.
Vollhardt, G. A. de Wijs, R. Wiesendanger, A. Yamasaki.

This work was supported by the Stichting voor Fundamenteel
Onderzoek der Materie (FOM), the Netherlands, and by the
Netherlands Organization for Scientific Research (Grant NWO
047.016.005). V. I. acknowledges support from the Russian Basic
Research Foundation (Grant No. 4640.2006.2) \ and A. L. - support
from the DFG (Grants No. SFB 668-A3). L. C. acknowledges financial
support offered by the Austrian Science Foundation FWF project nr.
P18505-N16 and the Romanian CNCSIS project nr.96/2006.

\bibliographystyle{apsrmp}
\bibliography{hmf}

\begin{thebibliography}{357}
\expandafter\ifx\csname natexlab\endcsname\relax\def\natexlab#1{#1}\fi
\expandafter\ifx\csname bibnamefont\endcsname\relax
  \def\bibnamefont#1{#1}\fi
\expandafter\ifx\csname bibfnamefont\endcsname\relax
  \def\bibfnamefont#1{#1}\fi
\expandafter\ifx\csname citenamefont\endcsname\relax
  \def\citenamefont#1{#1}\fi
\expandafter\ifx\csname url\endcsname\relax
  \def\url#1{\texttt{#1}}\fi
\expandafter\ifx\csname urlprefix\endcsname\relax\def\urlprefix{URL }\fi
\providecommand{\bibinfo}[2]{#2}
\providecommand{\eprint}[2][]{\url{#2}}

\bibitem[{\citenamefont{Abragam}(1961)}]{Abragam:1961}
\bibinfo{author}{\bibnamefont{Abragam}, \bibfnamefont{A.}},
  \bibinfo{year}{1961}, \emph{\bibinfo{title}{The Principles of Nuclear
  Magnetism}} (\bibinfo{publisher}{Clarendon, Oxford}).

\bibitem[{\citenamefont{Akinaga} \emph{et~al.}(2000)\citenamefont{Akinaga,
  Magano, and Shirai}}]{akinaga:L1118}
\bibinfo{author}{\bibnamefont{Akinaga}, \bibfnamefont{H.}},
  \bibinfo{author}{\bibfnamefont{T.}~\bibnamefont{Magano}}, and
  \bibinfo{author}{\bibfnamefont{M.}~\bibnamefont{Shirai}},
  \bibinfo{year}{2000}, \bibinfo{journal}{Jpn. J. Appl. Phys.}
  \textbf{\bibinfo{volume}{39}}(\bibinfo{number}{11B}), \bibinfo{pages}{L1118}.

\bibitem[{\citenamefont{Almbladh} \emph{et~al.}(1999)\citenamefont{Almbladh,
  Barth, and Leeuwen}}]{Almbladh:535}
\bibinfo{author}{\bibnamefont{Almbladh}, \bibfnamefont{C.-O.}},
  \bibinfo{author}{\bibfnamefont{U.~V.} \bibnamefont{Barth}}, and
  \bibinfo{author}{\bibfnamefont{R.~V.} \bibnamefont{Leeuwen}},
  \bibinfo{year}{1999}, \bibinfo{journal}{Int. J. Mod. Phys. B}
  \textbf{\bibinfo{volume}{13}}(\bibinfo{number}{5-6}), \bibinfo{pages}{535}.

\bibitem[{\citenamefont{Andersen}(1975)}]{Andersen:3060}
\bibinfo{author}{\bibnamefont{Andersen}, \bibfnamefont{O.~K.}},
  \bibinfo{year}{1975}, \bibinfo{journal}{Phys. Rev. B}
  \textbf{\bibinfo{volume}{12}}(\bibinfo{number}{8}), \bibinfo{pages}{3060}.

\bibitem[{\citenamefont{Andersen and Jepsen}(1984)}]{Andersen:2571}
\bibinfo{author}{\bibnamefont{Andersen}, \bibfnamefont{O.~K.}}, and
  \bibinfo{author}{\bibfnamefont{O.}~\bibnamefont{Jepsen}},
  \bibinfo{year}{1984}, \bibinfo{journal}{Phys. Rev. Lett.}
  \textbf{\bibinfo{volume}{53}}(\bibinfo{number}{27}), \bibinfo{pages}{2571}.

\bibitem[{\citenamefont{Andersen and Saha-Dasgupta}(2000)}]{Andersen:16219}
\bibinfo{author}{\bibnamefont{Andersen}, \bibfnamefont{O.~K.}}, and
  \bibinfo{author}{\bibfnamefont{T.}~\bibnamefont{Saha-Dasgupta}},
  \bibinfo{year}{2000}, \bibinfo{journal}{Phys. Rev. B}
  \textbf{\bibinfo{volume}{62}}(\bibinfo{number}{24}), \bibinfo{pages}{R16219}.

\bibitem[{\citenamefont{Anisimov}
  \emph{et~al.}(1997{\natexlab{a}})\citenamefont{Anisimov, Aryasetiawan, and
  Lichtenstein}}]{Anisimov:767}
\bibinfo{author}{\bibnamefont{Anisimov}, \bibfnamefont{V.~I.}},
  \bibinfo{author}{\bibfnamefont{F.}~\bibnamefont{Aryasetiawan}}, and
  \bibinfo{author}{\bibfnamefont{A.~I.} \bibnamefont{Lichtenstein}},
  \bibinfo{year}{1997}{\natexlab{a}}, \bibinfo{journal}{J. Phys.: Condens.
  Matter} \textbf{\bibinfo{volume}{9}}, \bibinfo{pages}{767}.

\bibitem[{\citenamefont{Anisimov and Gunnarsson}(1991)}]{anisimov:7570}
\bibinfo{author}{\bibnamefont{Anisimov}, \bibfnamefont{V.~I.}}, and
  \bibinfo{author}{\bibfnamefont{O.}~\bibnamefont{Gunnarsson}},
  \bibinfo{year}{1991}, \bibinfo{journal}{Phys. Rev. B}
  \textbf{\bibinfo{volume}{43}}(\bibinfo{number}{10}), \bibinfo{pages}{7570}.

\bibitem[{\citenamefont{Anisimov}
  \emph{et~al.}(1997{\natexlab{b}})\citenamefont{Anisimov, Poteryaev, Korotin,
  Anokhin, and Kotliar}}]{anisimov:7359}
\bibinfo{author}{\bibnamefont{Anisimov}, \bibfnamefont{V.~I.}},
  \bibinfo{author}{\bibfnamefont{A.~I.} \bibnamefont{Poteryaev}},
  \bibinfo{author}{\bibfnamefont{M.~A.} \bibnamefont{Korotin}},
  \bibinfo{author}{\bibfnamefont{A.~O.} \bibnamefont{Anokhin}}, and
  \bibinfo{author}{\bibfnamefont{G.}~\bibnamefont{Kotliar}},
  \bibinfo{year}{1997}{\natexlab{b}}, \bibinfo{journal}{J. Phys.: Condens.
  Matter} \textbf{\bibinfo{volume}{9}}(\bibinfo{number}{35}),
  \bibinfo{pages}{7359}.

\bibitem[{\citenamefont{Anisimov} \emph{et~al.}(1993)\citenamefont{Anisimov,
  Solovyev, Korotin, Czy\ifmmode~\dot{z}\else \.{z}\fi{}yk, and
  Sawatzky}}]{Anisimov:16929}
\bibinfo{author}{\bibnamefont{Anisimov}, \bibfnamefont{V.~I.}},
  \bibinfo{author}{\bibfnamefont{I.~V.} \bibnamefont{Solovyev}},
  \bibinfo{author}{\bibfnamefont{M.~A.} \bibnamefont{Korotin}},
  \bibinfo{author}{\bibfnamefont{M.~T.} \bibnamefont{Czy\ifmmode~\dot{z}\else
  \.{z}\fi{}yk}}, and \bibinfo{author}{\bibfnamefont{G.~A.}
  \bibnamefont{Sawatzky}}, \bibinfo{year}{1993}, \bibinfo{journal}{Phys. Rev.
  B} \textbf{\bibinfo{volume}{48}}(\bibinfo{number}{23}),
  \bibinfo{pages}{16929}.

\bibitem[{\citenamefont{Anokhin and Katsnelson}(1996)}]{Anokhin:2468}
\bibinfo{author}{\bibnamefont{Anokhin}, \bibfnamefont{A.~O.}}, and
  \bibinfo{author}{\bibfnamefont{M.~I.} \bibnamefont{Katsnelson}},
  \bibinfo{year}{1996}, \bibinfo{journal}{Int. J. Mod. Phys. B}
  \textbf{\bibinfo{volume}{10}}, \bibinfo{pages}{2468}.

\bibitem[{\citenamefont{Antonov} \emph{et~al.}(1999)\citenamefont{Antonov,
  Antropov, Harmon, Yaresko, and Perlov}}]{Antonov:14552}
\bibinfo{author}{\bibnamefont{Antonov}, \bibfnamefont{V.~N.}},
  \bibinfo{author}{\bibfnamefont{V.~P.} \bibnamefont{Antropov}},
  \bibinfo{author}{\bibfnamefont{B.~N.} \bibnamefont{Harmon}},
  \bibinfo{author}{\bibfnamefont{A.~N.} \bibnamefont{Yaresko}}, and
  \bibinfo{author}{\bibfnamefont{A.~Y.} \bibnamefont{Perlov}},
  \bibinfo{year}{1999}, \bibinfo{journal}{Phys. Rev. B}
  \textbf{\bibinfo{volume}{59}}(\bibinfo{number}{22}), \bibinfo{pages}{14552}.

\bibitem[{\citenamefont{Antonov} \emph{et~al.}(1997)\citenamefont{Antonov,
  Oppeneer, Yaresko, Perlov, and Kraft}}]{Antonov:13012}
\bibinfo{author}{\bibnamefont{Antonov}, \bibfnamefont{V.~N.}},
  \bibinfo{author}{\bibfnamefont{P.~M.} \bibnamefont{Oppeneer}},
  \bibinfo{author}{\bibfnamefont{A.~N.} \bibnamefont{Yaresko}},
  \bibinfo{author}{\bibfnamefont{A.~Y.} \bibnamefont{Perlov}}, and
  \bibinfo{author}{\bibfnamefont{T.}~\bibnamefont{Kraft}},
  \bibinfo{year}{1997}, \bibinfo{journal}{Phys. Rev. B}
  \textbf{\bibinfo{volume}{56}}(\bibinfo{number}{20}), \bibinfo{pages}{13012}.

\bibitem[{\citenamefont{Aryasetiawan}
  \emph{et~al.}(2004)\citenamefont{Aryasetiawan, Imada, Georges, Kotliar,
  Biermann, and Lichtenstein}}]{aryasetiawan:195104}
\bibinfo{author}{\bibnamefont{Aryasetiawan}, \bibfnamefont{F.}},
  \bibinfo{author}{\bibfnamefont{M.}~\bibnamefont{Imada}},
  \bibinfo{author}{\bibfnamefont{A.}~\bibnamefont{Georges}},
  \bibinfo{author}{\bibfnamefont{G.}~\bibnamefont{Kotliar}},
  \bibinfo{author}{\bibfnamefont{S.}~\bibnamefont{Biermann}}, and
  \bibinfo{author}{\bibfnamefont{A.~I.} \bibnamefont{Lichtenstein}},
  \bibinfo{year}{2004}, \bibinfo{journal}{Phys. Rev. B}
  \textbf{\bibinfo{volume}{70}}(\bibinfo{number}{19}), \bibinfo{eid}{195104}.

\bibitem[{\citenamefont{Attema} \emph{et~al.}(2004)\citenamefont{Attema, Fang,
  Chioncel, de~Wijs, Lichtenstein, and de~Groot}}]{attema:S5517}
\bibinfo{author}{\bibnamefont{Attema}, \bibfnamefont{J.~J.}},
  \bibinfo{author}{\bibfnamefont{C.~M.} \bibnamefont{Fang}},
  \bibinfo{author}{\bibfnamefont{L.}~\bibnamefont{Chioncel}},
  \bibinfo{author}{\bibfnamefont{G.~A.} \bibnamefont{de~Wijs}},
  \bibinfo{author}{\bibfnamefont{A.~I.} \bibnamefont{Lichtenstein}}, and
  \bibinfo{author}{\bibfnamefont{R.~A.} \bibnamefont{de~Groot}},
  \bibinfo{year}{2004}, \bibinfo{journal}{J. Phys.: Condens. Matter}
  \textbf{\bibinfo{volume}{16}}, \bibinfo{pages}{S5517}.

\bibitem[{\citenamefont{Attema} \emph{et~al.}(2005)\citenamefont{Attema,
  de~Wijs, Blake, and de~Groot}}]{Attema:16325}
\bibinfo{author}{\bibnamefont{Attema}, \bibfnamefont{J.~J.}},
  \bibinfo{author}{\bibfnamefont{G.}~\bibnamefont{de~Wijs}},
  \bibinfo{author}{\bibfnamefont{G.~R.} \bibnamefont{Blake}}, and
  \bibinfo{author}{\bibfnamefont{R.~A.} \bibnamefont{de~Groot}},
  \bibinfo{year}{2005}, \bibinfo{journal}{Journal of the American Chemical
  Society} \textbf{\bibinfo{volume}{127}}(\bibinfo{number}{46}),
  \bibinfo{pages}{16325}.

\bibitem[{\citenamefont{Attema} \emph{et~al.}(2006)\citenamefont{Attema,
  de~Wijs, and de~Groot}}]{Attema:793}
\bibinfo{author}{\bibnamefont{Attema}, \bibfnamefont{J.~J.}},
  \bibinfo{author}{\bibfnamefont{G.~A.} \bibnamefont{de~Wijs}}, and
  \bibinfo{author}{\bibfnamefont{R.~A.} \bibnamefont{de~Groot}},
  \bibinfo{year}{2006}, \bibinfo{journal}{J. Phys. D: Appl. Phys.}
  \textbf{\bibinfo{volume}{39}}, \bibinfo{pages}{793}.

\bibitem[{\citenamefont{Auslender and
  Irkhin}(1984{\natexlab{a}})}]{Auslender:301}
\bibinfo{author}{\bibnamefont{Auslender}, \bibfnamefont{M.~I.}}, and
  \bibinfo{author}{\bibfnamefont{V.~Y.} \bibnamefont{Irkhin}},
  \bibinfo{year}{1984}{\natexlab{a}}, \bibinfo{journal}{Z. Phys. B}
  \textbf{\bibinfo{volume}{56}}, \bibinfo{pages}{301}.

\bibitem[{\citenamefont{Auslender and
  Irkhin}(1984{\natexlab{b}})}]{Auslender:1003}
\bibinfo{author}{\bibnamefont{Auslender}, \bibfnamefont{M.~I.}}, and
  \bibinfo{author}{\bibfnamefont{V.~Y.} \bibnamefont{Irkhin}},
  \bibinfo{year}{1984}{\natexlab{b}}, \bibinfo{journal}{Solid State Commun.}
  \textbf{\bibinfo{volume}{50}}, \bibinfo{pages}{1003}.

\bibitem[{\citenamefont{Auslender and
  Irkhin}(1985{\natexlab{a}})}]{Auslender:129}
\bibinfo{author}{\bibnamefont{Auslender}, \bibfnamefont{M.~I.}}, and
  \bibinfo{author}{\bibfnamefont{V.~Y.} \bibnamefont{Irkhin}},
  \bibinfo{year}{1985}{\natexlab{a}}, \bibinfo{journal}{Z. Phys. B}
  \textbf{\bibinfo{volume}{61}}, \bibinfo{pages}{129}.

\bibitem[{\citenamefont{Auslender and
  Irkhin}(1985{\natexlab{b}})}]{auslender:701}
\bibinfo{author}{\bibnamefont{Auslender}, \bibfnamefont{M.~I.}}, and
  \bibinfo{author}{\bibfnamefont{V.~Y.} \bibnamefont{Irkhin}},
  \bibinfo{year}{1985}{\natexlab{b}}, \bibinfo{journal}{Solid State Commun.}
  \textbf{\bibinfo{volume}{56}}(\bibinfo{number}{8}), \bibinfo{pages}{701}.

\bibitem[{\citenamefont{Auslender} \emph{et~al.}(1988)\citenamefont{Auslender,
  Irkhin, and Katsnelson}}]{Auslender:5521}
\bibinfo{author}{\bibnamefont{Auslender}, \bibfnamefont{M.~I.}},
  \bibinfo{author}{\bibfnamefont{V.~Y.} \bibnamefont{Irkhin}}, and
  \bibinfo{author}{\bibfnamefont{M.~I.} \bibnamefont{Katsnelson}},
  \bibinfo{year}{1988}, \bibinfo{journal}{J. Phys. C}
  \textbf{\bibinfo{volume}{21}}, \bibinfo{pages}{5521}.

\bibitem[{\citenamefont{Auslender and Katsnelson}(1982)}]{Auslender:436}
\bibinfo{author}{\bibnamefont{Auslender}, \bibfnamefont{M.~I.}}, and
  \bibinfo{author}{\bibfnamefont{M.~I.} \bibnamefont{Katsnelson}},
  \bibinfo{year}{1982}, \bibinfo{journal}{Teor. Mat. Fiz.}
  \textbf{\bibinfo{volume}{51}}, \bibinfo{pages}{436}.

\bibitem[{\citenamefont{Auslender} \emph{et~al.}(1983)\citenamefont{Auslender,
  Katsnelson, and Irkhin}}]{Auslender:309}
\bibinfo{author}{\bibnamefont{Auslender}, \bibfnamefont{M.~I.}},
  \bibinfo{author}{\bibfnamefont{M.~I.} \bibnamefont{Katsnelson}}, and
  \bibinfo{author}{\bibfnamefont{V.~Y.} \bibnamefont{Irkhin}},
  \bibinfo{year}{1983}, \bibinfo{journal}{Physica B}
  \textbf{\bibinfo{volume}{119}}, \bibinfo{pages}{309}.

\bibitem[{\citenamefont{Auth} \emph{et~al.}(2003)\citenamefont{Auth, Jakob,
  Block, and Felser}}]{auth:024403}
\bibinfo{author}{\bibnamefont{Auth}, \bibfnamefont{N.}},
  \bibinfo{author}{\bibfnamefont{G.}~\bibnamefont{Jakob}},
  \bibinfo{author}{\bibfnamefont{T.}~\bibnamefont{Block}}, and
  \bibinfo{author}{\bibfnamefont{C.}~\bibnamefont{Felser}},
  \bibinfo{year}{2003}, \bibinfo{journal}{Phys. Rev. B}
  \textbf{\bibinfo{volume}{68}}(\bibinfo{number}{2}), \bibinfo{eid}{024403}.

\bibitem[{\citenamefont{Baym and Kadanoff}(1961)}]{baym:287}
\bibinfo{author}{\bibnamefont{Baym}, \bibfnamefont{G.}}, and
  \bibinfo{author}{\bibfnamefont{L.~P.} \bibnamefont{Kadanoff}},
  \bibinfo{year}{1961}, \bibinfo{journal}{Phys. Rev.}
  \textbf{\bibinfo{volume}{124}}(\bibinfo{number}{2}), \bibinfo{pages}{287}.

\bibitem[{\citenamefont{Bickers and Scalapino}(1989)}]{bickers:206}
\bibinfo{author}{\bibnamefont{Bickers}, \bibfnamefont{N.~E.}}, and
  \bibinfo{author}{\bibfnamefont{D.~J.} \bibnamefont{Scalapino}},
  \bibinfo{year}{1989}, \bibinfo{journal}{Ann. Phys.}
  \textbf{\bibinfo{volume}{193}}, \bibinfo{pages}{206}.

\bibitem[{\citenamefont{Biermann} \emph{et~al.}(2003)\citenamefont{Biermann,
  Aryasetiawan, and Georges}}]{biermann:086402}
\bibinfo{author}{\bibnamefont{Biermann}, \bibfnamefont{S.}},
  \bibinfo{author}{\bibfnamefont{F.}~\bibnamefont{Aryasetiawan}}, and
  \bibinfo{author}{\bibfnamefont{A.}~\bibnamefont{Georges}},
  \bibinfo{year}{2003}, \bibinfo{journal}{Phys. Rev. Lett.}
  \textbf{\bibinfo{volume}{90}}(\bibinfo{number}{8}), \bibinfo{eid}{086402}.

\bibitem[{\citenamefont{de~Boer and Groot}(1999)}]{Boer:10758}
\bibinfo{author}{\bibnamefont{de~Boer}, \bibfnamefont{P.~K.}}, and
  \bibinfo{author}{\bibfnamefont{R.~A.~d.} \bibnamefont{Groot}},
  \bibinfo{year}{1999}, \bibinfo{journal}{Phys. Rev. B}
  \textbf{\bibinfo{volume}{60}}(\bibinfo{number}{15}), \bibinfo{pages}{10758}.

\bibitem[{\citenamefont{Bona} \emph{et~al.}(1985)\citenamefont{Bona, Meier,
  Taborelli, Bucher, and Schmidt}}]{bona:391}
\bibinfo{author}{\bibnamefont{Bona}, \bibfnamefont{G.~L.}},
  \bibinfo{author}{\bibfnamefont{F.}~\bibnamefont{Meier}},
  \bibinfo{author}{\bibfnamefont{M.}~\bibnamefont{Taborelli}},
  \bibinfo{author}{\bibfnamefont{E.}~\bibnamefont{Bucher}}, and
  \bibinfo{author}{\bibfnamefont{P.~H.} \bibnamefont{Schmidt}},
  \bibinfo{year}{1985}, \bibinfo{journal}{Solid State Commun.}
  \textbf{\bibinfo{volume}{56}}(\bibinfo{number}{4}), \bibinfo{pages}{391}.

\bibitem[{\citenamefont{Borca} \emph{et~al.}(2001)\citenamefont{Borca, Komesu,
  Jeong, Dowben, Ristoiu, Hordequin, Nozi\`eres, Pierre, Stadler, and
  Idzerda}}]{Borca:052409}
\bibinfo{author}{\bibnamefont{Borca}, \bibfnamefont{C.~N.}},
  \bibinfo{author}{\bibfnamefont{T.}~\bibnamefont{Komesu}},
  \bibinfo{author}{\bibfnamefont{H.-K.} \bibnamefont{Jeong}},
  \bibinfo{author}{\bibfnamefont{P.~A.} \bibnamefont{Dowben}},
  \bibinfo{author}{\bibfnamefont{D.}~\bibnamefont{Ristoiu}},
  \bibinfo{author}{\bibfnamefont{C.}~\bibnamefont{Hordequin}},
  \bibinfo{author}{\bibfnamefont{J.~P.} \bibnamefont{Nozi\`eres}},
  \bibinfo{author}{\bibfnamefont{J.}~\bibnamefont{Pierre}},
  \bibinfo{author}{\bibfnamefont{S.}~\bibnamefont{Stadler}}, and
  \bibinfo{author}{\bibfnamefont{Y.~U.} \bibnamefont{Idzerda}},
  \bibinfo{year}{2001}, \bibinfo{journal}{Phys. Rev. B}
  \textbf{\bibinfo{volume}{64}}(\bibinfo{number}{5}), \bibinfo{pages}{052409}.

\bibitem[{\citenamefont{Bowen} \emph{et~al.}(2003)\citenamefont{Bowen, Bibes,
  Barthelemy, Contour, Anane, Lemaitre, and Fert}}]{bowen:233}
\bibinfo{author}{\bibnamefont{Bowen}, \bibfnamefont{M.}},
  \bibinfo{author}{\bibfnamefont{M.}~\bibnamefont{Bibes}},
  \bibinfo{author}{\bibfnamefont{A.}~\bibnamefont{Barthelemy}},
  \bibinfo{author}{\bibfnamefont{J.-P.} \bibnamefont{Contour}},
  \bibinfo{author}{\bibfnamefont{A.}~\bibnamefont{Anane}},
  \bibinfo{author}{\bibfnamefont{Y.}~\bibnamefont{Lemaitre}}, and
  \bibinfo{author}{\bibfnamefont{A.}~\bibnamefont{Fert}}, \bibinfo{year}{2003},
  \bibinfo{journal}{Appl. Phys. Lett.}
  \textbf{\bibinfo{volume}{82}}(\bibinfo{number}{2}), \bibinfo{pages}{233}.

\bibitem[{\citenamefont{Bratkovsky}(1997)}]{Bratkovsky:2344}
\bibinfo{author}{\bibnamefont{Bratkovsky}, \bibfnamefont{A.~M.}},
  \bibinfo{year}{1997}, \bibinfo{journal}{Phys. Rev. B}
  \textbf{\bibinfo{volume}{56}}(\bibinfo{number}{5}), \bibinfo{pages}{2344}.

\bibitem[{\citenamefont{Bratkovsky}(1998)}]{bratkovsky:2334}
\bibinfo{author}{\bibnamefont{Bratkovsky}, \bibfnamefont{A.~M.}},
  \bibinfo{year}{1998}, \bibinfo{journal}{Appl. Phys. Lett.}
  \textbf{\bibinfo{volume}{72}}(\bibinfo{number}{18}), \bibinfo{pages}{2334}.

\bibitem[{\citenamefont{Brown} \emph{et~al.}(2000)\citenamefont{Brown, Neumann,
  Webster, and Ziebeck}}]{br.ne.00}
\bibinfo{author}{\bibnamefont{Brown}, \bibfnamefont{P.~J.}},
  \bibinfo{author}{\bibfnamefont{K.~U.} \bibnamefont{Neumann}},
  \bibinfo{author}{\bibfnamefont{P.~J.} \bibnamefont{Webster}}, and
  \bibinfo{author}{\bibfnamefont{K.~R.~A.} \bibnamefont{Ziebeck}},
  \bibinfo{year}{2000}, \bibinfo{journal}{J. Phys.: Condens. Matter}
  \textbf{\bibinfo{volume}{12}}(\bibinfo{number}{8}), \bibinfo{pages}{1827}.

\bibitem[{\citenamefont{Carneiro and Pethick}(1975)}]{Carneiro:1106}
\bibinfo{author}{\bibnamefont{Carneiro}, \bibfnamefont{G.~M.}}, and
  \bibinfo{author}{\bibfnamefont{C.~J.} \bibnamefont{Pethick}},
  \bibinfo{year}{1975}, \bibinfo{journal}{Phys. Rev. B}
  \textbf{\bibinfo{volume}{11}}(\bibinfo{number}{3}), \bibinfo{pages}{1106}.

\bibitem[{\citenamefont{Ceperley and Alder}(1980)}]{ceperley:566}
\bibinfo{author}{\bibnamefont{Ceperley}, \bibfnamefont{D.~M.}}, and
  \bibinfo{author}{\bibfnamefont{B.~J.} \bibnamefont{Alder}},
  \bibinfo{year}{1980}, \bibinfo{journal}{Phys. Rev. Lett.}
  \textbf{\bibinfo{volume}{45}}(\bibinfo{number}{7}), \bibinfo{pages}{566}.

\bibitem[{\citenamefont{Chadov} \emph{et~al.}(2006)\citenamefont{Chadov, Minar,
  Ebert, Perlov, Chioncel, Katsnelson, and Lichtenstein}}]{chadov:140411}
\bibinfo{author}{\bibnamefont{Chadov}, \bibfnamefont{S.}},
  \bibinfo{author}{\bibfnamefont{J.}~\bibnamefont{Minar}},
  \bibinfo{author}{\bibfnamefont{H.}~\bibnamefont{Ebert}},
  \bibinfo{author}{\bibfnamefont{A.}~\bibnamefont{Perlov}},
  \bibinfo{author}{\bibfnamefont{L.}~\bibnamefont{Chioncel}},
  \bibinfo{author}{\bibfnamefont{M.~I.} \bibnamefont{Katsnelson}}, and
  \bibinfo{author}{\bibfnamefont{A.~I.} \bibnamefont{Lichtenstein}},
  \bibinfo{year}{2006}, \bibinfo{journal}{Phys. Rev. B}
  \textbf{\bibinfo{volume}{74}}(\bibinfo{number}{14}), \bibinfo{eid}{140411}.

\bibitem[{\citenamefont{Chamberland}(1977)}]{cham.77}
\bibinfo{author}{\bibnamefont{Chamberland}, \bibfnamefont{B.~L.}},
  \bibinfo{year}{1977}, \bibinfo{journal}{CRC Crit. Rev. Solid State Mater.
  Sci.} \textbf{\bibinfo{volume}{7}}, \bibinfo{pages}{1}.

\bibitem[{\citenamefont{Chiba} \emph{et~al.}(2003)\citenamefont{Chiba,
  Takamura, Matsukura, and Ohno}}]{chiba:3020}
\bibinfo{author}{\bibnamefont{Chiba}, \bibfnamefont{D.}},
  \bibinfo{author}{\bibfnamefont{K.}~\bibnamefont{Takamura}},
  \bibinfo{author}{\bibfnamefont{F.}~\bibnamefont{Matsukura}}, and
  \bibinfo{author}{\bibfnamefont{H.}~\bibnamefont{Ohno}}, \bibinfo{year}{2003},
  \bibinfo{journal}{Applied Physics Letters}
  \textbf{\bibinfo{volume}{82}}(\bibinfo{number}{18}), \bibinfo{pages}{3020}.

\bibitem[{\citenamefont{Chioncel}(2004)}]{Chioncel:thesis}
\bibinfo{author}{\bibnamefont{Chioncel}, \bibfnamefont{L.}},
  \bibinfo{year}{2004}, \emph{\bibinfo{title}{Finite Temperature Electronic
  Structure, beyond Local Density Approximation,}} (\bibinfo{publisher}{PhD
  Thesis, Radbout University Nijmegen}).

\bibitem[{\citenamefont{Chioncel}(2006)}]{Chioncel:private}
\bibinfo{author}{\bibnamefont{Chioncel}, \bibfnamefont{L.}},
  \bibinfo{year}{2006}, \bibinfo{journal}{unpublished} .

\bibitem[{\citenamefont{Chioncel} \emph{et~al.}(2007)\citenamefont{Chioncel,
  Allmaier, Arrigoni, Yamasaki, Daghofer, Katsnelson, and
  Lichtenstein}}]{chioncel:cro2}
\bibinfo{author}{\bibnamefont{Chioncel}, \bibfnamefont{L.}},
  \bibinfo{author}{\bibfnamefont{H.}~\bibnamefont{Allmaier}},
  \bibinfo{author}{\bibfnamefont{E.}~\bibnamefont{Arrigoni}},
  \bibinfo{author}{\bibfnamefont{A.}~\bibnamefont{Yamasaki}},
  \bibinfo{author}{\bibfnamefont{M.}~\bibnamefont{Daghofer}},
  \bibinfo{author}{\bibfnamefont{M.~I.} \bibnamefont{Katsnelson}}, and
  \bibinfo{author}{\bibfnamefont{A.~I.} \bibnamefont{Lichtenstein}},
  \bibinfo{year}{2007}, \bibinfo{journal}{Phys. Rev. B}
  \textbf{\bibinfo{volume}{75}}(\bibinfo{number}{14}), \bibinfo{eid}{140406}.

\bibitem[{\citenamefont{Chioncel}
  \emph{et~al.}(2006{\natexlab{a}})\citenamefont{Chioncel, Arrigoni,
  Katsnelson, and Lichtenstein}}]{chioncel:137203}
\bibinfo{author}{\bibnamefont{Chioncel}, \bibfnamefont{L.}},
  \bibinfo{author}{\bibfnamefont{E.}~\bibnamefont{Arrigoni}},
  \bibinfo{author}{\bibfnamefont{M.~I.} \bibnamefont{Katsnelson}}, and
  \bibinfo{author}{\bibfnamefont{A.~I.} \bibnamefont{Lichtenstein}},
  \bibinfo{year}{2006}{\natexlab{a}}, \bibinfo{journal}{Phys. Rev. Lett.}
  \textbf{\bibinfo{volume}{96}}(\bibinfo{number}{13}), \bibinfo{eid}{137203}.

\bibitem[{\citenamefont{Chioncel}
  \emph{et~al.}(2003{\natexlab{a}})\citenamefont{Chioncel, Katsnelson,
  de~Groot, and Lichtenstein}}]{chioncel:144425}
\bibinfo{author}{\bibnamefont{Chioncel}, \bibfnamefont{L.}},
  \bibinfo{author}{\bibfnamefont{M.~I.} \bibnamefont{Katsnelson}},
  \bibinfo{author}{\bibfnamefont{R.~A.} \bibnamefont{de~Groot}}, and
  \bibinfo{author}{\bibfnamefont{A.~I.} \bibnamefont{Lichtenstein}},
  \bibinfo{year}{2003}{\natexlab{a}}, \bibinfo{journal}{Phys. Rev. B}
  \textbf{\bibinfo{volume}{68}}(\bibinfo{number}{14}), \bibinfo{eid}{144425}.

\bibitem[{\citenamefont{Chioncel} \emph{et~al.}(2005)\citenamefont{Chioncel,
  Katsnelson, de~Wijs, de~Groot, and Lichtenstein}}]{chioncel:085111}
\bibinfo{author}{\bibnamefont{Chioncel}, \bibfnamefont{L.}},
  \bibinfo{author}{\bibfnamefont{M.~I.} \bibnamefont{Katsnelson}},
  \bibinfo{author}{\bibfnamefont{G.~A.} \bibnamefont{de~Wijs}},
  \bibinfo{author}{\bibfnamefont{R.~A.} \bibnamefont{de~Groot}}, and
  \bibinfo{author}{\bibfnamefont{A.~I.} \bibnamefont{Lichtenstein}},
  \bibinfo{year}{2005}, \bibinfo{journal}{Phys. Rev. B}
  \textbf{\bibinfo{volume}{71}}(\bibinfo{number}{8}), \bibinfo{eid}{085111}.

\bibitem[{\citenamefont{Chioncel}
  \emph{et~al.}(2006{\natexlab{b}})\citenamefont{Chioncel, Mavropoulos, Lezaic,
  Blugel, Arrigoni, Katsnelson, and Lichtenstein}}]{chioncel:197203}
\bibinfo{author}{\bibnamefont{Chioncel}, \bibfnamefont{L.}},
  \bibinfo{author}{\bibfnamefont{P.}~\bibnamefont{Mavropoulos}},
  \bibinfo{author}{\bibfnamefont{M.}~\bibnamefont{Lezaic}},
  \bibinfo{author}{\bibfnamefont{S.}~\bibnamefont{Blugel}},
  \bibinfo{author}{\bibfnamefont{E.}~\bibnamefont{Arrigoni}},
  \bibinfo{author}{\bibfnamefont{M.~I.} \bibnamefont{Katsnelson}}, and
  \bibinfo{author}{\bibfnamefont{A.~I.} \bibnamefont{Lichtenstein}},
  \bibinfo{year}{2006}{\natexlab{b}}, \bibinfo{journal}{Phys. Rev. Lett.}
  \textbf{\bibinfo{volume}{96}}(\bibinfo{number}{19}), \bibinfo{eid}{197203}.

\bibitem[{\citenamefont{Chioncel}
  \emph{et~al.}(2003{\natexlab{b}})\citenamefont{Chioncel, Vitos, Abrikosov,
  Kollar, Katsnelson, and Lichtenstein}}]{chioncel:235106}
\bibinfo{author}{\bibnamefont{Chioncel}, \bibfnamefont{L.}},
  \bibinfo{author}{\bibfnamefont{L.}~\bibnamefont{Vitos}},
  \bibinfo{author}{\bibfnamefont{I.~A.} \bibnamefont{Abrikosov}},
  \bibinfo{author}{\bibfnamefont{J.}~\bibnamefont{Kollar}},
  \bibinfo{author}{\bibfnamefont{M.~I.} \bibnamefont{Katsnelson}}, and
  \bibinfo{author}{\bibfnamefont{A.~I.} \bibnamefont{Lichtenstein}},
  \bibinfo{year}{2003}{\natexlab{b}}, \bibinfo{journal}{Phys. Rev. B}
  \textbf{\bibinfo{volume}{67}}(\bibinfo{number}{23}), \bibinfo{eid}{235106}.

\bibitem[{\citenamefont{Chitra and Kotliar}(2000)}]{chitra:12715}
\bibinfo{author}{\bibnamefont{Chitra}, \bibfnamefont{R.}}, and
  \bibinfo{author}{\bibfnamefont{G.}~\bibnamefont{Kotliar}},
  \bibinfo{year}{2000}, \bibinfo{journal}{Phys. Rev. B}
  \textbf{\bibinfo{volume}{62}}(\bibinfo{number}{19}), \bibinfo{pages}{12715}.

\bibitem[{\citenamefont{Chitra and Kotliar}(2001)}]{chitra:115110}
\bibinfo{author}{\bibnamefont{Chitra}, \bibfnamefont{R.}}, and
  \bibinfo{author}{\bibfnamefont{G.}~\bibnamefont{Kotliar}},
  \bibinfo{year}{2001}, \bibinfo{journal}{Phys. Rev. B}
  \textbf{\bibinfo{volume}{63}}(\bibinfo{number}{11}), \bibinfo{pages}{115110}.

\bibitem[{\citenamefont{Correa} \emph{et~al.}(2006)\citenamefont{Correa, Eibl,
  Rangelov, Braun, and Donath}}]{correa:125316}
\bibinfo{author}{\bibnamefont{Correa}, \bibfnamefont{J.~S.}},
  \bibinfo{author}{\bibfnamefont{C.}~\bibnamefont{Eibl}},
  \bibinfo{author}{\bibfnamefont{G.}~\bibnamefont{Rangelov}},
  \bibinfo{author}{\bibfnamefont{J.}~\bibnamefont{Braun}}, and
  \bibinfo{author}{\bibfnamefont{M.}~\bibnamefont{Donath}},
  \bibinfo{year}{2006}, \bibinfo{journal}{Phys. Rev. B}
  \textbf{\bibinfo{volume}{73}}(\bibinfo{number}{12}), \bibinfo{eid}{125316}.

\bibitem[{\citenamefont{Craco} \emph{et~al.}(2003)\citenamefont{Craco, Laad,
  and M{\"u}ller-Hartmann}}]{craco:237203}
\bibinfo{author}{\bibnamefont{Craco}, \bibfnamefont{L.}},
  \bibinfo{author}{\bibfnamefont{M.~S.} \bibnamefont{Laad}}, and
  \bibinfo{author}{\bibfnamefont{E.}~\bibnamefont{M{\"u}ller-Hartmann}},
  \bibinfo{year}{2003}, \bibinfo{journal}{Phys. Rev. Lett.}
  \textbf{\bibinfo{volume}{90}}(\bibinfo{number}{23}), \bibinfo{eid}{237203}.

\bibitem[{\citenamefont{Craco} \emph{et~al.}(2006)\citenamefont{Craco, Laad,
  and M{\"u}ller-Hartmann}}]{craco:064425}
\bibinfo{author}{\bibnamefont{Craco}, \bibfnamefont{L.}},
  \bibinfo{author}{\bibfnamefont{M.~S.} \bibnamefont{Laad}}, and
  \bibinfo{author}{\bibfnamefont{E.}~\bibnamefont{M{\"u}ller-Hartmann}},
  \bibinfo{year}{2006}, \bibinfo{journal}{Phys. Rev. B}
  \textbf{\bibinfo{volume}{74}}(\bibinfo{number}{6}), \bibinfo{eid}{064425}.

\bibitem[{\citenamefont{Dagotto}(2003)}]{Dagotto2003}
\bibinfo{author}{\bibnamefont{Dagotto}, \bibfnamefont{E.}},
  \bibinfo{year}{2003}, \emph{\bibinfo{title}{Nanoscale Phase Separation and
  Colossal Magnetoresistance: The Physics of Manganites and Related Compounds}}
  (\bibinfo{publisher}{Springer, Berlin}).

\bibitem[{\citenamefont{Dederichs} \emph{et~al.}(1984)\citenamefont{Dederichs,
  Bl{\"u}gel, Zeller, and Akai}}]{dederichs:2512}
\bibinfo{author}{\bibnamefont{Dederichs}, \bibfnamefont{P.~H.}},
  \bibinfo{author}{\bibfnamefont{S.}~\bibnamefont{Bl{\"u}gel}},
  \bibinfo{author}{\bibfnamefont{R.}~\bibnamefont{Zeller}}, and
  \bibinfo{author}{\bibfnamefont{H.}~\bibnamefont{Akai}}, \bibinfo{year}{1984},
  \bibinfo{journal}{Phys. Rev. Lett.}
  \textbf{\bibinfo{volume}{53}}(\bibinfo{number}{26}), \bibinfo{pages}{2512}.

\bibitem[{\citenamefont{Delves and Lewis}(1963)}]{Delves:549}
\bibinfo{author}{\bibnamefont{Delves}, \bibfnamefont{R.}}, and
  \bibinfo{author}{\bibfnamefont{B.}~\bibnamefont{Lewis}},
  \bibinfo{year}{1963}, \bibinfo{journal}{J. Phys. Chem. Solids}
  \textbf{\bibinfo{volume}{28}}, \bibinfo{pages}{549}.

\bibitem[{\citenamefont{Dowben and Skomski}(2003)}]{dowben:7948}
\bibinfo{author}{\bibnamefont{Dowben}, \bibfnamefont{P.~A.}}, and
  \bibinfo{author}{\bibfnamefont{R.}~\bibnamefont{Skomski}},
  \bibinfo{year}{2003}, \bibinfo{journal}{J. Appl. Phys.}
  \textbf{\bibinfo{volume}{93}}(\bibinfo{number}{10}), \bibinfo{pages}{7948}.

\bibitem[{\citenamefont{Dowben and Skomski}(2004)}]{dowben:7453}
\bibinfo{author}{\bibnamefont{Dowben}, \bibfnamefont{P.~A.}}, and
  \bibinfo{author}{\bibfnamefont{R.}~\bibnamefont{Skomski}},
  \bibinfo{year}{2004}, \textbf{\bibinfo{volume}{95}}(\bibinfo{number}{11}),
  \bibinfo{pages}{7453}.

\bibitem[{\citenamefont{Ebert}(1996)}]{ebert:1665}
\bibinfo{author}{\bibnamefont{Ebert}, \bibfnamefont{H.}}, \bibinfo{year}{1996},
  \bibinfo{journal}{Reports on Progress in Physics}
  \textbf{\bibinfo{volume}{59}}(\bibinfo{number}{12}), \bibinfo{pages}{1665}.

\bibitem[{\citenamefont{Ebert and Schutz}(1991)}]{ebert:4627}
\bibinfo{author}{\bibnamefont{Ebert}, \bibfnamefont{H.}}, and
  \bibinfo{author}{\bibfnamefont{G.}~\bibnamefont{Schutz}},
  \bibinfo{year}{1991}, \bibinfo{journal}{J. Appl. Phys.}
  \textbf{\bibinfo{volume}{69}}(\bibinfo{number}{8}), \bibinfo{pages}{4627}.

\bibitem[{\citenamefont{Edwards}(1983)}]{Edwards:L327}
\bibinfo{author}{\bibnamefont{Edwards}, \bibfnamefont{D.~M.}},
  \bibinfo{year}{1983}, \bibinfo{journal}{J. Phys. C}
  \textbf{\bibinfo{volume}{16}}, \bibinfo{pages}{L327}.

\bibitem[{\citenamefont{Edwards}(2002)}]{Edwards:1259}
\bibinfo{author}{\bibnamefont{Edwards}, \bibfnamefont{D.~M.}},
  \bibinfo{year}{2002}, \bibinfo{journal}{Adv. Phys.}
  \textbf{\bibinfo{volume}{51}}, \bibinfo{pages}{1259}.

\bibitem[{\citenamefont{Edwards and Hertz}(1973)}]{edwards:2191}
\bibinfo{author}{\bibnamefont{Edwards}, \bibfnamefont{D.~M.}}, and
  \bibinfo{author}{\bibfnamefont{J.~A.} \bibnamefont{Hertz}},
  \bibinfo{year}{1973}, \bibinfo{journal}{Journal of Physics F-Metal Physics}
  \textbf{\bibinfo{volume}{3}}(\bibinfo{number}{12}), \bibinfo{pages}{2191}.

\bibitem[{\citenamefont{Edwards and Katsnelson}(2006)}]{Edwards:7209}
\bibinfo{author}{\bibnamefont{Edwards}, \bibfnamefont{D.~M.}}, and
  \bibinfo{author}{\bibfnamefont{M.~I.} \bibnamefont{Katsnelson}},
  \bibinfo{year}{2006}, \bibinfo{journal}{J. Phys.: Condens. Matter}
  \textbf{\bibinfo{volume}{18}}, \bibinfo{pages}{7209}.

\bibitem[{\citenamefont{Eerenstein}
  \emph{et~al.}(2002)\citenamefont{Eerenstein, Palstra, Saxena, and
  Hibma}}]{Eerenstein:247204}
\bibinfo{author}{\bibnamefont{Eerenstein}, \bibfnamefont{W.}},
  \bibinfo{author}{\bibfnamefont{T.~T.~M.} \bibnamefont{Palstra}},
  \bibinfo{author}{\bibfnamefont{S.~S.} \bibnamefont{Saxena}}, and
  \bibinfo{author}{\bibfnamefont{T.}~\bibnamefont{Hibma}},
  \bibinfo{year}{2002}, \bibinfo{journal}{Phys. Rev. Lett.}
  \textbf{\bibinfo{volume}{88}}(\bibinfo{number}{24}), \bibinfo{pages}{247204}.

\bibitem[{\citenamefont{Egorushkin}
  \emph{et~al.}(1983)\citenamefont{Egorushkin, Kulkov, and
  Kulkova}}]{Egorushkin:61}
\bibinfo{author}{\bibnamefont{Egorushkin}, \bibfnamefont{V.}},
  \bibinfo{author}{\bibfnamefont{S.}~\bibnamefont{Kulkov}}, and
  \bibinfo{author}{\bibfnamefont{S.}~\bibnamefont{Kulkova}},
  \bibinfo{year}{1983}, \bibinfo{journal}{Physica}
  \textbf{\bibinfo{volume}{123B}}, \bibinfo{pages}{61}.

\bibitem[{\citenamefont{van Engen} \emph{et~al.}(1983)\citenamefont{van Engen,
  Buschow, Jongebreur, and Erman}}]{engen:202}
\bibinfo{author}{\bibnamefont{van Engen}, \bibfnamefont{P.~G.}},
  \bibinfo{author}{\bibfnamefont{K.~H.~J.} \bibnamefont{Buschow}},
  \bibinfo{author}{\bibfnamefont{R.}~\bibnamefont{Jongebreur}}, and
  \bibinfo{author}{\bibfnamefont{M.}~\bibnamefont{Erman}},
  \bibinfo{year}{1983}, \bibinfo{journal}{Appl. Phys. Lett.}
  \textbf{\bibinfo{volume}{42}}(\bibinfo{number}{2}), \bibinfo{pages}{202}.

\bibitem[{\citenamefont{Etgens} \emph{et~al.}(2004)\citenamefont{Etgens,
  de~Camargo, Eddrief, Mattana, George, and Garreau}}]{etgens:2004}
\bibinfo{author}{\bibnamefont{Etgens}, \bibfnamefont{V.~H.}},
  \bibinfo{author}{\bibfnamefont{P.~C.} \bibnamefont{de~Camargo}},
  \bibinfo{author}{\bibfnamefont{M.}~\bibnamefont{Eddrief}},
  \bibinfo{author}{\bibfnamefont{R.}~\bibnamefont{Mattana}},
  \bibinfo{author}{\bibfnamefont{J.~M.} \bibnamefont{George}}, and
  \bibinfo{author}{\bibfnamefont{Y.}~\bibnamefont{Garreau}},
  \bibinfo{year}{2004}, \bibinfo{journal}{Phys. Rev. Lett.}
  \textbf{\bibinfo{volume}{92}}(\bibinfo{number}{16}), \bibinfo{eid}{167205}.

\bibitem[{\citenamefont{Fazekas} \emph{et~al.}(1990)\citenamefont{Fazekas,
  Menge, and M{\"u}ller-Hartmann}}]{Fazekas:69}
\bibinfo{author}{\bibnamefont{Fazekas}, \bibfnamefont{P.}},
  \bibinfo{author}{\bibfnamefont{B.}~\bibnamefont{Menge}}, and
  \bibinfo{author}{\bibfnamefont{E.}~\bibnamefont{M{\"u}ller-Hartmann}},
  \bibinfo{year}{1990}, \bibinfo{journal}{Z. Phys. B}
  \textbf{\bibinfo{volume}{78}}, \bibinfo{pages}{69}.

\bibitem[{\citenamefont{Fecher} \emph{et~al.}(2006)\citenamefont{Fecher,
  Kandpal, Wurmehl, Felser, and Schonhense}}]{fecher:08J106}
\bibinfo{author}{\bibnamefont{Fecher}, \bibfnamefont{G.~H.}},
  \bibinfo{author}{\bibfnamefont{H.~C.} \bibnamefont{Kandpal}},
  \bibinfo{author}{\bibfnamefont{S.}~\bibnamefont{Wurmehl}},
  \bibinfo{author}{\bibfnamefont{C.}~\bibnamefont{Felser}}, and
  \bibinfo{author}{\bibfnamefont{G.}~\bibnamefont{Schonhense}},
  \bibinfo{year}{2006}, \bibinfo{journal}{J. Appl. Phys.}
  \textbf{\bibinfo{volume}{99}}(\bibinfo{number}{8}), \bibinfo{eid}{08J106}.

\bibitem[{\citenamefont{Flatte and Vignale}(2001)}]{flatte:1273}
\bibinfo{author}{\bibnamefont{Flatte}, \bibfnamefont{M.~E.}}, and
  \bibinfo{author}{\bibfnamefont{G.}~\bibnamefont{Vignale}},
  \bibinfo{year}{2001}, \bibinfo{journal}{Appl. Phys. Lett.}
  \textbf{\bibinfo{volume}{78}}(\bibinfo{number}{9}), \bibinfo{pages}{1273}.

\bibitem[{\citenamefont{Folkerts} \emph{et~al.}(1987)\citenamefont{Folkerts,
  Sawatzky, Haas, de~Groot, and Hillebrecht}}]{Folkerts:4135}
\bibinfo{author}{\bibnamefont{Folkerts}, \bibfnamefont{W.}},
  \bibinfo{author}{\bibfnamefont{G.}~\bibnamefont{Sawatzky}},
  \bibinfo{author}{\bibfnamefont{C.}~\bibnamefont{Haas}},
  \bibinfo{author}{\bibfnamefont{R.}~\bibnamefont{de~Groot}}, and
  \bibinfo{author}{\bibfnamefont{F.}~\bibnamefont{Hillebrecht}},
  \bibinfo{year}{1987}, \bibinfo{journal}{Journal of Physics C - Solid State
  Physics} \textbf{\bibinfo{volume}{20}}(\bibinfo{number}{26}),
  \bibinfo{pages}{4135}.

\bibitem[{\citenamefont{Fong} \emph{et~al.}(2004)\citenamefont{Fong, Qian,
  Pask, Yang, and Dag}}]{fong:239}
\bibinfo{author}{\bibnamefont{Fong}, \bibfnamefont{C.~Y.}},
  \bibinfo{author}{\bibfnamefont{M.~C.} \bibnamefont{Qian}},
  \bibinfo{author}{\bibfnamefont{J.~E.} \bibnamefont{Pask}},
  \bibinfo{author}{\bibfnamefont{L.~H.} \bibnamefont{Yang}}, and
  \bibinfo{author}{\bibfnamefont{S.}~\bibnamefont{Dag}}, \bibinfo{year}{2004},
  \bibinfo{journal}{Appl. Phys. Lett.}
  \textbf{\bibinfo{volume}{84}}(\bibinfo{number}{2}), \bibinfo{pages}{239}.

\bibitem[{\citenamefont{Fujii} \emph{et~al.}(1995)\citenamefont{Fujii, Ishida,
  and Asano}}]{fu.is.85}
\bibinfo{author}{\bibnamefont{Fujii}, \bibfnamefont{S.}},
  \bibinfo{author}{\bibfnamefont{S.}~\bibnamefont{Ishida}}, and
  \bibinfo{author}{\bibfnamefont{S.}~\bibnamefont{Asano}},
  \bibinfo{year}{1995}, \bibinfo{journal}{J. Phys. Soc. Jpn.}
  \textbf{\bibinfo{volume}{64}}, \bibinfo{pages}{184}.

\bibitem[{\citenamefont{Fukuda} \emph{et~al.}(1994)\citenamefont{Fukuda,
  Kotani, Suzuki, and Yokojima}}]{Fukuda:865}
\bibinfo{author}{\bibnamefont{Fukuda}, \bibfnamefont{R.}},
  \bibinfo{author}{\bibfnamefont{T.}~\bibnamefont{Kotani}},
  \bibinfo{author}{\bibfnamefont{Y.}~\bibnamefont{Suzuki}}, and
  \bibinfo{author}{\bibfnamefont{S.}~\bibnamefont{Yokojima}},
  \bibinfo{year}{1994}, \bibinfo{journal}{Prog. Theor. Phys.}
  \textbf{\bibinfo{volume}{92}}, \bibinfo{pages}{833}.

\bibitem[{\citenamefont{Furukawa}(2000)}]{Furukawa:1954}
\bibinfo{author}{\bibnamefont{Furukawa}, \bibfnamefont{N.}},
  \bibinfo{year}{2000}, \bibinfo{journal}{J. Phys. Soc. Jpn.}
  \textbf{\bibinfo{volume}{69}}, \bibinfo{pages}{1954}.

\bibitem[{\citenamefont{Galanakis}(2003)}]{Galanakis:6329}
\bibinfo{author}{\bibnamefont{Galanakis}, \bibfnamefont{I.}},
  \bibinfo{year}{2003}, \bibinfo{journal}{J. Phys.: Condens. Matter}
  \textbf{\bibinfo{volume}{14}}, \bibinfo{pages}{6329}.

\bibitem[{\citenamefont{Galanakis}(2004)}]{Galanakis:3089}
\bibinfo{author}{\bibnamefont{Galanakis}, \bibfnamefont{I.}},
  \bibinfo{year}{2004}, \bibinfo{journal}{J. Phys.: Condens. Matter}
  \textbf{\bibinfo{volume}{16}}, \bibinfo{pages}{3089}.

\bibitem[{\citenamefont{Galanakis}
  \emph{et~al.}(2002{\natexlab{a}})\citenamefont{Galanakis, Dederichs, and
  Papanikolaou}}]{galanakis:134428}
\bibinfo{author}{\bibnamefont{Galanakis}, \bibfnamefont{I.}},
  \bibinfo{author}{\bibfnamefont{P.~H.} \bibnamefont{Dederichs}}, and
  \bibinfo{author}{\bibfnamefont{N.}~\bibnamefont{Papanikolaou}},
  \bibinfo{year}{2002}{\natexlab{a}}, \bibinfo{journal}{Phys. Rev. B}
  \textbf{\bibinfo{volume}{66}}(\bibinfo{number}{13}), \bibinfo{pages}{134428}.

\bibitem[{\citenamefont{Galanakis}
  \emph{et~al.}(2002{\natexlab{b}})\citenamefont{Galanakis, Dederichs, and
  Papanikolaou}}]{ga.de.02}
\bibinfo{author}{\bibnamefont{Galanakis}, \bibfnamefont{I.}},
  \bibinfo{author}{\bibfnamefont{P.~H.} \bibnamefont{Dederichs}}, and
  \bibinfo{author}{\bibfnamefont{N.}~\bibnamefont{Papanikolaou}},
  \bibinfo{year}{2002}{\natexlab{b}}, \bibinfo{journal}{Phys. Rev. B}
  \textbf{\bibinfo{volume}{66}}(\bibinfo{number}{17}), \bibinfo{pages}{174429}.

\bibitem[{\citenamefont{Galanakis and Dederichs(eds.)}(2005)}]{Dederichs2005}
\bibinfo{author}{\bibnamefont{Galanakis}, \bibfnamefont{I.}}, and
  \bibinfo{author}{\bibfnamefont{P.~H.} \bibnamefont{Dederichs(eds.)}},
  \bibinfo{year}{2005}, \emph{\bibinfo{title}{Lecture Notes in Physics}}
  (\bibinfo{publisher}{Springer, Berlin Heidelberg}).

\bibitem[{\citenamefont{Galanakis and Mavropoulos}(2003)}]{galanakis:104417}
\bibinfo{author}{\bibnamefont{Galanakis}, \bibfnamefont{I.}}, and
  \bibinfo{author}{\bibfnamefont{P.}~\bibnamefont{Mavropoulos}},
  \bibinfo{year}{2003}, \bibinfo{journal}{Phys. Rev. B}
  \textbf{\bibinfo{volume}{67}}(\bibinfo{number}{10}), \bibinfo{eid}{104417}.

\bibitem[{\citenamefont{Galanakis and Mavropoulos}(2007)}]{Galanakis:315213}
\bibinfo{author}{\bibnamefont{Galanakis}, \bibfnamefont{I.}}, and
  \bibinfo{author}{\bibfnamefont{P.}~\bibnamefont{Mavropoulos}},
  \bibinfo{year}{2007}, \bibinfo{journal}{J. Phys.: Condens. Matter}
  \textbf{\bibinfo{volume}{19}}, \bibinfo{pages}{315213}.

\bibitem[{\citenamefont{Galanakis} \emph{et~al.}(2006)\citenamefont{Galanakis,
  Mavropoulos, and Dederichs}}]{Galanakis:765}
\bibinfo{author}{\bibnamefont{Galanakis}, \bibfnamefont{I.}},
  \bibinfo{author}{\bibfnamefont{P.}~\bibnamefont{Mavropoulos}}, and
  \bibinfo{author}{\bibfnamefont{P.~H.} \bibnamefont{Dederichs}},
  \bibinfo{year}{2006}, \bibinfo{journal}{J. Phys. D: Appl. Phys.}
  \textbf{\bibinfo{volume}{39}}, \bibinfo{pages}{765}.

\bibitem[{\citenamefont{Galitski}(1958)}]{galitski:1011}
\bibinfo{author}{\bibnamefont{Galitski}, \bibfnamefont{V.~M.}},
  \bibinfo{year}{1958}, \bibinfo{journal}{Zh. Eksper. Teor. Fiz.}
  \textbf{\bibinfo{volume}{34}}, \bibinfo{pages}{1011}.

\bibitem[{\citenamefont{Garsia and Sabias}(2004)}]{Garsia:R145}
\bibinfo{author}{\bibnamefont{Garsia}, \bibfnamefont{J.}}, and
  \bibinfo{author}{\bibfnamefont{G.}~\bibnamefont{Sabias}},
  \bibinfo{year}{2004}, \bibinfo{journal}{J. Phys.: Condens. Matter}
  \textbf{\bibinfo{volume}{16}}, \bibinfo{pages}{R145}.

\bibitem[{\citenamefont{Georges} \emph{et~al.}(1996)\citenamefont{Georges,
  Kotliar, Krauth, and Rozenberg}}]{georges:13}
\bibinfo{author}{\bibnamefont{Georges}, \bibfnamefont{A.}},
  \bibinfo{author}{\bibfnamefont{G.}~\bibnamefont{Kotliar}},
  \bibinfo{author}{\bibfnamefont{W.}~\bibnamefont{Krauth}}, and
  \bibinfo{author}{\bibfnamefont{M.~J.} \bibnamefont{Rozenberg}},
  \bibinfo{year}{1996}, \bibinfo{journal}{Rev. Mod. Phys.}
  \textbf{\bibinfo{volume}{68}}(\bibinfo{number}{1}), \bibinfo{pages}{13}.

\bibitem[{\citenamefont{Gercsi} \emph{et~al.}(2006)\citenamefont{Gercsi,
  Rajanikanth, Takahashi, Hono, Kikuchi, Tezuka, and Inomata}}]{gercsi:082512}
\bibinfo{author}{\bibnamefont{Gercsi}, \bibfnamefont{Z.}},
  \bibinfo{author}{\bibfnamefont{A.}~\bibnamefont{Rajanikanth}},
  \bibinfo{author}{\bibfnamefont{Y.~K.} \bibnamefont{Takahashi}},
  \bibinfo{author}{\bibfnamefont{K.}~\bibnamefont{Hono}},
  \bibinfo{author}{\bibfnamefont{M.}~\bibnamefont{Kikuchi}},
  \bibinfo{author}{\bibfnamefont{N.}~\bibnamefont{Tezuka}}, and
  \bibinfo{author}{\bibfnamefont{K.}~\bibnamefont{Inomata}},
  \bibinfo{year}{2006}, \bibinfo{journal}{Appl. Phys. Lett.}
  \textbf{\bibinfo{volume}{89}}(\bibinfo{number}{8}), \bibinfo{eid}{082512}.

\bibitem[{\citenamefont{Goering} \emph{et~al.}(2006)\citenamefont{Goering,
  Gold, Lafkioti, and Sch{\"u}tz}}]{Goering:97}
\bibinfo{author}{\bibnamefont{Goering}, \bibfnamefont{E.}},
  \bibinfo{author}{\bibfnamefont{S.}~\bibnamefont{Gold}},
  \bibinfo{author}{\bibfnamefont{M.}~\bibnamefont{Lafkioti}}, and
  \bibinfo{author}{\bibfnamefont{G.}~\bibnamefont{Sch{\"u}tz}},
  \bibinfo{year}{2006}, \bibinfo{journal}{Europhys. Lett.}
  \textbf{\bibinfo{volume}{73}}(\bibinfo{number}{1}), \bibinfo{pages}{97}.

\bibitem[{\citenamefont{Golosov}(2000)}]{Golosov:3974}
\bibinfo{author}{\bibnamefont{Golosov}, \bibfnamefont{D.~I.}},
  \bibinfo{year}{2000}, \bibinfo{journal}{Phys. Rev. Lett.}
  \textbf{\bibinfo{volume}{84}}(\bibinfo{number}{17}), \bibinfo{pages}{3974}.

\bibitem[{\citenamefont{Grigin and Nagaev}(1974)}]{Grigin:65}
\bibinfo{author}{\bibnamefont{Grigin}, \bibfnamefont{A.~P.}}, and
  \bibinfo{author}{\bibfnamefont{E.~L.} \bibnamefont{Nagaev}},
  \bibinfo{year}{1974}, \bibinfo{journal}{Phys. Stat. Sol. (b)}
  \textbf{\bibinfo{volume}{61}}, \bibinfo{pages}{65}.

\bibitem[{\citenamefont{de~Groot} \emph{et~al.}(1990)\citenamefont{de~Groot,
  Fuggle, Thole, and Sawatzky}}]{deGroot:5459}
\bibinfo{author}{\bibnamefont{de~Groot}, \bibfnamefont{F.~M.~F.}},
  \bibinfo{author}{\bibfnamefont{J.~C.} \bibnamefont{Fuggle}},
  \bibinfo{author}{\bibfnamefont{B.~T.} \bibnamefont{Thole}}, and
  \bibinfo{author}{\bibfnamefont{G.~A.} \bibnamefont{Sawatzky}},
  \bibinfo{year}{1990}, \bibinfo{journal}{Phys. Rev. B}
  \textbf{\bibinfo{volume}{42}}(\bibinfo{number}{9}), \bibinfo{pages}{5459}.

\bibitem[{\citenamefont{de~Groot}(1991)}]{deGroot:45}
\bibinfo{author}{\bibnamefont{de~Groot}, \bibfnamefont{R.~A.}},
  \bibinfo{year}{1991}, \bibinfo{journal}{Physica B}
  \textbf{\bibinfo{volume}{172}}(\bibinfo{number}{1-2}), \bibinfo{pages}{45}.

\bibitem[{\citenamefont{de~Groot}
  \emph{et~al.}(1983{\natexlab{a}})\citenamefont{de~Groot, Janner, and
  Mueller}}]{Rob83}
\bibinfo{author}{\bibnamefont{de~Groot}, \bibfnamefont{R.~A.}},
  \bibinfo{author}{\bibfnamefont{A.~G.~M.} \bibnamefont{Janner}}, and
  \bibinfo{author}{\bibfnamefont{F.~M.} \bibnamefont{Mueller}},
  \bibinfo{year}{1983}{\natexlab{a}}, \bibinfo{journal}{patents NL 19830000602,
  EP 198402000215} .

\bibitem[{\citenamefont{de~Groot} \emph{et~al.}(1986)\citenamefont{de~Groot,
  van~der Kraan, and Buschow}}]{deGroot:330}
\bibinfo{author}{\bibnamefont{de~Groot}, \bibfnamefont{R.~A.}},
  \bibinfo{author}{\bibfnamefont{A.~M.} \bibnamefont{van~der Kraan}}, and
  \bibinfo{author}{\bibfnamefont{K.~H.~J.} \bibnamefont{Buschow}},
  \bibinfo{year}{1986}, \bibinfo{journal}{J. Magn. Magn. Mater.}
  \textbf{\bibinfo{volume}{61}}, \bibinfo{pages}{330}.

\bibitem[{\citenamefont{de~Groot}
  \emph{et~al.}(1983{\natexlab{b}})\citenamefont{de~Groot, Mueller, Engen, and
  Buschow}}]{deGroot:2024}
\bibinfo{author}{\bibnamefont{de~Groot}, \bibfnamefont{R.~A.}},
  \bibinfo{author}{\bibfnamefont{F.~M.} \bibnamefont{Mueller}},
  \bibinfo{author}{\bibfnamefont{P.~G.~v.} \bibnamefont{Engen}}, and
  \bibinfo{author}{\bibfnamefont{K.~H.~J.} \bibnamefont{Buschow}},
  \bibinfo{year}{1983}{\natexlab{b}}, \bibinfo{journal}{Phys. Rev. Lett.}
  \textbf{\bibinfo{volume}{50}}(\bibinfo{number}{25}), \bibinfo{pages}{2024}.

\bibitem[{\citenamefont{Gunnarsson}
  \emph{et~al.}(1989)\citenamefont{Gunnarsson, Andersen, Jepsen, and
  Zaanen}}]{gunnarsson:1708}
\bibinfo{author}{\bibnamefont{Gunnarsson}, \bibfnamefont{O.}},
  \bibinfo{author}{\bibfnamefont{O.~K.} \bibnamefont{Andersen}},
  \bibinfo{author}{\bibfnamefont{O.}~\bibnamefont{Jepsen}}, and
  \bibinfo{author}{\bibfnamefont{J.}~\bibnamefont{Zaanen}},
  \bibinfo{year}{1989}, \bibinfo{journal}{Phys. Rev. B}
  \textbf{\bibinfo{volume}{39}}(\bibinfo{number}{3}), \bibinfo{pages}{1708}.

\bibitem[{\citenamefont{Gyorffy} \emph{et~al.}(1985)\citenamefont{Gyorffy,
  Pindor, Staunton, Stocks, and Winter}}]{gy.pi.85}
\bibinfo{author}{\bibnamefont{Gyorffy}, \bibfnamefont{B.~L.}},
  \bibinfo{author}{\bibfnamefont{A.~J.} \bibnamefont{Pindor}},
  \bibinfo{author}{\bibfnamefont{J.}~\bibnamefont{Staunton}},
  \bibinfo{author}{\bibfnamefont{G.~M.} \bibnamefont{Stocks}}, and
  \bibinfo{author}{\bibfnamefont{H.}~\bibnamefont{Winter}},
  \bibinfo{year}{1985}, \bibinfo{journal}{J. Phys. F: Met. Phys}
  \textbf{\bibinfo{volume}{15}}, \bibinfo{pages}{1337}.

\bibitem[{\citenamefont{Halilov} \emph{et~al.}(1998)\citenamefont{Halilov,
  Eschrig, Perlov, and Oppeneer}}]{halilov:293}
\bibinfo{author}{\bibnamefont{Halilov}, \bibfnamefont{S.~V.}},
  \bibinfo{author}{\bibfnamefont{H.}~\bibnamefont{Eschrig}},
  \bibinfo{author}{\bibfnamefont{A.~Y.} \bibnamefont{Perlov}}, and
  \bibinfo{author}{\bibfnamefont{P.~M.} \bibnamefont{Oppeneer}},
  \bibinfo{year}{1998}, \bibinfo{journal}{Phys. Rev. B}
  \textbf{\bibinfo{volume}{58}}(\bibinfo{number}{1}), \bibinfo{pages}{293}.

\bibitem[{\citenamefont{Hanssen and Mijnarends}(1986)}]{Hanssen:5009}
\bibinfo{author}{\bibnamefont{Hanssen}, \bibfnamefont{K.~E. H.~M.}}, and
  \bibinfo{author}{\bibfnamefont{P.~E.} \bibnamefont{Mijnarends}},
  \bibinfo{year}{1986}, \bibinfo{journal}{Phys. Rev. B}
  \textbf{\bibinfo{volume}{34}}(\bibinfo{number}{8}), \bibinfo{pages}{5009}.

\bibitem[{\citenamefont{Hanssen} \emph{et~al.}(1990)\citenamefont{Hanssen,
  Mijnarends, Rabou, and Buschow}}]{Hanssen:1533}
\bibinfo{author}{\bibnamefont{Hanssen}, \bibfnamefont{K.~E. H.~M.}},
  \bibinfo{author}{\bibfnamefont{P.~E.} \bibnamefont{Mijnarends}},
  \bibinfo{author}{\bibfnamefont{L.~P. L.~M.} \bibnamefont{Rabou}}, and
  \bibinfo{author}{\bibfnamefont{K.~H.~J.} \bibnamefont{Buschow}},
  \bibinfo{year}{1990}, \bibinfo{journal}{Phys. Rev. B}
  \textbf{\bibinfo{volume}{42}}(\bibinfo{number}{3}), \bibinfo{pages}{1533}.

\bibitem[{\citenamefont{Harland and Woolley}(1966)}]{harland:2715}
\bibinfo{author}{\bibnamefont{Harland}, \bibfnamefont{H.}}, and
  \bibinfo{author}{\bibfnamefont{J.~C.} \bibnamefont{Woolley}},
  \bibinfo{year}{1966}, \bibinfo{journal}{Can. J. of Phys.}
  \textbf{\bibinfo{volume}{44}}, \bibinfo{pages}{2715}.

\bibitem[{\citenamefont{Harris and Jones}(1974)}]{Harris:1170}
\bibinfo{author}{\bibnamefont{Harris}, \bibfnamefont{J.}}, and
  \bibinfo{author}{\bibfnamefont{R.}~\bibnamefont{Jones}},
  \bibinfo{year}{1974}, \bibinfo{journal}{Journal of Physics F - Metal Physics}
  \textbf{\bibinfo{volume}{4}}(\bibinfo{number}{8}), \bibinfo{pages}{1170}.

\bibitem[{\citenamefont{Hartman-Boutron}(1965)}]{Hartman:114}
\bibinfo{author}{\bibnamefont{Hartman-Boutron}, \bibfnamefont{F.}},
  \bibinfo{year}{1965}, \bibinfo{journal}{Phys. Kond. Mat.}
  \textbf{\bibinfo{volume}{4}}, \bibinfo{pages}{114}.

\bibitem[{\citenamefont{Hashemifar}
  \emph{et~al.}(2005)\citenamefont{Hashemifar, Kratzer, and
  Scheffler}}]{hashemifar:096402}
\bibinfo{author}{\bibnamefont{Hashemifar}, \bibfnamefont{S.~J.}},
  \bibinfo{author}{\bibfnamefont{P.}~\bibnamefont{Kratzer}}, and
  \bibinfo{author}{\bibfnamefont{M.}~\bibnamefont{Scheffler}},
  \bibinfo{year}{2005}, \bibinfo{journal}{Phys. Rev. Lett.}
  \textbf{\bibinfo{volume}{94}}(\bibinfo{number}{9}), \bibinfo{eid}{096402}.

\bibitem[{\citenamefont{Hedin}(1965)}]{Hedin:A796}
\bibinfo{author}{\bibnamefont{Hedin}, \bibfnamefont{L.}}, \bibinfo{year}{1965},
  \bibinfo{journal}{Phys. Rev.}
  \textbf{\bibinfo{volume}{139}}(\bibinfo{number}{3A}), \bibinfo{pages}{A796}.

\bibitem[{\citenamefont{Heinze} \emph{et~al.}(2000)\citenamefont{Heinze, Bode,
  Kubetzka, Pietzsch, Nie, Bl\"{u}gel, and Wiesendanger}}]{Heinze:1805}
\bibinfo{author}{\bibnamefont{Heinze}, \bibfnamefont{S.}},
  \bibinfo{author}{\bibfnamefont{M.}~\bibnamefont{Bode}},
  \bibinfo{author}{\bibfnamefont{A.}~\bibnamefont{Kubetzka}},
  \bibinfo{author}{\bibfnamefont{O.}~\bibnamefont{Pietzsch}},
  \bibinfo{author}{\bibfnamefont{X.}~\bibnamefont{Nie}},
  \bibinfo{author}{\bibfnamefont{S.}~\bibnamefont{Bl\"{u}gel}}, and
  \bibinfo{author}{\bibfnamefont{R.}~\bibnamefont{Wiesendanger}},
  \bibinfo{year}{2000}, \bibinfo{journal}{Science}
  \textbf{\bibinfo{volume}{288}}, \bibinfo{pages}{1805}.

\bibitem[{\citenamefont{Helmholdt} \emph{et~al.}(1984)\citenamefont{Helmholdt,
  de~Groot, van Engen, and Buschow}}]{Helmholdt:249}
\bibinfo{author}{\bibnamefont{Helmholdt}, \bibfnamefont{R.~B.}},
  \bibinfo{author}{\bibfnamefont{R.~A.} \bibnamefont{de~Groot}},
  \bibinfo{author}{\bibfnamefont{F.~M. M.~G.} \bibnamefont{van Engen}}, and
  \bibinfo{author}{\bibfnamefont{K.~H.~J.} \bibnamefont{Buschow}},
  \bibinfo{year}{1984}, \bibinfo{journal}{J. Magn. Magn. Mater.}
  \textbf{\bibinfo{volume}{43}}, \bibinfo{pages}{249}.

\bibitem[{\citenamefont{Herring}(1966)}]{Herring:1966}
\bibinfo{author}{\bibnamefont{Herring}, \bibfnamefont{C.}},
  \bibinfo{year}{1966}, \emph{\bibinfo{title}{Magnetism, vol. 4}}
  (\bibinfo{publisher}{New York, Academic Press}).

\bibitem[{\citenamefont{Heusler}(1903)}]{heus.03}
\bibinfo{author}{\bibnamefont{Heusler}, \bibfnamefont{F.}},
  \bibinfo{year}{1903}, \bibinfo{journal}{Verhandlungen der Deutschen
  Physikalischen Gesellschaft} \textbf{\bibinfo{volume}{5}},
  \bibinfo{pages}{219}.

\bibitem[{\citenamefont{Hewson}(1993)}]{Hewson:1993}
\bibinfo{author}{\bibnamefont{Hewson}, \bibfnamefont{A.~C.}},
  \bibinfo{year}{1993}, \emph{\bibinfo{title}{The Kondo Problem to Heavy
  Fermions}} (\bibinfo{publisher}{Cambridge University Press, Cambridge}).

\bibitem[{\citenamefont{Hirsch}(1983)}]{hirsch:4059}
\bibinfo{author}{\bibnamefont{Hirsch}, \bibfnamefont{J.~E.}},
  \bibinfo{year}{1983}, \bibinfo{journal}{Phys. Rev. B}
  \textbf{\bibinfo{volume}{28}}(\bibinfo{number}{7}), \bibinfo{pages}{4059}.

\bibitem[{\citenamefont{Hohenberg and Kohn}(1964)}]{Hohenberg:B864}
\bibinfo{author}{\bibnamefont{Hohenberg}, \bibfnamefont{P.}}, and
  \bibinfo{author}{\bibfnamefont{W.}~\bibnamefont{Kohn}}, \bibinfo{year}{1964},
  \bibinfo{journal}{Phys. Rev.}
  \textbf{\bibinfo{volume}{136}}(\bibinfo{number}{3B}), \bibinfo{pages}{B864}.

\bibitem[{\citenamefont{Hordequin}
  \emph{et~al.}(1997{\natexlab{a}})\citenamefont{Hordequin, Lelièvre-Bernab,
  and Pierre}}]{Hordequin:602}
\bibinfo{author}{\bibnamefont{Hordequin}, \bibfnamefont{C.}},
  \bibinfo{author}{\bibfnamefont{E.}~\bibnamefont{Lelièvre-Bernab}}, and
  \bibinfo{author}{\bibfnamefont{J.}~\bibnamefont{Pierre}},
  \bibinfo{year}{1997}{\natexlab{a}}, \bibinfo{journal}{Physica B}
  \textbf{\bibinfo{volume}{234-236}}, \bibinfo{pages}{602}.

\bibitem[{\citenamefont{Hordequin} \emph{et~al.}(1998)\citenamefont{Hordequin,
  Nozieres, and Pierre}}]{Hordequin:225}
\bibinfo{author}{\bibnamefont{Hordequin}, \bibfnamefont{C.}},
  \bibinfo{author}{\bibfnamefont{J.~P.} \bibnamefont{Nozieres}}, and
  \bibinfo{author}{\bibfnamefont{J.}~\bibnamefont{Pierre}},
  \bibinfo{year}{1998}, \bibinfo{journal}{J. Magn. Magn. Mater.}
  \textbf{\bibinfo{volume}{183}}, \bibinfo{pages}{225}.

\bibitem[{\citenamefont{Hordequin}
  \emph{et~al.}(1997{\natexlab{b}})\citenamefont{Hordequin, Pierre, and
  Currat}}]{Hordequin:605}
\bibinfo{author}{\bibnamefont{Hordequin}, \bibfnamefont{C.}},
  \bibinfo{author}{\bibfnamefont{J.}~\bibnamefont{Pierre}}, and
  \bibinfo{author}{\bibfnamefont{R.}~\bibnamefont{Currat}},
  \bibinfo{year}{1997}{\natexlab{b}}, \bibinfo{journal}{Physica B}
  \textbf{\bibinfo{volume}{234-236}}, \bibinfo{pages}{605}.

\bibitem[{\citenamefont{Hordequin} \emph{et~al.}(2000)\citenamefont{Hordequin,
  Ristoiu, Ranno, and Pierre}}]{Hordequin:287}
\bibinfo{author}{\bibnamefont{Hordequin}, \bibfnamefont{C.}},
  \bibinfo{author}{\bibfnamefont{D.}~\bibnamefont{Ristoiu}},
  \bibinfo{author}{\bibfnamefont{L.}~\bibnamefont{Ranno}}, and
  \bibinfo{author}{\bibfnamefont{J.}~\bibnamefont{Pierre}},
  \bibinfo{year}{2000}, \bibinfo{journal}{Eur. Phys. J. B}
  \textbf{\bibinfo{volume}{16}}(\bibinfo{number}{2}), \bibinfo{pages}{287}.

\bibitem[{\citenamefont{Hoshino} \emph{et~al.}(1995)\citenamefont{Hoshino,
  Kurikawa, Takeda, Nakajima, and Kaya}}]{Hoshino:3053}
\bibinfo{author}{\bibnamefont{Hoshino}, \bibfnamefont{K.}},
  \bibinfo{author}{\bibfnamefont{T.}~\bibnamefont{Kurikawa}},
  \bibinfo{author}{\bibfnamefont{H.}~\bibnamefont{Takeda}},
  \bibinfo{author}{\bibfnamefont{A.}~\bibnamefont{Nakajima}}, and
  \bibinfo{author}{\bibfnamefont{K.}~\bibnamefont{Kaya}}, \bibinfo{year}{1995},
  \bibinfo{journal}{J. Phys.Chem.}
  \textbf{\bibinfo{volume}{99}}(\bibinfo{number}{10}), \bibinfo{pages}{3053}.

\bibitem[{\citenamefont{Huang} \emph{et~al.}(2004)\citenamefont{Huang, Chang,
  Jeng, Guo, Lin, Wu, Ku, Fujimori, Takahashi, and Chen}}]{huang:077204}
\bibinfo{author}{\bibnamefont{Huang}, \bibfnamefont{D.~J.}},
  \bibinfo{author}{\bibfnamefont{C.~F.} \bibnamefont{Chang}},
  \bibinfo{author}{\bibfnamefont{H.~T.} \bibnamefont{Jeng}},
  \bibinfo{author}{\bibfnamefont{G.~Y.} \bibnamefont{Guo}},
  \bibinfo{author}{\bibfnamefont{H.~J.} \bibnamefont{Lin}},
  \bibinfo{author}{\bibfnamefont{W.~B.} \bibnamefont{Wu}},
  \bibinfo{author}{\bibfnamefont{H.~C.} \bibnamefont{Ku}},
  \bibinfo{author}{\bibfnamefont{A.}~\bibnamefont{Fujimori}},
  \bibinfo{author}{\bibfnamefont{Y.}~\bibnamefont{Takahashi}}, and
  \bibinfo{author}{\bibfnamefont{C.~T.} \bibnamefont{Chen}},
  \bibinfo{year}{2004}, \bibinfo{journal}{Phys. Rev. Lett.}
  \textbf{\bibinfo{volume}{93}}(\bibinfo{number}{7}), \bibinfo{eid}{077204}.

\bibitem[{\citenamefont{Huang} \emph{et~al.}(2003)\citenamefont{Huang, Tjeng,
  Chen, Chang, Wu, Chung, Tanaka, Guo, Lin, Shyu, Wu, and Chen}}]{huang:214419}
\bibinfo{author}{\bibnamefont{Huang}, \bibfnamefont{D.~J.}},
  \bibinfo{author}{\bibfnamefont{L.~H.} \bibnamefont{Tjeng}},
  \bibinfo{author}{\bibfnamefont{J.}~\bibnamefont{Chen}},
  \bibinfo{author}{\bibfnamefont{C.~F.} \bibnamefont{Chang}},
  \bibinfo{author}{\bibfnamefont{W.~P.} \bibnamefont{Wu}},
  \bibinfo{author}{\bibfnamefont{S.~C.} \bibnamefont{Chung}},
  \bibinfo{author}{\bibfnamefont{A.}~\bibnamefont{Tanaka}},
  \bibinfo{author}{\bibfnamefont{G.~Y.} \bibnamefont{Guo}},
  \bibinfo{author}{\bibfnamefont{H.~J.} \bibnamefont{Lin}},
  \bibinfo{author}{\bibfnamefont{S.~G.} \bibnamefont{Shyu}},
  \bibinfo{author}{\bibfnamefont{C.~C.} \bibnamefont{Wu}}, and
  \bibinfo{author}{\bibfnamefont{C.~T.} \bibnamefont{Chen}},
  \bibinfo{year}{2003}, \bibinfo{journal}{Phys. Rev. B}
  \textbf{\bibinfo{volume}{67}}(\bibinfo{number}{21}), \bibinfo{eid}{214419}.

\bibitem[{\citenamefont{Hubbard}(1963)}]{Hubbard:238}
\bibinfo{author}{\bibnamefont{Hubbard}, \bibfnamefont{J.}},
  \bibinfo{year}{1963}, \bibinfo{journal}{Proc. Roy. Soc.}
  \textbf{\bibinfo{volume}{A276}}, \bibinfo{pages}{238}.

\bibitem[{\citenamefont{Hybertsen} \emph{et~al.}(1989)\citenamefont{Hybertsen,
  Schl{\"u}ter, and Christensen}}]{hybertsen:9028}
\bibinfo{author}{\bibnamefont{Hybertsen}, \bibfnamefont{M.~S.}},
  \bibinfo{author}{\bibfnamefont{M.}~\bibnamefont{Schl{\"u}ter}}, and
  \bibinfo{author}{\bibfnamefont{N.~E.} \bibnamefont{Christensen}},
  \bibinfo{year}{1989}, \bibinfo{journal}{Phys. Rev. B}
  \textbf{\bibinfo{volume}{39}}(\bibinfo{number}{13}), \bibinfo{pages}{9028}.

\bibitem[{\citenamefont{Irkhin}(1987)}]{irkh.41}
\bibinfo{author}{\bibnamefont{Irkhin}, \bibfnamefont{V.~Y.}},
  \bibinfo{year}{1987}, \bibinfo{journal}{Fiz. Metalov Metalloved.}
  \textbf{\bibinfo{volume}{64}}, \bibinfo{pages}{260}.

\bibitem[{\citenamefont{Irkhin and Irkhin}(1994)}]{Irkhin:9}
\bibinfo{author}{\bibnamefont{Irkhin}, \bibfnamefont{V.~Y.}}, and
  \bibinfo{author}{\bibfnamefont{Y.~P.} \bibnamefont{Irkhin}},
  \bibinfo{year}{1994}, \bibinfo{journal}{Phys. Stat. Sol.(b)}
  \textbf{\bibinfo{volume}{183}}, \bibinfo{pages}{9}.

\bibitem[{\citenamefont{Irkhin and Irkhin}(2007)}]{Irkhin:2007}
\bibinfo{author}{\bibnamefont{Irkhin}, \bibfnamefont{V.~Y.}}, and
  \bibinfo{author}{\bibfnamefont{Y.~P.} \bibnamefont{Irkhin}},
  \bibinfo{year}{2007}, \emph{\bibinfo{title}{Electronic Structure, Correlation
  Effects and Physical Properties of d- and f-Metals and their Compound}}
  (\bibinfo{publisher}{Cambridge International Science Publishing}),
  \bibinfo{note}{cond-mat/9812072}.

\bibitem[{\citenamefont{Irkhin and Katsnelson}(1983)}]{Irkhin:1947}
\bibinfo{author}{\bibnamefont{Irkhin}, \bibfnamefont{V.~Y.}}, and
  \bibinfo{author}{\bibfnamefont{M.~I.} \bibnamefont{Katsnelson}},
  \bibinfo{year}{1983}, \bibinfo{journal}{Sov. Phys. - Solid State}
  \textbf{\bibinfo{volume}{25}}, \bibinfo{pages}{1947}.

\bibitem[{\citenamefont{Irkhin and Katsnelson}(1984)}]{Irkhin:3055}
\bibinfo{author}{\bibnamefont{Irkhin}, \bibfnamefont{V.~Y.}}, and
  \bibinfo{author}{\bibfnamefont{M.~I.} \bibnamefont{Katsnelson}},
  \bibinfo{year}{1984}, \bibinfo{journal}{Fiz. Tverd. Tela}
  \textbf{\bibinfo{volume}{26}}, \bibinfo{pages}{3055}.

\bibitem[{\citenamefont{Irkhin and
  Katsnelson}(1985{\natexlab{a}})}]{Irkhin:522}
\bibinfo{author}{\bibnamefont{Irkhin}, \bibfnamefont{V.~Y.}}, and
  \bibinfo{author}{\bibfnamefont{M.~I.} \bibnamefont{Katsnelson}},
  \bibinfo{year}{1985}{\natexlab{a}}, \bibinfo{journal}{Zh. Eksp. Teor. Fiz.}
  \textbf{\bibinfo{volume}{88}}, \bibinfo{pages}{522}, \bibinfo{note}{[Engl.
  Transl.: Sov. Phys. JETP \textbf{61}, 306 (1985)]}.

\bibitem[{\citenamefont{Irkhin and
  Katsnelson}(1985{\natexlab{b}})}]{Irkhin:4173}
\bibinfo{author}{\bibnamefont{Irkhin}, \bibfnamefont{V.~Y.}}, and
  \bibinfo{author}{\bibfnamefont{M.~I.} \bibnamefont{Katsnelson}},
  \bibinfo{year}{1985}{\natexlab{b}}, \bibinfo{journal}{J. Phys. C}
  \textbf{\bibinfo{volume}{18}}, \bibinfo{pages}{4173}.

\bibitem[{\citenamefont{Irkhin and Katsnelson}(1988)}]{Irkhin:41}
\bibinfo{author}{\bibnamefont{Irkhin}, \bibfnamefont{V.~Y.}}, and
  \bibinfo{author}{\bibfnamefont{M.~I.} \bibnamefont{Katsnelson}},
  \bibinfo{year}{1988}, \bibinfo{journal}{Fiz. Metalov Metalloved.}
  \textbf{\bibinfo{volume}{66}}, \bibinfo{pages}{41}.

\bibitem[{\citenamefont{Irkhin and Katsnelson}(1990)}]{irkhin:7151}
\bibinfo{author}{\bibnamefont{Irkhin}, \bibfnamefont{V.~Y.}}, and
  \bibinfo{author}{\bibfnamefont{M.~I.} \bibnamefont{Katsnelson}},
  \bibinfo{year}{1990}, \bibinfo{journal}{J. Phys.: Condens. Matter}
  \textbf{\bibinfo{volume}{2}}, \bibinfo{pages}{7151}.

\bibitem[{\citenamefont{Irkhin and Katsnelson}(1994)}]{Irkhin:705}
\bibinfo{author}{\bibnamefont{Irkhin}, \bibfnamefont{V.~Y.}}, and
  \bibinfo{author}{\bibfnamefont{M.~I.} \bibnamefont{Katsnelson}},
  \bibinfo{year}{1994}, \bibinfo{journal}{Uspekhi Fiz. Nauk Phys. Usp.}
  \textbf{\bibinfo{volume}{164}}, \bibinfo{pages}{705}, \bibinfo{note}{[Engl.
  Transl: Physics - Uspekhi 37, 659]}.

\bibitem[{\citenamefont{Irkhin and Katsnelson}(1996)}]{Irkhin:14008}
\bibinfo{author}{\bibnamefont{Irkhin}, \bibfnamefont{V.~Y.}}, and
  \bibinfo{author}{\bibfnamefont{M.~I.} \bibnamefont{Katsnelson}},
  \bibinfo{year}{1996}, \bibinfo{journal}{Phys. Rev. B}
  \textbf{\bibinfo{volume}{53}}(\bibinfo{number}{21}), \bibinfo{pages}{14008}.

\bibitem[{\citenamefont{Irkhin and Katsnelson}(2001)}]{irkhin:401}
\bibinfo{author}{\bibnamefont{Irkhin}, \bibfnamefont{V.~Y.}}, and
  \bibinfo{author}{\bibfnamefont{M.~I.} \bibnamefont{Katsnelson}},
  \bibinfo{year}{2001}, \bibinfo{journal}{Eur. Phys. J. B}
  \textbf{\bibinfo{volume}{19}}, \bibinfo{pages}{401}.

\bibitem[{\citenamefont{Irkhin and Katsnelson}(2002)}]{irkhin:481}
\bibinfo{author}{\bibnamefont{Irkhin}, \bibfnamefont{V.~Y.}}, and
  \bibinfo{author}{\bibfnamefont{M.~I.} \bibnamefont{Katsnelson}},
  \bibinfo{year}{2002}, \bibinfo{journal}{Eur. Phys. J. B}
  \textbf{\bibinfo{volume}{30}}, \bibinfo{pages}{481}.

\bibitem[{\citenamefont{Irkhin and
  Katsnelson}(2005{\natexlab{a}})}]{Irkhin:479}
\bibinfo{author}{\bibnamefont{Irkhin}, \bibfnamefont{V.~Y.}}, and
  \bibinfo{author}{\bibfnamefont{M.~I.} \bibnamefont{Katsnelson}},
  \bibinfo{year}{2005}{\natexlab{a}}, \bibinfo{journal}{Eur. Phys. J. B}
  \textbf{\bibinfo{volume}{43}}, \bibinfo{pages}{479}.

\bibitem[{\citenamefont{Irkhin and
  Katsnelson}(2005{\natexlab{b}})}]{irkhin:054421}
\bibinfo{author}{\bibnamefont{Irkhin}, \bibfnamefont{V.~Y.}}, and
  \bibinfo{author}{\bibfnamefont{M.~I.} \bibnamefont{Katsnelson}},
  \bibinfo{year}{2005}{\natexlab{b}}, \bibinfo{journal}{Phys. Rev. B}
  \textbf{\bibinfo{volume}{72}}(\bibinfo{number}{5}), \bibinfo{eid}{054421}.

\bibitem[{\citenamefont{Irkhin and Katsnelson}(2006)}]{irkhin:104429}
\bibinfo{author}{\bibnamefont{Irkhin}, \bibfnamefont{V.~Y.}}, and
  \bibinfo{author}{\bibfnamefont{M.~I.} \bibnamefont{Katsnelson}},
  \bibinfo{year}{2006}, \bibinfo{journal}{Phys. Rev. B}
  \textbf{\bibinfo{volume}{73}}(\bibinfo{number}{10}), \bibinfo{eid}{104429}.

\bibitem[{\citenamefont{Irkhin} \emph{et~al.}(1989)\citenamefont{Irkhin,
  Katsnelson, and Trefilov}}]{Irkhin:397}
\bibinfo{author}{\bibnamefont{Irkhin}, \bibfnamefont{V.~Y.}},
  \bibinfo{author}{\bibfnamefont{M.~I.} \bibnamefont{Katsnelson}}, and
  \bibinfo{author}{\bibfnamefont{A.~V.} \bibnamefont{Trefilov}},
  \bibinfo{year}{1989}, \bibinfo{journal}{Physica C}
  \textbf{\bibinfo{volume}{160}}, \bibinfo{pages}{397}.

\bibitem[{\citenamefont{Irkhin} \emph{et~al.}(1994)\citenamefont{Irkhin,
  Katsnelson, and Trefilov}}]{Irkhin:1733}
\bibinfo{author}{\bibnamefont{Irkhin}, \bibfnamefont{V.~Y.}},
  \bibinfo{author}{\bibfnamefont{M.~I.} \bibnamefont{Katsnelson}}, and
  \bibinfo{author}{\bibfnamefont{A.~V.} \bibnamefont{Trefilov}},
  \bibinfo{year}{1994}, \bibinfo{journal}{Zh. Eksp. Theor. Fiz.}
  \textbf{\bibinfo{volume}{105}}, \bibinfo{pages}{1733}, \bibinfo{note}{[Sov.
  Phys. JETP 78, 936 (1994)]}.

\bibitem[{\citenamefont{Irkhin and Zarubin}(2000)}]{Irkhin:463}
\bibinfo{author}{\bibnamefont{Irkhin}, \bibfnamefont{V.~Y.}}, and
  \bibinfo{author}{\bibfnamefont{A.~V.} \bibnamefont{Zarubin}},
  \bibinfo{year}{2000}, \bibinfo{journal}{Eur. Phys. J. B}
  \textbf{\bibinfo{volume}{16}}, \bibinfo{pages}{463}.

\bibitem[{\citenamefont{Irkhin and Zarubin}(2004)}]{irkhin:035116}
\bibinfo{author}{\bibnamefont{Irkhin}, \bibfnamefont{V.~Y.}}, and
  \bibinfo{author}{\bibfnamefont{A.~V.} \bibnamefont{Zarubin}},
  \bibinfo{year}{2004}, \bibinfo{journal}{Phys. Rev. B}
  \textbf{\bibinfo{volume}{70}}(\bibinfo{number}{3}), \bibinfo{eid}{035116}.

\bibitem[{\citenamefont{Irkhin and Zarubin}(2006)}]{irkhin:246}
\bibinfo{author}{\bibnamefont{Irkhin}, \bibfnamefont{V.~Y.}}, and
  \bibinfo{author}{\bibfnamefont{A.~V.} \bibnamefont{Zarubin}},
  \bibinfo{year}{2006}, \bibinfo{journal}{J. Magn. Magn. Mat.}
  \textbf{\bibinfo{volume}{300}}, \bibinfo{pages}{246}.

\bibitem[{\citenamefont{Ishida} \emph{et~al.}(1982)\citenamefont{Ishida,
  Akazawa, Kubo, and Ishida}}]{Ishida:1111}
\bibinfo{author}{\bibnamefont{Ishida}, \bibfnamefont{S.}},
  \bibinfo{author}{\bibfnamefont{S.}~\bibnamefont{Akazawa}},
  \bibinfo{author}{\bibfnamefont{Y.}~\bibnamefont{Kubo}}, and
  \bibinfo{author}{\bibfnamefont{J.}~\bibnamefont{Ishida}},
  \bibinfo{year}{1982}, \bibinfo{journal}{J. Phys. F: Met. Phys.}
  \textbf{\bibinfo{volume}{12}}, \bibinfo{pages}{1111}.

\bibitem[{\citenamefont{Ishida} \emph{et~al.}(1995)\citenamefont{Ishida, Fujii,
  Kashiwagi, and Asano}}]{Ishida:2152}
\bibinfo{author}{\bibnamefont{Ishida}, \bibfnamefont{S.}},
  \bibinfo{author}{\bibfnamefont{S.}~\bibnamefont{Fujii}},
  \bibinfo{author}{\bibfnamefont{S.}~\bibnamefont{Kashiwagi}}, and
  \bibinfo{author}{\bibfnamefont{S.}~\bibnamefont{Asano}},
  \bibinfo{year}{1995}, \bibinfo{journal}{J. Phys. Soc. Jpn.}
  \textbf{\bibinfo{volume}{64}}, \bibinfo{pages}{2152}.

\bibitem[{\citenamefont{Ishida}
  \emph{et~al.}(1976{\natexlab{a}})\citenamefont{Ishida, Ishida, Asano, and
  Yamashita}}]{Ishida:1239}
\bibinfo{author}{\bibnamefont{Ishida}, \bibfnamefont{S.}},
  \bibinfo{author}{\bibfnamefont{J.}~\bibnamefont{Ishida}},
  \bibinfo{author}{\bibfnamefont{S.}~\bibnamefont{Asano}}, and
  \bibinfo{author}{\bibfnamefont{J.}~\bibnamefont{Yamashita}},
  \bibinfo{year}{1976}{\natexlab{a}}, \bibinfo{journal}{J. Phys. Soc. Jpn.}
  \textbf{\bibinfo{volume}{41}}(\bibinfo{number}{5}), \bibinfo{pages}{1570}.

\bibitem[{\citenamefont{Ishida}
  \emph{et~al.}(1976{\natexlab{b}})\citenamefont{Ishida, Ishida, Asano, and
  Yamashita}}]{Ishida:1570}
\bibinfo{author}{\bibnamefont{Ishida}, \bibfnamefont{S.}},
  \bibinfo{author}{\bibfnamefont{J.}~\bibnamefont{Ishida}},
  \bibinfo{author}{\bibfnamefont{S.}~\bibnamefont{Asano}}, and
  \bibinfo{author}{\bibfnamefont{J.}~\bibnamefont{Yamashita}},
  \bibinfo{year}{1976}{\natexlab{b}}, \bibinfo{journal}{J. Phys. Soc. Jpn.}
  \textbf{\bibinfo{volume}{41}}(\bibinfo{number}{5}), \bibinfo{pages}{1570}.

\bibitem[{\citenamefont{Ishida} \emph{et~al.}(1980)\citenamefont{Ishida, Kubo,
  Ishida, and Asano}}]{Ishida:814}
\bibinfo{author}{\bibnamefont{Ishida}, \bibfnamefont{S.}},
  \bibinfo{author}{\bibfnamefont{Y.}~\bibnamefont{Kubo}},
  \bibinfo{author}{\bibfnamefont{J.}~\bibnamefont{Ishida}}, and
  \bibinfo{author}{\bibfnamefont{S.}~\bibnamefont{Asano}},
  \bibinfo{year}{1980}, \bibinfo{journal}{J. Phys. Soc. Jpn.}
  \textbf{\bibinfo{volume}{48}}, \bibinfo{pages}{814}.

\bibitem[{\citenamefont{Ishigaki and Moriya}(1996)}]{Ishigaki:3402}
\bibinfo{author}{\bibnamefont{Ishigaki}, \bibfnamefont{A.}}, and
  \bibinfo{author}{\bibfnamefont{T.}~\bibnamefont{Moriya}},
  \bibinfo{year}{1996}, \bibinfo{journal}{J. Phys. Soc. Jpn}
  \textbf{\bibinfo{volume}{65}}, \bibinfo{pages}{3402}.

\bibitem[{\citenamefont{Itoh} \emph{et~al.}(2000)\citenamefont{Itoh, Ohsawa,
  and Inoue}}]{it.oh.00}
\bibinfo{author}{\bibnamefont{Itoh}, \bibfnamefont{H.}},
  \bibinfo{author}{\bibfnamefont{T.}~\bibnamefont{Ohsawa}}, and
  \bibinfo{author}{\bibfnamefont{J.}~\bibnamefont{Inoue}},
  \bibinfo{year}{2000}, \bibinfo{journal}{Phys. Rev. Lett.}
  \textbf{\bibinfo{volume}{84}}(\bibinfo{number}{11}), \bibinfo{pages}{2501}.

\bibitem[{\citenamefont{Jansen} \emph{et~al.}(1999)\citenamefont{Jansen,
  Hagenmayer, and Korber}}]{Jansen:591}
\bibinfo{author}{\bibnamefont{Jansen}, \bibfnamefont{M.}},
  \bibinfo{author}{\bibfnamefont{R.}~\bibnamefont{Hagenmayer}}, and
  \bibinfo{author}{\bibfnamefont{N.}~\bibnamefont{Korber}},
  \bibinfo{year}{1999}, \bibinfo{journal}{Comptes Rendus}
  \textbf{\bibinfo{volume}{2}}(\bibinfo{number}{11-13}), \bibinfo{pages}{591}.

\bibitem[{\citenamefont{Jarrell}(1992)}]{Jarrell:168}
\bibinfo{author}{\bibnamefont{Jarrell}, \bibfnamefont{M.}},
  \bibinfo{year}{1992}, \bibinfo{journal}{Phys. Rev. Lett.}
  \textbf{\bibinfo{volume}{69}}(\bibinfo{number}{1}), \bibinfo{pages}{168}.

\bibitem[{\citenamefont{Jarrell and Gubernatis}(1996)}]{Jarrell:134}
\bibinfo{author}{\bibnamefont{Jarrell}, \bibfnamefont{M.}}, and
  \bibinfo{author}{\bibfnamefont{J.~E.} \bibnamefont{Gubernatis}},
  \bibinfo{year}{1996}, \bibinfo{journal}{Physics Reports}
  \textbf{\bibinfo{volume}{269}}(\bibinfo{number}{3}), \bibinfo{pages}{134}.

\bibitem[{\citenamefont{Jarrett} \emph{et~al.}(1968)\citenamefont{Jarrett,
  Cloud, Bouchard, Butler, Frederick, and Gillson}}]{Jarrett:617}
\bibinfo{author}{\bibnamefont{Jarrett}, \bibfnamefont{H.~S.}},
  \bibinfo{author}{\bibfnamefont{W.~H.} \bibnamefont{Cloud}},
  \bibinfo{author}{\bibfnamefont{R.~J.} \bibnamefont{Bouchard}},
  \bibinfo{author}{\bibfnamefont{S.~R.} \bibnamefont{Butler}},
  \bibinfo{author}{\bibfnamefont{C.~G.} \bibnamefont{Frederick}}, and
  \bibinfo{author}{\bibfnamefont{J.~L.} \bibnamefont{Gillson}},
  \bibinfo{year}{1968}, \bibinfo{journal}{Phys. Rev. Lett.}
  \textbf{\bibinfo{volume}{21}}(\bibinfo{number}{9}), \bibinfo{pages}{617}.

\bibitem[{\citenamefont{Ji} \emph{et~al.}(2001)\citenamefont{Ji, Strijkers,
  Yang, Chien, Byers, Anguelouch, Xiao, and Gupta}}]{ji:5585}
\bibinfo{author}{\bibnamefont{Ji}, \bibfnamefont{Y.}},
  \bibinfo{author}{\bibfnamefont{G.~J.} \bibnamefont{Strijkers}},
  \bibinfo{author}{\bibfnamefont{F.~Y.} \bibnamefont{Yang}},
  \bibinfo{author}{\bibfnamefont{C.~L.} \bibnamefont{Chien}},
  \bibinfo{author}{\bibfnamefont{J.~M.} \bibnamefont{Byers}},
  \bibinfo{author}{\bibfnamefont{A.}~\bibnamefont{Anguelouch}},
  \bibinfo{author}{\bibfnamefont{G.}~\bibnamefont{Xiao}}, and
  \bibinfo{author}{\bibfnamefont{A.}~\bibnamefont{Gupta}},
  \bibinfo{year}{2001}, \bibinfo{journal}{Phys. Rev. Lett.}
  \textbf{\bibinfo{volume}{86}}(\bibinfo{number}{24}), \bibinfo{pages}{5585}.

\bibitem[{\citenamefont{Jo} \emph{et~al.}(2000{\natexlab{a}})\citenamefont{Jo,
  Mathur, Evetts, and Blamire}}]{jo:3803}
\bibinfo{author}{\bibnamefont{Jo}, \bibfnamefont{M.-H.}},
  \bibinfo{author}{\bibfnamefont{N.~D.} \bibnamefont{Mathur}},
  \bibinfo{author}{\bibfnamefont{J.~E.} \bibnamefont{Evetts}}, and
  \bibinfo{author}{\bibfnamefont{M.~G.} \bibnamefont{Blamire}},
  \bibinfo{year}{2000}{\natexlab{a}}, \bibinfo{journal}{Appl. Phys. Lett.}
  \textbf{\bibinfo{volume}{77}}(\bibinfo{number}{23}), \bibinfo{pages}{3803}.

\bibitem[{\citenamefont{Jo} \emph{et~al.}(2000{\natexlab{b}})\citenamefont{Jo,
  Mathur, Todd, and Blamire}}]{jo:R14905}
\bibinfo{author}{\bibnamefont{Jo}, \bibfnamefont{M.-H.}},
  \bibinfo{author}{\bibfnamefont{N.~D.} \bibnamefont{Mathur}},
  \bibinfo{author}{\bibfnamefont{N.~K.} \bibnamefont{Todd}}, and
  \bibinfo{author}{\bibfnamefont{M.~G.} \bibnamefont{Blamire}},
  \bibinfo{year}{2000}{\natexlab{b}}, \bibinfo{journal}{Phys. Rev. B}
  \textbf{\bibinfo{volume}{61}}(\bibinfo{number}{22}), \bibinfo{pages}{R14905}.

\bibitem[{\citenamefont{Jonker and Santen}(1950)}]{jonker:337}
\bibinfo{author}{\bibnamefont{Jonker}, \bibfnamefont{J.}}, and
  \bibinfo{author}{\bibfnamefont{G.~H.~V.} \bibnamefont{Santen}},
  \bibinfo{year}{1950}, \bibinfo{journal}{J. Magn. Magn. Mater.}
  \textbf{\bibinfo{volume}{16}}, \bibinfo{pages}{337}.

\bibitem[{\citenamefont{Joss} \emph{et~al.}(1984)\citenamefont{Joss, Hall,
  Crabtree, and Vuillemin}}]{Joss:5637}
\bibinfo{author}{\bibnamefont{Joss}, \bibfnamefont{W.}},
  \bibinfo{author}{\bibfnamefont{L.~N.} \bibnamefont{Hall}},
  \bibinfo{author}{\bibfnamefont{G.~W.} \bibnamefont{Crabtree}}, and
  \bibinfo{author}{\bibfnamefont{J.~J.} \bibnamefont{Vuillemin}},
  \bibinfo{year}{1984}, \bibinfo{journal}{Phys. Rev. B}
  \textbf{\bibinfo{volume}{30}}(\bibinfo{number}{10}), \bibinfo{pages}{5637}.

\bibitem[{\citenamefont{Judd}(1963)}]{JUDD:1963}
\bibinfo{author}{\bibnamefont{Judd}, \bibfnamefont{B.~R.}},
  \bibinfo{year}{1963}, \emph{\bibinfo{title}{Operator Techniques in Atomic
  Spectroscopy}} (\bibinfo{publisher}{McGrow-Hill, New York}).

\bibitem[{\citenamefont{Kajueter and Kotliar}(1996)}]{kajueter:131}
\bibinfo{author}{\bibnamefont{Kajueter}, \bibfnamefont{H.}}, and
  \bibinfo{author}{\bibfnamefont{G.}~\bibnamefont{Kotliar}},
  \bibinfo{year}{1996}, \bibinfo{journal}{Phys. Rev. Lett.}
  \textbf{\bibinfo{volume}{77}}(\bibinfo{number}{1}), \bibinfo{pages}{131}.

\bibitem[{\citenamefont{K\"amper} \emph{et~al.}(1987)\citenamefont{K\"amper,
  Schmitt, G\"untherodt, Gambino, and Ruf}}]{Kamper:2788}
\bibinfo{author}{\bibnamefont{K\"amper}, \bibfnamefont{K.~P.}},
  \bibinfo{author}{\bibfnamefont{W.}~\bibnamefont{Schmitt}},
  \bibinfo{author}{\bibfnamefont{G.}~\bibnamefont{G\"untherodt}},
  \bibinfo{author}{\bibfnamefont{R.~J.} \bibnamefont{Gambino}}, and
  \bibinfo{author}{\bibfnamefont{R.}~\bibnamefont{Ruf}}, \bibinfo{year}{1987},
  \bibinfo{journal}{Phys. Rev. Lett.}
  \textbf{\bibinfo{volume}{59}}(\bibinfo{number}{24}), \bibinfo{pages}{2788}.

\bibitem[{\citenamefont{Kanamori}(1963)}]{kanamori:276}
\bibinfo{author}{\bibnamefont{Kanamori}, \bibfnamefont{J.}},
  \bibinfo{year}{1963}, \bibinfo{journal}{Prog. Theor, Phys.}
  \textbf{\bibinfo{volume}{30}}, \bibinfo{pages}{276}.

\bibitem[{\citenamefont{Kaplan} \emph{et~al.}(2001)\citenamefont{Kaplan,
  Mahanti, and Su}}]{Kaplan:3634}
\bibinfo{author}{\bibnamefont{Kaplan}, \bibfnamefont{T.~A.}},
  \bibinfo{author}{\bibfnamefont{S.~D.} \bibnamefont{Mahanti}}, and
  \bibinfo{author}{\bibfnamefont{Y.-S.} \bibnamefont{Su}},
  \bibinfo{year}{2001}, \bibinfo{journal}{Phys. Rev. Lett.}
  \textbf{\bibinfo{volume}{86}}(\bibinfo{number}{16}), \bibinfo{pages}{3634}.

\bibitem[{\citenamefont{Karla} \emph{et~al.}(1999)\citenamefont{Karla, Pierre,
  Murani, and Neumann}}]{Karla:294}
\bibinfo{author}{\bibnamefont{Karla}, \bibfnamefont{I.}},
  \bibinfo{author}{\bibfnamefont{J.}~\bibnamefont{Pierre}},
  \bibinfo{author}{\bibfnamefont{A.~P.} \bibnamefont{Murani}}, and
  \bibinfo{author}{\bibfnamefont{M.}~\bibnamefont{Neumann}},
  \bibinfo{year}{1999}, \bibinfo{journal}{Physica B}
  \textbf{\bibinfo{volume}{271}}, \bibinfo{pages}{294}.

\bibitem[{\citenamefont{Karla}
  \emph{et~al.}(1998{\natexlab{a}})\citenamefont{Karla, Pierre, and
  Ouladdiaf}}]{Karla:215}
\bibinfo{author}{\bibnamefont{Karla}, \bibfnamefont{I.}},
  \bibinfo{author}{\bibfnamefont{J.}~\bibnamefont{Pierre}}, and
  \bibinfo{author}{\bibfnamefont{B.}~\bibnamefont{Ouladdiaf}},
  \bibinfo{year}{1998}{\natexlab{a}}, \bibinfo{journal}{Physica B}
  \textbf{\bibinfo{volume}{253}}, \bibinfo{pages}{215}.

\bibitem[{\citenamefont{Karla}
  \emph{et~al.}(1998{\natexlab{b}})\citenamefont{Karla, Pierre, and
  Skolozdra}}]{Karla:42}
\bibinfo{author}{\bibnamefont{Karla}, \bibfnamefont{I.}},
  \bibinfo{author}{\bibfnamefont{J.}~\bibnamefont{Pierre}}, and
  \bibinfo{author}{\bibfnamefont{R.~V.} \bibnamefont{Skolozdra}},
  \bibinfo{year}{1998}{\natexlab{b}}, \bibinfo{journal}{J. Alloys Compound}
  \textbf{\bibinfo{volume}{265}}, \bibinfo{pages}{42}.

\bibitem[{\citenamefont{Kato} \emph{et~al.}(2002)\citenamefont{Kato, Okuda,
  Okimoto, Tomioka, Takenoya, Ohkubo, Kawasaki, and Tokura}}]{kato:328}
\bibinfo{author}{\bibnamefont{Kato}, \bibfnamefont{H.}},
  \bibinfo{author}{\bibfnamefont{T.}~\bibnamefont{Okuda}},
  \bibinfo{author}{\bibfnamefont{Y.}~\bibnamefont{Okimoto}},
  \bibinfo{author}{\bibfnamefont{Y.}~\bibnamefont{Tomioka}},
  \bibinfo{author}{\bibfnamefont{Y.}~\bibnamefont{Takenoya}},
  \bibinfo{author}{\bibfnamefont{A.}~\bibnamefont{Ohkubo}},
  \bibinfo{author}{\bibfnamefont{M.}~\bibnamefont{Kawasaki}}, and
  \bibinfo{author}{\bibfnamefont{Y.}~\bibnamefont{Tokura}},
  \bibinfo{year}{2002}, \bibinfo{journal}{Appl. Phys. Lett.}
  \textbf{\bibinfo{volume}{81}}(\bibinfo{number}{2}), \bibinfo{pages}{328}.

\bibitem[{\citenamefont{Katsnelson and Edwards}(1992)}]{Katsnelson:3289}
\bibinfo{author}{\bibnamefont{Katsnelson}, \bibfnamefont{M.~I.}}, and
  \bibinfo{author}{\bibfnamefont{D.~M.} \bibnamefont{Edwards}},
  \bibinfo{year}{1992}, \bibinfo{journal}{J. Phys.: Condens. Matter}
  \textbf{\bibinfo{volume}{4}}, \bibinfo{pages}{3289}.

\bibitem[{\citenamefont{Katsnelson and Lichtenstein}(1999)}]{katsnelson:1037}
\bibinfo{author}{\bibnamefont{Katsnelson}, \bibfnamefont{M.~I.}}, and
  \bibinfo{author}{\bibfnamefont{A.~I.} \bibnamefont{Lichtenstein}},
  \bibinfo{year}{1999}, \bibinfo{journal}{J. Phys.: Condens. Matter}
  \textbf{\bibinfo{volume}{11}}, \bibinfo{pages}{1037}.

\bibitem[{\citenamefont{Katsnelson and Lichtenstein}(2000)}]{Katsnelson:8906}
\bibinfo{author}{\bibnamefont{Katsnelson}, \bibfnamefont{M.~I.}}, and
  \bibinfo{author}{\bibfnamefont{A.~I.} \bibnamefont{Lichtenstein}},
  \bibinfo{year}{2000}, \bibinfo{journal}{Phys. Rev. B}
  \textbf{\bibinfo{volume}{61}}(\bibinfo{number}{13}), \bibinfo{pages}{8906}.

\bibitem[{\citenamefont{Katsnelson and Lichtenstein}(2002)}]{katsnelson:9}
\bibinfo{author}{\bibnamefont{Katsnelson}, \bibfnamefont{M.~I.}}, and
  \bibinfo{author}{\bibfnamefont{A.~I.} \bibnamefont{Lichtenstein}},
  \bibinfo{year}{2002}, \bibinfo{journal}{Eur. Phys. J. B}
  \textbf{\bibinfo{volume}{30}}, \bibinfo{pages}{9}.

\bibitem[{\citenamefont{Katsnelson and Trefilov}(1990)}]{katsnelson:63}
\bibinfo{author}{\bibnamefont{Katsnelson}, \bibfnamefont{M.~I.}}, and
  \bibinfo{author}{\bibfnamefont{A.~V.} \bibnamefont{Trefilov}},
  \bibinfo{year}{1990}, \bibinfo{journal}{Z. Phys. B}
  \textbf{\bibinfo{volume}{80}}, \bibinfo{pages}{63}.

\bibitem[{\citenamefont{Keizer} \emph{et~al.}(2006)\citenamefont{Keizer,
  Goennenwein, Klapwijk, Miao, Xiao, and Gupta}}]{keizer:825}
\bibinfo{author}{\bibnamefont{Keizer}, \bibfnamefont{R.~S.}},
  \bibinfo{author}{\bibfnamefont{S.~T.~B.} \bibnamefont{Goennenwein}},
  \bibinfo{author}{\bibfnamefont{T.~M.} \bibnamefont{Klapwijk}},
  \bibinfo{author}{\bibfnamefont{G.}~\bibnamefont{Miao}},
  \bibinfo{author}{\bibfnamefont{G.}~\bibnamefont{Xiao}}, and
  \bibinfo{author}{\bibfnamefont{A.}~\bibnamefont{Gupta}},
  \bibinfo{year}{2006}, \bibinfo{journal}{Nature}
  \textbf{\bibinfo{volume}{439}}, \bibinfo{pages}{825}.

\bibitem[{\citenamefont{Kino} \emph{et~al.}(2003)\citenamefont{Kino,
  Aryasetiawanx, Solovyev, Miyake, Ohno, and Terakura}}]{Kino:858}
\bibinfo{author}{\bibnamefont{Kino}, \bibfnamefont{H.}},
  \bibinfo{author}{\bibfnamefont{F.}~\bibnamefont{Aryasetiawanx}},
  \bibinfo{author}{\bibfnamefont{I.}~\bibnamefont{Solovyev}},
  \bibinfo{author}{\bibfnamefont{T.}~\bibnamefont{Miyake}},
  \bibinfo{author}{\bibfnamefont{T.}~\bibnamefont{Ohno}}, and
  \bibinfo{author}{\bibfnamefont{K.}~\bibnamefont{Terakura}},
  \bibinfo{year}{2003}, \bibinfo{journal}{Physica B}
  \textbf{\bibinfo{volume}{329-333}}, \bibinfo{pages}{858}.

\bibitem[{\citenamefont{Kisker} \emph{et~al.}(1978)\citenamefont{Kisker, Baum,
  Mahan, Raith, and Reihl}}]{kisker:2256}
\bibinfo{author}{\bibnamefont{Kisker}, \bibfnamefont{E.}},
  \bibinfo{author}{\bibfnamefont{G.}~\bibnamefont{Baum}},
  \bibinfo{author}{\bibfnamefont{A.~H.} \bibnamefont{Mahan}},
  \bibinfo{author}{\bibfnamefont{W.}~\bibnamefont{Raith}}, and
  \bibinfo{author}{\bibfnamefont{B.}~\bibnamefont{Reihl}},
  \bibinfo{year}{1978}, \bibinfo{journal}{Phys. Rev. B}
  \textbf{\bibinfo{volume}{18}}(\bibinfo{number}{5}), \bibinfo{pages}{2256}.

\bibitem[{\citenamefont{Kisker} \emph{et~al.}(1987)\citenamefont{Kisker,
  Carbone, Flipse, and Wassermann}}]{Kisker:21}
\bibinfo{author}{\bibnamefont{Kisker}, \bibfnamefont{E.}},
  \bibinfo{author}{\bibfnamefont{C.}~\bibnamefont{Carbone}},
  \bibinfo{author}{\bibfnamefont{C.~F.} \bibnamefont{Flipse}}, and
  \bibinfo{author}{\bibfnamefont{E.~F.} \bibnamefont{Wassermann}},
  \bibinfo{year}{1987}, \bibinfo{journal}{J. Magn. Magn. Mater.}
  \textbf{\bibinfo{volume}{70}}(\bibinfo{number}{1-3}), \bibinfo{pages}{21}.

\bibitem[{\citenamefont{Kleiber} \emph{et~al.}(2000)\citenamefont{Kleiber,
  Bode, Ravli\ifmmode~\acute{c}\else \'{c}\fi{}, and
  Wiesendanger}}]{Kleiber:4606}
\bibinfo{author}{\bibnamefont{Kleiber}, \bibfnamefont{M.}},
  \bibinfo{author}{\bibfnamefont{M.}~\bibnamefont{Bode}},
  \bibinfo{author}{\bibfnamefont{R.}~\bibnamefont{Ravli\ifmmode~\acute{c}\else
  \'{c}\fi{}}}, and
  \bibinfo{author}{\bibfnamefont{R.}~\bibnamefont{Wiesendanger}},
  \bibinfo{year}{2000}, \bibinfo{journal}{Phys. Rev. Lett.}
  \textbf{\bibinfo{volume}{85}}(\bibinfo{number}{21}), \bibinfo{pages}{4606}.

\bibitem[{\citenamefont{Kobayashi} \emph{et~al.}(1998)\citenamefont{Kobayashi,
  Kimura, Sawada, Terakura, and Tokura}}]{Kobayashi:677}
\bibinfo{author}{\bibnamefont{Kobayashi}, \bibfnamefont{K.~I.}},
  \bibinfo{author}{\bibfnamefont{T.}~\bibnamefont{Kimura}},
  \bibinfo{author}{\bibfnamefont{H.}~\bibnamefont{Sawada}},
  \bibinfo{author}{\bibfnamefont{K.}~\bibnamefont{Terakura}}, and
  \bibinfo{author}{\bibfnamefont{Y.}~\bibnamefont{Tokura}},
  \bibinfo{year}{1998}, \bibinfo{journal}{Nature}
  \textbf{\bibinfo{volume}{395}}, \bibinfo{pages}{677}.

\bibitem[{\citenamefont{Kobayashi} \emph{et~al.}(1999)\citenamefont{Kobayashi,
  Kimura, Tomioka, Sawada, Terakura, and Tokura}}]{Kobayashi:11159}
\bibinfo{author}{\bibnamefont{Kobayashi}, \bibfnamefont{K.-I.}},
  \bibinfo{author}{\bibfnamefont{T.}~\bibnamefont{Kimura}},
  \bibinfo{author}{\bibfnamefont{Y.}~\bibnamefont{Tomioka}},
  \bibinfo{author}{\bibfnamefont{H.}~\bibnamefont{Sawada}},
  \bibinfo{author}{\bibfnamefont{K.}~\bibnamefont{Terakura}}, and
  \bibinfo{author}{\bibfnamefont{Y.}~\bibnamefont{Tokura}},
  \bibinfo{year}{1999}, \bibinfo{journal}{Phys. Rev. B}
  \textbf{\bibinfo{volume}{59}}(\bibinfo{number}{17}), \bibinfo{pages}{11159}.

\bibitem[{\citenamefont{Kohn and Sham}(1965)}]{kohn:A1133}
\bibinfo{author}{\bibnamefont{Kohn}, \bibfnamefont{W.}}, and
  \bibinfo{author}{\bibfnamefont{L.~J.} \bibnamefont{Sham}},
  \bibinfo{year}{1965}, \bibinfo{journal}{Phys. Rev.}
  \textbf{\bibinfo{volume}{140}}(\bibinfo{number}{4A}), \bibinfo{pages}{A1133}.

\bibitem[{\citenamefont{Korotin} \emph{et~al.}(1998)\citenamefont{Korotin,
  Anisimov, Khomskii, and Sawatzky}}]{korotin:4305}
\bibinfo{author}{\bibnamefont{Korotin}, \bibfnamefont{M.~A.}},
  \bibinfo{author}{\bibfnamefont{V.~I.} \bibnamefont{Anisimov}},
  \bibinfo{author}{\bibfnamefont{D.~I.} \bibnamefont{Khomskii}}, and
  \bibinfo{author}{\bibfnamefont{G.~A.} \bibnamefont{Sawatzky}},
  \bibinfo{year}{1998}, \bibinfo{journal}{Phys. Rev. Lett.}
  \textbf{\bibinfo{volume}{80}}(\bibinfo{number}{19}), \bibinfo{pages}{4305}.

\bibitem[{\citenamefont{Kotliar} \emph{et~al.}(2006)\citenamefont{Kotliar,
  Savrasov, Haule, Oudovenko, Parcollet, and Marianetti}}]{kotliar:865}
\bibinfo{author}{\bibnamefont{Kotliar}, \bibfnamefont{G.}},
  \bibinfo{author}{\bibfnamefont{S.~Y.} \bibnamefont{Savrasov}},
  \bibinfo{author}{\bibfnamefont{K.}~\bibnamefont{Haule}},
  \bibinfo{author}{\bibfnamefont{V.~S.} \bibnamefont{Oudovenko}},
  \bibinfo{author}{\bibfnamefont{O.}~\bibnamefont{Parcollet}}, and
  \bibinfo{author}{\bibfnamefont{C.~A.} \bibnamefont{Marianetti}},
  \bibinfo{year}{2006}, \bibinfo{journal}{Reviews of Modern Physics}
  \textbf{\bibinfo{volume}{78}}(\bibinfo{number}{3}), \bibinfo{eid}{865}.

\bibitem[{\citenamefont{K{\"u}bler}(1984)}]{Kubler:257}
\bibinfo{author}{\bibnamefont{K{\"u}bler}, \bibfnamefont{J.}},
  \bibinfo{year}{1984}, \bibinfo{journal}{Physica} \textbf{\bibinfo{volume}{127
  B}}, \bibinfo{pages}{257}.

\bibitem[{\citenamefont{K{\"u}bler}(2000)}]{Kuebler2000}
\bibinfo{author}{\bibnamefont{K{\"u}bler}, \bibfnamefont{J.}},
  \bibinfo{year}{2000}, \emph{\bibinfo{title}{Theory of Itinerant Electron
  Magnetism}} (\bibinfo{publisher}{Calderon Press: Oxford}).

\bibitem[{\citenamefont{K\"ubler}(2003)}]{kubl.03}
\bibinfo{author}{\bibnamefont{K\"ubler}, \bibfnamefont{J.}},
  \bibinfo{year}{2003}, \bibinfo{journal}{Phys. Rev. B}
  \textbf{\bibinfo{volume}{67}}(\bibinfo{number}{22}), \bibinfo{pages}{220403}.

\bibitem[{\citenamefont{K\"ubler} \emph{et~al.}(1983)\citenamefont{K\"ubler,
  William, and Sommers}}]{ku.wi.83}
\bibinfo{author}{\bibnamefont{K\"ubler}, \bibfnamefont{J.}},
  \bibinfo{author}{\bibfnamefont{A.~R.} \bibnamefont{William}}, and
  \bibinfo{author}{\bibfnamefont{C.~B.} \bibnamefont{Sommers}},
  \bibinfo{year}{1983}, \bibinfo{journal}{Phys. Rev. B}
  \textbf{\bibinfo{volume}{28}}(\bibinfo{number}{4}), \bibinfo{pages}{1745}.

\bibitem[{\citenamefont{Kubo and Ohata}(1972)}]{kubo:21}
\bibinfo{author}{\bibnamefont{Kubo}, \bibfnamefont{K.}}, and
  \bibinfo{author}{\bibfnamefont{N.}~\bibnamefont{Ohata}},
  \bibinfo{year}{1972}, \bibinfo{journal}{J. Phys. Soc. Jpn.}
  \textbf{\bibinfo{volume}{33}}, \bibinfo{pages}{21}.

\bibitem[{\citenamefont{Kubo}(1957)}]{kubo:570}
\bibinfo{author}{\bibnamefont{Kubo}, \bibfnamefont{R.}}, \bibinfo{year}{1957},
  \bibinfo{journal}{J. Phys. Soc. Jpn.}
  \textbf{\bibinfo{volume}{12}}(\bibinfo{number}{6}), \bibinfo{pages}{570}.

\bibitem[{\citenamefont{Kulatov and Mazin}(2003)}]{kulatov:343}
\bibinfo{author}{\bibnamefont{Kulatov}, \bibfnamefont{E.}}, and
  \bibinfo{author}{\bibfnamefont{I.~I.} \bibnamefont{Mazin}},
  \bibinfo{year}{2003}, \bibinfo{journal}{J. Phys.: Condens. Matter}
  \textbf{\bibinfo{volume}{2}}, \bibinfo{pages}{343}.

\bibitem[{\citenamefont{Kurmaev} \emph{et~al.}(2003)\citenamefont{Kurmaev,
  Moewes, Butorin, Katsnelson, Finkelstein, Nordgren, and
  Tedrow}}]{kurmaev:155105}
\bibinfo{author}{\bibnamefont{Kurmaev}, \bibfnamefont{E.~Z.}},
  \bibinfo{author}{\bibfnamefont{A.}~\bibnamefont{Moewes}},
  \bibinfo{author}{\bibfnamefont{S.~M.} \bibnamefont{Butorin}},
  \bibinfo{author}{\bibfnamefont{M.~I.} \bibnamefont{Katsnelson}},
  \bibinfo{author}{\bibfnamefont{L.~D.} \bibnamefont{Finkelstein}},
  \bibinfo{author}{\bibfnamefont{J.}~\bibnamefont{Nordgren}}, and
  \bibinfo{author}{\bibfnamefont{P.~M.} \bibnamefont{Tedrow}},
  \bibinfo{year}{2003}, \bibinfo{journal}{Phys. Rev. B}
  \textbf{\bibinfo{volume}{67}}(\bibinfo{number}{15}), \bibinfo{eid}{155105}.

\bibitem[{\citenamefont{Kurtulus} \emph{et~al.}(2005)\citenamefont{Kurtulus,
  Dronskowski, Samolyuk, and Antropov}}]{kurtulus:014425}
\bibinfo{author}{\bibnamefont{Kurtulus}, \bibfnamefont{Y.}},
  \bibinfo{author}{\bibfnamefont{R.}~\bibnamefont{Dronskowski}},
  \bibinfo{author}{\bibfnamefont{G.~D.} \bibnamefont{Samolyuk}}, and
  \bibinfo{author}{\bibfnamefont{V.~P.} \bibnamefont{Antropov}},
  \bibinfo{year}{2005}, \bibinfo{journal}{Phys. Rev. B}
  \textbf{\bibinfo{volume}{71}}(\bibinfo{number}{1}), \bibinfo{eid}{014425}
  (pages~\bibinfo{numpages}{12}).

\bibitem[{\citenamefont{Laad} \emph{et~al.}(2001)\citenamefont{Laad, Craco, and
  M{\"u}ller-Hartmann}}]{laad:214421}
\bibinfo{author}{\bibnamefont{Laad}, \bibfnamefont{M.~S.}},
  \bibinfo{author}{\bibfnamefont{L.}~\bibnamefont{Craco}}, and
  \bibinfo{author}{\bibfnamefont{E.}~\bibnamefont{M{\"u}ller-Hartmann}},
  \bibinfo{year}{2001}, \bibinfo{journal}{Phys. Rev. B}
  \textbf{\bibinfo{volume}{64}}(\bibinfo{number}{21}), \bibinfo{pages}{214421}.

\bibitem[{\citenamefont{Leighton} \emph{et~al.}(2007)\citenamefont{Leighton,
  Manno, Cady, Freeland, Wang, Umemoto, Wentzcovitch, Chen, Chien, Kuhns, Hoch,
  Reyes} \emph{et~al.}}]{Leighton:315219}
\bibinfo{author}{\bibnamefont{Leighton}, \bibfnamefont{C.}},
  \bibinfo{author}{\bibfnamefont{M.}~\bibnamefont{Manno}},
  \bibinfo{author}{\bibfnamefont{A.}~\bibnamefont{Cady}},
  \bibinfo{author}{\bibfnamefont{J.~W.} \bibnamefont{Freeland}},
  \bibinfo{author}{\bibfnamefont{L.}~\bibnamefont{Wang}},
  \bibinfo{author}{\bibfnamefont{K.}~\bibnamefont{Umemoto}},
  \bibinfo{author}{\bibfnamefont{R.~M.} \bibnamefont{Wentzcovitch}},
  \bibinfo{author}{\bibfnamefont{T.~Y.} \bibnamefont{Chen}},
  \bibinfo{author}{\bibfnamefont{C.~L.} \bibnamefont{Chien}},
  \bibinfo{author}{\bibfnamefont{P.~L.} \bibnamefont{Kuhns}},
  \bibinfo{author}{\bibfnamefont{M.~J.~R.} \bibnamefont{Hoch}},
  \bibinfo{author}{\bibfnamefont{A.~P.} \bibnamefont{Reyes}}, \emph{et~al.},
  \bibinfo{year}{2007}, \bibinfo{journal}{J. Phys.: Condens. Matter}
  \textbf{\bibinfo{volume}{19}}, \bibinfo{pages}{315219}.

\bibitem[{\citenamefont{Leonov} \emph{et~al.}(2006)\citenamefont{Leonov,
  Yaresko, Antonov, and Anisimov}}]{leonov:165117}
\bibinfo{author}{\bibnamefont{Leonov}, \bibfnamefont{I.}},
  \bibinfo{author}{\bibfnamefont{A.~N.} \bibnamefont{Yaresko}},
  \bibinfo{author}{\bibfnamefont{V.~N.} \bibnamefont{Antonov}}, and
  \bibinfo{author}{\bibfnamefont{V.~I.} \bibnamefont{Anisimov}},
  \bibinfo{year}{2006}, \bibinfo{journal}{Phys. Rev. B}
  \textbf{\bibinfo{volume}{74}}(\bibinfo{number}{16}), \bibinfo{eid}{165117}.

\bibitem[{\citenamefont{van Leuken and de~Groot}(1995)}]{vanLeuken:7176}
\bibinfo{author}{\bibnamefont{van Leuken}, \bibfnamefont{H.}}, and
  \bibinfo{author}{\bibfnamefont{R.~A.} \bibnamefont{de~Groot}},
  \bibinfo{year}{1995}, \bibinfo{journal}{Phys. Rev. B}
  \textbf{\bibinfo{volume}{51}}(\bibinfo{number}{11}), \bibinfo{pages}{7176}.

\bibitem[{\citenamefont{Lewis} \emph{et~al.}(1997)\citenamefont{Lewis, Allen,
  and Sasaki}}]{levis:10253}
\bibinfo{author}{\bibnamefont{Lewis}, \bibfnamefont{S.~P.}},
  \bibinfo{author}{\bibfnamefont{P.~B.} \bibnamefont{Allen}}, and
  \bibinfo{author}{\bibfnamefont{T.}~\bibnamefont{Sasaki}},
  \bibinfo{year}{1997}, \bibinfo{journal}{Phys. Rev. B}
  \textbf{\bibinfo{volume}{55}}(\bibinfo{number}{16}), \bibinfo{pages}{10253}.

\bibitem[{\citenamefont{Lezaic}(2006)}]{Lezaic:private}
\bibinfo{author}{\bibnamefont{Lezaic}, \bibfnamefont{M.}},
  \bibinfo{year}{2006}, \bibinfo{journal}{unpublished} .

\bibitem[{\citenamefont{Lezaic} \emph{et~al.}(2006)\citenamefont{Lezaic,
  Mavropoulos, Enkovaara, Bihlmayer, and Blugel}}]{lezaic:026404}
\bibinfo{author}{\bibnamefont{Lezaic}, \bibfnamefont{M.}},
  \bibinfo{author}{\bibfnamefont{P.}~\bibnamefont{Mavropoulos}},
  \bibinfo{author}{\bibfnamefont{J.}~\bibnamefont{Enkovaara}},
  \bibinfo{author}{\bibfnamefont{G.}~\bibnamefont{Bihlmayer}}, and
  \bibinfo{author}{\bibfnamefont{S.}~\bibnamefont{Blugel}},
  \bibinfo{year}{2006}, \bibinfo{journal}{Phys. Rev. Lett.}
  \textbf{\bibinfo{volume}{97}}(\bibinfo{number}{2}), \bibinfo{eid}{026404}.

\bibitem[{\citenamefont{Liang and Pang}(1995)}]{Liang:173}
\bibinfo{author}{\bibnamefont{Liang}, \bibfnamefont{S.}}, and
  \bibinfo{author}{\bibfnamefont{H.}~\bibnamefont{Pang}}, \bibinfo{year}{1995},
  \bibinfo{journal}{Europhys. Lett.} \textbf{\bibinfo{volume}{32}},
  \bibinfo{pages}{173}.

\bibitem[{\citenamefont{Lichtenstein and Katsnelson}(1998)}]{lichtenstein:6884}
\bibinfo{author}{\bibnamefont{Lichtenstein}, \bibfnamefont{A.~I.}}, and
  \bibinfo{author}{\bibfnamefont{M.~I.} \bibnamefont{Katsnelson}},
  \bibinfo{year}{1998}, \bibinfo{journal}{Phys. Rev. B}
  \textbf{\bibinfo{volume}{57}}(\bibinfo{number}{12}), \bibinfo{pages}{6884}.

\bibitem[{\citenamefont{Lichtenstein}
  \emph{et~al.}(2001)\citenamefont{Lichtenstein, Katsnelson, and
  Kotliar}}]{lichtenstein:067205}
\bibinfo{author}{\bibnamefont{Lichtenstein}, \bibfnamefont{A.~I.}},
  \bibinfo{author}{\bibfnamefont{M.~I.} \bibnamefont{Katsnelson}}, and
  \bibinfo{author}{\bibfnamefont{G.}~\bibnamefont{Kotliar}},
  \bibinfo{year}{2001}, \bibinfo{journal}{Phys. Rev. Lett.}
  \textbf{\bibinfo{volume}{87}}(\bibinfo{number}{6}), \bibinfo{pages}{067205}.

\bibitem[{\citenamefont{Liechtenstein}
  \emph{et~al.}(1987)\citenamefont{Liechtenstein, Katsnelson, Antropov, and
  Gubanov}}]{li.ka.87}
\bibinfo{author}{\bibnamefont{Liechtenstein}, \bibfnamefont{A.~I.}},
  \bibinfo{author}{\bibfnamefont{M.~I.} \bibnamefont{Katsnelson}},
  \bibinfo{author}{\bibfnamefont{V.~P.} \bibnamefont{Antropov}}, and
  \bibinfo{author}{\bibfnamefont{V.~A.} \bibnamefont{Gubanov}},
  \bibinfo{year}{1987}, \bibinfo{journal}{J. Mag. Mag. Matt.}
  \textbf{\bibinfo{volume}{67}}(\bibinfo{number}{1}), \bibinfo{pages}{65}.

\bibitem[{\citenamefont{von~der Linden and Edwards}(1991)}]{Linden:4917}
\bibinfo{author}{\bibnamefont{von~der Linden}, \bibfnamefont{W.}}, and
  \bibinfo{author}{\bibfnamefont{D.~M.} \bibnamefont{Edwards}},
  \bibinfo{year}{1991}, \bibinfo{journal}{J. Phys.: Condens. Matter}
  \textbf{\bibinfo{volume}{3}}, \bibinfo{pages}{4917}.

\bibitem[{\citenamefont{Liu}(2003)}]{liu:172411}
\bibinfo{author}{\bibnamefont{Liu}, \bibfnamefont{B.-G.}},
  \bibinfo{year}{2003}, \bibinfo{journal}{Phys. Rev. B}
  \textbf{\bibinfo{volume}{67}}(\bibinfo{number}{17}), \bibinfo{eid}{172411}.

\bibitem[{\citenamefont{Lutovinov and Reizer}(1979)}]{lutovinov:707}
\bibinfo{author}{\bibnamefont{Lutovinov}, \bibfnamefont{V.~S.}}, and
  \bibinfo{author}{\bibfnamefont{M.~Y.} \bibnamefont{Reizer}},
  \bibinfo{year}{1979}, \bibinfo{journal}{Zh. Eksp. Theor. Fiz.}
  \textbf{\bibinfo{volume}{77}}, \bibinfo{pages}{707}.

\bibitem[{\citenamefont{Luttinger and Ward}(1960)}]{luttinger:1417}
\bibinfo{author}{\bibnamefont{Luttinger}, \bibfnamefont{J.~M.}}, and
  \bibinfo{author}{\bibfnamefont{J.~C.} \bibnamefont{Ward}},
  \bibinfo{year}{1960}, \bibinfo{journal}{Phys. Rev.}
  \textbf{\bibinfo{volume}{118}}(\bibinfo{number}{5}), \bibinfo{pages}{1417}.

\bibitem[{\citenamefont{MacDonald} \emph{et~al.}(1998)\citenamefont{MacDonald,
  Jungwirth, and Kasner}}]{ma.ju.98}
\bibinfo{author}{\bibnamefont{MacDonald}, \bibfnamefont{A.~H.}},
  \bibinfo{author}{\bibfnamefont{T.}~\bibnamefont{Jungwirth}}, and
  \bibinfo{author}{\bibfnamefont{M.}~\bibnamefont{Kasner}},
  \bibinfo{year}{1998}, \bibinfo{journal}{Phys. Rev. Lett.}
  \textbf{\bibinfo{volume}{81}}(\bibinfo{number}{3}), \bibinfo{pages}{705}.

\bibitem[{\citenamefont{Machintosh and Andersen}(1980)}]{ma.an.80}
\bibinfo{author}{\bibnamefont{Machintosh}, \bibfnamefont{A.~R.}}, and
  \bibinfo{author}{\bibfnamefont{O.~K.} \bibnamefont{Andersen}},
  \bibinfo{year}{1980} (\bibinfo{publisher}{Ed. by M. Springford, Cambridge
  Univ. Press.}, \bibinfo{address}{London}).

\bibitem[{\citenamefont{Maeno} \emph{et~al.}(1994)\citenamefont{Maeno,
  Hashimoto, Yoshida, Nishizaki, Fujita, Bednorz, and Lichtenberg}}]{Maeno:532}
\bibinfo{author}{\bibnamefont{Maeno}, \bibfnamefont{Y.}},
  \bibinfo{author}{\bibfnamefont{H.}~\bibnamefont{Hashimoto}},
  \bibinfo{author}{\bibfnamefont{K.}~\bibnamefont{Yoshida}},
  \bibinfo{author}{\bibfnamefont{S.}~\bibnamefont{Nishizaki}},
  \bibinfo{author}{\bibfnamefont{T.}~\bibnamefont{Fujita}},
  \bibinfo{author}{\bibfnamefont{J.~G.} \bibnamefont{Bednorz}}, and
  \bibinfo{author}{\bibfnamefont{F.}~\bibnamefont{Lichtenberg}},
  \bibinfo{year}{1994}, \bibinfo{journal}{Nature}
  \textbf{\bibinfo{volume}{372}}, \bibinfo{pages}{532}.

\bibitem[{\citenamefont{Mahan}(1990)}]{mahan:1990}
\bibinfo{author}{\bibnamefont{Mahan}, \bibfnamefont{G.~D.}},
  \bibinfo{year}{1990}, \emph{\bibinfo{title}{Many-Particle Physics}}
  (\bibinfo{publisher}{Plenum Press, New York}).

\bibitem[{\citenamefont{van~der Marel and Sawatzky}(1988)}]{Sawatzky:10674}
\bibinfo{author}{\bibnamefont{van~der Marel}, \bibfnamefont{D.}}, and
  \bibinfo{author}{\bibfnamefont{G.~A.} \bibnamefont{Sawatzky}},
  \bibinfo{year}{1988}, \bibinfo{journal}{Phys. Rev. B}
  \textbf{\bibinfo{volume}{37}}(\bibinfo{number}{18}), \bibinfo{pages}{10674}.

\bibitem[{\citenamefont{Matar} \emph{et~al.}(1992)\citenamefont{Matar,
  Demazeau, Sticht, Eyert, and K{\"u}bler}}]{Matar:315}
\bibinfo{author}{\bibnamefont{Matar}, \bibfnamefont{S.}},
  \bibinfo{author}{\bibfnamefont{G.}~\bibnamefont{Demazeau}},
  \bibinfo{author}{\bibfnamefont{J.}~\bibnamefont{Sticht}},
  \bibinfo{author}{\bibfnamefont{V.}~\bibnamefont{Eyert}}, and
  \bibinfo{author}{\bibfnamefont{J.}~\bibnamefont{K{\"u}bler}},
  \bibinfo{year}{1992}, \bibinfo{journal}{Journal de Physique I}
  \textbf{\bibinfo{volume}{2}}, \bibinfo{pages}{315}.

\bibitem[{\citenamefont{Mavropoulos}
  \emph{et~al.}(2005)\citenamefont{Mavropoulos, Le\ifmmode \check{z}\else
  \v{z}\fi{}ai\ifmmode~\acute{c}\else \'{c}\fi{}, and
  Bl\"ugel}}]{mavropoulos:174428}
\bibinfo{author}{\bibnamefont{Mavropoulos}, \bibfnamefont{P.}},
  \bibinfo{author}{\bibfnamefont{M.}~\bibnamefont{Le\ifmmode \check{z}\else
  \v{z}\fi{}ai\ifmmode~\acute{c}\else \'{c}\fi{}}}, and
  \bibinfo{author}{\bibfnamefont{S.}~\bibnamefont{Bl\"ugel}},
  \bibinfo{year}{2005}, \bibinfo{journal}{Phys. Rev. B}
  \textbf{\bibinfo{volume}{72}}(\bibinfo{number}{17}), \bibinfo{pages}{174428}.

\bibitem[{\citenamefont{Mavropoulos}
  \emph{et~al.}(2004)\citenamefont{Mavropoulos, Galanakis, Popescu, and
  Dederichs}}]{Mavropoulos:S5759}
\bibinfo{author}{\bibnamefont{Mavropoulos}, \bibfnamefont{P.}},
  \bibinfo{author}{\bibfnamefont{I.}~\bibnamefont{Galanakis}},
  \bibinfo{author}{\bibfnamefont{V.}~\bibnamefont{Popescu}}, and
  \bibinfo{author}{\bibfnamefont{P.~H.} \bibnamefont{Dederichs}},
  \bibinfo{year}{2004}, \bibinfo{journal}{J. Phys.: Condens. Matter}
  \textbf{\bibinfo{volume}{16}}, \bibinfo{pages}{S5759}.

\bibitem[{\citenamefont{Mazin}(2000)}]{mazin:3000}
\bibinfo{author}{\bibnamefont{Mazin}, \bibfnamefont{I.~I.}},
  \bibinfo{year}{2000}, \bibinfo{journal}{Appl. Phys. Lett.}
  \textbf{\bibinfo{volume}{77}}(\bibinfo{number}{19}), \bibinfo{pages}{3000}.

\bibitem[{\citenamefont{Mazin} \emph{et~al.}(1999)\citenamefont{Mazin, Singh,
  and Ambrosch-Draxl}}]{mazin:411}
\bibinfo{author}{\bibnamefont{Mazin}, \bibfnamefont{I.~I.}},
  \bibinfo{author}{\bibfnamefont{D.~J.} \bibnamefont{Singh}}, and
  \bibinfo{author}{\bibfnamefont{C.}~\bibnamefont{Ambrosch-Draxl}},
  \bibinfo{year}{1999}, \bibinfo{journal}{Phys. Rev. B}
  \textbf{\bibinfo{volume}{59}}(\bibinfo{number}{1}), \bibinfo{pages}{411}.

\bibitem[{\citenamefont{McCann and Fal\char39{}ko}(2002)}]{McCann:134424}
\bibinfo{author}{\bibnamefont{McCann}, \bibfnamefont{E.}}, and
  \bibinfo{author}{\bibfnamefont{V.~I.} \bibnamefont{Fal\char39{}ko}},
  \bibinfo{year}{2002}, \bibinfo{journal}{Phys. Rev. B}
  \textbf{\bibinfo{volume}{66}}(\bibinfo{number}{13}), \bibinfo{pages}{134424}.

\bibitem[{\citenamefont{McCann and Fal'ko}(2003)}]{mccann:172404}
\bibinfo{author}{\bibnamefont{McCann}, \bibfnamefont{E.}}, and
  \bibinfo{author}{\bibfnamefont{V.~I.} \bibnamefont{Fal'ko}},
  \bibinfo{year}{2003}, \bibinfo{journal}{Phys. Rev. B}
  \textbf{\bibinfo{volume}{68}}(\bibinfo{number}{17}), \bibinfo{eid}{172404}.

\bibitem[{\citenamefont{McMahan} \emph{et~al.}(1988)\citenamefont{McMahan,
  Martin, and Satpathy}}]{mcmahan:6650}
\bibinfo{author}{\bibnamefont{McMahan}, \bibfnamefont{A.~K.}},
  \bibinfo{author}{\bibfnamefont{R.~M.} \bibnamefont{Martin}}, and
  \bibinfo{author}{\bibfnamefont{S.}~\bibnamefont{Satpathy}},
  \bibinfo{year}{1988}, \bibinfo{journal}{Phys. Rev. B}
  \textbf{\bibinfo{volume}{38}}(\bibinfo{number}{10}), \bibinfo{pages}{6650}.

\bibitem[{\citenamefont{Miao} \emph{et~al.}(2006)\citenamefont{Miao, LeClair,
  Gupta, Xiao, Varela, and Pennycook}}]{miao:022511}
\bibinfo{author}{\bibnamefont{Miao}, \bibfnamefont{G.~X.}},
  \bibinfo{author}{\bibfnamefont{P.}~\bibnamefont{LeClair}},
  \bibinfo{author}{\bibfnamefont{A.}~\bibnamefont{Gupta}},
  \bibinfo{author}{\bibfnamefont{G.}~\bibnamefont{Xiao}},
  \bibinfo{author}{\bibfnamefont{M.}~\bibnamefont{Varela}}, and
  \bibinfo{author}{\bibfnamefont{S.}~\bibnamefont{Pennycook}},
  \bibinfo{year}{2006}, \bibinfo{journal}{Appl. Phys. Lett.}
  \textbf{\bibinfo{volume}{89}}(\bibinfo{number}{2}), \bibinfo{eid}{022511}.

\bibitem[{\citenamefont{Millis} \emph{et~al.}(1990)\citenamefont{Millis,
  Monien, and Pines}}]{Millis:167}
\bibinfo{author}{\bibnamefont{Millis}, \bibfnamefont{A.~J.}},
  \bibinfo{author}{\bibfnamefont{H.}~\bibnamefont{Monien}}, and
  \bibinfo{author}{\bibfnamefont{D.}~\bibnamefont{Pines}},
  \bibinfo{year}{1990}, \bibinfo{journal}{Phys. Rev. B}
  \textbf{\bibinfo{volume}{42}}(\bibinfo{number}{1}), \bibinfo{pages}{167}.

\bibitem[{\citenamefont{Min} \emph{et~al.}(1986)\citenamefont{Min, Oguchi,
  Jansen, and Freeman}}]{Min:324}
\bibinfo{author}{\bibnamefont{Min}, \bibfnamefont{B.~I.}},
  \bibinfo{author}{\bibfnamefont{T.}~\bibnamefont{Oguchi}},
  \bibinfo{author}{\bibfnamefont{H.~J.~F.} \bibnamefont{Jansen}}, and
  \bibinfo{author}{\bibfnamefont{A.~J.} \bibnamefont{Freeman}},
  \bibinfo{year}{1986}, \bibinfo{journal}{Phys. Rev. B}
  \textbf{\bibinfo{volume}{33}}(\bibinfo{number}{1}), \bibinfo{pages}{324}.

\bibitem[{\citenamefont{Min} \emph{et~al.}(2004)\citenamefont{Min, Park, and
  Park}}]{Min:S5509}
\bibinfo{author}{\bibnamefont{Min}, \bibfnamefont{B.~I.}},
  \bibinfo{author}{\bibfnamefont{M.~S.} \bibnamefont{Park}}, and
  \bibinfo{author}{\bibfnamefont{J.~H.} \bibnamefont{Park}},
  \bibinfo{year}{2004}, \bibinfo{journal}{J. Phys.: Condens. Matter.}
  \textbf{\bibinfo{volume}{16}}, \bibinfo{pages}{S5509}.

\bibitem[{\citenamefont{Mizuguchi} \emph{et~al.}(2002)\citenamefont{Mizuguchi,
  Akinaga, Manago, Ono, Oshima, Shirai, Yuri, Lin, Hsieh, and
  Chen}}]{mizuguchi:7917}
\bibinfo{author}{\bibnamefont{Mizuguchi}, \bibfnamefont{M.}},
  \bibinfo{author}{\bibfnamefont{H.}~\bibnamefont{Akinaga}},
  \bibinfo{author}{\bibfnamefont{T.}~\bibnamefont{Manago}},
  \bibinfo{author}{\bibfnamefont{K.}~\bibnamefont{Ono}},
  \bibinfo{author}{\bibfnamefont{M.}~\bibnamefont{Oshima}},
  \bibinfo{author}{\bibfnamefont{M.}~\bibnamefont{Shirai}},
  \bibinfo{author}{\bibfnamefont{M.}~\bibnamefont{Yuri}},
  \bibinfo{author}{\bibfnamefont{H.~J.} \bibnamefont{Lin}},
  \bibinfo{author}{\bibfnamefont{H.~H.} \bibnamefont{Hsieh}}, and
  \bibinfo{author}{\bibfnamefont{C.~T.} \bibnamefont{Chen}},
  \bibinfo{year}{2002}, \textbf{\bibinfo{volume}{91}}(\bibinfo{number}{10}),
  \bibinfo{pages}{7917}.

\bibitem[{\citenamefont{Moodera and Mootoo}(1994)}]{moodera:6101}
\bibinfo{author}{\bibnamefont{Moodera}, \bibfnamefont{J.~S.}}, and
  \bibinfo{author}{\bibfnamefont{D.~M.} \bibnamefont{Mootoo}},
  \bibinfo{year}{1994}, \bibinfo{journal}{J. Appl. Phys}
  \textbf{\bibinfo{volume}{76}}(\bibinfo{number}{10}), \bibinfo{pages}{6101}.

\bibitem[{\citenamefont{Mori}(1965)}]{mori:399}
\bibinfo{author}{\bibnamefont{Mori}, \bibfnamefont{H.}}, \bibinfo{year}{1965},
  \bibinfo{journal}{Prog. Theor. Phys.} \textbf{\bibinfo{volume}{34}},
  \bibinfo{pages}{399}.

\bibitem[{\citenamefont{Moriya}(1963)}]{Moriya:516}
\bibinfo{author}{\bibnamefont{Moriya}, \bibfnamefont{T.}},
  \bibinfo{year}{1963}, \bibinfo{journal}{J. Phys. Soc. Jpn.}
  \textbf{\bibinfo{volume}{18}}, \bibinfo{pages}{516}.

\bibitem[{\citenamefont{Moriya}(1985)}]{Moriya:1985}
\bibinfo{author}{\bibnamefont{Moriya}, \bibfnamefont{T.}},
  \bibinfo{year}{1985}, \emph{\bibinfo{title}{Spin Fluctuations in Itinerant
  Electron Magnetism}} (\bibinfo{publisher}{Springer, Berlin}).

\bibitem[{\citenamefont{Moriya}(1994)}]{Moriya:1994}
\bibinfo{author}{\bibnamefont{Moriya}, \bibfnamefont{T.}},
  \bibinfo{year}{1994}, \emph{\bibinfo{title}{Spectroscopy of Mott Insulators
  and Correlated Metals}} (\bibinfo{publisher}{Springer, Berlin}).

\bibitem[{\citenamefont{Mortonx} \emph{et~al.}(2002)\citenamefont{Mortonx,
  Waddill, Kim, Schuller, Chambers, and Tobin}}]{Mortonx:L451}
\bibinfo{author}{\bibnamefont{Mortonx}, \bibfnamefont{S.~A.}},
  \bibinfo{author}{\bibfnamefont{G.~D.} \bibnamefont{Waddill}},
  \bibinfo{author}{\bibfnamefont{S.}~\bibnamefont{Kim}},
  \bibinfo{author}{\bibfnamefont{I.~K.} \bibnamefont{Schuller}},
  \bibinfo{author}{\bibfnamefont{S.~A.} \bibnamefont{Chambers}}, and
  \bibinfo{author}{\bibfnamefont{J.~G.} \bibnamefont{Tobin}},
  \bibinfo{year}{2002}, \bibinfo{journal}{Surface Science}
  \textbf{\bibinfo{volume}{513}}(\bibinfo{number}{3}), \bibinfo{pages}{L451}.

\bibitem[{\citenamefont{Mott}(1974)}]{Mott1974}
\bibinfo{author}{\bibnamefont{Mott}, \bibfnamefont{N.~F.}},
  \bibinfo{year}{1974}, \emph{\bibinfo{title}{Metal-Insulator Transitions}}
  (\bibinfo{publisher}{Taylor and Francis, London}).

\bibitem[{\citenamefont{Mott}(1980)}]{Mott:327}
\bibinfo{author}{\bibnamefont{Mott}, \bibfnamefont{N.~F.}},
  \bibinfo{year}{1980}, \bibinfo{journal}{Philos. Mag.}
  \textbf{\bibinfo{volume}{B42}}, \bibinfo{pages}{327}.

\bibitem[{\citenamefont{Nadgorny}(2007)}]{Nadgorny:315209}
\bibinfo{author}{\bibnamefont{Nadgorny}, \bibfnamefont{B.}},
  \bibinfo{year}{2007}, \bibinfo{journal}{J. Phys.: Condens. Matter}
  \textbf{\bibinfo{volume}{19}}, \bibinfo{pages}{315209}.

\bibitem[{\citenamefont{Nadgorny} \emph{et~al.}(2001)\citenamefont{Nadgorny,
  Mazin, Osofsky, Soulen, Broussard, Stroud, Singh, Harris, Arsenov, and
  Mukovskii}}]{Nadgorny:184433}
\bibinfo{author}{\bibnamefont{Nadgorny}, \bibfnamefont{B.}},
  \bibinfo{author}{\bibfnamefont{I.~I.} \bibnamefont{Mazin}},
  \bibinfo{author}{\bibfnamefont{M.}~\bibnamefont{Osofsky}},
  \bibinfo{author}{\bibfnamefont{R.~J.} \bibnamefont{Soulen}},
  \bibinfo{author}{\bibfnamefont{P.}~\bibnamefont{Broussard}},
  \bibinfo{author}{\bibfnamefont{R.~M.} \bibnamefont{Stroud}},
  \bibinfo{author}{\bibfnamefont{D.~J.} \bibnamefont{Singh}},
  \bibinfo{author}{\bibfnamefont{V.~G.} \bibnamefont{Harris}},
  \bibinfo{author}{\bibfnamefont{A.}~\bibnamefont{Arsenov}}, and
  \bibinfo{author}{\bibfnamefont{Y.}~\bibnamefont{Mukovskii}},
  \bibinfo{year}{2001}, \bibinfo{journal}{Phys. Rev. B}
  \textbf{\bibinfo{volume}{63}}(\bibinfo{number}{18}), \bibinfo{pages}{184433}.

\bibitem[{\citenamefont{Nagaev}(1983)}]{nagaev:1983}
\bibinfo{author}{\bibnamefont{Nagaev}, \bibfnamefont{E.~L.}},
  \bibinfo{year}{1983}, \emph{\bibinfo{title}{Physics of Magnetic
  Semiconductors}} (\bibinfo{publisher}{Mir, Moscow}).

\bibitem[{\citenamefont{Nagaev}(2001)}]{nagaev:387}
\bibinfo{author}{\bibnamefont{Nagaev}, \bibfnamefont{E.~L.}},
  \bibinfo{year}{2001}, \bibinfo{journal}{Physics Reports}
  \textbf{\bibinfo{volume}{346}}(\bibinfo{number}{6}), \bibinfo{pages}{387}.

\bibitem[{\citenamefont{Nagao} \emph{et~al.}(2004)\citenamefont{Nagao, Shirai,
  and Miura}}]{na.ma.04}
\bibinfo{author}{\bibnamefont{Nagao}, \bibfnamefont{K.}},
  \bibinfo{author}{\bibfnamefont{M.}~\bibnamefont{Shirai}}, and
  \bibinfo{author}{\bibfnamefont{Y.}~\bibnamefont{Miura}},
  \bibinfo{year}{2004}, \bibinfo{journal}{J. Phys.: Condens. Matter}
  \textbf{\bibinfo{volume}{16}}, \bibinfo{pages}{S5725}.

\bibitem[{\citenamefont{Nagaoka}(1966)}]{Nagaoka:392}
\bibinfo{author}{\bibnamefont{Nagaoka}, \bibfnamefont{Y.}},
  \bibinfo{year}{1966}, \bibinfo{journal}{Phys. Rev.}
  \textbf{\bibinfo{volume}{147}}(\bibinfo{number}{1}), \bibinfo{pages}{392}.

\bibitem[{\citenamefont{Nakano}(1957)}]{nakano:145}
\bibinfo{author}{\bibnamefont{Nakano}, \bibfnamefont{H.}},
  \bibinfo{year}{1957}, \bibinfo{journal}{Prog. Theor. Phys.}
  \textbf{\bibinfo{volume}{17}}, \bibinfo{pages}{145}.

\bibitem[{\citenamefont{Nanda and Dasgupta}(2003)}]{nanda:7307}
\bibinfo{author}{\bibnamefont{Nanda}, \bibfnamefont{B.~R.~K.}}, and
  \bibinfo{author}{\bibfnamefont{I.}~\bibnamefont{Dasgupta}},
  \bibinfo{year}{2003}, \bibinfo{journal}{J. Phys.: Condens. Matter}
  \textbf{\bibinfo{volume}{15}}, \bibinfo{pages}{7307}.

\bibitem[{\citenamefont{Nazmul} \emph{et~al.}(2002)\citenamefont{Nazmul,
  Sugahara, and Tanaka}}]{nazmul:3120}
\bibinfo{author}{\bibnamefont{Nazmul}, \bibfnamefont{A.~M.}},
  \bibinfo{author}{\bibfnamefont{S.}~\bibnamefont{Sugahara}}, and
  \bibinfo{author}{\bibfnamefont{M.}~\bibnamefont{Tanaka}},
  \bibinfo{year}{2002}, \bibinfo{journal}{Appl. Phys. Lett.}
  \textbf{\bibinfo{volume}{80}}(\bibinfo{number}{17}), \bibinfo{pages}{3120}.

\bibitem[{\citenamefont{Norman and Freeman}(1986)}]{norman:8896}
\bibinfo{author}{\bibnamefont{Norman}, \bibfnamefont{M.~R.}}, and
  \bibinfo{author}{\bibfnamefont{A.~J.} \bibnamefont{Freeman}},
  \bibinfo{year}{1986}, \bibinfo{journal}{Phys. Rev. B}
  \textbf{\bibinfo{volume}{33}}(\bibinfo{number}{12}), \bibinfo{pages}{8896}.

\bibitem[{\citenamefont{Nozieres}(1964)}]{Nozieres:1964}
\bibinfo{author}{\bibnamefont{Nozieres}, \bibfnamefont{P.}},
  \bibinfo{year}{1964}, \emph{\bibinfo{title}{Theory of Interacting Fermi
  Systems}} (\bibinfo{publisher}{Benjamin, New York}).

\bibitem[{\citenamefont{Obermeier} \emph{et~al.}(1997)\citenamefont{Obermeier,
  Pruschke, and Keller}}]{Obermeier:8479}
\bibinfo{author}{\bibnamefont{Obermeier}, \bibfnamefont{T.}},
  \bibinfo{author}{\bibfnamefont{T.}~\bibnamefont{Pruschke}}, and
  \bibinfo{author}{\bibfnamefont{J.}~\bibnamefont{Keller}},
  \bibinfo{year}{1997}, \bibinfo{journal}{Phys. Rev. B}
  \textbf{\bibinfo{volume}{56}}(\bibinfo{number}{14}), \bibinfo{pages}{R8479}.

\bibitem[{\citenamefont{Ohno}(1998{\natexlab{a}})}]{ohno:951}
\bibinfo{author}{\bibnamefont{Ohno}, \bibfnamefont{H.}},
  \bibinfo{year}{1998}{\natexlab{a}}, \bibinfo{journal}{Science}
  \textbf{\bibinfo{volume}{281}}, \bibinfo{pages}{951}.

\bibitem[{\citenamefont{Ohno}(1998{\natexlab{b}})}]{ohno:110}
\bibinfo{author}{\bibnamefont{Ohno}, \bibfnamefont{H.}},
  \bibinfo{year}{1998}{\natexlab{b}}, \bibinfo{journal}{J. Magn. Magn. Mater.}
  \textbf{\bibinfo{volume}{200}}, \bibinfo{pages}{110}.

\bibitem[{\citenamefont{Okabayashi}
  \emph{et~al.}(2004)\citenamefont{Okabayashi, Mizuguchi, Ono, Oshima,
  Fujimori, Kuramochi, and Akinaga}}]{okabayashi:233305}
\bibinfo{author}{\bibnamefont{Okabayashi}, \bibfnamefont{J.}},
  \bibinfo{author}{\bibfnamefont{M.}~\bibnamefont{Mizuguchi}},
  \bibinfo{author}{\bibfnamefont{K.}~\bibnamefont{Ono}},
  \bibinfo{author}{\bibfnamefont{M.}~\bibnamefont{Oshima}},
  \bibinfo{author}{\bibfnamefont{A.}~\bibnamefont{Fujimori}},
  \bibinfo{author}{\bibfnamefont{H.}~\bibnamefont{Kuramochi}}, and
  \bibinfo{author}{\bibfnamefont{H.}~\bibnamefont{Akinaga}},
  \bibinfo{year}{2004}, \bibinfo{journal}{Phys. Rev. B}
  \textbf{\bibinfo{volume}{70}}(\bibinfo{number}{23}), \bibinfo{eid}{233305}.

\bibitem[{\citenamefont{Orgassa} \emph{et~al.}(1999)\citenamefont{Orgassa,
  Fujiwara, Schulthess, and Butler}}]{Orgassa:13237}
\bibinfo{author}{\bibnamefont{Orgassa}, \bibfnamefont{D.}},
  \bibinfo{author}{\bibfnamefont{H.}~\bibnamefont{Fujiwara}},
  \bibinfo{author}{\bibfnamefont{T.~C.} \bibnamefont{Schulthess}}, and
  \bibinfo{author}{\bibfnamefont{W.~H.} \bibnamefont{Butler}},
  \bibinfo{year}{1999}, \bibinfo{journal}{Phys. Rev. B}
  \textbf{\bibinfo{volume}{60}}(\bibinfo{number}{19}), \bibinfo{pages}{13237}.

\bibitem[{\citenamefont{Orgassa} \emph{et~al.}(2000)\citenamefont{Orgassa,
  Fujiwara, Schulthess, and Butler}}]{orgassa:5870}
\bibinfo{author}{\bibnamefont{Orgassa}, \bibfnamefont{D.}},
  \bibinfo{author}{\bibfnamefont{H.}~\bibnamefont{Fujiwara}},
  \bibinfo{author}{\bibfnamefont{T.~C.} \bibnamefont{Schulthess}}, and
  \bibinfo{author}{\bibfnamefont{W.~H.} \bibnamefont{Butler}},
  \bibinfo{year}{2000}, \bibinfo{journal}{J. Appl. Phys.}
  \textbf{\bibinfo{volume}{87}}(\bibinfo{number}{9}), \bibinfo{pages}{5870}.

\bibitem[{\citenamefont{Otto} \emph{et~al.}(1989)\citenamefont{Otto, van
  Woerden, van~der Valk, Wijngaard, van Bruggen, and Haas}}]{otto:2351}
\bibinfo{author}{\bibnamefont{Otto}, \bibfnamefont{M.~J.}},
  \bibinfo{author}{\bibfnamefont{R.~A.~M.} \bibnamefont{van Woerden}},
  \bibinfo{author}{\bibfnamefont{P.~J.} \bibnamefont{van~der Valk}},
  \bibinfo{author}{\bibfnamefont{J.}~\bibnamefont{Wijngaard}},
  \bibinfo{author}{\bibfnamefont{C.~F.} \bibnamefont{van Bruggen}}, and
  \bibinfo{author}{\bibfnamefont{C.}~\bibnamefont{Haas}}, \bibinfo{year}{1989},
  \bibinfo{journal}{J. Phys.: Condens. Matter} \textbf{\bibinfo{volume}{1}},
  \bibinfo{pages}{2351}.

\bibitem[{\citenamefont{Ozdogan} \emph{et~al.}(2006)\citenamefont{Ozdogan,
  Galanakis, Sasioglu, and Aktas}}]{Ozdogan:2905}
\bibinfo{author}{\bibnamefont{Ozdogan}, \bibfnamefont{K.}},
  \bibinfo{author}{\bibfnamefont{I.}~\bibnamefont{Galanakis}},
  \bibinfo{author}{\bibfnamefont{E.}~\bibnamefont{Sasioglu}}, and
  \bibinfo{author}{\bibfnamefont{B.}~\bibnamefont{Aktas}},
  \bibinfo{year}{2006}, \bibinfo{journal}{J. Phys.: Condens. Matter}
  \textbf{\bibinfo{volume}{18}}, \bibinfo{pages}{2905}.

\bibitem[{\citenamefont{Paessler}(1999)}]{paes.99}
\bibinfo{author}{\bibnamefont{Paessler}, \bibfnamefont{R.}},
  \bibinfo{year}{1999}, \bibinfo{journal}{Phys. Stat. Sol. (b)}
  \textbf{\bibinfo{volume}{216}}, \bibinfo{pages}{975}.

\bibitem[{\citenamefont{Park} \emph{et~al.}(1998)\citenamefont{Park, Vescovo,
  Kim, Kwon, Ramesh, and Venkatesan}}]{Park:794}
\bibinfo{author}{\bibnamefont{Park}, \bibfnamefont{J.~H.}},
  \bibinfo{author}{\bibfnamefont{E.}~\bibnamefont{Vescovo}},
  \bibinfo{author}{\bibfnamefont{H.~J.} \bibnamefont{Kim}},
  \bibinfo{author}{\bibfnamefont{C.}~\bibnamefont{Kwon}},
  \bibinfo{author}{\bibfnamefont{R.}~\bibnamefont{Ramesh}}, and
  \bibinfo{author}{\bibfnamefont{T.}~\bibnamefont{Venkatesan}},
  \bibinfo{year}{1998}, \bibinfo{journal}{Nature}
  \textbf{\bibinfo{volume}{392}}, \bibinfo{pages}{794}.

\bibitem[{\citenamefont{Park} \emph{et~al.}(2001)\citenamefont{Park, Kwon, and
  Min}}]{Park:100403}
\bibinfo{author}{\bibnamefont{Park}, \bibfnamefont{M.~S.}},
  \bibinfo{author}{\bibfnamefont{S.~K.} \bibnamefont{Kwon}}, and
  \bibinfo{author}{\bibfnamefont{B.~I.} \bibnamefont{Min}},
  \bibinfo{year}{2001}, \bibinfo{journal}{Phys. Rev. B}
  \textbf{\bibinfo{volume}{64}}(\bibinfo{number}{10}), \bibinfo{pages}{100403}.

\bibitem[{\citenamefont{Pickett}(1998)}]{Pickett:10613}
\bibinfo{author}{\bibnamefont{Pickett}, \bibfnamefont{W.~E.}},
  \bibinfo{year}{1998}, \bibinfo{journal}{Phys. Rev. B}
  \textbf{\bibinfo{volume}{57}}(\bibinfo{number}{17}), \bibinfo{pages}{10613}.

\bibitem[{\citenamefont{Pickett} \emph{et~al.}(1998)\citenamefont{Pickett,
  Erwin, and Ethridge}}]{pickett:1201}
\bibinfo{author}{\bibnamefont{Pickett}, \bibfnamefont{W.~E.}},
  \bibinfo{author}{\bibfnamefont{S.~C.} \bibnamefont{Erwin}}, and
  \bibinfo{author}{\bibfnamefont{E.~C.} \bibnamefont{Ethridge}},
  \bibinfo{year}{1998}, \bibinfo{journal}{Phys. Rev. B}
  \textbf{\bibinfo{volume}{58}}(\bibinfo{number}{3}), \bibinfo{pages}{1201}.

\bibitem[{\citenamefont{Pickett and Eschrig}(2007)}]{Pickett:315203}
\bibinfo{author}{\bibnamefont{Pickett}, \bibfnamefont{W.~E.}}, and
  \bibinfo{author}{\bibfnamefont{H.}~\bibnamefont{Eschrig}},
  \bibinfo{year}{2007}, \bibinfo{journal}{J. Phys.: Condens. Matter}
  \textbf{\bibinfo{volume}{19}}, \bibinfo{pages}{315203}.

\bibitem[{\citenamefont{Pickett and Singh}(1996)}]{Pickett:1146}
\bibinfo{author}{\bibnamefont{Pickett}, \bibfnamefont{W.~E.}}, and
  \bibinfo{author}{\bibfnamefont{D.~J.} \bibnamefont{Singh}},
  \bibinfo{year}{1996}, \bibinfo{journal}{Phys. Rev. B}
  \textbf{\bibinfo{volume}{53}}(\bibinfo{number}{3}), \bibinfo{pages}{1146}.

\bibitem[{\citenamefont{Pierre and Karla}(2000)}]{Pierre:74}
\bibinfo{author}{\bibnamefont{Pierre}, \bibfnamefont{J.}}, and
  \bibinfo{author}{\bibfnamefont{I.}~\bibnamefont{Karla}},
  \bibinfo{year}{2000}, \bibinfo{journal}{J. Magn. Magn. Mater.}
  \textbf{\bibinfo{volume}{217}}, \bibinfo{pages}{74}.

\bibitem[{\citenamefont{Pierre} \emph{et~al.}(1999)\citenamefont{Pierre, Karla,
  and Kaczmarska}}]{Pierre:845}
\bibinfo{author}{\bibnamefont{Pierre}, \bibfnamefont{J.}},
  \bibinfo{author}{\bibfnamefont{I.}~\bibnamefont{Karla}}, and
  \bibinfo{author}{\bibfnamefont{K.}~\bibnamefont{Kaczmarska}},
  \bibinfo{year}{1999}, \bibinfo{journal}{Physica B}
  \textbf{\bibinfo{volume}{261}}, \bibinfo{pages}{845}.

\bibitem[{\citenamefont{Porta} \emph{et~al.}(1972)\citenamefont{Porta, Marezio,
  Remeika, and Dernier}}]{porta:157}
\bibinfo{author}{\bibnamefont{Porta}, \bibfnamefont{P.}},
  \bibinfo{author}{\bibfnamefont{M.}~\bibnamefont{Marezio}},
  \bibinfo{author}{\bibfnamefont{J.~P.} \bibnamefont{Remeika}}, and
  \bibinfo{author}{\bibfnamefont{P.~D.} \bibnamefont{Dernier}},
  \bibinfo{year}{1972}, \bibinfo{journal}{Mater. Res. Bull.}
  \textbf{\bibinfo{volume}{7}}, \bibinfo{pages}{157}.

\bibitem[{\citenamefont{Pourovskii}
  \emph{et~al.}(2005)\citenamefont{Pourovskii, Katsnelson, and
  Lichtenstein}}]{pourovskii:115106}
\bibinfo{author}{\bibnamefont{Pourovskii}, \bibfnamefont{L.~V.}},
  \bibinfo{author}{\bibfnamefont{M.~I.} \bibnamefont{Katsnelson}}, and
  \bibinfo{author}{\bibfnamefont{A.~I.} \bibnamefont{Lichtenstein}},
  \bibinfo{year}{2005}, \bibinfo{journal}{Phys. Rev. B}
  \textbf{\bibinfo{volume}{72}}(\bibinfo{number}{11}), \bibinfo{eid}{115106}.

\bibitem[{\citenamefont{Pourovskii}
  \emph{et~al.}(2006)\citenamefont{Pourovskii, Katsnelson, and
  Lichtenstein}}]{pourovskii:060506}
\bibinfo{author}{\bibnamefont{Pourovskii}, \bibfnamefont{L.~V.}},
  \bibinfo{author}{\bibfnamefont{M.~I.} \bibnamefont{Katsnelson}}, and
  \bibinfo{author}{\bibfnamefont{A.~I.} \bibnamefont{Lichtenstein}},
  \bibinfo{year}{2006}, \bibinfo{journal}{Phys. Rev. B}
  \textbf{\bibinfo{volume}{73}}(\bibinfo{number}{6}), \bibinfo{eid}{060506}.

\bibitem[{\citenamefont{Prinz}(1998)}]{Prinz:1660}
\bibinfo{author}{\bibnamefont{Prinz}, \bibfnamefont{G.~A.}},
  \bibinfo{year}{1998}, \bibinfo{journal}{Science}
  \textbf{\bibinfo{volume}{282}}, \bibinfo{pages}{1660}.

\bibitem[{\citenamefont{Rabe} \emph{et~al.}(2002)\citenamefont{Rabe, Pommer,
  Samm, Oezyilmaz, Koenig, Fraune, Ruediger, Guentherodt, Senz, and
  Hesse}}]{rabe:7}
\bibinfo{author}{\bibnamefont{Rabe}, \bibfnamefont{M.}},
  \bibinfo{author}{\bibfnamefont{J.}~\bibnamefont{Pommer}},
  \bibinfo{author}{\bibfnamefont{K.}~\bibnamefont{Samm}},
  \bibinfo{author}{\bibfnamefont{B.}~\bibnamefont{Oezyilmaz}},
  \bibinfo{author}{\bibfnamefont{C.}~\bibnamefont{Koenig}},
  \bibinfo{author}{\bibfnamefont{M.}~\bibnamefont{Fraune}},
  \bibinfo{author}{\bibfnamefont{U.}~\bibnamefont{Ruediger}},
  \bibinfo{author}{\bibfnamefont{G.}~\bibnamefont{Guentherodt}},
  \bibinfo{author}{\bibfnamefont{S.}~\bibnamefont{Senz}}, and
  \bibinfo{author}{\bibfnamefont{D.}~\bibnamefont{Hesse}},
  \bibinfo{year}{2002}, \bibinfo{journal}{J. Phys.: Condens. Matter}
  \textbf{\bibinfo{volume}{14}}, \bibinfo{pages}{7}.

\bibitem[{\citenamefont{Rahman} \emph{et~al.}(2004)\citenamefont{Rahman,
  Kisaku, Kishi, Matsunaka, Dino, Nakanishi, and Kasai}}]{Rahman:S5755}
\bibinfo{author}{\bibnamefont{Rahman}, \bibfnamefont{M.~M.}},
  \bibinfo{author}{\bibfnamefont{M.}~\bibnamefont{Kisaku}},
  \bibinfo{author}{\bibfnamefont{T.}~\bibnamefont{Kishi}},
  \bibinfo{author}{\bibfnamefont{D.}~\bibnamefont{Matsunaka}},
  \bibinfo{author}{\bibfnamefont{W.~A.} \bibnamefont{Dino}},
  \bibinfo{author}{\bibfnamefont{H.}~\bibnamefont{Nakanishi}}, and
  \bibinfo{author}{\bibfnamefont{H.}~\bibnamefont{Kasai}},
  \bibinfo{year}{2004}, \bibinfo{journal}{J. Phys.: Condens. Matter.}
  \textbf{\bibinfo{volume}{16}}, \bibinfo{pages}{S5755}.

\bibitem[{\citenamefont{Ramesha} \emph{et~al.}(2004)\citenamefont{Ramesha,
  Seshadri, Ederer, He, and Subramanian}}]{ramesha:214409}
\bibinfo{author}{\bibnamefont{Ramesha}, \bibfnamefont{K.}},
  \bibinfo{author}{\bibfnamefont{R.}~\bibnamefont{Seshadri}},
  \bibinfo{author}{\bibfnamefont{C.}~\bibnamefont{Ederer}},
  \bibinfo{author}{\bibfnamefont{T.}~\bibnamefont{He}}, and
  \bibinfo{author}{\bibfnamefont{M.~A.} \bibnamefont{Subramanian}},
  \bibinfo{year}{2004}, \bibinfo{journal}{Phys. Rev. B}
  \textbf{\bibinfo{volume}{70}}(\bibinfo{number}{21}), \bibinfo{eid}{214409}.

\bibitem[{\citenamefont{Rengade}(1907)}]{Rengade:348}
\bibinfo{author}{\bibnamefont{Rengade}, \bibfnamefont{M.~E.}},
  \bibinfo{year}{1907}, \bibinfo{journal}{Ann. Chim. Phys.}
  \textbf{\bibinfo{volume}{11}}, \bibinfo{pages}{348}.

\bibitem[{\citenamefont{Ristoiu} \emph{et~al.}(2000)\citenamefont{Ristoiu,
  Nozieres, Borca, Borca, and Dowben}}]{ristoiu:2349}
\bibinfo{author}{\bibnamefont{Ristoiu}, \bibfnamefont{D.}},
  \bibinfo{author}{\bibfnamefont{J.~P.} \bibnamefont{Nozieres}},
  \bibinfo{author}{\bibfnamefont{C.~N.} \bibnamefont{Borca}},
  \bibinfo{author}{\bibfnamefont{B.}~\bibnamefont{Borca}}, and
  \bibinfo{author}{\bibfnamefont{P.~A.} \bibnamefont{Dowben}},
  \bibinfo{year}{2000}, \bibinfo{journal}{Appl. Phys. Lett.}
  \textbf{\bibinfo{volume}{76}}(\bibinfo{number}{17}), \bibinfo{pages}{2349}.

\bibitem[{\citenamefont{Roesler}(1965)}]{Roesler:K31}
\bibinfo{author}{\bibnamefont{Roesler}, \bibfnamefont{M.}},
  \bibinfo{year}{1965}, \bibinfo{journal}{Phys. Stat. Sol.}
  \textbf{\bibinfo{volume}{8}}, \bibinfo{pages}{K31}.

\bibitem[{\citenamefont{Roth}(1969{\natexlab{a}})}]{Roth:451}
\bibinfo{author}{\bibnamefont{Roth}, \bibfnamefont{L.~M.}},
  \bibinfo{year}{1969}{\natexlab{a}}, \bibinfo{journal}{Phys. Rev.}
  \textbf{\bibinfo{volume}{184}}(\bibinfo{number}{2}), \bibinfo{pages}{451}.

\bibitem[{\citenamefont{Roth}(1969{\natexlab{b}})}]{Roth:428}
\bibinfo{author}{\bibnamefont{Roth}, \bibfnamefont{L.~M.}},
  \bibinfo{year}{1969}{\natexlab{b}}, \bibinfo{journal}{Phys. Rev.}
  \textbf{\bibinfo{volume}{186}}(\bibinfo{number}{2}), \bibinfo{pages}{428}.

\bibitem[{\citenamefont{Rozenberg}(1997)}]{Rozenberg:R4855}
\bibinfo{author}{\bibnamefont{Rozenberg}, \bibfnamefont{M.~J.}},
  \bibinfo{year}{1997}, \bibinfo{journal}{Phys. Rev. B}
  \textbf{\bibinfo{volume}{55}}(\bibinfo{number}{8}), \bibinfo{pages}{R4855}.

\bibitem[{\citenamefont{Rubtsov} \emph{et~al.}(2005)\citenamefont{Rubtsov,
  Savkin, and Lichtenstein}}]{rubtsov:035122}
\bibinfo{author}{\bibnamefont{Rubtsov}, \bibfnamefont{A.~N.}},
  \bibinfo{author}{\bibfnamefont{V.~V.} \bibnamefont{Savkin}}, and
  \bibinfo{author}{\bibfnamefont{A.~I.} \bibnamefont{Lichtenstein}},
  \bibinfo{year}{2005}, \bibinfo{journal}{Phys. Rev. B}
  \textbf{\bibinfo{volume}{72}}(\bibinfo{number}{3}), \bibinfo{eid}{035122}.

\bibitem[{\citenamefont{Rusz} \emph{et~al.}(2006)\citenamefont{Rusz, Bergqvist,
  Kudrnovsky, and Turek}}]{rusz:214412}
\bibinfo{author}{\bibnamefont{Rusz}, \bibfnamefont{J.}},
  \bibinfo{author}{\bibfnamefont{L.}~\bibnamefont{Bergqvist}},
  \bibinfo{author}{\bibfnamefont{J.}~\bibnamefont{Kudrnovsky}}, and
  \bibinfo{author}{\bibfnamefont{I.}~\bibnamefont{Turek}},
  \bibinfo{year}{2006}, \bibinfo{journal}{Phys. Rev. B}
  \textbf{\bibinfo{volume}{73}}(\bibinfo{number}{21}), \bibinfo{eid}{214412}.

\bibitem[{\citenamefont{Rybchenko} \emph{et~al.}(2006)\citenamefont{Rybchenko,
  Fujishiro, Takagi, and Awano}}]{rybchenko:132509}
\bibinfo{author}{\bibnamefont{Rybchenko}, \bibfnamefont{S.~I.}},
  \bibinfo{author}{\bibfnamefont{Y.}~\bibnamefont{Fujishiro}},
  \bibinfo{author}{\bibfnamefont{H.}~\bibnamefont{Takagi}}, and
  \bibinfo{author}{\bibfnamefont{M.}~\bibnamefont{Awano}},
  \bibinfo{year}{2006}, \bibinfo{journal}{Appl. Phys. Lett.}
  \textbf{\bibinfo{volume}{89}}(\bibinfo{number}{13}), \bibinfo{eid}{132509}.

\bibitem[{\citenamefont{Sakuma}(2002)}]{saku.02}
\bibinfo{author}{\bibnamefont{Sakuma}, \bibfnamefont{A.}},
  \bibinfo{year}{2002}, \bibinfo{journal}{J. Phys. Soc. Japan}
  \textbf{\bibinfo{volume}{71}}, \bibinfo{pages}{2534}.

\bibitem[{\citenamefont{Sakuraba}
  \emph{et~al.}(2006{\natexlab{a}})\citenamefont{Sakuraba, Hattori, Oogane,
  Ando, Kato, Sakuma, Miyazaki, and Kubota}}]{sa.ha.06}
\bibinfo{author}{\bibnamefont{Sakuraba}, \bibfnamefont{Y.}},
  \bibinfo{author}{\bibfnamefont{M.}~\bibnamefont{Hattori}},
  \bibinfo{author}{\bibfnamefont{M.}~\bibnamefont{Oogane}},
  \bibinfo{author}{\bibfnamefont{Y.}~\bibnamefont{Ando}},
  \bibinfo{author}{\bibfnamefont{H.}~\bibnamefont{Kato}},
  \bibinfo{author}{\bibfnamefont{A.}~\bibnamefont{Sakuma}},
  \bibinfo{author}{\bibfnamefont{T.}~\bibnamefont{Miyazaki}}, and
  \bibinfo{author}{\bibfnamefont{H.}~\bibnamefont{Kubota}},
  \bibinfo{year}{2006}{\natexlab{a}}, \bibinfo{journal}{Appl. Phys. Lett.}
  \textbf{\bibinfo{volume}{88}}(\bibinfo{number}{19}), \bibinfo{pages}{192508}.

\bibitem[{\citenamefont{Sakuraba} \emph{et~al.}(2007)\citenamefont{Sakuraba,
  Hattori, Oogane, Kubota, Ando, Sakuma, and Miyazaki1}}]{sa.ha.07}
\bibinfo{author}{\bibnamefont{Sakuraba}, \bibfnamefont{Y.}},
  \bibinfo{author}{\bibfnamefont{M.}~\bibnamefont{Hattori}},
  \bibinfo{author}{\bibfnamefont{M.}~\bibnamefont{Oogane}},
  \bibinfo{author}{\bibfnamefont{H.}~\bibnamefont{Kubota}},
  \bibinfo{author}{\bibfnamefont{Y.}~\bibnamefont{Ando}},
  \bibinfo{author}{\bibfnamefont{A.}~\bibnamefont{Sakuma}}, and
  \bibinfo{author}{\bibfnamefont{T.}~\bibnamefont{Miyazaki1}},
  \bibinfo{year}{2007}, \bibinfo{journal}{J. Phys. D: Appl. Phys.}
  \textbf{\bibinfo{volume}{40}}(\bibinfo{number}{5}), \bibinfo{pages}{1221}.

\bibitem[{\citenamefont{Sakuraba}
  \emph{et~al.}(2006{\natexlab{b}})\citenamefont{Sakuraba, Miyakoshi, Oogane,
  Ando, Sakuma, Miyazaki, and Kubota}}]{sakuraba:052508}
\bibinfo{author}{\bibnamefont{Sakuraba}, \bibfnamefont{Y.}},
  \bibinfo{author}{\bibfnamefont{T.}~\bibnamefont{Miyakoshi}},
  \bibinfo{author}{\bibfnamefont{M.}~\bibnamefont{Oogane}},
  \bibinfo{author}{\bibfnamefont{Y.}~\bibnamefont{Ando}},
  \bibinfo{author}{\bibfnamefont{A.}~\bibnamefont{Sakuma}},
  \bibinfo{author}{\bibfnamefont{T.}~\bibnamefont{Miyazaki}}, and
  \bibinfo{author}{\bibfnamefont{H.}~\bibnamefont{Kubota}},
  \bibinfo{year}{2006}{\natexlab{b}}, \bibinfo{journal}{Appl. Phys. Lett.}
  \textbf{\bibinfo{volume}{89}}(\bibinfo{number}{5}), \bibinfo{eid}{052508}.

\bibitem[{\citenamefont{Salamon and Jaime}(2001)}]{Salamon:583}
\bibinfo{author}{\bibnamefont{Salamon}, \bibfnamefont{M.~B.}}, and
  \bibinfo{author}{\bibfnamefont{M.}~\bibnamefont{Jaime}},
  \bibinfo{year}{2001}, \bibinfo{journal}{Rev. Mod. Phys.}
  \textbf{\bibinfo{volume}{73}}(\bibinfo{number}{3}), \bibinfo{pages}{583}.

\bibitem[{\citenamefont{Sandratskii}(1998)}]{sand.98}
\bibinfo{author}{\bibnamefont{Sandratskii}, \bibfnamefont{L.~M.}},
  \bibinfo{year}{1998}, \bibinfo{journal}{Advances in Physics}
  \textbf{\bibinfo{volume}{47}}(\bibinfo{number}{1}), \bibinfo{pages}{91}.

\bibitem[{\citenamefont{Sandratskii}(2001)}]{sand.01}
\bibinfo{author}{\bibnamefont{Sandratskii}, \bibfnamefont{L.~M.}},
  \bibinfo{year}{2001}, \bibinfo{journal}{Phys. Rev. B}
  \textbf{\bibinfo{volume}{64}}(\bibinfo{number}{13}), \bibinfo{pages}{134402}.

\bibitem[{\citenamefont{Sandratskii and Bruno}(2003)}]{sa.br.03}
\bibinfo{author}{\bibnamefont{Sandratskii}, \bibfnamefont{L.~M.}}, and
  \bibinfo{author}{\bibfnamefont{P.}~\bibnamefont{Bruno}},
  \bibinfo{year}{2003}, \bibinfo{journal}{Phys. Rev. B}
  \textbf{\bibinfo{volume}{67}}(\bibinfo{number}{21}), \bibinfo{pages}{214402}.

\bibitem[{\citenamefont{Sanyal} \emph{et~al.}(2003)\citenamefont{Sanyal,
  Bergqvist, and Eriksson}}]{sanyal:054417}
\bibinfo{author}{\bibnamefont{Sanyal}, \bibfnamefont{B.}},
  \bibinfo{author}{\bibfnamefont{L.}~\bibnamefont{Bergqvist}}, and
  \bibinfo{author}{\bibfnamefont{O.}~\bibnamefont{Eriksson}},
  \bibinfo{year}{2003}, \bibinfo{journal}{Phys. Rev. B}
  \textbf{\bibinfo{volume}{68}}(\bibinfo{number}{5}), \bibinfo{eid}{054417}.

\bibitem[{\citenamefont{Sasioglu}
  \emph{et~al.}(2005{\natexlab{a}})\citenamefont{Sasioglu, Galanakis,
  Sandratskii, and Bruno}}]{sa.ga.05}
\bibinfo{author}{\bibnamefont{Sasioglu}, \bibfnamefont{E.}},
  \bibinfo{author}{\bibfnamefont{I.}~\bibnamefont{Galanakis}},
  \bibinfo{author}{\bibfnamefont{L.~M.} \bibnamefont{Sandratskii}}, and
  \bibinfo{author}{\bibfnamefont{P.}~\bibnamefont{Bruno}},
  \bibinfo{year}{2005}{\natexlab{a}}, \bibinfo{journal}{J. Phys.: Condens.
  Matter} \textbf{\bibinfo{volume}{17}}, \bibinfo{pages}{3915}.

\bibitem[{\citenamefont{Sasioglu}
  \emph{et~al.}(2005{\natexlab{b}})\citenamefont{Sasioglu, Sandratskii, and
  Bruno}}]{sasioglu:063523}
\bibinfo{author}{\bibnamefont{Sasioglu}, \bibfnamefont{E.}},
  \bibinfo{author}{\bibfnamefont{L.~M.} \bibnamefont{Sandratskii}}, and
  \bibinfo{author}{\bibfnamefont{P.}~\bibnamefont{Bruno}},
  \bibinfo{year}{2005}{\natexlab{b}}, \bibinfo{journal}{Journal of Applied
  Physics} \textbf{\bibinfo{volume}{98}}(\bibinfo{number}{6}),
  \bibinfo{eid}{063523} (pages~\bibinfo{numpages}{6}).

\bibitem[{\citenamefont{Sasioglu}
  \emph{et~al.}(2005{\natexlab{c}})\citenamefont{Sasioglu, Sandratskii, Bruno,
  and Galanakis}}]{sasioglu:184415}
\bibinfo{author}{\bibnamefont{Sasioglu}, \bibfnamefont{E.}},
  \bibinfo{author}{\bibfnamefont{L.~M.} \bibnamefont{Sandratskii}},
  \bibinfo{author}{\bibfnamefont{P.}~\bibnamefont{Bruno}}, and
  \bibinfo{author}{\bibfnamefont{I.}~\bibnamefont{Galanakis}},
  \bibinfo{year}{2005}{\natexlab{c}}, \bibinfo{journal}{Phys. Rev. B}
  \textbf{\bibinfo{volume}{72}}(\bibinfo{number}{18}), \bibinfo{eid}{184415}.

\bibitem[{\citenamefont{Savkin} \emph{et~al.}(2005)\citenamefont{Savkin,
  Rubtsov, Katsnelson, and Lichtenstein}}]{savkin:026402}
\bibinfo{author}{\bibnamefont{Savkin}, \bibfnamefont{V.~V.}},
  \bibinfo{author}{\bibfnamefont{A.~N.} \bibnamefont{Rubtsov}},
  \bibinfo{author}{\bibfnamefont{M.~I.} \bibnamefont{Katsnelson}}, and
  \bibinfo{author}{\bibfnamefont{A.~I.} \bibnamefont{Lichtenstein}},
  \bibinfo{year}{2005}, \bibinfo{journal}{Phys. Rev. Lett.}
  \textbf{\bibinfo{volume}{94}}(\bibinfo{number}{2}), \bibinfo{eid}{026402}.

\bibitem[{\citenamefont{Savrasov and Kotliar}(2004)}]{savrasov:245101}
\bibinfo{author}{\bibnamefont{Savrasov}, \bibfnamefont{S.~Y.}}, and
  \bibinfo{author}{\bibfnamefont{G.}~\bibnamefont{Kotliar}},
  \bibinfo{year}{2004}, \bibinfo{journal}{Phys. Rev. B}
  \textbf{\bibinfo{volume}{69}}(\bibinfo{number}{24}), \bibinfo{eid}{245101}.

\bibitem[{\citenamefont{Schaf} \emph{et~al.}(1983)\citenamefont{Schaf, Dang,
  Veillet, and Campbell}}]{sc.da.83}
\bibinfo{author}{\bibnamefont{Schaf}, \bibfnamefont{J.}},
  \bibinfo{author}{\bibfnamefont{K.~L.} \bibnamefont{Dang}},
  \bibinfo{author}{\bibfnamefont{P.}~\bibnamefont{Veillet}}, and
  \bibinfo{author}{\bibfnamefont{I.~A.} \bibnamefont{Campbell}},
  \bibinfo{year}{1983}, \bibinfo{journal}{J. Phys. F: Met. Phys}
  \textbf{\bibinfo{volume}{13}}, \bibinfo{pages}{1311}.

\bibitem[{\citenamefont{Schlottmann}(2003)}]{schlottmann:174419}
\bibinfo{author}{\bibnamefont{Schlottmann}, \bibfnamefont{P.}},
  \bibinfo{year}{2003}, \bibinfo{journal}{Phys. Rev. B}
  \textbf{\bibinfo{volume}{67}}(\bibinfo{number}{17}), \bibinfo{eid}{174419}.

\bibitem[{\citenamefont{Schwarz}(1986)}]{schwarz:L211}
\bibinfo{author}{\bibnamefont{Schwarz}, \bibfnamefont{K.}},
  \bibinfo{year}{1986}, \bibinfo{journal}{Journal of Physics F-Metal Physics}
  \textbf{\bibinfo{volume}{16}}(\bibinfo{number}{9}), \bibinfo{pages}{L211}.

\bibitem[{\citenamefont{Senateur} \emph{et~al.}(1972)\citenamefont{Senateur,
  Ronault, Fruchart, and Fruchart}}]{Senateur:226}
\bibinfo{author}{\bibnamefont{Senateur}, \bibfnamefont{J.~P.}},
  \bibinfo{author}{\bibfnamefont{A.}~\bibnamefont{Ronault}},
  \bibinfo{author}{\bibfnamefont{R.}~\bibnamefont{Fruchart}}, and
  \bibinfo{author}{\bibfnamefont{D.}~\bibnamefont{Fruchart}},
  \bibinfo{year}{1972}, \bibinfo{journal}{J. Sol. St. Chem.}
  \textbf{\bibinfo{volume}{5}}, \bibinfo{pages}{226}.

\bibitem[{\citenamefont{Shirai}(2003)}]{shirai:6844}
\bibinfo{author}{\bibnamefont{Shirai}, \bibfnamefont{M.}},
  \bibinfo{year}{2003}, \bibinfo{journal}{J. Appl. Phys.}
  \textbf{\bibinfo{volume}{93}}(\bibinfo{number}{10}), \bibinfo{pages}{6844}.

\bibitem[{\citenamefont{Shirai} \emph{et~al.}(1998)\citenamefont{Shirai, Ogawa,
  Kitagawa, and Suzuki}}]{shirai:1383}
\bibinfo{author}{\bibnamefont{Shirai}, \bibfnamefont{M.}},
  \bibinfo{author}{\bibfnamefont{T.}~\bibnamefont{Ogawa}},
  \bibinfo{author}{\bibfnamefont{I.}~\bibnamefont{Kitagawa}}, and
  \bibinfo{author}{\bibfnamefont{N.}~\bibnamefont{Suzuki}},
  \bibinfo{year}{1998}, \bibinfo{journal}{J. Magn. Magn. Mater.}
  \textbf{\bibinfo{volume}{177-181}}, \bibinfo{pages}{1383}.

\bibitem[{\citenamefont{Shirakawa} \emph{et~al.}(1977)\citenamefont{Shirakawa,
  Louis, MacDiarmid, Chiang, and Heeger}}]{Shirakawa:578}
\bibinfo{author}{\bibnamefont{Shirakawa}, \bibfnamefont{H.}},
  \bibinfo{author}{\bibfnamefont{E.~J.} \bibnamefont{Louis}},
  \bibinfo{author}{\bibfnamefont{A.~G.} \bibnamefont{MacDiarmid}},
  \bibinfo{author}{\bibfnamefont{C.~K.} \bibnamefont{Chiang}}, and
  \bibinfo{author}{\bibfnamefont{A.~J.} \bibnamefont{Heeger}},
  \bibinfo{year}{1977}, \bibinfo{journal}{J. Chem. Soc., Chem. Commun.}
  \textbf{\bibinfo{volume}{16}}, \bibinfo{pages}{578}.

\bibitem[{\citenamefont{Silin and Solontsov}(1984)}]{Silin:1080}
\bibinfo{author}{\bibnamefont{Silin}, \bibfnamefont{V.~P.}}, and
  \bibinfo{author}{\bibfnamefont{A.~Z.} \bibnamefont{Solontsov}},
  \bibinfo{year}{1984}, \bibinfo{journal}{Fiz. Metallov. Metalloved.}
  \textbf{\bibinfo{volume}{58}}, \bibinfo{pages}{1080}.

\bibitem[{\citenamefont{Singley} \emph{et~al.}(1999)\citenamefont{Singley,
  Weber, Basov, Barry, and Coey}}]{singley:4126}
\bibinfo{author}{\bibnamefont{Singley}, \bibfnamefont{E.~J.}},
  \bibinfo{author}{\bibfnamefont{C.~P.} \bibnamefont{Weber}},
  \bibinfo{author}{\bibfnamefont{D.~N.} \bibnamefont{Basov}},
  \bibinfo{author}{\bibfnamefont{A.}~\bibnamefont{Barry}}, and
  \bibinfo{author}{\bibfnamefont{J.~M.~D.} \bibnamefont{Coey}},
  \bibinfo{year}{1999}, \bibinfo{journal}{Phys. Rev. B}
  \textbf{\bibinfo{volume}{60}}(\bibinfo{number}{6}), \bibinfo{pages}{4126}.

\bibitem[{\citenamefont{Skomski}(2007)}]{Skomski:315202}
\bibinfo{author}{\bibnamefont{Skomski}, \bibfnamefont{R.}},
  \bibinfo{year}{2007}, \bibinfo{journal}{J. Phys.: Condens. Matter}
  \textbf{\bibinfo{volume}{19}}, \bibinfo{pages}{315202}.

\bibitem[{\citenamefont{Skomski and Dowben}(2002)}]{skomski:544}
\bibinfo{author}{\bibnamefont{Skomski}, \bibfnamefont{R.}}, and
  \bibinfo{author}{\bibfnamefont{P.~A.} \bibnamefont{Dowben}},
  \bibinfo{year}{2002}, \bibinfo{journal}{Europhys. Lett.,}
  \textbf{\bibinfo{volume}{58}}, \bibinfo{pages}{544}.

\bibitem[{\citenamefont{Sokolov} \emph{et~al.}(1977)\citenamefont{Sokolov,
  Grebennikov, and Turov}}]{Sokolov:383}
\bibinfo{author}{\bibnamefont{Sokolov}, \bibfnamefont{O.~B.}},
  \bibinfo{author}{\bibfnamefont{V.~I.} \bibnamefont{Grebennikov}}, and
  \bibinfo{author}{\bibfnamefont{E.~A.} \bibnamefont{Turov}},
  \bibinfo{year}{1977}, \bibinfo{journal}{Phys. Stat. Sol. (b)}
  \textbf{\bibinfo{volume}{83}}, \bibinfo{pages}{383}.

\bibitem[{\citenamefont{Solovyev} \emph{et~al.}(1996)\citenamefont{Solovyev,
  Hamada, and Terakura}}]{solovyev:7158}
\bibinfo{author}{\bibnamefont{Solovyev}, \bibfnamefont{I.}},
  \bibinfo{author}{\bibfnamefont{N.}~\bibnamefont{Hamada}}, and
  \bibinfo{author}{\bibfnamefont{K.}~\bibnamefont{Terakura}},
  \bibinfo{year}{1996}, \bibinfo{journal}{Phys. Rev. B}
  \textbf{\bibinfo{volume}{53}}(\bibinfo{number}{11}), \bibinfo{pages}{7158}.

\bibitem[{\citenamefont{Solovyev and Imada}(2005)}]{solovyev:045103}
\bibinfo{author}{\bibnamefont{Solovyev}, \bibfnamefont{I.~V.}}, and
  \bibinfo{author}{\bibfnamefont{M.}~\bibnamefont{Imada}},
  \bibinfo{year}{2005}, \bibinfo{journal}{Phys. Rev. B}
  \textbf{\bibinfo{volume}{71}}(\bibinfo{number}{4}), \bibinfo{eid}{045103}.

\bibitem[{\citenamefont{Sorantin and Schwarz}(1992)}]{sorantin:567}
\bibinfo{author}{\bibnamefont{Sorantin}, \bibfnamefont{P.~I.}}, and
  \bibinfo{author}{\bibfnamefont{K.}~\bibnamefont{Schwarz}},
  \bibinfo{year}{1992}, \bibinfo{journal}{Inorg. Chem.}
  \textbf{\bibinfo{volume}{31}}, \bibinfo{pages}{567}.

\bibitem[{\citenamefont{Soulen} \emph{et~al.}(1998)\citenamefont{Soulen, Byers,
  Osofsky, Nadgorny, Ambrose, Cheng, Broussard, Tanaka, Nowak, Moodera, Barry,
  and Coey}}]{soulen:85}
\bibinfo{author}{\bibnamefont{Soulen}, \bibfnamefont{R.~J.}},
  \bibinfo{author}{\bibfnamefont{J.~M.} \bibnamefont{Byers}},
  \bibinfo{author}{\bibfnamefont{M.~S.} \bibnamefont{Osofsky}},
  \bibinfo{author}{\bibfnamefont{B.}~\bibnamefont{Nadgorny}},
  \bibinfo{author}{\bibfnamefont{T.}~\bibnamefont{Ambrose}},
  \bibinfo{author}{\bibfnamefont{S.~F.} \bibnamefont{Cheng}},
  \bibinfo{author}{\bibfnamefont{P.~R.} \bibnamefont{Broussard}},
  \bibinfo{author}{\bibfnamefont{C.~T.} \bibnamefont{Tanaka}},
  \bibinfo{author}{\bibfnamefont{J.}~\bibnamefont{Nowak}},
  \bibinfo{author}{\bibfnamefont{J.~S.} \bibnamefont{Moodera}},
  \bibinfo{author}{\bibfnamefont{A.}~\bibnamefont{Barry}}, and
  \bibinfo{author}{\bibfnamefont{J.~M.~D.} \bibnamefont{Coey}},
  \bibinfo{year}{1998}, \bibinfo{journal}{Science}
  \textbf{\bibinfo{volume}{282}}(\bibinfo{number}{5386}), \bibinfo{pages}{85}.

\bibitem[{\citenamefont{Soulen} \emph{et~al.}(1999)\citenamefont{Soulen,
  Osofsky, Nadgorny, Ambrose, Broussard, Cheng, Byers, Tanaka, Nowack, Moodera,
  Laprade, Barry} \emph{et~al.}}]{soulen:4589}
\bibinfo{author}{\bibnamefont{Soulen}, \bibfnamefont{R.~J.}},
  \bibinfo{author}{\bibfnamefont{M.~S.} \bibnamefont{Osofsky}},
  \bibinfo{author}{\bibfnamefont{B.}~\bibnamefont{Nadgorny}},
  \bibinfo{author}{\bibfnamefont{T.}~\bibnamefont{Ambrose}},
  \bibinfo{author}{\bibfnamefont{P.}~\bibnamefont{Broussard}},
  \bibinfo{author}{\bibfnamefont{S.~F.} \bibnamefont{Cheng}},
  \bibinfo{author}{\bibfnamefont{J.}~\bibnamefont{Byers}},
  \bibinfo{author}{\bibfnamefont{C.~T.} \bibnamefont{Tanaka}},
  \bibinfo{author}{\bibfnamefont{J.}~\bibnamefont{Nowack}},
  \bibinfo{author}{\bibfnamefont{J.~S.} \bibnamefont{Moodera}},
  \bibinfo{author}{\bibfnamefont{G.}~\bibnamefont{Laprade}},
  \bibinfo{author}{\bibfnamefont{A.}~\bibnamefont{Barry}}, \emph{et~al.},
  \bibinfo{year}{1999}, \bibinfo{journal}{J. Appl. Phys.}
  \textbf{\bibinfo{volume}{85}}(\bibinfo{number}{8}), \bibinfo{pages}{4589}.

\bibitem[{\citenamefont{Springford}(1980)}]{Lonzarich:225}
\bibinfo{author}{\bibnamefont{Springford}, \bibfnamefont{M.}},
  \bibinfo{year}{1980}, \emph{\bibinfo{title}{Electrons at the Fermi Surface}}
  (\bibinfo{publisher}{Cambridge Univ. Press, Cambridge}).

\bibitem[{\citenamefont{Stagarescu}
  \emph{et~al.}(2000)\citenamefont{Stagarescu, Su, Eastman, Altmann, Himpsel,
  and Gupta}}]{stagarescu:9233}
\bibinfo{author}{\bibnamefont{Stagarescu}, \bibfnamefont{C.~B.}},
  \bibinfo{author}{\bibfnamefont{X.}~\bibnamefont{Su}},
  \bibinfo{author}{\bibfnamefont{D.~E.} \bibnamefont{Eastman}},
  \bibinfo{author}{\bibfnamefont{K.~N.} \bibnamefont{Altmann}},
  \bibinfo{author}{\bibfnamefont{F.~J.} \bibnamefont{Himpsel}}, and
  \bibinfo{author}{\bibfnamefont{A.}~\bibnamefont{Gupta}},
  \bibinfo{year}{2000}, \bibinfo{journal}{Phys. Rev. B}
  \textbf{\bibinfo{volume}{61}}(\bibinfo{number}{14}), \bibinfo{pages}{R9233}.

\bibitem[{\citenamefont{Suzuki and Tedrow}(1998)}]{suzuki:11597}
\bibinfo{author}{\bibnamefont{Suzuki}, \bibfnamefont{K.}}, and
  \bibinfo{author}{\bibfnamefont{P.~M.} \bibnamefont{Tedrow}},
  \bibinfo{year}{1998}, \bibinfo{journal}{Phys. Rev. B}
  \textbf{\bibinfo{volume}{58}}(\bibinfo{number}{17}), \bibinfo{pages}{11597}.

\bibitem[{\citenamefont{Tkachov} \emph{et~al.}(2001)\citenamefont{Tkachov,
  McCann, and Fal\char39{}ko}}]{Tkachov:024519}
\bibinfo{author}{\bibnamefont{Tkachov}, \bibfnamefont{G.}},
  \bibinfo{author}{\bibfnamefont{E.}~\bibnamefont{McCann}}, and
  \bibinfo{author}{\bibfnamefont{V.~I.} \bibnamefont{Fal\char39{}ko}},
  \bibinfo{year}{2001}, \bibinfo{journal}{Phys. Rev. B}
  \textbf{\bibinfo{volume}{65}}(\bibinfo{number}{2}), \bibinfo{pages}{024519}.

\bibitem[{\citenamefont{Tomioka} \emph{et~al.}(2000)\citenamefont{Tomioka,
  Okuda, Okimoto, Kumai, Kobayashi, and Tokura}}]{Tomioka:422}
\bibinfo{author}{\bibnamefont{Tomioka}, \bibfnamefont{Y.}},
  \bibinfo{author}{\bibfnamefont{T.}~\bibnamefont{Okuda}},
  \bibinfo{author}{\bibfnamefont{Y.}~\bibnamefont{Okimoto}},
  \bibinfo{author}{\bibfnamefont{R.}~\bibnamefont{Kumai}},
  \bibinfo{author}{\bibfnamefont{K.~I.} \bibnamefont{Kobayashi}}, and
  \bibinfo{author}{\bibfnamefont{Y.}~\bibnamefont{Tokura}},
  \bibinfo{year}{2000}, \bibinfo{journal}{Phys. Rev. B}
  \textbf{\bibinfo{volume}{61}}(\bibinfo{number}{1}), \bibinfo{pages}{422}.

\bibitem[{\citenamefont{Toropova} \emph{et~al.}(2005)\citenamefont{Toropova,
  Kotliar, Savrasov, and Oudovenko}}]{to.ko.05}
\bibinfo{author}{\bibnamefont{Toropova}, \bibfnamefont{A.}},
  \bibinfo{author}{\bibfnamefont{G.}~\bibnamefont{Kotliar}},
  \bibinfo{author}{\bibfnamefont{S.~Y.} \bibnamefont{Savrasov}}, and
  \bibinfo{author}{\bibfnamefont{V.~S.} \bibnamefont{Oudovenko}},
  \bibinfo{year}{2005}, \bibinfo{journal}{Phys. Rev. B}
  \textbf{\bibinfo{volume}{71}}(\bibinfo{number}{17}), \bibinfo{eid}{172403}.

\bibitem[{\citenamefont{Troyer and Wiese}(2005)}]{troyer:170201}
\bibinfo{author}{\bibnamefont{Troyer}, \bibfnamefont{M.}}, and
  \bibinfo{author}{\bibfnamefont{U.-J.} \bibnamefont{Wiese}},
  \bibinfo{year}{2005}, \bibinfo{journal}{Phys. Rev. Lett.}
  \textbf{\bibinfo{volume}{94}}(\bibinfo{number}{17}), \bibinfo{eid}{170201}.

\bibitem[{\citenamefont{Tsujioka} \emph{et~al.}(1997)\citenamefont{Tsujioka,
  Mizokawa, Okamoto, Fujimori, Nohara, Takagi, Yamaura, and
  Takano}}]{tsujioka:15509}
\bibinfo{author}{\bibnamefont{Tsujioka}, \bibfnamefont{T.}},
  \bibinfo{author}{\bibfnamefont{T.}~\bibnamefont{Mizokawa}},
  \bibinfo{author}{\bibfnamefont{J.}~\bibnamefont{Okamoto}},
  \bibinfo{author}{\bibfnamefont{A.}~\bibnamefont{Fujimori}},
  \bibinfo{author}{\bibfnamefont{M.}~\bibnamefont{Nohara}},
  \bibinfo{author}{\bibfnamefont{H.}~\bibnamefont{Takagi}},
  \bibinfo{author}{\bibfnamefont{K.}~\bibnamefont{Yamaura}}, and
  \bibinfo{author}{\bibfnamefont{M.}~\bibnamefont{Takano}},
  \bibinfo{year}{1997}, \bibinfo{journal}{Phys. Rev. B}
  \textbf{\bibinfo{volume}{56}}(\bibinfo{number}{24}), \bibinfo{pages}{R15509}.

\bibitem[{\citenamefont{Turzhevskii}
  \emph{et~al.}(1990)\citenamefont{Turzhevskii, Lichtenstein, and
  Katsnelson}}]{Turzhevskii:1952}
\bibinfo{author}{\bibnamefont{Turzhevskii}, \bibfnamefont{S.~A.}},
  \bibinfo{author}{\bibfnamefont{A.~I.} \bibnamefont{Lichtenstein}}, and
  \bibinfo{author}{\bibfnamefont{M.~I.} \bibnamefont{Katsnelson}},
  \bibinfo{year}{1990}, \bibinfo{journal}{Fizika Tverdogo Tela [Engl. transl.:
  Sov. Phys.: Solid State 32, 1138 (1990)]} \textbf{\bibinfo{volume}{32}},
  \bibinfo{pages}{1952}.

\bibitem[{\citenamefont{Ueda and Moriya}(1975)}]{Ueda:32}
\bibinfo{author}{\bibnamefont{Ueda}, \bibfnamefont{K.}}, and
  \bibinfo{author}{\bibfnamefont{T.}~\bibnamefont{Moriya}},
  \bibinfo{year}{1975}, \bibinfo{journal}{J. Phys. Soc. Jpn}
  \textbf{\bibinfo{volume}{38}}, \bibinfo{pages}{32}.

\bibitem[{\citenamefont{\"O\ifmmode~\breve{g}\else \u{g}\fi{}\"ut and
  Rabe}(1995)}]{ogut:10443}
\bibinfo{author}{\bibnamefont{\"O\ifmmode~\breve{g}\else \u{g}\fi{}\"ut},
  \bibfnamefont{S.}}, and \bibinfo{author}{\bibfnamefont{K.~M.}
  \bibnamefont{Rabe}}, \bibinfo{year}{1995}, \bibinfo{journal}{Phys. Rev. B}
  \textbf{\bibinfo{volume}{51}}(\bibinfo{number}{16}), \bibinfo{pages}{10443}.

\bibitem[{\citenamefont{Ukraintsev}(1996)}]{Ukraintsev:11176}
\bibinfo{author}{\bibnamefont{Ukraintsev}, \bibfnamefont{V.~A.}},
  \bibinfo{year}{1996}, \bibinfo{journal}{Phys. Rev. B}
  \textbf{\bibinfo{volume}{53}}(\bibinfo{number}{16}), \bibinfo{pages}{11176}.

\bibitem[{\citenamefont{Ulmke}(1998)}]{ulmke:301}
\bibinfo{author}{\bibnamefont{Ulmke}, \bibfnamefont{M.}}, \bibinfo{year}{1998},
  \bibinfo{journal}{Eur. Phys. J. B} \textbf{\bibinfo{volume}{1}},
  \bibinfo{pages}{301}.

\bibitem[{\citenamefont{Vaitheeswaran}
  \emph{et~al.}(2005)\citenamefont{Vaitheeswaran, Kanchana, and
  Delin}}]{vaitheeswaran:032513}
\bibinfo{author}{\bibnamefont{Vaitheeswaran}, \bibfnamefont{G.}},
  \bibinfo{author}{\bibfnamefont{V.}~\bibnamefont{Kanchana}}, and
  \bibinfo{author}{\bibfnamefont{A.}~\bibnamefont{Delin}},
  \bibinfo{year}{2005}, \bibinfo{journal}{Appl. Phys. Lett.}
  \textbf{\bibinfo{volume}{86}}(\bibinfo{number}{3}), \bibinfo{eid}{032513}.

\bibitem[{\citenamefont{Verwey}(1939)}]{Verwey:327}
\bibinfo{author}{\bibnamefont{Verwey}, \bibfnamefont{E.}},
  \bibinfo{year}{1939}, \bibinfo{journal}{Nature}
  \textbf{\bibinfo{volume}{144}}, \bibinfo{pages}{327}.

\bibitem[{\citenamefont{Viret} \emph{et~al.}(1997)\citenamefont{Viret, Ranno,
  and Coey}}]{Viret:8067}
\bibinfo{author}{\bibnamefont{Viret}, \bibfnamefont{M.}},
  \bibinfo{author}{\bibfnamefont{L.}~\bibnamefont{Ranno}}, and
  \bibinfo{author}{\bibfnamefont{J.~M.~D.} \bibnamefont{Coey}},
  \bibinfo{year}{1997}, \bibinfo{journal}{Phys. Rev. B}
  \textbf{\bibinfo{volume}{55}}(\bibinfo{number}{13}), \bibinfo{pages}{8067}.

\bibitem[{\citenamefont{Vollhardt} \emph{et~al.}(1999)\citenamefont{Vollhardt,
  Blumer, Held, and Kollar}}]{Vollhardt:383}
\bibinfo{author}{\bibnamefont{Vollhardt}, \bibfnamefont{D.}},
  \bibinfo{author}{\bibfnamefont{N.}~\bibnamefont{Blumer}},
  \bibinfo{author}{\bibfnamefont{K.}~\bibnamefont{Held}}, and
  \bibinfo{author}{\bibfnamefont{M.}~\bibnamefont{Kollar}},
  \bibinfo{year}{1999}, \bibinfo{journal}{Advances In Solid State Physics}
  \textbf{\bibinfo{volume}{38}}, \bibinfo{pages}{383}.

\bibitem[{\citenamefont{Vonsovsky}(1974)}]{Vonsovsky:1974}
\bibinfo{author}{\bibnamefont{Vonsovsky}, \bibfnamefont{S.~V.}},
  \bibinfo{year}{1974}, \emph{\bibinfo{title}{Magnetism}}
  (\bibinfo{publisher}{New York, Wiley}).

\bibitem[{\citenamefont{Vonsovsky and Katsnelson}(1989)}]{Vonsovsky:1989}
\bibinfo{author}{\bibnamefont{Vonsovsky}, \bibfnamefont{S.~V.}}, and
  \bibinfo{author}{\bibfnamefont{M.~I.} \bibnamefont{Katsnelson}},
  \bibinfo{year}{1989}, \emph{\bibinfo{title}{Quantum Solid State Physics}}
  (\bibinfo{publisher}{Springer, Berlin}).

\bibitem[{\citenamefont{Walz}(2002)}]{Walz:R285}
\bibinfo{author}{\bibnamefont{Walz}, \bibfnamefont{F.}}, \bibinfo{year}{2002},
  \bibinfo{journal}{J. Phys.: Condens. Matter} \textbf{\bibinfo{volume}{14}},
  \bibinfo{pages}{R285}.

\bibitem[{\citenamefont{Wang} \emph{et~al.}(2005)\citenamefont{Wang, Umemoto,
  Wentzcovitch, Chen, Chien, Checkelsky, Eckert, Dahlberg, and
  Leighton}}]{wang:056602}
\bibinfo{author}{\bibnamefont{Wang}, \bibfnamefont{L.}},
  \bibinfo{author}{\bibfnamefont{K.}~\bibnamefont{Umemoto}},
  \bibinfo{author}{\bibfnamefont{R.~M.} \bibnamefont{Wentzcovitch}},
  \bibinfo{author}{\bibfnamefont{T.~Y.} \bibnamefont{Chen}},
  \bibinfo{author}{\bibfnamefont{C.~L.} \bibnamefont{Chien}},
  \bibinfo{author}{\bibfnamefont{J.~G.} \bibnamefont{Checkelsky}},
  \bibinfo{author}{\bibfnamefont{J.~C.} \bibnamefont{Eckert}},
  \bibinfo{author}{\bibfnamefont{E.~D.} \bibnamefont{Dahlberg}}, and
  \bibinfo{author}{\bibfnamefont{C.}~\bibnamefont{Leighton}},
  \bibinfo{year}{2005}, \bibinfo{journal}{Phys. Rev. Lett.}
  \textbf{\bibinfo{volume}{94}}(\bibinfo{number}{5}), \bibinfo{eid}{056602}.

\bibitem[{\citenamefont{Watts} \emph{et~al.}(2000)\citenamefont{Watts, Wirth,
  von Moln{\'a}r, Barry, and Coey}}]{wa.wi.00}
\bibinfo{author}{\bibnamefont{Watts}, \bibfnamefont{S.~M.}},
  \bibinfo{author}{\bibfnamefont{S.}~\bibnamefont{Wirth}},
  \bibinfo{author}{\bibfnamefont{S.}~\bibnamefont{von Moln{\'a}r}},
  \bibinfo{author}{\bibfnamefont{A.}~\bibnamefont{Barry}}, and
  \bibinfo{author}{\bibfnamefont{J.~M.~D.} \bibnamefont{Coey}},
  \bibinfo{year}{2000}, \bibinfo{journal}{Phys. Rev. B}
  \textbf{\bibinfo{volume}{61}}(\bibinfo{number}{14}), \bibinfo{pages}{9621}.

\bibitem[{\citenamefont{Weht and Pickett}(1999)}]{ru.pi.99}
\bibinfo{author}{\bibnamefont{Weht}, \bibfnamefont{R.}}, and
  \bibinfo{author}{\bibfnamefont{W.~E.} \bibnamefont{Pickett}},
  \bibinfo{year}{1999}, \bibinfo{journal}{Phys. Rev. B}
  \textbf{\bibinfo{volume}{60}}(\bibinfo{number}{18}), \bibinfo{pages}{13006}.

\bibitem[{\citenamefont{Wessely} \emph{et~al.}(2003)\citenamefont{Wessely, Roy,
  Aberg, Andersson, Edvardsson, Karis, Sanyal, Svedlindh, Katsnelson,
  Gunnarsson, Arvanitis, Bengone} \emph{et~al.}}]{wessely:235109}
\bibinfo{author}{\bibnamefont{Wessely}, \bibfnamefont{O.}},
  \bibinfo{author}{\bibfnamefont{P.}~\bibnamefont{Roy}},
  \bibinfo{author}{\bibfnamefont{D.}~\bibnamefont{Aberg}},
  \bibinfo{author}{\bibfnamefont{C.}~\bibnamefont{Andersson}},
  \bibinfo{author}{\bibfnamefont{S.}~\bibnamefont{Edvardsson}},
  \bibinfo{author}{\bibfnamefont{O.}~\bibnamefont{Karis}},
  \bibinfo{author}{\bibfnamefont{B.}~\bibnamefont{Sanyal}},
  \bibinfo{author}{\bibfnamefont{P.}~\bibnamefont{Svedlindh}},
  \bibinfo{author}{\bibfnamefont{M.~I.} \bibnamefont{Katsnelson}},
  \bibinfo{author}{\bibfnamefont{R.}~\bibnamefont{Gunnarsson}},
  \bibinfo{author}{\bibfnamefont{D.}~\bibnamefont{Arvanitis}},
  \bibinfo{author}{\bibfnamefont{O.}~\bibnamefont{Bengone}}, \emph{et~al.},
  \bibinfo{year}{2003}, \bibinfo{journal}{Phys. Rev. B}
  \textbf{\bibinfo{volume}{68}}(\bibinfo{number}{23}), \bibinfo{eid}{235109}.

\bibitem[{\citenamefont{Wiesendanger}
  \emph{et~al.}(1990)\citenamefont{Wiesendanger, G\"untherodt, G\"untherodt,
  Gambino, and Ruf}}]{Wiesendanger:247}
\bibinfo{author}{\bibnamefont{Wiesendanger}, \bibfnamefont{R.}},
  \bibinfo{author}{\bibfnamefont{H.-J.} \bibnamefont{G\"untherodt}},
  \bibinfo{author}{\bibfnamefont{G.}~\bibnamefont{G\"untherodt}},
  \bibinfo{author}{\bibfnamefont{R.~J.} \bibnamefont{Gambino}}, and
  \bibinfo{author}{\bibfnamefont{R.}~\bibnamefont{Ruf}}, \bibinfo{year}{1990},
  \bibinfo{journal}{Phys. Rev. Lett.}
  \textbf{\bibinfo{volume}{65}}(\bibinfo{number}{2}), \bibinfo{pages}{247}.

\bibitem[{\citenamefont{Wijngaard} \emph{et~al.}(1989)\citenamefont{Wijngaard,
  Haas, and de~Groot}}]{Wijngaard:9318}
\bibinfo{author}{\bibnamefont{Wijngaard}, \bibfnamefont{J.~H.}},
  \bibinfo{author}{\bibfnamefont{C.}~\bibnamefont{Haas}}, and
  \bibinfo{author}{\bibfnamefont{R.~A.} \bibnamefont{de~Groot}},
  \bibinfo{year}{1989}, \bibinfo{journal}{Phys. Rev. B}
  \textbf{\bibinfo{volume}{40}}(\bibinfo{number}{13}), \bibinfo{pages}{9318}.

\bibitem[{\citenamefont{Wijngaard} \emph{et~al.}(1992)\citenamefont{Wijngaard,
  Haas, and de~Groot}}]{Wijngaard:5395}
\bibinfo{author}{\bibnamefont{Wijngaard}, \bibfnamefont{J.~H.}},
  \bibinfo{author}{\bibfnamefont{C.}~\bibnamefont{Haas}}, and
  \bibinfo{author}{\bibfnamefont{R.~A.} \bibnamefont{de~Groot}},
  \bibinfo{year}{1992}, \bibinfo{journal}{Phys. Rev. B}
  \textbf{\bibinfo{volume}{45}}(\bibinfo{number}{10}), \bibinfo{pages}{5395}.

\bibitem[{\citenamefont{de~Wijs and de~Groot}(2001)}]{dewijs:020402}
\bibinfo{author}{\bibnamefont{de~Wijs}, \bibfnamefont{G.~A.}}, and
  \bibinfo{author}{\bibfnamefont{R.~A.} \bibnamefont{de~Groot}},
  \bibinfo{year}{2001}, \bibinfo{journal}{Phys. Rev. B}
  \textbf{\bibinfo{volume}{64}}(\bibinfo{number}{2}), \bibinfo{pages}{020402}.

\bibitem[{\citenamefont{Wurmehl} \emph{et~al.}(2005)\citenamefont{Wurmehl,
  Fecher, Kandpal, Ksenofontov, Felser, Lin, and Morais}}]{wurmehl:184434}
\bibinfo{author}{\bibnamefont{Wurmehl}, \bibfnamefont{S.}},
  \bibinfo{author}{\bibfnamefont{G.~H.} \bibnamefont{Fecher}},
  \bibinfo{author}{\bibfnamefont{H.~C.} \bibnamefont{Kandpal}},
  \bibinfo{author}{\bibfnamefont{V.}~\bibnamefont{Ksenofontov}},
  \bibinfo{author}{\bibfnamefont{C.}~\bibnamefont{Felser}},
  \bibinfo{author}{\bibfnamefont{H.-J.} \bibnamefont{Lin}}, and
  \bibinfo{author}{\bibfnamefont{J.}~\bibnamefont{Morais}},
  \bibinfo{year}{2005}, \bibinfo{journal}{Phys. Rev. B}
  \textbf{\bibinfo{volume}{72}}(\bibinfo{number}{18}), \bibinfo{eid}{184434}.

\bibitem[{\citenamefont{Xiang} \emph{et~al.}(2005)\citenamefont{Xiang, Yang,
  Hou, and Zhu}}]{xiang:243113}
\bibinfo{author}{\bibnamefont{Xiang}, \bibfnamefont{H.~J.}},
  \bibinfo{author}{\bibfnamefont{J.}~\bibnamefont{Yang}},
  \bibinfo{author}{\bibfnamefont{J.~G.} \bibnamefont{Hou}}, and
  \bibinfo{author}{\bibfnamefont{Q.}~\bibnamefont{Zhu}}, \bibinfo{year}{2005},
  \bibinfo{journal}{Appl. Phys. Lett.}
  \textbf{\bibinfo{volume}{87}}(\bibinfo{number}{24}), \bibinfo{eid}{243113}
  (pages~\bibinfo{numpages}{3}).

\bibitem[{\citenamefont{Xie}
  \emph{et~al.}(2003{\natexlab{a}})\citenamefont{Xie, Liu, and
  Pettifor}}]{xie:134407}
\bibinfo{author}{\bibnamefont{Xie}, \bibfnamefont{W.-H.}},
  \bibinfo{author}{\bibfnamefont{B.-G.} \bibnamefont{Liu}}, and
  \bibinfo{author}{\bibfnamefont{D.~G.} \bibnamefont{Pettifor}},
  \bibinfo{year}{2003}{\natexlab{a}}, \bibinfo{journal}{Phys. Rev. B}
  \textbf{\bibinfo{volume}{68}}(\bibinfo{number}{13}), \bibinfo{eid}{134407}.

\bibitem[{\citenamefont{Xie}
  \emph{et~al.}(2003{\natexlab{b}})\citenamefont{Xie, Xu, Liu, and
  Pettifor}}]{xie:2003}
\bibinfo{author}{\bibnamefont{Xie}, \bibfnamefont{W.-H.}},
  \bibinfo{author}{\bibfnamefont{Y.-Q.} \bibnamefont{Xu}},
  \bibinfo{author}{\bibfnamefont{B.-G.} \bibnamefont{Liu}}, and
  \bibinfo{author}{\bibfnamefont{D.~G.} \bibnamefont{Pettifor}},
  \bibinfo{year}{2003}{\natexlab{b}}, \bibinfo{journal}{Phys. Rev. Lett.}
  \textbf{\bibinfo{volume}{91}}(\bibinfo{number}{3}), \bibinfo{eid}{037204}.

\bibitem[{\citenamefont{Yablonskikh}
  \emph{et~al.}(2001)\citenamefont{Yablonskikh, Yarmoshenko, Grebennikov,
  Kurmaev, Butorin, Duda, Nordgren, Plogmann, and
  Neumann}}]{Yablonskikh:235117}
\bibinfo{author}{\bibnamefont{Yablonskikh}, \bibfnamefont{M.~V.}},
  \bibinfo{author}{\bibfnamefont{Y.~M.} \bibnamefont{Yarmoshenko}},
  \bibinfo{author}{\bibfnamefont{V.~I.} \bibnamefont{Grebennikov}},
  \bibinfo{author}{\bibfnamefont{E.~Z.} \bibnamefont{Kurmaev}},
  \bibinfo{author}{\bibfnamefont{S.~M.} \bibnamefont{Butorin}},
  \bibinfo{author}{\bibfnamefont{L.-C.} \bibnamefont{Duda}},
  \bibinfo{author}{\bibfnamefont{J.}~\bibnamefont{Nordgren}},
  \bibinfo{author}{\bibfnamefont{S.}~\bibnamefont{Plogmann}}, and
  \bibinfo{author}{\bibfnamefont{M.}~\bibnamefont{Neumann}},
  \bibinfo{year}{2001}, \bibinfo{journal}{Phys. Rev. B}
  \textbf{\bibinfo{volume}{63}}(\bibinfo{number}{23}), \bibinfo{pages}{235117}.

\bibitem[{\citenamefont{Yamasaki}(2006)}]{Yamasaki:private}
\bibinfo{author}{\bibnamefont{Yamasaki}, \bibfnamefont{A.}},
  \bibinfo{year}{2006}, \bibinfo{journal}{unpublished} .

\bibitem[{\citenamefont{Yamasaki} \emph{et~al.}(2006)\citenamefont{Yamasaki,
  Chioncel, Lichtenstein, and Andersen}}]{yamasaki:024419}
\bibinfo{author}{\bibnamefont{Yamasaki}, \bibfnamefont{A.}},
  \bibinfo{author}{\bibfnamefont{L.}~\bibnamefont{Chioncel}},
  \bibinfo{author}{\bibfnamefont{A.~I.} \bibnamefont{Lichtenstein}}, and
  \bibinfo{author}{\bibfnamefont{O.~K.} \bibnamefont{Andersen}},
  \bibinfo{year}{2006}, \bibinfo{journal}{Phys. Rev. B}
  \textbf{\bibinfo{volume}{74}}(\bibinfo{number}{2}), \bibinfo{eid}{024419}.

\bibitem[{\citenamefont{Yanase and Siratori}(1984)}]{Yanase:312}
\bibinfo{author}{\bibnamefont{Yanase}, \bibfnamefont{A.}}, and
  \bibinfo{author}{\bibfnamefont{K.}~\bibnamefont{Siratori}},
  \bibinfo{year}{1984}, \bibinfo{journal}{J. Phys. Soc. Jpn}
  \textbf{\bibinfo{volume}{53}}, \bibinfo{pages}{312}.

\bibitem[{\citenamefont{Yarmoshenko}
  \emph{et~al.}(1998)\citenamefont{Yarmoshenko, Katsnelson, Shreder, Kurmaev,
  Slebarski, Plogmann, Schlathoelter, Braun, and Neumann}}]{Yarmoshenko:1}
\bibinfo{author}{\bibnamefont{Yarmoshenko}, \bibfnamefont{Y.~M.}},
  \bibinfo{author}{\bibfnamefont{M.~I.} \bibnamefont{Katsnelson}},
  \bibinfo{author}{\bibfnamefont{E.~I.} \bibnamefont{Shreder}},
  \bibinfo{author}{\bibfnamefont{E.~Z.} \bibnamefont{Kurmaev}},
  \bibinfo{author}{\bibfnamefont{A.}~\bibnamefont{Slebarski}},
  \bibinfo{author}{\bibfnamefont{S.}~\bibnamefont{Plogmann}},
  \bibinfo{author}{\bibfnamefont{T.}~\bibnamefont{Schlathoelter}},
  \bibinfo{author}{\bibfnamefont{J.}~\bibnamefont{Braun}}, and
  \bibinfo{author}{\bibfnamefont{M.}~\bibnamefont{Neumann}},
  \bibinfo{year}{1998}, \bibinfo{journal}{Eur. Phys. J. B}
  \textbf{\bibinfo{volume}{2}}, \bibinfo{pages}{1}.

\bibitem[{\citenamefont{Zener}(1951)}]{Zener:403}
\bibinfo{author}{\bibnamefont{Zener}, \bibfnamefont{C.}}, \bibinfo{year}{1951},
  \bibinfo{journal}{Phys. Rev.}
  \textbf{\bibinfo{volume}{82}}(\bibinfo{number}{3}), \bibinfo{pages}{403}.

\bibitem[{\citenamefont{Zhang} \emph{et~al.}(2007)\citenamefont{Zhang, Ye, Sha,
  Dai, Fernandez-Baca, and Plummer}}]{Zhang:315204}
\bibinfo{author}{\bibnamefont{Zhang}, \bibfnamefont{J.}},
  \bibinfo{author}{\bibfnamefont{F.}~\bibnamefont{Ye}},
  \bibinfo{author}{\bibfnamefont{H.}~\bibnamefont{Sha}},
  \bibinfo{author}{\bibfnamefont{P.}~\bibnamefont{Dai}},
  \bibinfo{author}{\bibfnamefont{J.~A.} \bibnamefont{Fernandez-Baca}}, and
  \bibinfo{author}{\bibfnamefont{E.~W.} \bibnamefont{Plummer}},
  \bibinfo{year}{2007}, \bibinfo{journal}{J. Phys.: Condens. Matter}
  \textbf{\bibinfo{volume}{19}}, \bibinfo{pages}{315202}.

\bibitem[{\citenamefont{Zhao} \emph{et~al.}(1987)\citenamefont{Zhao, Nie,
  Zhang, Chao, and Micnas}}]{Zhao:2321}
\bibinfo{author}{\bibnamefont{Zhao}, \bibfnamefont{B.-H.}},
  \bibinfo{author}{\bibfnamefont{H.-Q.} \bibnamefont{Nie}},
  \bibinfo{author}{\bibfnamefont{K.-Y.} \bibnamefont{Zhang}},
  \bibinfo{author}{\bibfnamefont{K.~A.} \bibnamefont{Chao}}, and
  \bibinfo{author}{\bibfnamefont{R.}~\bibnamefont{Micnas}},
  \bibinfo{year}{1987}, \bibinfo{journal}{Phys. Rev. B}
  \textbf{\bibinfo{volume}{36}}(\bibinfo{number}{4}), \bibinfo{pages}{2321}.

\bibitem[{\citenamefont{Zhao} \emph{et~al.}(1993)\citenamefont{Zhao, Callaway,
  and Hayashibara}}]{Zhao:15781}
\bibinfo{author}{\bibnamefont{Zhao}, \bibfnamefont{G.~L.}},
  \bibinfo{author}{\bibfnamefont{J.}~\bibnamefont{Callaway}}, and
  \bibinfo{author}{\bibfnamefont{M.}~\bibnamefont{Hayashibara}},
  \bibinfo{year}{1993}, \bibinfo{journal}{Phys. Rev. B}
  \textbf{\bibinfo{volume}{48}}(\bibinfo{number}{21}), \bibinfo{pages}{15781}.

\bibitem[{\citenamefont{Zhao} \emph{et~al.}(2001)\citenamefont{Zhao, Matsukura,
  Takamura, Abe, Chiba, and Ohno}}]{zhao:2776}
\bibinfo{author}{\bibnamefont{Zhao}, \bibfnamefont{J.~H.}},
  \bibinfo{author}{\bibfnamefont{F.}~\bibnamefont{Matsukura}},
  \bibinfo{author}{\bibfnamefont{K.}~\bibnamefont{Takamura}},
  \bibinfo{author}{\bibfnamefont{E.}~\bibnamefont{Abe}},
  \bibinfo{author}{\bibfnamefont{D.}~\bibnamefont{Chiba}}, and
  \bibinfo{author}{\bibfnamefont{H.}~\bibnamefont{Ohno}}, \bibinfo{year}{2001},
  \bibinfo{journal}{Appl. Phys. Lett.}
  \textbf{\bibinfo{volume}{79}}(\bibinfo{number}{17}), \bibinfo{pages}{2776}.

\bibitem[{\citenamefont{Zhao} \emph{et~al.}(2006)\citenamefont{Zhao, Feng,
  Huang, gao Zhao, Lu, Han, and Zhan}}]{zhao:052506}
\bibinfo{author}{\bibnamefont{Zhao}, \bibfnamefont{K.}},
  \bibinfo{author}{\bibfnamefont{J.}~\bibnamefont{Feng}},
  \bibinfo{author}{\bibfnamefont{Y.}~\bibnamefont{Huang}},
  \bibinfo{author}{\bibfnamefont{J.}~\bibnamefont{gao Zhao}},
  \bibinfo{author}{\bibfnamefont{H.}~\bibnamefont{Lu}},
  \bibinfo{author}{\bibfnamefont{X.}~\bibnamefont{Han}}, and
  \bibinfo{author}{\bibfnamefont{W.}~\bibnamefont{Zhan}}, \bibinfo{year}{2006},
  \bibinfo{journal}{Appl. Phys. Lett.}
  \textbf{\bibinfo{volume}{88}}(\bibinfo{number}{5}), \bibinfo{eid}{052506}.

\bibitem[{\citenamefont{Zhao} \emph{et~al.}(2005)\citenamefont{Zhao, Feng,
  Huang, Zhao, Lu, Han, and Zhan}}]{Zhao:2595}
\bibinfo{author}{\bibnamefont{Zhao}, \bibfnamefont{K.}},
  \bibinfo{author}{\bibfnamefont{J.~F.} \bibnamefont{Feng}},
  \bibinfo{author}{\bibfnamefont{Y.~H.} \bibnamefont{Huang}},
  \bibinfo{author}{\bibfnamefont{J.~G.} \bibnamefont{Zhao}},
  \bibinfo{author}{\bibfnamefont{H.~B.} \bibnamefont{Lu}},
  \bibinfo{author}{\bibfnamefont{X.~F.} \bibnamefont{Han}}, and
  \bibinfo{author}{\bibfnamefont{W.~S.} \bibnamefont{Zhan}},
  \bibinfo{year}{2005}, \bibinfo{journal}{Chinese Physics}
  \textbf{\bibinfo{volume}{14}}(\bibinfo{number}{12}), \bibinfo{pages}{2595}.

\bibitem[{\citenamefont{Zhao and Zunger}(2005)}]{zhao:132403}
\bibinfo{author}{\bibnamefont{Zhao}, \bibfnamefont{Y.-J.}}, and
  \bibinfo{author}{\bibfnamefont{A.}~\bibnamefont{Zunger}},
  \bibinfo{year}{2005}, \bibinfo{journal}{Phys. Rev. B}
  \textbf{\bibinfo{volume}{71}}(\bibinfo{number}{13}), \bibinfo{eid}{132403}.

\bibitem[{\citenamefont{Ziebeck and Webster}(1974)}]{zi.we.74}
\bibinfo{author}{\bibnamefont{Ziebeck}, \bibfnamefont{K.~R.~A.}}, and
  \bibinfo{author}{\bibfnamefont{P.~J.} \bibnamefont{Webster}},
  \bibinfo{year}{1974}, \bibinfo{journal}{Journal of Physics and Chemistry of
  Solids} \textbf{\bibinfo{volume}{35}}(\bibinfo{number}{1}),
  \bibinfo{pages}{1}.

\bibitem[{\citenamefont{Ziese}(2002)}]{ziese:143}
\bibinfo{author}{\bibnamefont{Ziese}, \bibfnamefont{M.}}, \bibinfo{year}{2002},
  \bibinfo{journal}{Reports on Progress in Physics}
  \textbf{\bibinfo{volume}{65}}(\bibinfo{number}{2}), \bibinfo{pages}{143}.

\bibitem[{\citenamefont{Zurek} \emph{et~al.}(2005)\citenamefont{Zurek, Jepsen,
  and Andersen}}]{zurek:1934}
\bibinfo{author}{\bibnamefont{Zurek}, \bibfnamefont{E.}},
  \bibinfo{author}{\bibfnamefont{O.}~\bibnamefont{Jepsen}}, and
  \bibinfo{author}{\bibfnamefont{O.~K.} \bibnamefont{Andersen}},
  \bibinfo{year}{2005}, \bibinfo{journal}{Chem. Phys. Chem}
  \textbf{\bibinfo{volume}{6}}, \bibinfo{pages}{1934}.

\bibitem[{\citenamefont{Zutic} \emph{et~al.}(2004)\citenamefont{Zutic, Fabian,
  and Sarma}}]{zutic:323}
\bibinfo{author}{\bibnamefont{Zutic}, \bibfnamefont{I.}},
  \bibinfo{author}{\bibfnamefont{J.}~\bibnamefont{Fabian}}, and
  \bibinfo{author}{\bibfnamefont{S.~D.} \bibnamefont{Sarma}},
  \bibinfo{year}{2004}, \bibinfo{journal}{Reviews of Modern Physics}
  \textbf{\bibinfo{volume}{76}}(\bibinfo{number}{2}), \bibinfo{eid}{323}.

\end{thebibliography}
\newpage

\end{document}